\documentclass[a4paper,11pt]{article}
\usepackage{jheppub}
\usepackage{upgreek}
\usepackage{subcaption}
\usepackage{tikz-feynman}
\usepackage{amsmath}
\usepackage{slashed}
\usepackage{cleveref}
\usepackage{multirow}
\usepackage{tcolorbox}
\usepackage{adjustbox}
\usepackage{orcidlink}
\usepackage{pifont}
\usepackage{siunitx,booktabs,graphicx}
\usepackage[table]{xcolor}
\usepackage[dvipsnames]{xcolor}
\usepackage[normalem]{ulem}
\usepackage[font=small]{caption}
\usepackage[font={small},labelfont={bf}]{caption}
\usepackage{hyperref}
\usepackage{url}
\newcommand{\cmark}{\ding{51}}
\newcommand{\xmark}{{\color{gray}\ding{55}}}
\newcommand{\psibar}{\bar{\psi}}
\newcommand{\chibar}{\bar{\chi}}
\newcommand{\sphiN}{\sigma_{\rm \Phi \mathbb{N}}^{\rm eff}}

\newcommand{\TFO}{{\rm T_{FO}}}
\newcommand{\TEW}{{\rm T_{EW}}}
\newcommand{\yq}{\mathtt{y}_{\rm q}^{}}
\newcommand{\yl}{\mathtt{y}_{\ell}^{}}
\newcommand{\yd}{\mathtt{y}_{\rm d}^{}}
\newcommand{\ys}{\mathtt{y}_{\rm s}^{}}
\newcommand{\yb}{\mathtt{y}_{\rm b}^{}}
\newcommand{\ybs}{\mathtt{y}_{\rm bs}^{}}
\newcommand{\ye}{\mathtt{y}_{\rm e}^{}}
\newcommand{\ymu}{\mathtt{y}_{\mu}^{}}
\newcommand{\yta}{\mathtt{y}_{\tau}^{}}
\newcommand{\mpsi}{m_{\psi}^{}}
\newcommand{\omgdm}{{\Omega_{\rm DM}}}
\newcommand{\mchi}{m_{\chi}^{}}
\newcommand{\mphi}{m_{\Phi}^{}}
\newcommand{\TeV}{\text{TeV}}
\newcommand{\GeV}{\text{GeV}}
\newcommand{\MeV}{\text{MeV}}
\newcommand{\VLF}{\text{VLF}}
\newcommand{\BR}{{\mathcal{B}}}

\newcommand{\lphiH}{\lambda_{\Phi\rm H}^{}}
\newcommand{\D}{\mathcal{D}}

\newcommand{\qR}{q_{\mathtt{R}}^{}}
\newcommand{\dR}{\rm d_{\mathtt{R}}^{}}
\newcommand{\sR}{\rm s_{\mathtt{R}}^{}}
\newcommand{\bR}{\rm b_{\mathtt{R}}^{}}
\newcommand{\lR}{\ell_{\mathtt{R}}^{}}
\newcommand{\eR}{\rm e_{\mathtt{R}}^{}}
\newcommand{\muR}{\rm \mu_{\mathtt{R}}^{}}
\newcommand{\tauR}{\rm \tau_{\mathtt{R}}^{}}
\newcommand{\Eq}{Eq\,.~}
\newcommand{\eq}{Eq\,.~}
\newcommand{\Eqs}{Eqs\,.~}

\newcommand{\fig}{Fig\,.~}
\newcommand{\figs}{Figs\,.~}
\newcommand{\Fig}{Fig\,.~}

\newcommand{\Zthree}{\mathcal{Z}_3}
\newcommand{\sveff}{\langle\sigma v\rangle^{\rm eff}}
\newcommand{\Delpsi}{\Delta_{\psi\Phi}}
\newcommand{\Delchi}{\Delta_{\chi\Phi}}
\definecolor{deepmagenta}{rgb}{0.8, 0.0, 0.8}
\definecolor{mediumtealblue}{rgb}{0.0, 0.33, 0.71}
\definecolor{warmblack}{rgb}{0.0, 0.26, 0.26}
\definecolor{bostonuniversityred}{rgb}{0.8, 0.0, 0.0}
\definecolor{junglegreen}{rgb}{0.16, 0.67, 0.53}
\definecolor{lightcornflowerblue}{rgb}{0.6, 0.81, 0.93}
\definecolor{mypink1}{rgb}{0.858, 0.188, 0.478}
\definecolor{mypink2}{RGB}{219, 48, 122}
\definecolor{mypink3}{cmyk}{0, 0.7808, 0.4429, 0.1412}
\definecolor{mygray}{gray}{0.2}
\definecolor{ForestGreen}{RGB}{34,139,34}
\definecolor{MyDarkBlue}{rgb}{0.1, 0.1, 0.8}
\definecolor{SBlue}{rgb}{0.2, 0.4, 0.7}
\definecolor{MyLightBlue}{rgb}{0.22,0.51,0.9}
\definecolor{MyGreen}{rgb}{0.0, 0.5, 0.0}
\definecolor{BrickRed}{rgb}{0.8, 0.25, 0.33}
\hypersetup{colorlinks, citecolor=BrickRed,linkcolor=MyDarkBlue, urlcolor=MyGreen}

\title{Complex Scalar Dark Matter with a Vector-Like Quark and Lepton: Precision, Flavor, and HL-LHC}
\author[a]{Lipika Kolay\orcidlink{0000-0003-0715-6668}, }
\emailAdd{lipika.kolay@iitgn.ac.in}
\author[a]{Rusa Mandal\orcidlink{0000-0001-8053-6881}, }
\emailAdd{rusa.mandal@iitgn.ac.in}
\author[b,c]{Manimala Mitra\orcidlink{0000-0002-8032-5125}, }
\emailAdd{manimala@iopb.res.in}
\author[b,c]{Dipankar Pradhan\orcidlink{0000-0002-2450-6677}, }
\emailAdd{dipankar.pradhan@iopb.res.in}
\author[b,c]{Subham Saha\orcidlink{0009-0009-1183-3271}}
\emailAdd{subham.saha@iopb.res.in}
\affiliation[a]{Department of Physics, Indian Institute of Technology Gandhinagar, Palaj, Gujarat 382355, India}
\affiliation[b]{Institute of Physics, Sachivalaya Marg, Bhubaneswar, Odisha 751005, India}
\affiliation[c]{Homi Bhabha National Institute, BARC Training School Complex, Anushakti Nagar, Mumbai 400094, India}
\abstract{We investigate a minimal extension of the SM consisting of a $\Zthree$-stabilized complex scalar dark matter (CSDM) candidate, a down-type vector-like quark (VLQ), and a charged vector-like lepton (VLL). The additional vector-like fermions not only enable the CSDM to reproduce the observed relic abundance beyond the Higgs-resonance region through semi-annihilation and co-annihilation processes, but also induce correlated signatures across flavor, electroweak precision, dark matter, and collider observables. We perform a comprehensive one-loop analysis of neutral meson mixing, rare meson decays, charged lepton flavor violation, anomalous magnetic moments of charged leptons, and $Z$-pole observables. We find that neutral meson mixing and rare meson decays provide the dominant constraints on the VLQ sector, while charged lepton flavor-violating processes strongly restrict the VLL Yukawa couplings. Current direct-detection limits require Higgs-DM coupling $\lphiH \lesssim5\times10^{-3}$ and the VLQ Yukawa coupling $\yd\lesssim0.05$ for TeV-scale VLQ masses, whereas present indirect-detection searches impose no additional constraints. Combining all flavor, electroweak precision, dark matter, and collider constraints, we identify viable parameter regions with CSDM masses above approximately $1.0~\TeV$ and VLQ masses above about $1.5~\TeV$.  We also find that the LHC can exclude VLQs (VLLs) with masses up to approximately $1.6~(0.38)$ TeV, depending on the CSDM mass, at the $2\sigma$ confidence level. The HL-LHC can further probe an extended region of the parameter space, with discovery prospects at the $3\sigma$ level. Finally, we demonstrate the complementarity of flavor, dark matter, and collider searches in probing this framework.
}
\begin{document}
\maketitle
\flushbottom
\section{Introduction}
\label{sec:intro}
The Standard Model (SM) has been remarkably successful in describing particle interactions over a wide range of energies. Nevertheless, several outstanding questions, including the particle nature of dark matter (DM), the origin of possible new flavor dynamics, and the possible existence of new particles beyond the electroweak scale, provide compelling evidence for physics beyond the SM (BSM). Among the well-motivated extensions of the SM, models containing vector-like fermions have attracted considerable attention, as they arise naturally in a variety of theoretical frameworks and can induce observable effects in flavor physics while remaining free from gauge anomalies. At the same time, the introduction of an extended dark sector can accommodate a viable DM candidate, whose relic abundance and detection prospects are closely intertwined with the flavor structure of the model. Such scenarios can also contribute to several flavor observables associated with the current flavor anomalies, while predicting a rich phenomenology across flavor, dark matter, and collider experiments. Motivated by these considerations, we investigate a minimal extension of the SM consisting of a complex scalar dark matter (CSDM) candidate together with a vector-like quark (VLQ) and a vector-like lepton (VLL), stabilized by a discrete $\Zthree$ symmetry.

A minimal Higgs-portal CSDM model stabilized by a discrete $\Zthree$ symmetry~\cite{Barger:2008jx, Rindler-Daller:2013zxa, McDonald:1993ex} is severely constrained by current direct-detection (DD) experiments when the observed relic abundance is generated through the standard thermal freeze-out mechanism in a radiation-dominated Universe~\cite{Belanger:2012zr, Hektor:2019ote}. Although several alternative scenarios have been proposed to evade these constraints (see, e.g., Refs\,.~\cite{Giacchino:2015hvk, Acaroglu:2022hrm, Bai:2014osa, Biondini:2026ryb, Ishiwata:2013gma, Falkowski:2013jya, Barducci:2018esg, CMS:2024bni}), including modified dark sectors, non-standard cosmological histories~\cite{Mitra:2025cmo}, or multi-component DM setups~\cite{Bhattacharya:2024nla}, in this work we pursue a different possibility by extending the dark sector with a down-type VLQ~\cite{Aguilar-Saavedra:2013qpa} and a charged VLL, in addition to the CSDM candidate~\cite{Barger:2008jx}. A similar extension involving a Dirac or Majorana DM candidate together with two charged scalar particles carrying the same quantum numbers as the VLQ and VLL has been studied in Refs\,.~\cite{Acaroglu:2021qae, Acaroglu:2023phy}. All three new particles carry the same non-trivial charge under the discrete $\Zthree$ symmetry, which guarantees the stability of the CSDM candidate while allowing the trilinear scalar interaction $\Phi^3$. This interaction opens up semi-annihilation and co-annihilation channels involving the vector-like fermions, thereby providing additional mechanisms for reproducing the observed relic abundance while remaining consistent with current direct-detection limits~\cite{Benincasa:2023vyp, Aranda:2019vda, Bhattacharya:2025mlg, DiazSaez:2022nhp, Lahiri:2024rxc}. The efficiency of these processes depends sensitively on the mass hierarchy among the dark-sector particles, making the relic abundance strongly correlated with the underlying model parameters.

The remaining free parameters of the model, in particular the six Yukawa couplings, play a central role not only in determining the dark matter phenomenology but also in generating correlated signatures in both the quark and lepton flavor sectors. In the quark sector, they induce new contributions to flavor-changing neutral current processes, including neutral meson mixing, leptonic and semileptonic $B$-meson decays, and rare kaon decays. In the lepton sector, the same Yukawa interactions contribute to the anomalous magnetic moments of the charged leptons and to charged lepton flavor-violating (LFV) processes such as $\mu\to e\gamma$, $\mu\to3e$, and other LFV observables. Furthermore, the direct couplings of the charged leptons to the vector-like fermions and the complex scalar dark matter particle generate one-loop corrections to the $Z\ell\bar{\ell}$ vertex, thereby modifying the precisely measured $Z$-pole observables. Consequently, precision measurements of $R_\ell$, $R_b$, and the forward--backward asymmetries $A_\ell$, $A_b$, and $A_s$ provide important complementary constraints on the model parameter space. In contrast, since the vector-like fermions are singlets under $\rm SU(2)_\mathtt{L}$, their contributions to the oblique electroweak parameters $S$, $T$, and $U$ are highly suppressed and remain negligible throughout the parameter space considered in this work.

Furthermore, the model predicts a rich collider phenomenology via the production and decay of vector-like fermions. Their signatures closely resemble those of supersymmetric particles at the LHC, allowing existing LHC searches to be reinterpreted within the present framework. Owing to the dominant decay modes $\psi \rightarrow \Phi\,b$ and $\chi \rightarrow \Phi \, \ell$, the resulting signatures closely resemble those predicted in supersymmetric scenarios featuring heavy-flavor jets or leptons accompanied by missing transverse momentum ($E_{\mathrm{T}}^{\mathrm{miss}}$). As a result, existing searches for stops, squarks, sleptons and charginos set stringent constraints on the model parameter space spanned by the masses and couplings of the new particles.

An equally important question is whether the parameter space capable of reproducing the observed dark matter relic abundance while satisfying all current experimental constraints remains testable in future experiments. In this regard, the High-Luminosity LHC (HL-LHC) will substantially extend the discovery reach for the new vector-like states through direct searches, while next-generation direct- and indirect-detection experiments, such as XLZD, Fermi-LAT, and cosmic microwave background (CMB) observations, will provide complementary probes of the dark matter sector. At the same time, future high-precision flavor experiments will further constrain the underlying Yukawa structure of the model through its virtual contributions to flavor observables. The complementarity among collider, dark matter, and flavor searches therefore offers a comprehensive strategy for testing this class of models across the Energy, Cosmic, and Intensity Frontiers.

The remainder of this paper is organised as follows. In \autoref{sec:model}, we introduce the model and discuss its salient features. The flavor phenomenology and electroweak precision constraints are presented in \autoref{sec:flavour}. The dark matter phenomenology is discussed in \autoref{sec:darkmatter}, while the collider prospects at the HL-LHC are investigated in \autoref{sec:collider}. A combined analysis incorporating all relevant experimental constraints is performed in \autoref{sec:combined}. Finally, we summarize our results and present our conclusions in \autoref{sec:summary}. Additional details and supplementary material are collected in Appendices~\ref{app:EFT}--\ref{app:feynman}.
\section{The Model}
\label{sec:model}
In this work, we extend the SM by introducing a complex scalar field, $\Phi$, together with two SM singlet vector-like fermions, $\psi$ and $\chi$, carrying hypercharges $\mathtt{Y}_\psi=-1/3$ and $\mathtt{Y}_\chi=-1$, respectively. All three BSM fields transform identically under a discrete $\Zthree$ symmetry. Their quantum number assignments are summarised in \autoref{tab:tab1}. Owing to the common $\Zthree$ charge of the new fields, the lightest electrically neutral BSM particle, namely the complex scalar $\Phi$, is stable against decay into SM particles and therefore constitutes a viable dark matter candidate\footnote{The stability of dark matter requires it to be the lightest particle charged under the same $\Zthree$ symmetry. This condition imposes the mass hierarchy $\mpsi > \mphi + m_{d_i}$ and $\mchi > \mphi + m_{e_i}$.}.
\begin{table}[htb!]
\begin{center}
\begin{tabular}{|c|c|c|c|c|}\hline
\rowcolor{gray!25}{\bf Fields}& $\rm SU(3)_{c}$ & $\rm SU(2)_{\mathtt{L}}$ & $\rm U(1)_{\mathtt{Y}}$   &  $\Zthree$ \\\hline\hline
\rowcolor{red!10} $\psi$    & 3 & 1 & $-1/3$   & $\omega$ \\
\rowcolor{green!10} $\chi$     & 1 & 1 & $-1$     & $\omega$ \\
\rowcolor{cyan!10} $\Phi$  & 1 & 1 & $0$    & $\omega$ \\\hline
\end{tabular}
\end{center}
\caption{The beyond-standard-model particle contents and their corresponding quantum numbers while $\omega=e^{-i (2\pi/3)}$.}
\label{tab:tab1}
\end{table}

The Lagrangian involving the new BSM fields, consistent with the SM gauge symmetry and the imposed $\Zthree$ symmetry, is given by
\begin{eqnarray}
\begin{split}
\mathcal{L}\supset &|\partial_{\mu}\Phi |^2-m_{\Phi}^2|\Phi|^2-\lambda_{\Phi}|\Phi|^4-\frac{\mu_{3}}{2}\left[\Phi^3+\mathrm{h.c.}\right]
-\lambda_{\Phi H}|\Phi|^2\left(H^{\dagger}H-\frac{v^2}{2}\right)\\
&+\psibar\left(i\gamma^{\mu}\mathcal{D}^{(\psi)}_{\mu}-\mpsi\right)\psi
+\chibar\left(i\gamma^{\mu}\mathcal{D}^{(\chi)}_{\mu}-\mchi\right)\chi \\
&-\left(\sum_{q=d,s,b}\mathtt{y}_{q}\psibar q_{\mathtt{R}}\Phi
+\sum_{\ell=e,\mu,\tau}\mathtt{y}_{\ell}\chibar\ell_{\mathtt{R}}\Phi
+\mathrm{h.c.}\right),
\end{split}
\label{eq:model}
\end{eqnarray}
where the covariant derivatives are
\begin{align}
\mathcal{D}^{(\psi)}_{\mu}
=\partial_{\mu}-ig_sT^aG^a_{\mu}-ig^{\prime} {\tt Y}_{\psi}B_{\mu},\qquad
\mathcal{D}^{(\chi)}_{\mu}
=\partial_{\mu}-ig^{\prime} {\tt Y}_{\chi}B_{\mu},
\end{align}
with $g_s$ and $g^{\prime}$ denoting the gauge couplings of $\mathrm{SU}(3)_c$ and $\mathrm{U}(1)_Y$, respectively. 
The first line of \eq(\ref{eq:model}) contains the scalar kinetic term together with the most general renormalizable scalar potential consistent with the imposed $\Zthree$ symmetry. In contrast to the well-studied real scalar Higgs-portal dark matter model~\cite{Burgess:2000yq}, the complex scalar field $\Phi$ admits the cubic self-interaction term proportional to $\mu_3$, which is a characteristic feature of $\Zthree$-symmetric dark matter scenarios. This interaction gives rise to semi-annihilation processes of the form $\Phi\Phi\rightarrow\Phi^* X_{\rm SM}$ and, in the presence of the vector-like fermions, co-annihilation channels involving the dark-sector particles. Consequently, the dark matter relic abundance is determined not only by the Higgs portal interaction but also by the interplay between the trilinear scalar coupling and the Yukawa interactions. These additional annihilation channels considerably enlarge the viable parameter space by allowing the observed relic abundance to be achieved even for relatively small Higgs portal and Yukawa couplings, thereby helping evade stringent constraints from collider and direct detection searches. In contrast, the corresponding real scalar Higgs-portal scenario generally relies more strongly on the Higgs portal interaction and may require alternative non-standard cosmological histories, such as freeze-out during the reheating era, to simultaneously satisfy the observed relic abundance and current experimental constraints~\cite{Mitra:2025cmo}. We write the Higgs portal interaction in the shifted form $(H^\dagger H-v^2/2)$ so that Electroweak Symmetry Breaking (EWSB) does not induce an additional contribution to the tree-level mass parameter $\mphi$.

The second line of \eq(\ref{eq:model}) describes the kinetic and Dirac mass terms of the vector-like fermions together with their Yukawa interactions with the SM fermions and the dark scalar. Since the left- and right-handed components of $\psi$ and $\chi$ transform identically under the SM gauge group, gauge-invariant Dirac mass terms can be written independently of EWSB. The hypercharge assignments of the new fermions permit the gauge-invariant Yukawa interactions $\psibar q_R\Phi$ and $\chibar\ell_R\Phi$, thereby coupling $\psi$ exclusively to the right-handed down-type quarks and $\chi$ to the right-handed charged leptons. Besides opening co-annihilation channels in the early Universe, the vector-like fermions mediate the interactions between the dark sector and the SM fermions, generating the one-loop contributions to flavor-changing neutral current processes and charged-lepton observables that will be discussed in the following sections.
Before discussing the phenomenological implications of the model, we first summarize the most relevant theoretical constraints that define the viable parameter space.
\begin{itemize}
\item \texttt{Perturbativity\\} 
To ensure the validity of perturbation theory, the loop corrections to the couplings must remain smaller than their corresponding tree-level values. This requirement leads to the following perturbativity bounds on the model parameters \cite{Lerner:2009xg}:
\begin{align}
|\lambda_{\rm H}|\leq 2\pi/3\,,\quad
|\lambda_{\Phi}|\leq \pi\,,\quad
|\lambda_{\rm\Phi H}|\leq 4\pi\,,\quad
|\yq|\leq\sqrt{4\pi}\,,\quad|\yl|\leq\sqrt{4\pi}\,.
\end{align}
\item \texttt{Tree-level unitarity\\}
The tree-level unitarity bounds, obtained in the high-energy scattering limit ($\texttt{s}\to \infty$) and considering only the quartic interaction terms, are given by \cite{Bhattacharya:2024nla},
\begin{eqnarray}
\begin{split}
|\lambda_{\rm H}|\leq 4\pi\,,\quad
|\lambda_{\Phi}|\leq 4\pi\,,\quad
|\lambda_{\rm\Phi H}|\leq 8\pi\,,\qquad\quad\\
|3\lambda_{\rm H}+2\lambda_\Phi \pm \sqrt{9\lambda^2_{\rm H}-12\lambda_{\rm H}\lambda_\Phi+4\lambda_\Phi^2+2\lambda_{\rm\Phi H}^2}|\leq 8\pi\,.
\end{split}
\end{eqnarray}
\item \texttt{Vacuum stability\\}
The necessary conditions required to stabilise the potential are given by \cite{Athron:2018ipf, McDonald:1993ex, Kannike:2012pe},
\begin{align}
\lambda_{\rm H}>0\,,\quad
\lambda_{\Phi}>0\,,\quad
\lambda_{\Phi{\rm H}}+2\sqrt{\lambda_{\Phi}\lambda_{\rm H}}>0\,.
\end{align}
Additionally, to ensure that the electroweak vacuum is sufficiently long-lived, the cubic coupling satisfies the approximate upper bound \cite{Belanger:2012zr},
\begin{align}
{\rm max}[\mu_3]\approx 2\sqrt{\lambda_\Phi}\,\mphi.
\end{align}
Therefore, within the perturbative limit, we consider a maximum value of $\mu_3 = 2\sqrt{\pi}\mphi$ for our phenomenological analysis.
\end{itemize}
\noindent
As evident from \eq(\ref{eq:model}), the relic abundance of the CSDM is naturally determined through the standard thermal freeze-out mechanism. The CSDM therefore behaves as a WIMP\footnote{An alternative possibility is that the CSDM is produced via the freeze-in mechanism if both the Yukawa couplings and the Higgs portal coupling are sufficiently feeble. In this regime, the dark matter population is generated not only through Higgs-portal interactions but also via Yukawa-mediated decays and scattering processes involving the vector-like fermions.}. This follows from the fact that the $\Zthree$-charged vector-like fermions, $\psi$ and $\chi$, remain in thermal equilibrium with the SM bath through their gauge interactions, while the CSDM is efficiently thermalized via its Yukawa interactions with these fermions (and, in general, through the Higgs portal interaction).

Before presenting the phenomenological analysis, we first discuss the parameter dependence of the key observables considered in this work and subsequently identify the regions of parameter space that can simultaneously accommodate all experimental constraints. For convenience, \autoref{tab:placeholder} provides a road map by summarising the model parameters relevant to each class of observables, namely dark matter searches (red), neutral meson mixing (green), leptonic meson decays (violet), semileptonic meson decays (yellow), charged-lepton anomalous magnetic moments and rare decays (blue), and electroweak $Z$-pole observables (pink). The symbols \cmark\ and \xmark\ denote, respectively, that an observable is sensitive or insensitive to a given model parameter.
\begin{table}[htb!]
\centering
\renewcommand{\arraystretch}{1.0}
\setlength{\tabcolsep}{6pt}
\setlength{\cmidrulewidth}{0.8pt}
\resizebox{1\linewidth}{!}{
\begin{tabular}{|c|ccccccccccc|}\hline
\rowcolor{lightgray} & \multicolumn{11}{c|}{Model parameters} \\ \cline{2-12}
\rowcolor{lightgray}\multirow{-2}{*}{Observables}& $\mphi$ & $\mpsi$ & $\mchi$ & $\lphiH$ & $\mu_3$ & $\ye$  & $\ymu$ & $\yta$ & $\yd$  & $\ys$  & $\yb$ \\\hline\hline 
\rowcolor{red!10}DM relic density   & \cmark & \cmark   & \cmark  & \cmark   & \cmark  & \cmark & \cmark & \cmark & \cmark & \cmark & \cmark\\
\rowcolor{red!10}DM direct search   & \cmark & \cmark   & \xmark  & \cmark   & \xmark  & \xmark & \xmark & \xmark & \cmark & \xmark & \xmark\\
\rowcolor{red!10}DM indirect search & \cmark & \cmark   & \cmark  & \cmark   & \xmark  & \cmark & \cmark & \cmark & \xmark & \xmark & \cmark\\
\rowcolor{red!10}DM collider search & \cmark & \cmark   & \cmark  & \cmark   & \xmark  & \cmark & \cmark & \cmark & \cmark & \cmark & \cmark\\
\rowcolor{green!10}$B_{s}^{0}-\bar{B}_{s}^{0}$ mixing& \cmark & \cmark   & \xmark  & \xmark   & \xmark  & \xmark & \xmark & \xmark & \xmark & \cmark & \cmark\\
\rowcolor{green!10}$B_{d}^{0}-\bar{B}_{d}^{0}$ mixing& \cmark & \cmark   & \xmark  & \xmark   & \xmark  & \xmark & \xmark & \xmark & \cmark & \xmark & \cmark\\
\rowcolor{green!10}$K_{}^{0}-\bar{K}^{0}$ mixing& \cmark & \cmark   & \xmark  & \xmark   & \xmark  & \xmark & \xmark & \xmark & \cmark & \cmark & \xmark\\
\rowcolor{blue!10}${\rm K_L^0\to\mu^-\mu^+}$& \cmark & \cmark   & \cmark  & \xmark   & \xmark  & \xmark & \cmark & \xmark & \cmark & \cmark & \xmark\\
\rowcolor{blue!10}${\rm K_L^0\to e^-e^+}$& \cmark & \cmark   & \cmark  & \xmark   & \xmark  & \cmark & \xmark & \xmark & \cmark & \cmark & \xmark\\
\rowcolor{blue!10}${\rm B^0\to \mu^-\mu^+}$& \cmark & \cmark   & \cmark  & \xmark   & \xmark  & \xmark & \cmark & \xmark & \cmark & \xmark & \cmark\\
\rowcolor{blue!10}${\rm B_s^0\to \mu^-\mu^+}$& \cmark & \cmark   & \cmark  & \xmark   & \xmark  & \xmark & \cmark & \xmark & \xmark & \cmark & \cmark\\
\rowcolor{yellow!10}$b \to s ~\mu^- ~\mu^+ $& \cmark & \cmark   & \cmark  & \xmark   & \xmark  & \xmark & \cmark & \xmark & \xmark & \cmark & \cmark\\
\rowcolor{yellow!10}$b \to s ~e^- ~e^+ $& \cmark & \cmark   & \cmark  & \xmark   & \xmark  & \cmark & \xmark & \xmark & \xmark & \cmark & \cmark\\
\rowcolor{yellow!10}$b \to d ~e^- ~e^+ $& \cmark & \cmark   & \cmark  & \xmark   & \xmark  & \cmark & \xmark & \xmark & \cmark & \xmark & \cmark\\
\rowcolor{cyan!10}$\Delta a_e$&\cmark & \xmark  & \cmark  & \xmark   & \xmark & \cmark  & \xmark & \xmark & \xmark & \xmark & \xmark\\
\rowcolor{cyan!10}$\mu \to e ~\gamma $& \cmark  & \xmark   & \cmark  & \xmark & \xmark  & \cmark & \cmark & \xmark & \xmark & \xmark    & \xmark\\
\rowcolor{cyan!10}$\tau \to e ~\gamma $& \cmark & \xmark   & \cmark  & \xmark & \xmark  & \cmark & \xmark & \cmark & \xmark & \xmark    & \xmark\\
\rowcolor{cyan!10}$\tau \to \mu ~\gamma $& \cmark& \xmark   & \cmark  & \xmark & \xmark  & \xmark & \cmark & \cmark & \xmark & \xmark    & \xmark\\
\rowcolor{cyan!10}$\mu \to 3e $& \cmark & \xmark& \cmark   & \xmark  & \xmark & \cmark  & \cmark & \xmark & \xmark & \xmark & \xmark  \\
\rowcolor{cyan!10}$\mu^- ~ \mathbb{N} \to e^{\mp}~\mathbb{N}$   & \cmark   & \cmark  & \cmark & \xmark  & \xmark & \cmark & \cmark & \xmark & \cmark & \xmark & \xmark\\
\rowcolor{cyan!5}$\tau \to \mu ~\pi^0 $& \cmark & \cmark   & \cmark  & \xmark & \xmark  & \xmark & \cmark & \cmark & \cmark & \xmark    & \xmark\\
\rowcolor{cyan!5}$\tau \to \mu ~K_S^0 $& \cmark& \cmark   & \cmark  & \xmark & \xmark  & \xmark & \cmark & \cmark & \cmark & \cmark    & \xmark\\
\rowcolor{cyan!5}$\tau \to \mu ~K^{*0} $& \cmark & \cmark   & \cmark  & \xmark & \xmark  & \xmark & \cmark & \cmark & \cmark & \cmark    & \xmark\\
\rowcolor{cyan!5}$\tau \to \mu ~\rho^0 $& \cmark& \cmark   & \cmark  & \xmark & \xmark  & \xmark & \cmark & \cmark & \cmark & \xmark    & \xmark\\
\rowcolor{cyan!5}$\tau \to \mu ~\phi $& \cmark & \cmark   & \cmark  & \xmark & \xmark  & \xmark & \cmark & \cmark & \xmark & \cmark    & \xmark\\
\rowcolor{magenta!10}$ R_\ell$&\cmark& \cmark   & \cmark  & \xmark   & \xmark  & \cmark & \cmark  & \cmark & \cmark & \cmark & \cmark\\
\rowcolor{magenta!10}$ R_b$&\cmark   & \cmark   & \xmark  & \xmark   & \xmark  & \xmark & \xmark  & \xmark & \cmark & \cmark & \cmark\\
\rowcolor{magenta!10}$A_\ell$&\cmark & \xmark   & \cmark  & \xmark   & \xmark  & \cmark & \cmark  & \cmark & \xmark & \xmark & \xmark\\
\rowcolor{magenta!10}$ A_s$&\cmark   & \cmark   & \xmark  & \xmark   & \xmark  & \xmark & \xmark  & \xmark & \xmark & \cmark & \xmark\\
\rowcolor{magenta!10}$ A_b$&\cmark   & \cmark   & \xmark  & \xmark   & \xmark  & \xmark & \xmark  & \xmark & \xmark & \xmark & \cmark\\
\hline\end{tabular}}
\caption{Summary of dependence of the observables on the model parameters considered in this work. The sign \cmark\ (\xmark) indicates that the corresponding observable is sensitive (insensitive) to a given model parameter.}
\label{tab:placeholder}
\end{table}

Notably, only the DM relic density is sensitive to all the model parameters, whereas the remaining observables are insensitive to at least one parameter, thereby leaving additional freedom in the parameter space. In particular, the trilinear coupling $\mu_3$ enters exclusively into the calculation of the DM relic density. Consequently, even when the remaining model parameters are tightly constrained by other observables, $\mu_3$ provides sufficient freedom to reproduce the observed relic abundance. Similarly, the portal coupling $\lphiH$ primarily affects the DM relic density and DM detection observables. Although it also contributes to flavor observables, its effect is highly suppressed compared to the gauge-mediated contributions and is therefore negligible in practice.

Before proceeding to the detailed phenomenological analysis, we remark that, in general, the Yukawa couplings and the Higgs portal coupling in \eq\eqref{eq:model} can be complex\footnote{Notably, an imaginary Yukawa coupling can address the matter-antimatter asymmetry problem, albeit in an extended scenario; see Ref\,.~\cite{Heisig:2024mwr, Belfatto:2026uze} for details.}, thereby introducing new sources of CP violation beyond the SM. Such complex phases can give rise to a variety of CP-violating observables in both the quark and lepton sectors, including electric dipole moments and CP asymmetries in flavor-changing processes. A comprehensive study of these effects is beyond the scope of the present work. We therefore restrict our analysis to the CP-conserving limit, in which all new couplings are taken to be real, and investigate the resulting phenomenological implications in the following sections.
\section{Constraints from Low-energy flavor Observables and Electroweak Precision Observables}
\label{sec:flavour}
The interactions introduced in \eq\eqref{eq:model} not only govern the dark matter phenomenology but also generate a rich set of low-energy observables through loop-induced effects involving the complex scalar $\Phi$ and the vector-like fermions. Since the new Yukawa interactions couple the dark sector simultaneously to the down-type quark and charged-lepton sectors, the model predicts correlated contributions to flavor-changing neutral current processes, charged-lepton observables, and electroweak precision measurements. Owing to the high precision achieved in many of these observables, they provide powerful and complementary probes of the model parameter space, often extending well beyond the direct reach of collider searches. In this section, we therefore investigate the constraints arising from low-energy flavor observables and electroweak precision measurements before discussing the dark matter and collider phenomenology.
\subsection{Meson Observables}
In contrast to the SM, the amplitudes for low-energy flavor transitions are determined by the new Yukawa couplings and are not necessarily subject to the generation-dependent suppressions arising from the CKM matrix. As a result, flavor observables involving all three generations of down-type quarks and charged leptons can provide stringent and complementary constraints on the model parameter space. For convenience, we discuss these observables in three categories: purely hadronic processes, purely leptonic processes, and semileptonic processes involving both quarks and leptons.
\subsubsection{Neutral Meson Mixing}
\label{subsec:meson_mixing}
\begin{figure}[htb!]
\centering
\begin{adjustbox}{width=0.9\textwidth}
\begin{tcolorbox}[
colback=gray!5, colframe=black!10, boxrule=0.8pt, arc=4mm, boxsep=0pt, left=0pt, right=0pt, top=0pt, bottom=0pt, width=0.75\textwidth, halign=center]
\begin{tikzpicture}
\begin{feynman}
\vertex (a1){\(d_i\)};
\vertex[right=1.3cm of a1](a2);
\vertex[above=2cm of a2](a3);
\vertex[left=1cm of a3](a4){\(d_j\)};
\vertex[right=2cm of a2](a5);
\vertex[right=2cm of a3](a6);
\vertex[right=1cm of a6](a7){\(d_i\)};
\vertex[right=1cm of a5](a8){\(d_j\)};
\diagram* { 
(a1) --[fermion,line width=0.35mm, arrow size=0.8pt](a2),
(a2) --[fermion, line width=0.35mm,arrow size=0.8pt,style=black!50,edge label=\(\color{black}\psi\)](a3),
(a3)--[fermion, line width=0.35mm,arrow size=0.8pt](a4),
(a5) --[charged scalar,line width=0.35mm,arrow size=0.8pt,style=black!50,edge label'=\(\color{black}\Phi\)](a2),
(a3) --[charged scalar,line width=0.35mm,style=black!50,arrow size=0.8pt,edge label'={\(\color{black}\Phi\)}](a6),
(a7) --[fermion,line width=0.35mm, arrow size=0.8pt](a6),
(a6) --[fermion,line width=0.35mm, arrow size=0.8pt,style=black!50,edge label=\(\color{black}\psi\)](a5),
(a5) --[fermion,line width=0.35mm, arrow size=0.8pt](a8)};
\node at (a2)[circle,fill,style=gray,inner sep=1pt]{};
\node at (a3)[circle,fill,style=gray,inner sep=1pt]{};
\node at (a5)[circle,fill,style=gray,inner sep=1pt]{};
\node at (a6)[circle,fill,style=gray,inner sep=1pt]{};
\end{feynman}
\end{tikzpicture}\qquad
\begin{tikzpicture}
\begin{feynman}
\vertex (a1){\(d_i\)};
\vertex[right=1.3cm of a1](a2);
\vertex[above=2cm of a2](a3);
\vertex[left=1cm of a3](a4){\(d_j\)};
\vertex[right=2cm of a2](a5);
\vertex[right=2cm of a3](a6);
\vertex[right=1cm of a6](a7){\(d_i\)};
\vertex[right=1cm of a5](a8){\(d_j\)};
\diagram* { 
(a1) --[fermion,line width=0.35mm, arrow size=0.8pt](a2),
(a3) --[charged scalar,style=black!50, arrow size=0.8pt,line width=0.35mm,edge label'=\(\color{black}\Phi\)](a2),
(a3) --[fermion, line width=0.35mm,arrow size=0.8pt](a4),
(a2) --[fermion,line width=0.35mm,arrow size=0.8pt,style=black!50,edge label=\(\color{black}\psi\)](a5),
(a6) --[fermion,line width=0.35mm,arrow size=0.8pt,style=black!50,edge label={\(\color{black}\psi\)}](a3),
(a7) --[fermion,line width=0.35mm, arrow size=0.8pt](a6),
(a5) --[charged scalar,line width=0.35mm, arrow size=0.8pt,style=black!50,edge label'=\(\color{black}\Phi\)](a6),
(a5) --[fermion,line width=0.35mm, arrow size=0.8pt](a8)};
\node at (a2)[circle,fill,style=gray,inner sep=1pt]{};
\node at (a3)[circle,fill,style=gray,inner sep=1pt]{};
\node at (a5)[circle,fill,style=gray,inner sep=1pt]{};
\node at (a6)[circle,fill,style=gray,inner sep=1pt]{};
\end{feynman}
\end{tikzpicture}
\end{tcolorbox}
\end{adjustbox}
\caption{The Feynman diagrams contributing to the neutral meson mixing process. Here, $\psi$ and $\Phi$ are the VLQ and DM, respectively.}
\label{fig:meson_mixing_Feyn}
\end{figure}
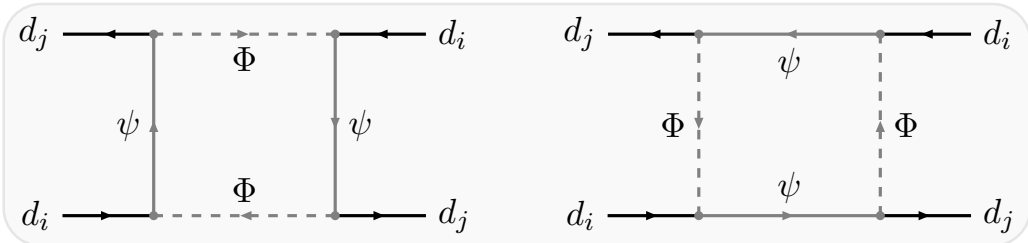

Neutral pseudoscalar meson mixing provides an important probe of NP models owing to the high precision achieved in experimental measurements. These processes are particularly sensitive to BSM physics since they first arise at the one-loop level in the SM. Moreover, they are theoretically relatively clean, as the relevant hadronic uncertainties are encapsulated primarily in the meson decay constants and bag parameters entering the hadronic matrix elements, which we introduce in the following. In the SM, the dominant contribution to neutral meson mixing arises from box diagrams involving the $W$ boson and the top quark (for down-type meson systems). In the present NP scenario, additional contributions arise from analogous box diagrams in which the VLQ and the dark matter particle propagate in the loop. The corresponding representative Feynman diagrams are shown in \fig\ref{fig:meson_mixing_Feyn}. Since the VLQ couples only to down-type quarks, all neutral meson systems involving down-type quarks, namely $B_s^0-\bar{B}_s^0$, $B^0-\bar{B}^0$, and $K^0-\bar{K}^0$, provide relevant constraints. In contrast, the model does not contribute to $D^0-\bar{D}^0$ mixing, while constraints on analogous models with up-type VLQs are discussed in Ref\,.~\cite{Bhattacharya:2025mlg}.

In the absence of CP-violating new physics effects, the mass difference, $\Delta M$, is the primary observable characterizing the neutral meson oscillation $P^{0}-\bar{P}^{0}$ and is given by
\begin{equation}\label{eq:mixing_deltaM}
\Delta M_P = 2 |M_{12}| = 2 \frac{|\mathcal{M}|}{2 m_{P}}\,,
\end{equation}
where $m_{P}$ is the mass of the pseudoscalar meson $P_0$ and $|\mathcal{M}|$ is the mixing amplitude obtained from the sum of the dispersive parts of the meson mixing diagrams in \fig\ref{fig:meson_mixing_Feyn}. We can express $\Delta M$ as the sum of contributions from the SM and the new physics scenarios:
\begin{equation}
\Delta M_{\rm tot}^{P} = \Delta M_{\rm SM}^{P} + \Delta M_{\rm NP}^{P}\,.
\end{equation}
As an amplitude-level observable, the NP contributions will be directly added at the observable level. 
It is convenient to quantify the size of the NP contribution relative to the SM prediction through the ratio~\cite{Kolay:2024wns, Kolay:2025jip, Kala:2025srq}
\begin{equation}
\Delta_{P} = \frac{\Delta M_{\rm NP}^{P}}{\Delta M_{\rm SM}^{P}} = \left(\frac{\Delta M_{\rm exp}^{P}}{\Delta M_{\rm SM}^{P}} - 1 \right)\,,
\end{equation}
where $\Delta M_{\rm exp}^{P}$ denotes the experimentally measured mass difference.
Expressing the mixing observable in this form helps reduce the impact of hadronic uncertainties by canceling common input parameters. The SM and experimental values of the observable $\Delta M_{P}$ are given in \autoref{tab:mixing_obs}, from which the bounds on the parameter $\Delta_{P}$ can easily be calculated.
\begin{table}[htb!]
\centering
\begin{tabular}{|c|c|c|c|}
\hline
\rowcolor{gray!40}Process& Observable & SM Prediction ($\rm ps^{-1}$) & Exp. Measurement ($\rm ps^{-1}$) \\\hline \hline
\rowcolor{red!10}$B^{0}_d-\bar{B}^{0}_d $ & $\Delta M_{B^0}$ & $0.535 \pm 0.021 $ \cite{Albrecht:2024oyn, Wang:2022lfq} & $0.5065\pm0.0019$ \cite{LHCb:2023sim, Belle-II:2023bps}  \\
\rowcolor{green!10}$B^{0}_s-\bar{B}^{0}_s $ & $\Delta M_{B_s}$ & $18.23\pm 0.63 $ \cite{Albrecht:2024oyn, Wang:2022lfq} & $17.765 \pm 0.006 $ \cite{LHCb:2023sim, Belle-II:2023bps} \\
\rowcolor{cyan!10}$K^{0}-\bar{K}^{0} $ & $\Delta M_{K}$ & $ 0.00580 \pm 0.00237 $ \cite{ParticleDataGroup:2024cfk} &  $0.005289 \pm 0.000010  $ \cite{KTeV:2010sng} \\
\hline
\end{tabular}
\caption{The SM prediction and experimental measurement of the neutral meson mixing observable $\Delta M_P$ for $B^0$, $B_s^0$, and $K^0$ meson systems. }
\label{tab:mixing_obs}
\end{table}
In the presence of the new particles, the neutral meson mixing amplitudes receive additional contributions from the box diagrams shown in \fig\ref{fig:meson_mixing_Feyn}. These diagrams generate the following effective four-fermion operator:
\begin{equation}\label{eq:mixing_loop}
\mathcal{L}_{\rm eff}=\mathcal{C}_{\rm RR}\left(\bar d_{j}\gamma_{\mu}\mathrm{P}_{\mathtt{R}}d_{i}\right)\left(\bar d_{j}\gamma^{\mu}\mathrm{P}_{\mathtt{R}}d_{i}\right)\,,
\end{equation}
where $\mathcal{C}_{\rm RR}$ denotes the corresponding Wilson coefficient, receiving contributions from both box diagrams in \fig\ref{fig:meson_mixing_Feyn}. Its explicit expression is collected in \eq\eqref{eq:mixing_loop_eff} in \autoref{app:loops}. Owing to the chiral structure of the Yukawa interactions in \eq(\ref{eq:model}), only the $(V+A)\times (V+A)$ four-fermion operator is generated at one loop. The corresponding NP contribution to the neutral meson mass difference can then be expressed as
\begin{equation}
\Delta M_{\rm NP} = \frac{2}{3} \, B_{P}\,  m_{P} \,  f_{P}^2 \, \mathcal{C}_{\rm RR} \,.
\end{equation}
In the above equation, $B_{P}$ is the bag factor related to the amplitude, defined as
\begin{equation}
\langle P_{0}|(\bar{d}_j\,\gamma_{\mu}\,\mathrm{P}_{\mathtt{R}}\,d_{i})(\bar{d}_j\,\gamma^{\mu}\,\mathrm{P}_{\mathtt{R}}\, d_{i})|D_{0}\rangle=\frac{2}{3}\,B_{P}\,f_{P}^2\, m_{P}^2\,,
\end{equation}
where $f_{P}$ is the decay constant of the meson $P$. We have used the following values of the inputs for $B$-meson~\cite{FlavourLatticeAveragingGroupFLAG:2024oxs}: $$B_{B_s} = 1.232 (53), \, B_{B_d} = 1.222(61), \, f_{B_s} = 230.3 (1.3) {~\rm MeV}, \, f_{B_d} = 190.0(1.3) ~{\rm MeV}.$$ For $K$-meson, the corresponding values are~\cite{FlavourLatticeAveragingGroupFLAG:2024oxs, ParticleDataGroup:2024cfk, Bazavov:2017lyh}: $$B_{B_K} = 0.717 (24),~f_K = \frac{f_{K^{\pm}}}{{\sqrt{1+\delta_{\rm SU(2)}}}} \,,{\rm with}~f_{K^{\pm}} = 155.7 (3)  {\rm\, MeV}, ~ \delta_{\rm SU(2)} = -0.0052(9). $$ 

\paragraph{\underline{Bounds on the Parameter Space from Meson Mixing:}} 
\begin{figure}[htb!]
\centering
\subfloat[]{\includegraphics[width=0.33\linewidth]{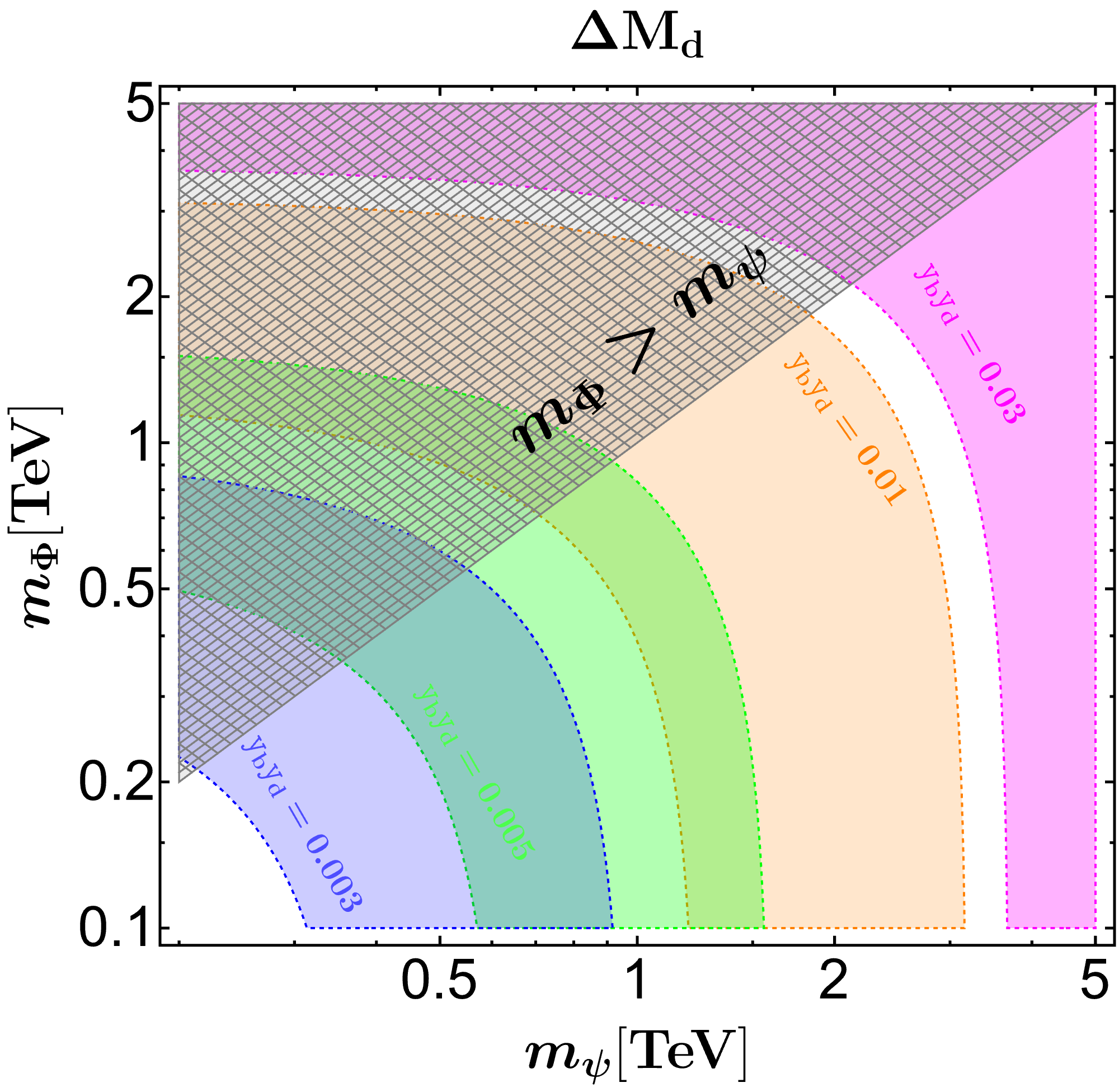}\label{fig:meson_mixing_DeltaMd}}~~
\subfloat[]{\includegraphics[width=0.33\linewidth]{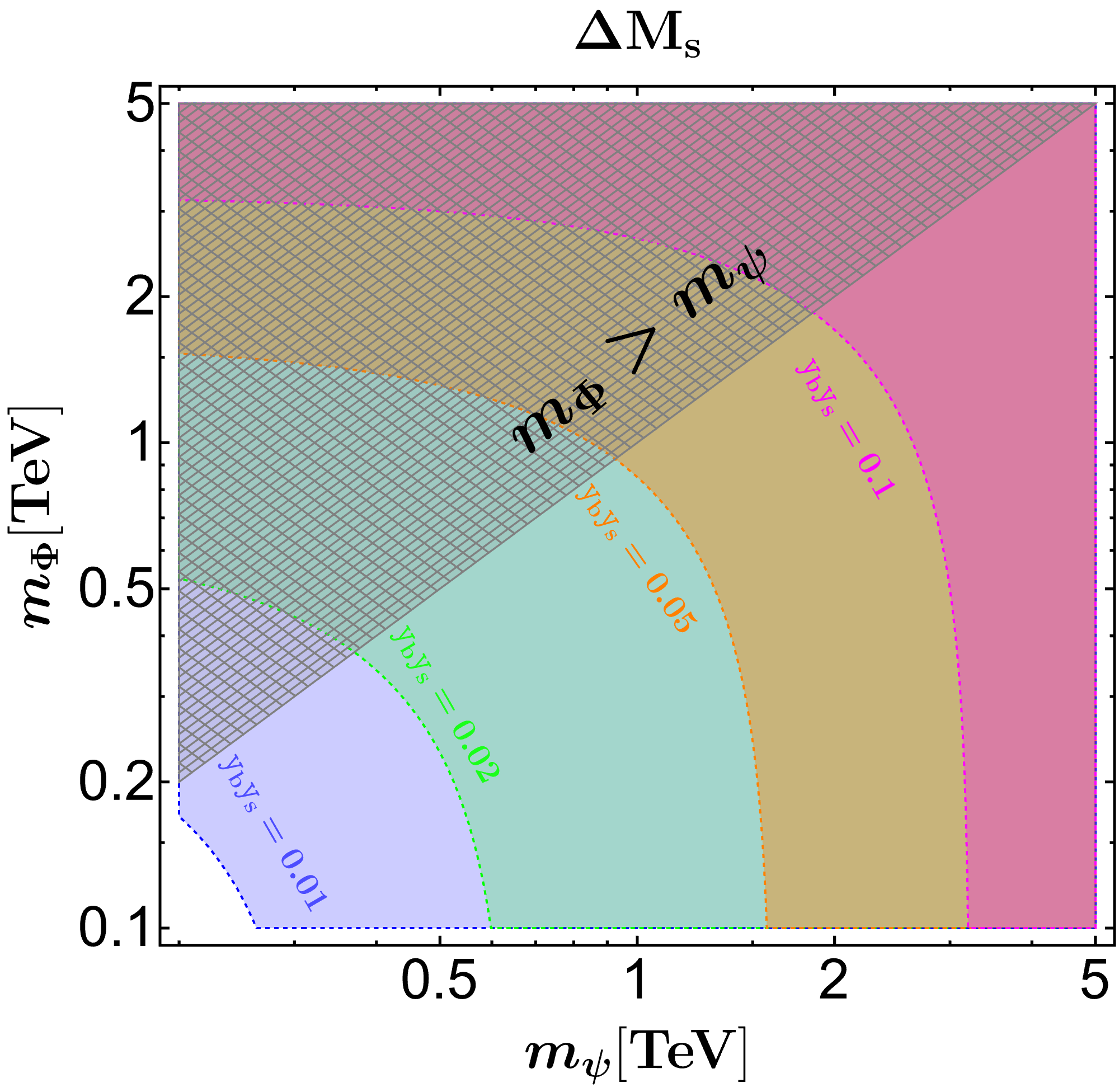}\label{fig:meson_mixing_DeltaMs}}~~
\subfloat[]{\includegraphics[width=0.33\linewidth]{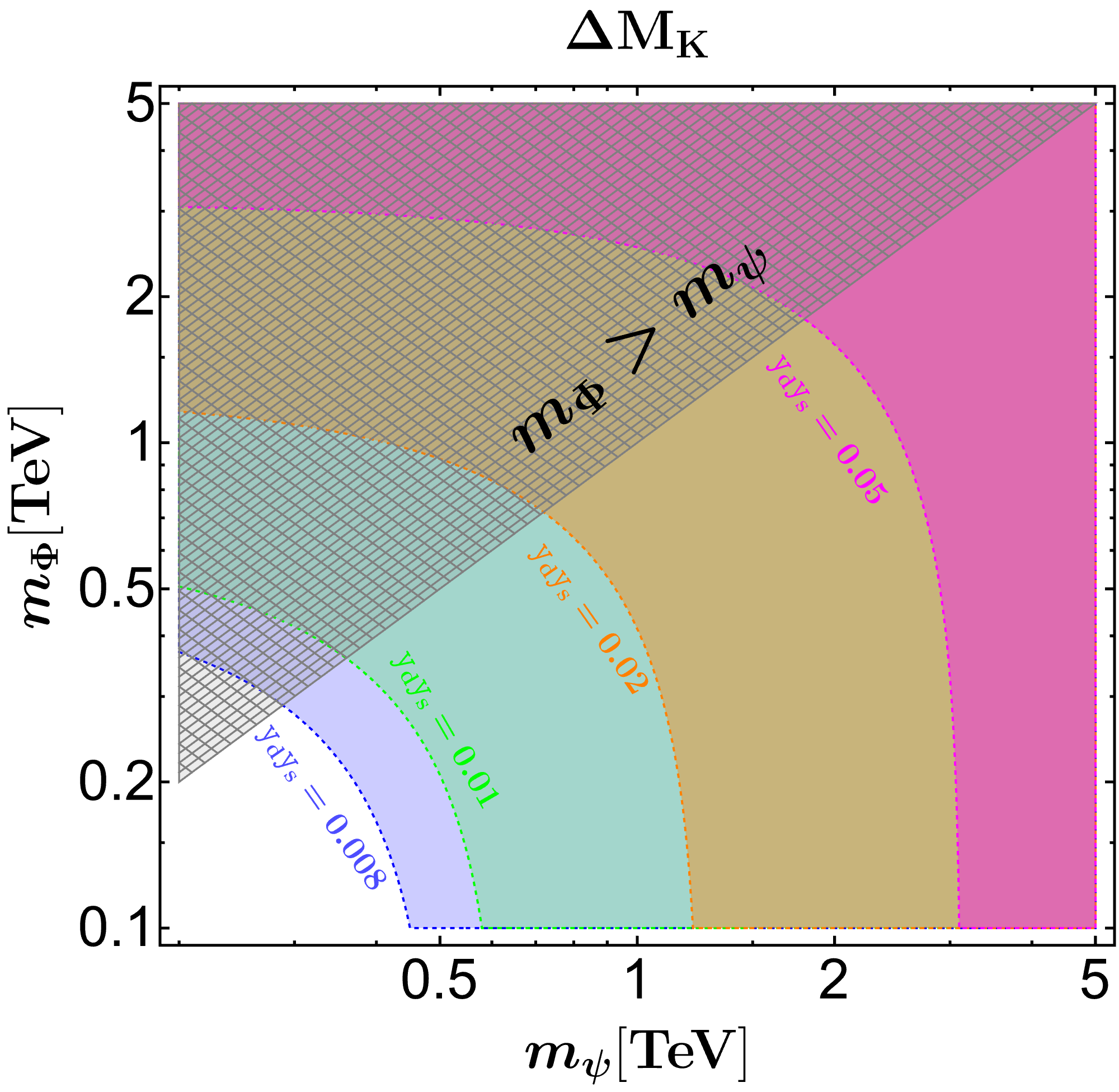}\label{fig:meson_mixing_DeltaMk}}
\caption{Allowed regions of the $(\mpsi\,,~\mphi)$ parameter space at the $68\%$ C.L. from the neutral meson mixing observables: (left) $B^{0}$--$\bar{B}^{0}$ mixing, (middle) $B_{s}^{0}$--$\bar{B}_{s}^{0}$ mixing, and (right) $K^{0}$--$\bar{K}^{0}$ mixing. The blue, green, orange, and magenta bands denote the parameter space allowed for different fixed values of the relevant Yukawa coupling combinations, as indicated in each panel. The hatched grey region, corresponding to $\mphi>\mpsi$, is excluded by the requirement that the complex scalar $\Phi$ remains the stable dark matter candidate. The $B^{0}$ mixing constraint results in allowed bands because both the upper and lower experimental limits contribute, whereas the $B_{s}^{0}$ and $K^{0}$ mixing observables yield one-sided lower-mass exclusions, allowing the entire higher-mass region.
}
\label{fig:mixing_allowed}
\end{figure}
As discussed in the previous subsection, neutral meson mixing constrains the masses of the VLQ and the DM particle, as well as the Yukawa couplings involving the VLQ. To minimize the impact of theoretical uncertainties, we use the ratio $\Delta_{P}$ as the relevant observable. The free parameters entering this analysis are $\mpsi$, $\mphi$, $\yd$, $\ys$, and $\yb$. Since the VLL does not participate in neutral meson mixing, neither its mass nor its Yukawa couplings are constrained in this sector. Figure\,.~\ref{fig:mixing_allowed} shows the parameter space allowed by the individual neutral meson mixing observables, with each panel corresponding to a specific neutral meson system. From the box diagrams, or equivalently from the expression for the Wilson coefficient $\mathcal{C}_{\rm RR}$, it follows that the NP contribution scales as $\Delta M_{P}^{\rm NP}\propto (\mathtt{y}_{i}\mathtt{y}_{j})^{2},$ where $P^{0}\equiv d_{i}\bar d_{j}$. Furthermore, the NP contribution decouples with increasing masses of the dark matter particle and the VLQ, leading to weaker constraints in the heavy-mass regime.

Figure\,.~\ref{fig:meson_mixing_DeltaMd} shows the regions of parameter space allowed by the $\Delta_d$ with $1\sigma$ uncertainty. The blue, green, orange, and magenta bands correspond to the coupling combinations $\yb\yd=\{0.003,\,0.005,\,0.01,\,0.03\}$, respectively. The hatched grey region denotes $\mphi>\mpsi$, which is disfavoured from a cosmological perspective but is otherwise unrelated to the meson-mixing constraint. As expected from the dependence of the Wilson coefficient $\mathcal{C}_{\rm RR}$ on the Yukawa couplings and the masses of the DM particle and the VLQ, smaller values of the coupling combination allow lighter masses, whereas larger couplings require heavier new particles to suppress the NP contribution. Furthermore, the allowed parameter space appears as narrow bands rather than one-sided exclusion regions. This is because the experimentally measured value of $\Delta M_d$ exhibits a tension of approximately $1.35\sigma$ with the SM prediction. Consequently, the $1\sigma$ allowed interval of $\Delta_d$ does not include zero, requiring a non-zero NP contribution. As a result, both the upper and lower bounds on $\Delta_d$, listed in \autoref{tab:mixing_obs}, simultaneously constrain the parameter space, giving rise to the band-like allowed regions.

The remaining panels of \Fig\ref{fig:mixing_allowed} show the parameter space allowed by the $B_s^0-\bar{B}_s^0$ and $K^0-\bar{K}^0$ mixing observables. In contrast to the $\Delta M_d$ case, these measurements yield only lower bounds on the allowed parameter space. Consequently, for a given Yukawa coupling combination, only the low-mass region is excluded, while the entire higher-mass region remains allowed. For example, in the case of $B_s^0-\bar{B}_s^0$ system, coupling combinations satisfying $\yb\ys \geq 0.02$ permit the region with $m_{\psi,\Phi}\gtrsim 500~\GeV$. Similarly, for $K^0-\bar{K}^0$ mixing, the corresponding parameter space is allowed for $\ys\yd \geq 0.01$.
\subsubsection{Rare Leptonic Decays of Mesons}
Rare dileptonic decays of neutral pseudoscalar mesons provide sensitive probes of physics beyond the SM, as they are highly suppressed in the SM by the GIM mechanism and arise only at the loop level. In the present model, the decays $P\to \ell^{+}\ell^{-}$, with $P=B^{0},\,B_{s}^{0},\,K_{L}^{0},$ and $K_{S}^{0}$, receive additional contributions from the exchange of the new particles introduced in \eq(\ref{eq:model}). The current experimental measurements of the corresponding branching ratios, together with the SM predictions, are summarized in \autoref{tab:rare_decays_val}.
The rare leptonic and semileptonic decays of $B$- and $K$-mesons are conveniently described within the framework of the weak effective theory, in which the heavy degrees of freedom are integrated out, and their effects are encoded in a set of local effective operators. The complete operator basis relevant for these processes can be found in Refs\,.~\cite{Bobeth:1999mk, Becirevic:2012fy, Altmannshofer:2008dz}. The corresponding Wilson coefficients parameterize the short-distance contributions, while the hadronic matrix elements are expressed in terms of meson decay constants for purely leptonic decays. In the present model, only a subset of these operators is generated at one loop. We therefore focus on the corresponding Wilson coefficients relevant for the observables considered below.
\begin{table}[htb!]
\centering
\renewcommand{\arraystretch}{1.3}
\setlength{\tabcolsep}{3pt}
\resizebox{0.9\textwidth}{!}{
\begin{tabular}{|c|c|c|}\hline
\rowcolor{gray!40}Processes & Experimental Measurement & SM predictions\\
\hline \hline 
\rowcolor{red!10}$B_{s}^{0} \to \mu^{+} \mu^{-}$ & $(3.83 \pm 0.43) \times 10^{-9}$ \cite{CMS:2022mgd}  & $(3.66 \pm 0.14) \times 10^{-9} $\cite{Beneke:2019slt} \\ 
\rowcolor{red!10}$B_{s}^{0} \to e^{+} e^{-}$ & $< 9.4 \times 10^{-9}$ \cite{LHCb:2020pcv} & $( 8.60 \pm 0.36) \times 10^{-14}$ \cite{Beneke:2019slt} \\
\rowcolor{red!10}$B_{s}^{0} \to \tau^{+} \tau^{-}$ & $<6.8 \times 10^{-3}$ \cite{LHCb:2017myy}& $(7.73 \pm 0.49) \times 10^{-7}$ \cite{Bobeth:2013uxa} \\
\rowcolor{green!10}& $(1.20 \pm 0.78) \times 10^{-10}  $ \cite{LHCb:2021vsc} & \\ 
\rowcolor{green!10}\multirow{-2}{*}{$B^{0} \to \mu^{+} \mu^{-}$}& $<1.5 \times 10^{-10}$ \cite{CMS:2022mgd} & \multirow{-2}{*}{$(1.03 \pm 0.05) \times 10^{-10} $\cite{Beneke:2019slt}} \\ 
\rowcolor{green!10}$B^{0} \to e^+  e^-$& $<2.5 \times 10^{-9}$ \cite{LHCb:2020pcv} & $(2.41 \pm 0.13) \times 10^{-15}$ \cite{Beneke:2019slt} \\
\rowcolor{green!10}$B^{0} \to \tau^{+} \tau^{-}$& $<2.1 \times 10^{-3} $ \cite{LHCb:2017myy}& $ (2.22 \pm 0.19) \times 10^{-8}$ \cite{Bobeth:2013uxa}\\ 
\rowcolor{cyan!10}$K_{L}^0 \to \mu^{+} \mu^{-} $ & $(6.84 \pm 0.11) \times 10^{-9}$ \cite{ParticleDataGroup:2024cfk} &  $(6.85 \pm  0.80_{\rm LD} \pm  0.06_{\rm SD}) \times 10^{-9}$ \cite{DAmbrosio:1994fgc} \\
\rowcolor{cyan!10}$K_{L}^0 \to e^{+} e^{-} $ & $(8.7 \pm 4.9) \times 10^{-12}$ \cite{BNLE871:1998bii}& $\sim 9.0\times 10^{-12}$ \cite{Valencia:1997xe}\\
\rowcolor{cyan!10}$K_{S}^0 \to \mu^{+} \mu^{-}$ & $<2.1 \times 10^{-10}$ \cite{LHCb:2020ycd}& $(5.18 \pm 1.50_{\rm LD} \pm 0.02_{\rm SD} ) \times 10^{-12} $\cite{DAmbrosio:2017klp, Buras:2022qip}\\
\rowcolor{cyan!10}$K_{S}^0 \to e^{+} e^{-}$ & $<9.0 \times 10^{-9}$ \cite{KLOE:2008acb}& $6.0\times 10^{-15}$ \cite{Sehgal:1969zok}\\\hline
\end{tabular}}
\caption{Experimental measurements and SM predictions for the branching ratios of the rare $B$- and $K$-meson decays considered in this work. For the kaon decay modes, the subscripts ``LD'' and ``SD'' denote the long-distance and short-distance contributions, respectively.}
\label{tab:rare_decays_val}
\end{table}
\begin{figure}[htb!]
\centering
\begin{adjustbox}{width=1\linewidth}
\begin{tcolorbox}[colback=gray!5, colframe=black!10, boxrule=0.8pt, arc=4mm, boxsep=0pt, left=0pt, right=0pt, top=0pt, bottom=0pt, width=1.2\linewidth, halign=center]
\subfloat[]{\begin{tikzpicture}
\begin{feynman}
\vertex (a2);
\vertex[left = 1.25cm of a2](a1){\(d_i\)};
\vertex[above = 2cm of a2](a3);
\vertex[left = 1.25cm of a3](a4){\(d_j\)};
\vertex[right = 2cm of a2](a5);
\vertex[right = 2cm of a3](a6);
\vertex[right = 1cm of a6](a7){\(\ell_n\)};
\vertex[right = 1cm of a5](a8){\(\ell_m\)};
\diagram*{ 
(a1) --[fermion,line width=0.35mm, arrow size=0.8pt](a2),
(a2) --[fermion, line width=0.35mm,arrow size=0.8pt,style=black!50,edge label=\(\color{black}\psi\)](a3),
(a3)--[fermion, line width=0.35mm,arrow size=0.8pt](a4),
(a5) --[charged scalar,line width=0.35mm,arrow size=0.8pt,style=black!50,edge label=\(\color{black}\Phi\)](a2),
(a3) --[charged scalar,line width=0.35mm,style=black!50,arrow size=0.8pt,edge label={\(\color{black}\Phi\)}](a6),
(a7) --[fermion,line width=0.35mm, arrow size=0.8pt](a6),
(a6) --[fermion,line width=0.35mm, arrow size=0.8pt,style=black!50,edge label=\(\color{black}\chi\)](a5),
(a5) --[fermion,line width=0.35mm, arrow size=0.8pt](a8)};
\node at (a2)[circle,fill,style=gray,inner sep=1pt]{};
\node at (a3)[circle,fill,style=gray,inner sep=1pt]{};
\node at (a5)[circle,fill,style=gray,inner sep=1pt]{};
\node at (a6)[circle,fill,style=gray,inner sep=1pt]{};
\end{feynman}
\end{tikzpicture}\label{fig:Feyn_didjlmln_box}}\quad
\subfloat[]{\begin{tikzpicture}
\begin{feynman}
\vertex(a);
\vertex[above =2.0cm of a](b);
\vertex[above right=1cm and 1cm of a](c);
\vertex[right =1.5cm of c](d);
\vertex[above right=1.0cm and 1.0cm of d](d1){\(\ell_{n}\)};
\vertex[below right=1.0cm and 1.0cm of d](d2){\(\ell_{m}\)};
\vertex[left=1.0cm of b](b1){\(d_j\)};
\vertex[left=1.0cm of a](a1){\(d_i\)};
\diagram*{
(a1) -- [fermion,line width=0.35mm,arrow size=1pt,style=black] (a), (b) -- [charged scalar,line width=0.35mm,arrow size=1pt,style=black!50,edge label'=\(\color{black}\Phi\)] (a), (b) -- [fermion,line width=0.35mm,arrow size=1pt, style=black] (b1), (a) -- [fermion,line width=0.35mm,arrow size=1pt, quarter right,style=black!50,edge label'=\(\color{black}\psi\)] (c) -- [fermion,line width=0.35mm,arrow size=1pt,style=black!50,quarter right,edge label'=\(\color{black}\psi\)](b), (c) -- [boson,line width=0.35mm,style=black!50,edge label'=\(\color{black}\gamma/Z\)](d), (d1) -- [fermion,line width=0.35mm,arrow size=1pt,style=black] (d) --[fermion,line width=0.35mm,arrow size=1pt,style=black](d2)};
\end{feynman}
\node at (a)[circle,fill,style=gray,inner sep=1pt]{};
\node at (b)[circle,fill,style=gray,inner sep=1pt]{};
\node at (c)[circle,fill,style=gray,inner sep=1pt]{};
\node at (d)[circle,fill,style=black,inner sep=1pt]{};
\end{tikzpicture}\label{fig:Feyn_didjlmln_penguin}}\quad
\subfloat[]{\begin{tikzpicture}
\begin{feynman}
\vertex(a);
\vertex[above =2.0cm of a](b);
\vertex[above right=1cm and 1cm of a](c);
\vertex[right =1.5cm of c](d);
\vertex[above right=1.0cm and 1.0cm of d](d1){\(\nu_\ell\)};
\vertex[below right=1.0cm and 1.0cm of d](d2){\(\nu_\ell\)};
\vertex[left= 1.0cm of b](b1){\(d_j\)};
\vertex[left= 1.0cm of a](a1){\(d_j\)};
\diagram* {
(a1) --[fermion,line width=0.35mm,arrow size=1pt,style=black] (a),
(b) -- [charged scalar,line width=0.35mm,arrow size=1pt,style=black!50,edge label'=\(\color{black}\Phi\)] (a),
(b) -- [fermion,line width=0.35mm,arrow size=1pt,style=black] (b1),
(a) --[fermion,line width=0.35mm,arrow size=1pt,quarter right,style=black!50,edge label'=\(\color{black}\psi\)](c),
(c) --[fermion,line width=0.35mm,arrow size=1pt,style=black!50,quarter right,edge label'=\(\color{black}\psi\)](b),
(c) --[boson,line width=0.35mm,style=black!50,edge label'=\(\color{black}Z\)](d),
(d2) --[fermion,line width=0.35mm,arrow size=1pt,style=black] (d) --[fermion,line width=0.35mm,arrow size=1pt,style=black](d1)};
\end{feynman}
\node at (a)[circle,fill,style=gray,inner sep=1pt]{};
\node at (b)[circle,fill,style=gray,inner sep=1pt]{};
\node at (c)[circle,fill,style=gray,inner sep=1pt]{};
\node at (d)[circle,fill,style=black,inner sep=1pt]{};
\end{tikzpicture}\label{fig:Feyn_di2djnunubar}}
\end{tcolorbox}
\end{adjustbox}
\caption{Feynman diagrams contributing to the flavor-changing neutral-current transition $d_i\to d_j\ell_m\bar{\ell}_n$: box diagrams and electroweak penguin diagrams are shown in Figs.~\ref{fig:Feyn_didjlmln_box} and \ref{fig:Feyn_didjlmln_penguin}, respectively. Figure\,.~\ref{fig:Feyn_di2djnunubar} shows the corresponding one-loop diagrams contributing to the invisible decay process with the underlying quark-level transition $d_i\to d_j\nu_\ell\bar{\nu}_\ell$. Higgs-penguin diagrams are also generated in the model; however, their contributions are suppressed by the external fermion masses and are therefore neglected.
}
\label{fig:Feyn_di2djll}
\end{figure}
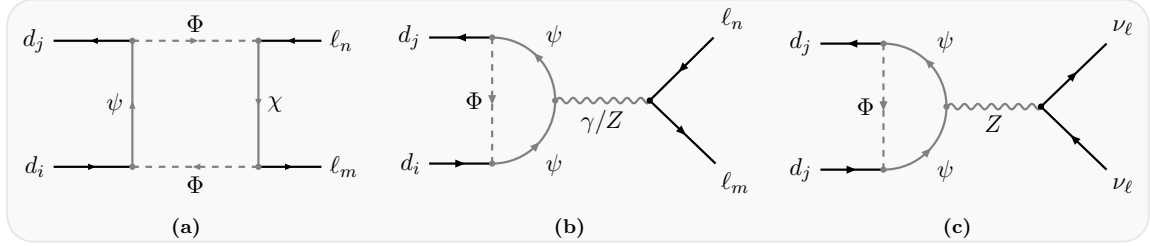

The expression for the branching ratio of the rare leptonic decays of the $B$-meson is given by:
\begin{eqnarray}
\begin{split} 
\label{eq:rare_BR_formula}
\BR(B_q \rightarrow \mu^+ \mu^-) = & \tau_{B_q} f_{B_q}^2  m_{B_q} \frac{G_F^2 \alpha^2}{64 \pi^3} |V^*_{tq}V_{tb}|^2 \beta_{\mu}(m_{B_q}^2) \left[   \frac{m_{B_q}^2}{m_b^2} |C_s - C'_s|^2 \left(1-\frac{4m_{\mu}^2}{m_{B_q}^2}\right) \right.\\& \left.  + \bigg|\frac{m_{B_q}}{m_b}(C_p - C'_p) + 2\frac{m_{\mu}}{m_{B_q}} (C_{10} - C'_{10})\bigg|^2 \right]\,,
\end{split}
\end{eqnarray}
where $f_{B_q}$ is the decay constant of the meson $B_{q}$. The branching ratios of $K_{L, S}^0 \to \ell^+\ell^-$ can be found from \cite{Kolay:2024wns}. In the model under our consideration, the rare dileptonic decays receive additional one-loop contributions mediated by a vector-like quark and lepton. The underlying flavor-changing neutral-current transitions are $b\to s\ell^{+}\ell^{-}$ and $s\to d\ell^{+}\ell^{-}$, corresponding to the decays of $B_s^0$, $B^0$, and $K_{L,S}^0$, respectively. The relevant one-loop Feynman diagrams are shown in \Fig\ref{fig:Feyn_di2djll}, consisting of box, photon-penguin, and $Z$-penguin topologies. The box contribution to the Wilson coefficients is proportional to the coupling combination $\mathtt{y}_{d_i}\mathtt{y}_{d_j}\mathtt{y}_{\ell}^{\,2}$, while the penguin diagrams generate additional contributions through virtual $\gamma$ and $Z$ exchange. As in the SM, these flavor-changing neutral-current transitions arise only at the loop level through box and electroweak penguin diagrams. A Higgs-penguin diagram is also generated in our model; however, its contribution is suppressed by the external fermion masses and is therefore neglected. The remaining diagrams correspond to external-leg (wave-function) renormalization. Altogether, the model generates contributions to the Wilson coefficients $C_{7}^{(\prime)}$, $C_{8}^{(\prime)}$, and $C_{9,10}^{(\prime)}$, whose explicit expressions are collected in \Eq\eqref{eq:wilson_expression} of \autoref{app:loops}. Finally, the branching ratio of $K_S^0\to\ell^{+}\ell^{-}$ depends on the imaginary part of the relevant coupling combination~\cite{Mandal:2019gff}. Since our analysis is restricted to the CP-conserving limit with real couplings, no new contribution arises to this decay.

\begin{figure}[htb!]
\centering
\subfloat[]{\includegraphics[width=0.475\linewidth]{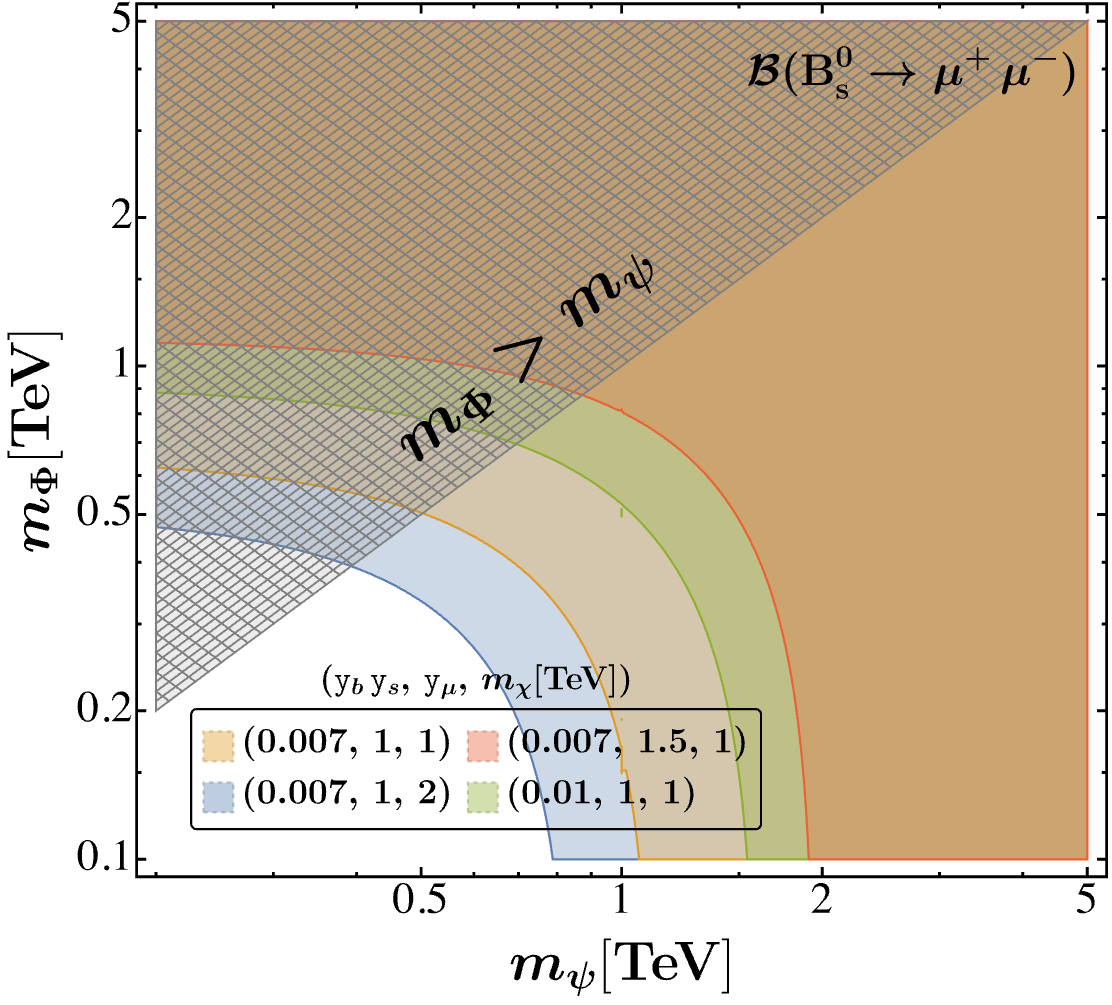}\label{fig:rare_bs0}}\quad
\subfloat[]{\includegraphics[width=0.475\linewidth]{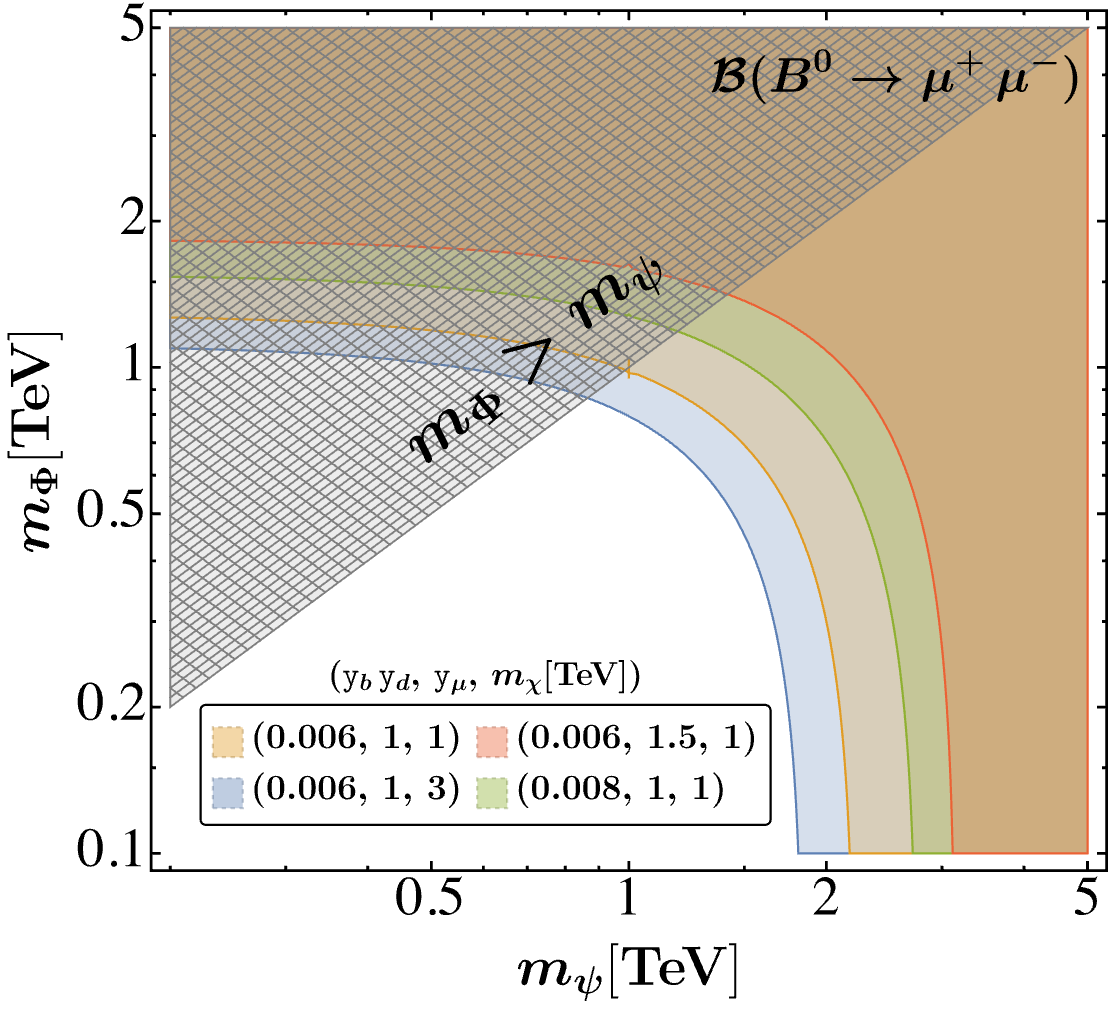}\label{fig:rare_bd0}}\\
\subfloat[]{\includegraphics[width=0.475\linewidth]{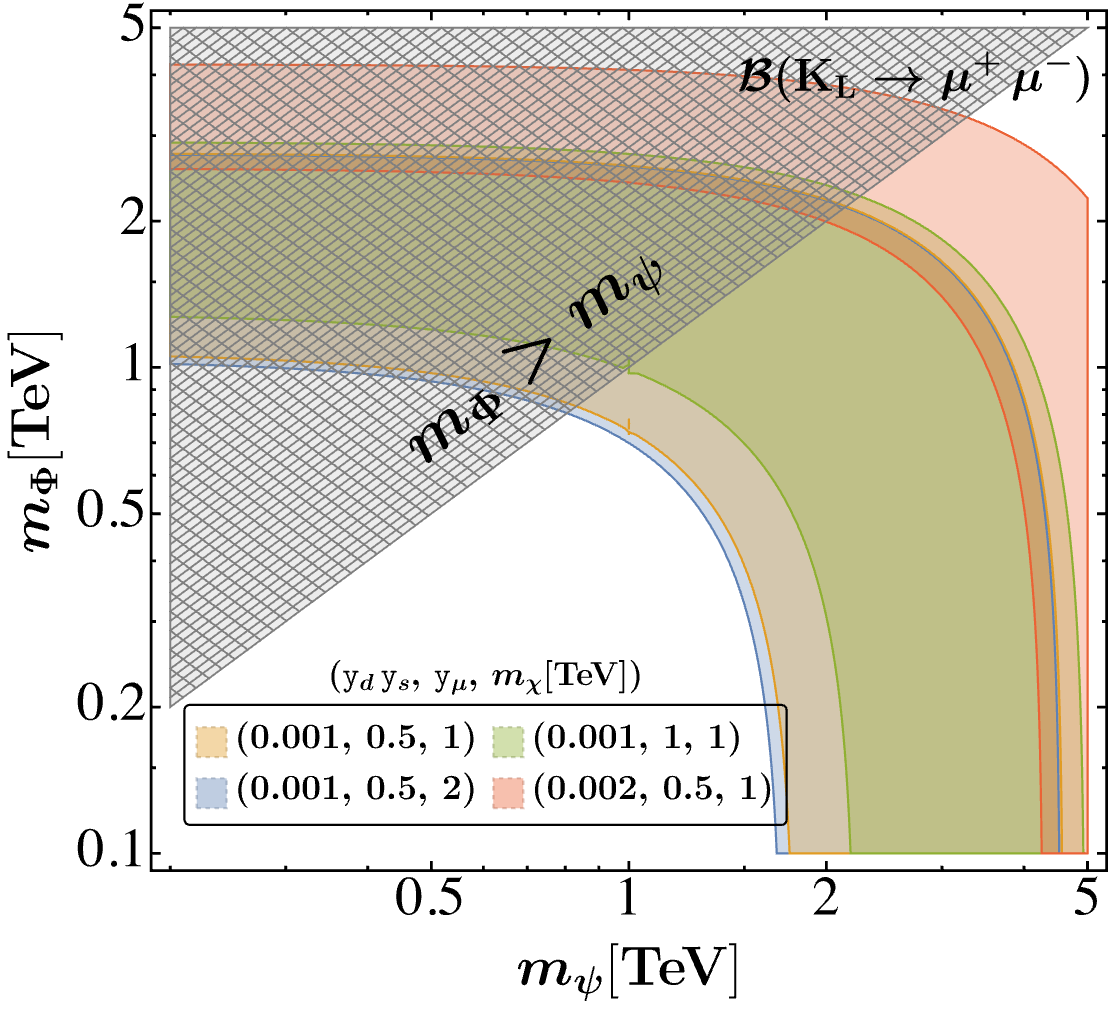}\label{fig:rare_KL}}\quad
\subfloat[]{\includegraphics[width=0.475\linewidth]{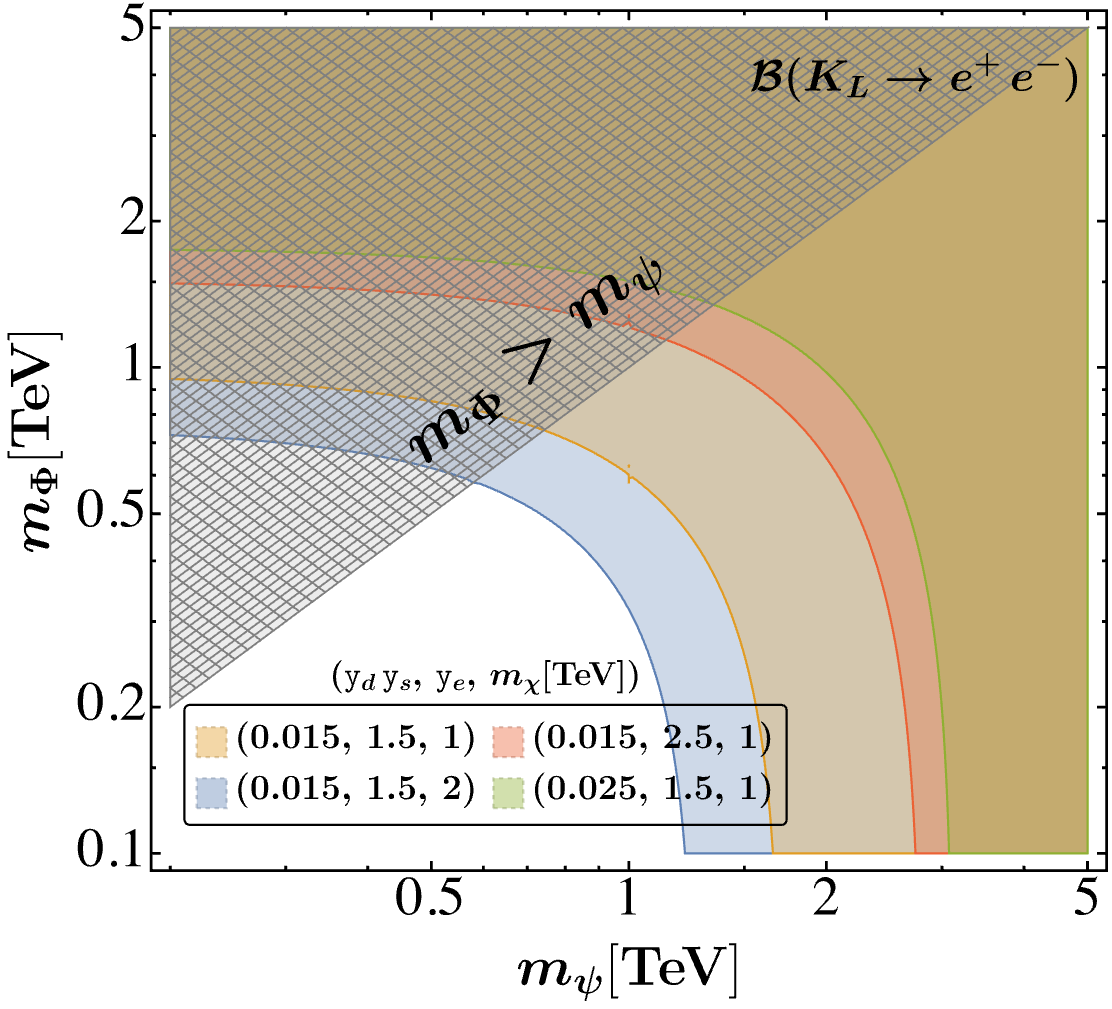}\label{fig:rare_KL_ee}}
\caption{Allowed regions of the $(\mpsi\,,~\mphi)$ parameter space at the $1\sigma$ confidence level from the rare leptonic decays: (a) $B^{0}\to\mu^{+}\mu^{-}$, (b) $B_{s}^{0}\to\mu^{+}\mu^{-}$, (c) $K_{L}^{0}\to\mu^{+}\mu^{-}$, and (d) $K_{L}^{0}\to e^{+}e^{-}$. The colored regions denote the parameter space consistent with the measured branching ratios for the benchmark points specified in the corresponding legends. The hatched grey region, corresponding to $\mphi>\mpsi$, is excluded by the requirement that the complex scalar $\Phi$ remains the stable dark matter candidate.}
\label{fig:rare_allowed_individual}
\end{figure}
\paragraph{\underline{Allowed Parameter Space from Rare Decays :}}
Figure\,.~\ref{fig:rare_allowed_individual} shows the regions of the $(\mpsi\,,~\mphi)$ parameter space allowed by the rare leptonic decays of $B$ and $K$ mesons at the $1\sigma$ confidence level. The colored regions correspond to the parameter space consistent with the measured branching ratios. Since the experimental measurements of the rare $B$-meson decays are in good agreement with the SM predictions at the $1\sigma$ level, these observables provide only lower bounds on the new-particle masses, leaving the higher-mass region allowed.
For the decay $B^{0}\to\mu^{+}\mu^{-}$, measurements are available from both the CMS and LHCb collaborations~\cite{CMS:2022mgd,LHCb:2021vsc}. As the CMS measurement provides the more stringent constraint, we use it to derive the allowed parameter space. The branching ratio receives contributions from both the box and electroweak penguin diagrams. While the penguin contribution depends only on the VLQ sector, the box contribution also involves the VLL mass and the lepton Yukawa coupling through the coupling combination $\mathtt{y}_{d_i}\mathtt{y}_{d_j}\mathtt{y}_{\ell}^{2}$. Consequently, the effects of $\mchi$ and $\ymu$ become significant only for sufficiently large Yukawa couplings. Increasing $\mchi$ suppresses the box contribution, thereby allowing lower values of $\mphi$ and $\mpsi$. Conversely, for fixed masses, a larger value of $\ymu$ enhances the new-physics contribution to the branching ratio, leading to a more restrictive allowed parameter space.

In contrast, the decay $K_{L}^{0}\to\mu^{+}\mu^{-}$ constrains the parameter space through both the upper and lower limits implied by the measured branching ratio. This decay is dominated by long-distance contributions, while the short-distance SM contribution is comparatively small with respect to the experimental measurement. We therefore determine the allowed parameter space by requiring the new-physics contribution to vary only within the experimental $1\sigma$ uncertainty. The colored regions in \Fig\ref{fig:rare_allowed_individual} correspond to four representative benchmark choices of the Yukawa couplings. The figure shows that coupling combinations of order $\mathcal{O}(10^{-3})$ allow CSDM and VLQ masses in the $(1\!-\!2)$~TeV range.
For the decay $K_{L}^{0}\to e^{+}e^{-}$, the branching ratio is almost entirely dominated by long-distance contributions and is in good agreement with the experimental measurement. Consequently, we allow the new-physics contribution to vary only within the experimental uncertainty. Since the resulting new-physics contribution is consistent with zero, this observable provides only a lower bound on the parameter space, leaving the higher-mass region allowed, similar to the rare $B$-meson decays. For both kaon decay modes, the sensitivity to the lepton Yukawa coupling $\yl$ is more pronounced than in the corresponding rare $B$-meson decays. The hatched grey region, corresponding to $\mphi>\mpsi$, is also shown, as in the previous section, and is excluded by the requirement that the complex scalar $\Phi$ remains the stable dark matter candidate.
\subsubsection{Rare Semileptonic Decays of Mesons to Charged Leptons}
In the previous section, we discussed the impact of our model parameters on rare leptonic decays of mesons with the underlying quark-level transition $d_{i} \to d_{j} \ell \ell$. However, several other observables share the same quark-level structure and can provide additional constraints on the NP parameter space. In this section, we focus on such observables. We consider three different quark-level transitions, which are presented below. 

\paragraph{\underline{Processes with $b \to s \ell \ell$ transition} :}The relevant meson-level processes for this quark transition include $B \to K^{(*)} \ell \ell$ and $B \to \phi \ell \ell$. Numerous experimental measurements are available in this sector, enabling comprehensive analysis. The key observables include branching ratios, angular observables, asymmetry observables, and lepton flavor universality violating (LFUV) ratios such as $R_{K^{(*)}}$. Most of these observables are consistent with the SM at the $1\sigma$ level and therefore place strong constraints on any NP contributions. However, a few observables, such as $P_{5}^{\prime}$, exhibit deviations exceeding $3\sigma$, making them particularly important probes of NP~\cite{LHCb:2020lmf}.
There is extensive literature on model-independent constraints on the Wilson coefficients obtained from global fits to all available data in the $b \to s \ell \ell$ sector \cite{Alguero:2023jeh, Alguero:2021anc, Wen:2023pfq, Biswas:2020uaq, Ciuchini:2022wbq, Hurth:2023jwr, London:2021lfn, Gubernari:2022hxn, Alguero:2022wkd}. These fits provide bounds on the relevant operator structures, which can then be translated into constraints on the parameter space of a given NP model through matching at the appropriate scale \cite{Alguero:2023jeh, Alguero:2021anc, Wen:2023pfq}. This approach provides a systematic, model-independent way to test the viability of NP scenarios. In our scenario, new-physics contributions are generated to the Wilson coefficients $(C_{7,8,9,10}^{(\prime)})$. However, the contributions to the dipole coefficients $(C_{7,8}^{(\prime)})$ are found to be negligible, and hence only the four semileptonic Wilson coefficients $(C_{9,10}^{(\prime)})$ are phenomenologically relevant. Accordingly, we employ the model-independent constraints on these four-operator scenarios obtained in Ref\,.~\cite{Alguero:2023jeh} for the case of muons in the final state.

\begin{figure}[htb!]
\centering
\subfloat[]{\includegraphics[width=0.475\linewidth]{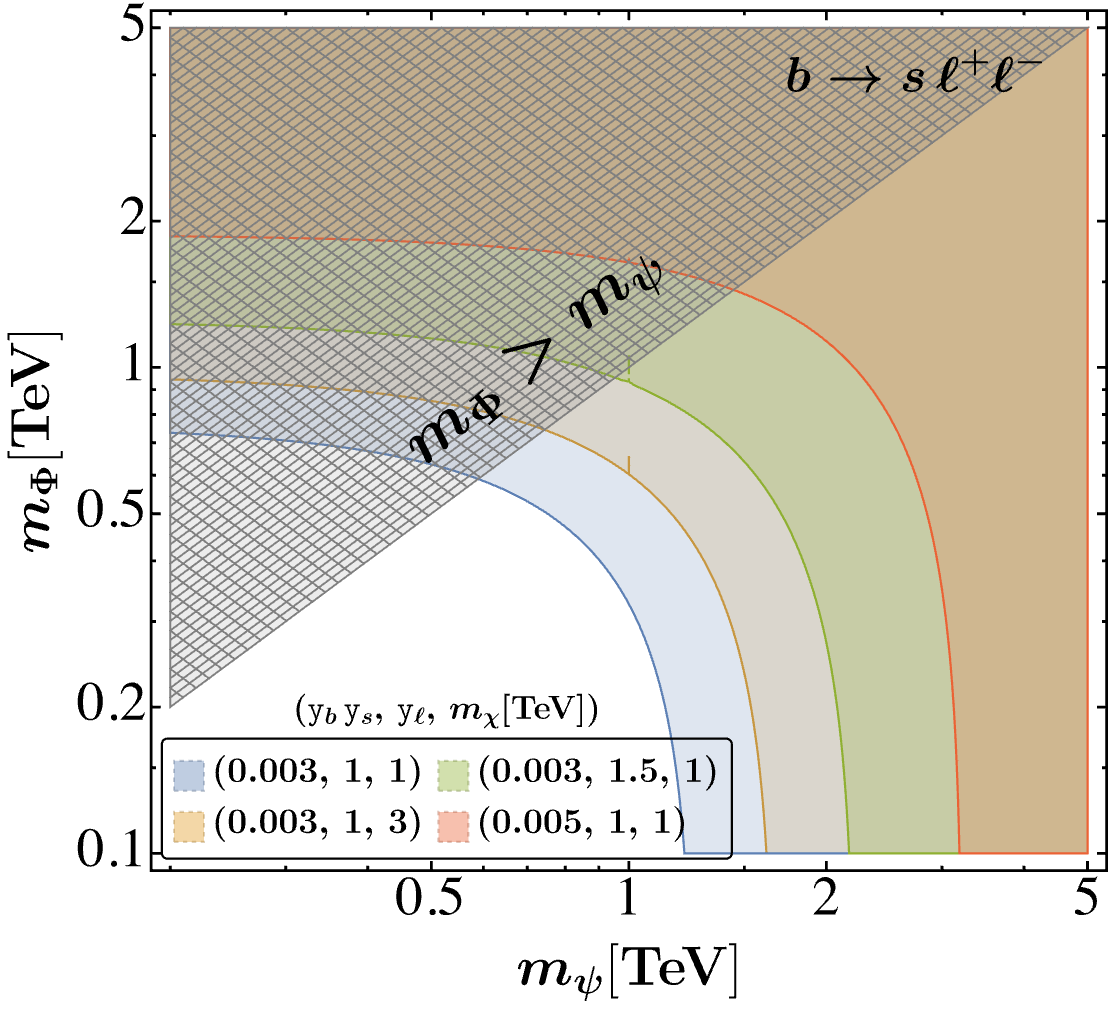}\label{fig:global_b2sll}}~~
\subfloat[]{\includegraphics[width=0.475\linewidth]{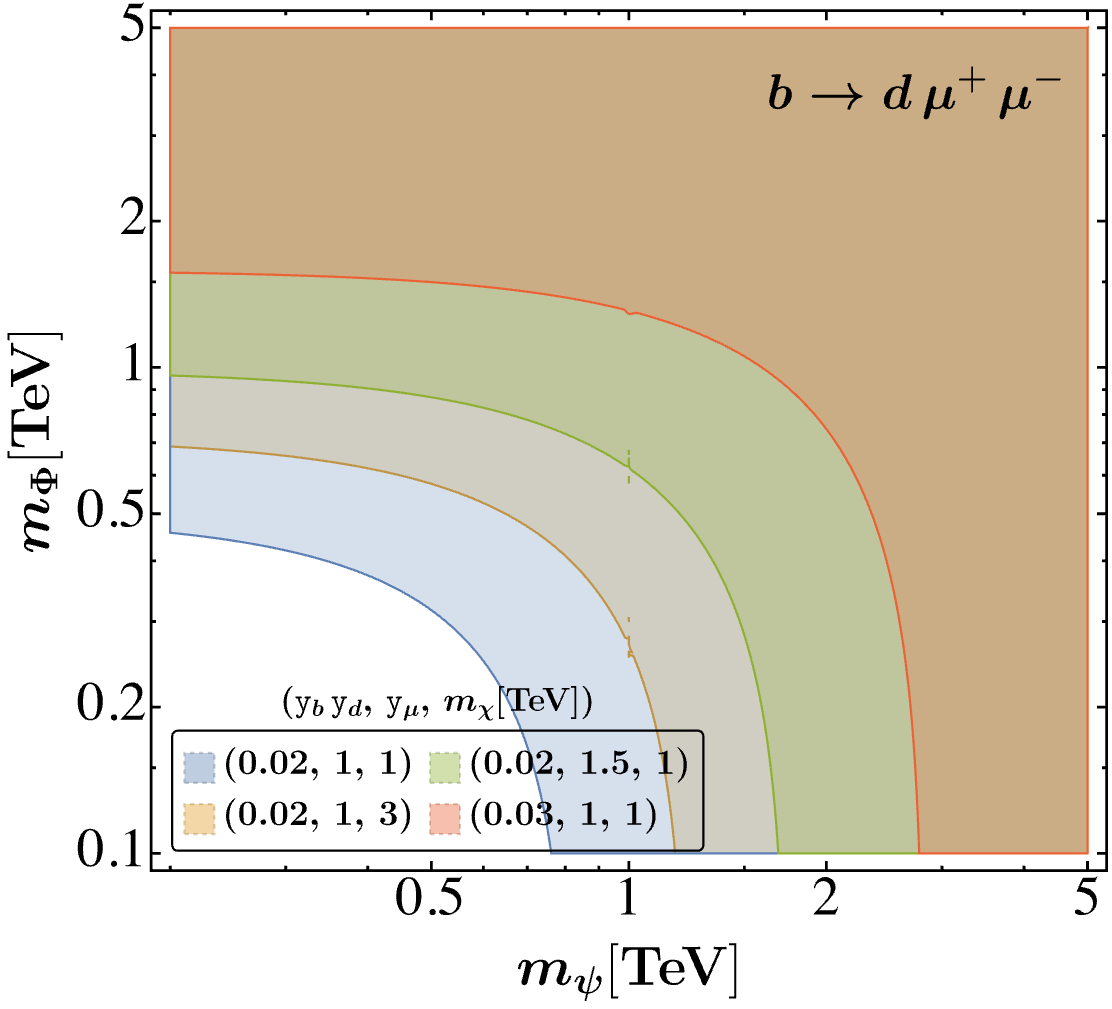}\label{fig:global_b2dll}}
\caption{Allowed regions of the $(\mpsi\,,~\mphi)$ parameter space at the $1\sigma$ confidence level obtained from constraints on the Wilson coefficients \cite{Alguero:2023jeh} governing the $b\to s\ell^+\ell^-$ (left) and $b\to d\mu^+\mu^-$ (right) transitions. The analysis incorporates all relevant observables within their corresponding $1\sigma$ experimental uncertainties.}
\end{figure}

Figure\,.~\ref{fig:global_b2sll} shows the allowed parameter space in the DM and VLQ mass plane, $(\mpsi\,,~\mphi)$, obtained using the $\pm 1\sigma$ allowed interval for the Wilson coefficients. The plots are generated for a few benchmark coupling values, chosen to cover the TeV mass range for both particles. Each coloured region represents the parameter space that simultaneously satisfies the $1\sigma$ constraints from the vector Wilson coefficients, $C_{9}^{(\prime)}$ and $C_{10}^{(\prime)}$, for the corresponding benchmark couplings. Among these, the strongest constraint comes from $C_{10}^{\prime}$. The dependence of the allowed regions on the VLL mass and the coupling $\yl$ is similar to that observed in the $B_s^0 \to \mu^+\mu^-$ process. However, the couplings are required to be smaller in order to satisfy the allowed ranges of the Wilson coefficients. As discussed in Ref\,.~\cite{Alguero:2023jeh}, the fit results depend strongly on the chosen fitting scenario, such as the number of Wilson coefficients included in the fit and whether lepton-flavor universality is assumed or not. For this reason, although we only show the $1\sigma$ allowed regions, the $2\sigma$ results are also important. At the $2\sigma$ level, the enlarged parameter space becomes allowed; hence, we do not show those regions separately.

\paragraph{\underline{Processes with $b \to d \, \ell \ell$:}} For the $b \to d \ell \ell$ transition, the relevant semi-leptonic decay modes are $B \to \pi \ell \ell$ and $B \to \rho\,\ell \ell$, with the branching ratios and angular observables providing the most important constraints. A detailed analysis of these observables can be found in Ref\,.~\cite{Biswas:2022lhu}. As in the $b \to s \ell \ell$ case, model-independent constraints on the Wilson coefficients have been obtained from global analyses of the available experimental data~\cite{Bause:2022rrs}. We use the $1\sigma$ ranges of these fit results to constrain the parameter space of our NP model.

Figure\,.~\ref{fig:global_b2dll} shows the allowed parameter space in the DM and VLQ mass plane obtained from the global-fit constraints on the $b \to d \ell \ell$ Wilson coefficients. We consider the four-parameter fit involving $C_{9,10}^{(\prime)}$, with the fit results taken from Ref\,.~\cite{Bause:2022rrs}. Each coloured region represents the parameter space that simultaneously satisfies the $1\sigma$ bounds on all the relevant Wilson coefficients for the corresponding benchmark couplings. Among these, the strongest constraint comes from $C_{10}$. Compared to the $b \to s \ell \ell$ transition, the available experimental data for $b \to d \ell \ell$ are much more limited. Consequently, the uncertainties in the fitted Wilson coefficients are larger, resulting in weaker constraints on our model and hence a larger allowed parameter space. We only show the $1\sigma$ regions, since extending the analysis to the $2\sigma$ ranges would allow almost the entire parameter space.

\paragraph{\underline{Processes with $s \to d \, \ell \ell$:}}
Unlike the $b\to s(d)\ell^+\ell^-$ transitions, no global fit to the Wilson coefficients is currently available for the $s\to d\ell^+\ell^-$ sector. Instead, the relevant constraints arise from rare kaon decays such as $K_{L,S}^{(+)}\to\pi^{(0,+)}\ell^+\ell^-$. Since our analysis assumes all new-physics couplings to be real, the CP-violating decay $K_L^0\to\pi^0\ell^+\ell^-$ does not receive any new-physics contribution. In contrast, the CP-conserving modes $K_{S}^{(+)}\to\pi^{0(+)}\ell^+\ell^-$ can receive non-zero contributions in the present model~\cite{Mandal:2019gff}. We have explicitly verified that the parameter space compatible with the constraints from $K_L^0\to\ell^+\ell^-$ remains consistent with the current bounds from these semileptonic kaon decays. Since the latter do not impose any additional significant constraints on the parameter space, we do not present the corresponding plots. Further discussions of these decay modes can be found in Refs\,.~\cite{Mescia:2006jd, DAmbrosio:2022kvb, DAmbrosio:2024ncc, Chen:2003nz, Gao:2003wy}.
\subsubsection{Rare Semileptonic Decays of Meson to Neutral Leptons}
Invisible decay modes provide important probes of new physics due to their clean theoretical structure and their loop-suppressed SM contributions. In SM, the invisible decays are the semileptonic decays of the mesons with a neutrino pair in the final state, i.e., decays like $ P \to M \nu \bar{\nu}$, with $P$ standing for a pseudoscalar meson and $M$ standing for the final pseudoscalar/vector meson. Recent data by the Belle-II collaboration on the branching fraction of $B^{+} \to K^{+} \nu \bar{\nu}$ decay shows a $2.7\sigma$ deviation from the SM prediction \cite{Belle-II:2023esi}. These channels are theoretically very clean to study any possible new physics effects in these processes. The corresponding SM predictions and the experimental measurements are listed in \autoref{tab:rareinputs_invisible}.  

In this NP scenario, we get contributions to the invisible decays via the Feynman diagrams of \fig\ref{fig:Feyn_di2djnunubar} via a penguin diagram, with $Z$ mediation. Since the neutrinos are left-handed massless particles in the SM, we only get contributions to the vector operators. To study the invisible decay, we use a similar operator basis as $b \to s \ell \ell$. The details of this operator basis can be found from \cite{Buras:2024ewl}. In our scenario, we get modifications to both left-handed and right-handed vector currents. In our model, we get the loop contribution as:
\begin{align}
\mathcal{L}_{\rm loop}^{d_{i} \to d_{j} \nu \bar{\nu}} = C_{L}^{\nu} (\bar{d}_{j} \gamma_{\mu} \mathrm{P}_{\mathtt{L}} d_{i})(\bar{\nu} \gamma^{\mu} \mathrm{P}_{\mathtt{L}} \nu) + C_{R}^{\nu} (\bar{d}_{j} \gamma_{\mu} \mathrm{P}_{\mathtt{R}} d_{i})(\bar{\nu} \gamma^{\mu} \mathrm{P}_{\mathtt{L}} \nu) \,. 
\end{align}
The Wilson coefficients $C_{L,R}^{\nu}$ corresponding to the $d_{i} \to d_{j} \nu \bar{\nu}$ transition can be expressed in NP scenarios as
\begin{align}
C_{L,R}^{\nu} = C_{L,R}^{\nu, \rm SM} + C_{L,R}^{\nu, \rm NP} \,.
\end{align}
The value of the Wilson coefficient in the SM is given by \cite{Chen:2024jlj, Buras:2014fpa}: 
\begin{equation}
C_{L}^{\nu, \rm SM} = - 6.32 \pm 0.07, \ \ \ \ C_{R}^{\nu, \rm SM} =0\,.
\end{equation} 
Since neutrinos do not couple to photons, the contribution to this decay in our case arises only from the $Z$-mediated penguin diagram. The dominant contribution to the loop amplitude enters the right-handed current, $C_{R}^{\nu,\rm SM}$, while the left-handed contribution is suppressed by the external particle mass. The corresponding loop contribution is given in \autoref{app:loops}. Since the same couplings contribute to both this decay and the $b \to d_{j}\ell\bar\ell$ processes, the resulting constraints on the mass-coupling parameter space are not independent. Moreover, the constraints obtained from this decay are weaker than those from $b \to d_{j}\ell\ell$. Therefore, we do not show the corresponding results explicitly.
\begin{table}[htb!]
\begin{center}
\rowcolors{1}{Emerald!15}{SpringGreen!10}
\renewcommand{\arraystretch}{1.3}
\begin{tabular}{|c|c|c|}
\hline
Branching Ratio & SM Value & Experimental Value \\
\hline
\hline
$ B^+ \to K^+ \nu\bar{\nu} $ & $ \left(5.06 \pm 0.14 \pm 0.28\right) \times 10^{-6} $ \cite{Becirevic:2023aov} & $ \left(2.3 \pm 0.5^{+0.5}_{-0.4}\right) \times 10^{-5} $ \cite{Belle-II:2023esi} \\
$ B_0 \to K_{S}^0 \nu\bar{\nu} $ & $ \left( 2.05 \pm 0.07 \pm 0.12 \right) \times 10^{-6} $ \cite{Becirevic:2023aov} & $ < 1.3 \times 10^{-5} $ \cite{Belle:2017oht}\\
$ B^+ \to K^{*+} \nu \bar{\nu} $ & $ \left( 10.86 \pm 1.30 \pm 0.59 \right) \times 10^{-6} $ \cite{Becirevic:2023aov} & $< 6.1 \times 10^{-5} $ \cite{Belle:2017oht} \\\rowcolor{SpringGreen!10}	
&  & $ \left( 3.8^{+2.9}_{-2.6} \right) \times 10^{-5} $  \cite{BaBar:2013npw} \\
\multirow{-2}{*}{$ B_0 \to K^{*0} \nu \bar{\nu} $} &  \multirow{-2}{*}{$ \left( 9.05 \pm  1.25 \pm  0.55 \right) \times 10^{-6} $ \cite{Becirevic:2023aov}}  &   $ < 1.8 \times 10^{-5} $ \cite{Belle:2017oht} \\
$ B^+ \to \pi^+ \nu\bar{\nu} $ & $ \left(1.57 \pm 0.44 \right) \times 10^{-7} $ \cite{Straub:2018kue} & $ < 1.4  \times 10^{-5} $ \cite{Belle:2017oht} \\
$ B_0 \to \pi^{0} \nu\bar{\nu} $ & $ (0.73\pm 0.21) \times 10^{-7} $ \cite{Straub:2018kue} & $ < 0.9 \times 10^{-5}  $ \cite{Belle:2017oht} \\
$ B^+ \to \rho^{+} \nu\bar{\nu} $ & $ (3.92\pm 0.79) \times 10^{-7} $ \cite{Straub:2018kue} & $ < 3 \times 10^{-5} $ \cite{Belle:2017oht} \\
$ B_0 \to \rho^0 \nu\bar{\nu} $ & $ \left(1.82 \pm 0.31 \right) \times 10^{-7} $ \cite{Straub:2018kue} & $ < 4 \times 10^{-5} $ \cite{Belle:2017oht}\\
$ K^+ \rightarrow \pi^+ \nu\bar{\nu}  $ & $ (8.60 \pm 0.42)\times 10^{-11} $ \cite{Buras:2024ewl} & $ \left(10.6^{+4.0}_{-3.4} \pm 0.9 \right) \times 10^{-11} $ \cite{NA62:2021zjw,NA62:2024pjp} \\
$ K_L^0 \to \pi^0 \nu\bar{\nu} $ & $ \left(2.94 \pm 0.15 \right) \times 10^{-11} $ \cite{Buras:2024ewl} & $ <2.2 \times 10^{-9} $ \cite{KOTO:2024zbl} \\\hline
\end{tabular}
\end{center}
\caption{The SM predictions and the experimentally measured values or upper limit on the branching ratios of the semileptonic decays of $B$ and $K$ mesons to neutral leptons.}\label{tab:rareinputs_invisible}
\end{table}

\subsection{Lepton Observables}
The Yukawa interactions among the VLL, the DM particle, and the SM charged leptons give rise to rich phenomenology in the charged-lepton sector. While the previous sections focused on rare semileptonic meson decays with lepton-flavor-conserving final states, the same interactions also induce one-loop contributions to the anomalous magnetic moments of charged leptons and charged LFV processes. The latter include radiative decays such as $\ell_{\alpha}\to\ell_{\beta}\gamma$, three-body decays $\ell_{\alpha}\to\ell_{\beta}\ell_{\delta}\bar{\ell}_{\rho}$, $\mu\to e$ conversion in nuclei, and muonium-antimuonium oscillation, as well as LFV meson decays such as $P\to M\ell_{1}\bar{\ell}_{2}$ and $P\to\ell_{1}\bar{\ell}_{2}$. Since charged LFV processes are absent in the SM and remain extremely suppressed, any observable signal would constitute a clear indication of physics beyond the SM. The absence of any experimental evidence for such processes therefore translates into stringent upper limits, providing powerful constraints on the model's parameter space. In this section, we derive the corresponding constraints from the anomalous magnetic moments and charged LFV observables.
\subsubsection{Anomalous Magnetic Moments}
\begin{figure}[htb!]
\begin{adjustbox}{width=1\textwidth}
\begin{tcolorbox}[colback=gray!5, colframe=black!10, boxrule=1pt, arc=4mm, boxsep=0pt, left=0pt, right=0pt, top=0pt, bottom=0pt, width=1.625\textwidth, halign=center]
\subfloat[]{\begin{tikzpicture}
\begin{feynman}
\vertex (a){\(\ell_{\alpha}\)};
\vertex[right=1.5cm of a](b);
\vertex[right=3.5cm of a](c);
\vertex[right=5.0cm of a](d){\(\ell_{\beta}\)};
\vertex[above right=1.0cm and 2.5cm of a](e);
\vertex[above =2.5cm of d](f){\(\gamma\)};
\vertex[below =1.0cm of d](g);
\diagram*{
(a) --[fermion,line width=0.35mm, arrow size=0.8pt] (b) --[fermion, line width=0.35mm, quarter left, arrow size=0.8pt,style=black!50,edge label=\(\color{black}\chi\)] (e) --[fermion, line width=0.35mm, quarter left, arrow size=0.8pt,style=black!50,edge label=\(\color{black}\chi\)] (c) --[fermion,line width=0.35mm, arrow size=0.8pt](d),
(c) --[charged scalar,half left,line width=0.35mm,style=black!50,arrow size=0.8pt,edge label={\(\color{black}\Phi\)}](b),
(e) --[boson,line width=0.35mm, arrow size=0.8pt](f),
(d) --[boson,line width=0.35mm, style=gray!5, arrow size=0.8pt](g)};
\node at (b)[circle,fill,style=gray,inner sep=1pt]{};
\node at (c)[circle,fill,style=gray,inner sep=1pt]{};
\node at (e)[circle,fill,style=gray,inner sep=1pt]{};
\end{feynman}
\end{tikzpicture}\label{feyn:mutoe}}\,
\subfloat[]{\begin{tikzpicture}
\begin{feynman}
\vertex (a){\(\ell_\alpha\)};
\vertex[right=1.5cm of a](b);
\vertex[above right=1.25cm and 1.25cm of b](b1);
\vertex[below right=1.25cm and 1.25cm of b](b3);
\vertex[above right=1.5cm and 3.5cm of b](b2){\(\ell_\beta\)};
\vertex[below right=1.5cm and 3.5cm of b](b4){\(\ell_\rho\)};
\vertex[right=2.5cm of b](b5);
\vertex[right=1.0cm of b5](b6){\(\ell_\delta\)};
\diagram* { 
(a) --[fermion,line width=0.35mm, arrow size=0.8pt]
(b) --[fermion, line width=0.35mm,arrow size=0.8pt,style=black!50,edge label=\(\color{black}\chi\)](b1) --[fermion, line width=0.35mm,arrow size=0.8pt](b2),
(b1) --[charged scalar,line width=0.35mm,arrow size=0.8pt,style=black!50,edge label=\(\color{black}\Phi\)](b5),
(b3) --[charged scalar,line width=0.35mm,arrow size=0.8pt,style=black!50,edge label=\(\color{black}\Phi\)](b),
(b6) --[fermion,line width=0.35mm, arrow size=0.8pt]
(b5) --[fermion,line width=0.35mm, arrow size=0.8pt,style=black!50,edge label=\(\color{black}\chi\)](b3) --[fermion,line width=0.35mm, arrow size=0.8pt](b4)};
\node at (b)[circle,fill,style=gray,inner sep=1pt]{};
\node at (b1)[circle,fill,style=gray,inner sep=1pt]{};
\node at (b5)[circle,fill,style=gray,inner sep=1pt]{};
\node at (b3)[circle,fill,style=gray,inner sep=1pt]{};
\end{feynman}
\end{tikzpicture}\label{feyn:muto3ea}}\,
\subfloat[]{\begin{tikzpicture}
\begin{feynman}
\vertex (a){\(\ell_\alpha\)};
\vertex[right=1.5cm of a](b);
\vertex[above right=1.25cm and 1.25cm of b](b1);
\vertex[below right=1.25cm and 1.25cm of b](b3);
\vertex[right=2.5cm of b](b5);
\vertex[above right=1.5cm and 3.5cm of b](b2){\(\ell_\beta\)};
\vertex[below right=1.5cm and 3.5cm of b](b4){\(\ell_\rho\)};
\vertex[right=1.0cm of b5](b6){\(\ell_\delta\)};
\diagram* { 
(a) --[fermion,line width=0.35mm, arrow size=0.8pt] (b),
(b1) --[charged scalar, line width=0.35mm,arrow size=0.8pt,style=black!50,edge label'=\(\color{black}\Phi\)] (b),
(b1)--[fermion, line width=0.35mm,arrow size=0.8pt] (b2),
(b5) --[fermion,line width=0.35mm,arrow size=0.8pt,style=black!50,edge label'=\(\color{black}\chi\)] (b1),
(b) --[fermion,line width=0.35mm,arrow size=0.8pt,style=black!50,edge label'=\(\color{black}\chi\)] (b3),
(b6) --[fermion,line width=0.35mm, arrow size=0.8pt] (b5),
(b3)--[charged scalar,line width=0.35mm, arrow size=0.8pt,style=black!50,edge label'=\(\color{black}\Phi\)] (b5),
(b3)--[fermion,line width=0.35mm, arrow size=0.8pt](b4)};
\node at (b)[circle,fill,style=gray,inner sep=1pt]{};
\node at (b1)[circle,fill,style=gray,inner sep=1pt]{};
\node at (b5)[circle,fill,style=gray,inner sep=1pt]{};
\node at (b3)[circle,fill,style=gray,inner sep=1pt]{};
\end{feynman}
\end{tikzpicture}\label{feyn:muto3eb}}\,
\subfloat[]{\begin{tikzpicture}
\begin{feynman}
\vertex (a){\(\ell_{\alpha}\)};
\vertex [right=1.5cm of a](b);
\vertex[above right=0.75cm and 1.5cm of b](b1);
\vertex[below right=0.75cm and 1.5cm of b](c0);
\vertex[right=1.5cm of c0](c1);
\vertex[above right=0.75cm and 0.75cm of c1](c2){\(\ell_{\delta}\)};
\vertex[below right=0.75cm and 0.75cm of c1](c3){\(\ell_{\delta}\)};
\vertex[above right=0.75cm and 2.25cm of b1](b2){\(\ell_{\beta}\)};
\diagram* {
(a) --[fermion,line width=0.35mm,arrow size=1pt,style=black](b),
(b) --[fermion,quarter right,line width=0.35mm,arrow size=1pt,style=black!50,edge label'=\(\color{black}\chi\)](c0),
(c0) --[fermion,line width=0.35mm,arrow size=1pt,style=black!50,edge label'=\(\color{black}\chi\)](b1),
(b1) --[fermion,line width=0.35mm,arrow size=1pt,style=black](b2),
(b1) --[charged scalar,quarter right,line width=0.35mm,arrow size=1pt,style=black!50,edge label'=\(\color{black}\Phi\)](b),
(c0) --[boson,line width=0.35mm,style=black!50,edge label'=\(\color{black}\gamma/Z\)](c1),
(c2) --[fermion,line width=0.35mm,arrow size=1pt,style=black](c1),
(c1) --[fermion,line width=0.35mm,arrow size=1pt,style=black](c3)};
\end{feynman}
\node at (b)[circle,fill,style=gray,inner sep=1pt]{};
\node at (b1)[circle,fill,style=gray,inner sep=1pt]{};
\node at (c0)[circle,fill,style=gray,inner sep=1pt]{};
\node at (c1)[circle,fill,style=black,inner sep=1pt]{};
\end{tikzpicture}\label{feyn:muto3ec}}
\end{tcolorbox}
\end{adjustbox}
\caption{One-loop Feynman diagrams contributing to the charged-lepton flavor-violating decays $\ell_\alpha\to\ell_\beta\gamma$ (\Fig\ref{feyn:mutoe}) and $\ell_\alpha\to\ell_\beta\bar{\ell}_\delta\ell_\rho$ (Figs.~\ref{feyn:muto3ea}, \ref{feyn:muto3eb}, and \ref{feyn:muto3ec}), mediated by the VLL and dark matter through penguin and box topologies.}
\label{feyn:LFV}
\end{figure}
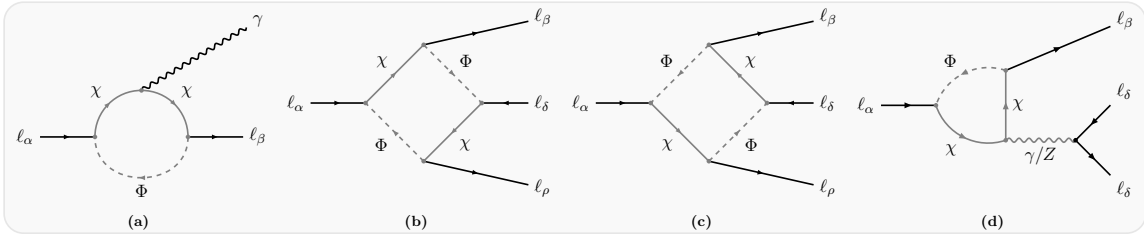

The anomalous magnetic moments (AMMs) of charged leptons provide some of the most precise tests of the SM and constitute sensitive probes of physics beyond the SM. In the present scenario, the Yukawa interactions involving the VLL, the dark matter particle, and the SM charged leptons generate one-loop contributions to the flavor-diagonal dipole operator, thereby modifying the lepton AMMs. The corresponding one-loop diagram is shown in \Fig\ref{feyn:mutoe}. Throughout this work, all couplings appearing in \eq\eqref{eq:model} are taken to be real. Consequently, the model does not induce electric dipole moments (EDMs), since non-zero EDMs require CP-violating interactions. We therefore restrict our analysis to the anomalous magnetic moments. The current experimental measurements and SM predictions for the charged-lepton AMMs are summarized in \autoref{tab:gminus2}.

The interpretation of the electron and muon anomalous magnetic moments depends on the corresponding SM predictions. In particular, different determinations of the fine-structure constant, $\alpha$, lead to either positive or negative values of the electron anomaly, $\Delta a_e$. Since the new-physics contribution in our model satisfies $\Delta a_e^{\rm NP}\geq 0$, we adopt the positive determination, $\Delta a_e=(3.38\pm1.61)\times10^{-13}$, in our numerical analysis. We do not, however, impose the electron AMM as a mandatory constraint, so parameter regions predicting a negligible new-physics contribution are also retained, even though they do not explain the observed central value. For the muon, the inferred new-physics contribution, $\Delta a_\mu$, is compatible with zero within the current experimental and theoretical uncertainties. Consequently, the muon AMM serves only to place an upper bound on the allowed new-physics contribution. In contrast, no direct measurement of the tau anomalous magnetic moment is available because of its extremely short lifetime. Existing indirect bounds are several orders of magnitude weaker than those for the electron and muon, and are therefore not included in our numerical analysis.

\begin{table}[htb!]
\centering
\begin{tabular}{|c|c|c|}\hline
\rowcolor{gray!40}Observables & SM Predictions & Experimental Values \\\hline\hline
\rowcolor{red!10} & $1159652181.547(229) \times 10^{-12}$ \cite{Volkov:2019phy} & \\
\rowcolor{red!10}\multirow{-2}{*}{$a_e$} & $1159652181.606(229) \times 10^{-12}$ \cite{Aoyama:2019ryr} & \multirow{-2}{*}{$1159652180.59(13) \times 10^{-12}$ \cite{Fan:2022eto}} \\
\rowcolor{green!10} &  & $-8.8(3.6)\times10^{-13}$ \cite{Parker:2018vye}\\
\rowcolor{green!10} & & $4.8(3.0)\times10^{-13}$ \cite{Morel:2020dww,Hanneke:2008tm}  \\
\rowcolor{green!10} & & $-9.57(2.64)\times10^{-13}$\\
\rowcolor{green!10} & & $-10.16(2.64)\times10^{-13}$ \cite{Volkov:2019phy,Fan:2022eto,Aoyama:2019ryr} \\
\rowcolor{green!10}\multirow{-5}{*}{$\Delta a_e$}& \multirow{-5}{*}{-} & $3.38 ( 1.61) \times 10^{-13}$ \cite{Morel:2020dww,Fan:2022eto} \\
\rowcolor{cyan!10}$\Delta a_\mu$ & - & $3.85 (6.37) \times 10^{-10}$ \cite{Aliberti:2025beg, Muong-2:2025xyk} \\ 
\rowcolor{lime!10} &  & $ -0.007 < a_\tau<0.005 $ \cite{Gonzalez-Sprinberg:2000lzf} \\
\rowcolor{lime!10} \multirow{-2}{*}{$a_{\tau}$}& \multirow{-2}{*}{$1.17721 (5) \times 10^{-3}$ \cite{Eidelman:2007sb}} & $-0.057 < a_\tau < 0.024 $ \cite{ATLAS:2022ryk}\\ \hline
\end{tabular}
\caption{The SM predictions and experimental measurements of the anomalous magnetic moments of the charged leptons.}
\label{tab:gminus2}
\end{table}

Assuming only conservation of charge and Lorentz invariance, the effective operators relevant for the on-shell transition $\ell_{i} \to \ell_{j} \gamma$ are given by:
\begin{equation}
\label{eq:gen_lagragian_magnetic_moment}
\mathcal{L}_{\ell_{i} \to \ell_{j} \gamma} = \frac{\mu_{ij}^M}{2} \bar{\ell}_{j} \sigma_{\mu \nu } \ell_{i} \, F^{\mu \nu} + \frac{\mu_{ij}^{E}}{2} \bar{\ell}_{j} \, i \gamma_{5} \sigma_{\mu \nu} \ell_{i} F^{\mu \nu}\,,
\end{equation}
where the diagonal elements ($i=j$) will generate the transition magnetic dipole moment $\mu^{M}$ for the leptons. The CP-odd part $\mu^{E}$ will generate a contribution to the electric dipole moments. The off-diagonal parts will contribute to the radiative LFV decays. The magnetic dipole moment can be written in terms of the magnetic dipole form factor as \cite{Lindner:2016bgg}:
\begin{equation}
\mu_{ij}^{M}  = e\, m_{i} \, A_{ii}^{M} /2\,.
\end{equation}
The AMM can be expressed as:
\begin{eqnarray}
\Delta a_{\ell_{i}} & = A^{M}_{ii} m_{i}^2= \displaystyle\frac{2 \, m_{i}}{e} \mu_{ii}^{M}\,.
\end{eqnarray}
The corresponding loop contributions are given in \Eq\eqref{eq:magnetic_moment_loop} in \autoref{app:loops}.
\begin{figure}[htb!]
\centering
\subfloat[]{ \includegraphics[width=0.475\linewidth]{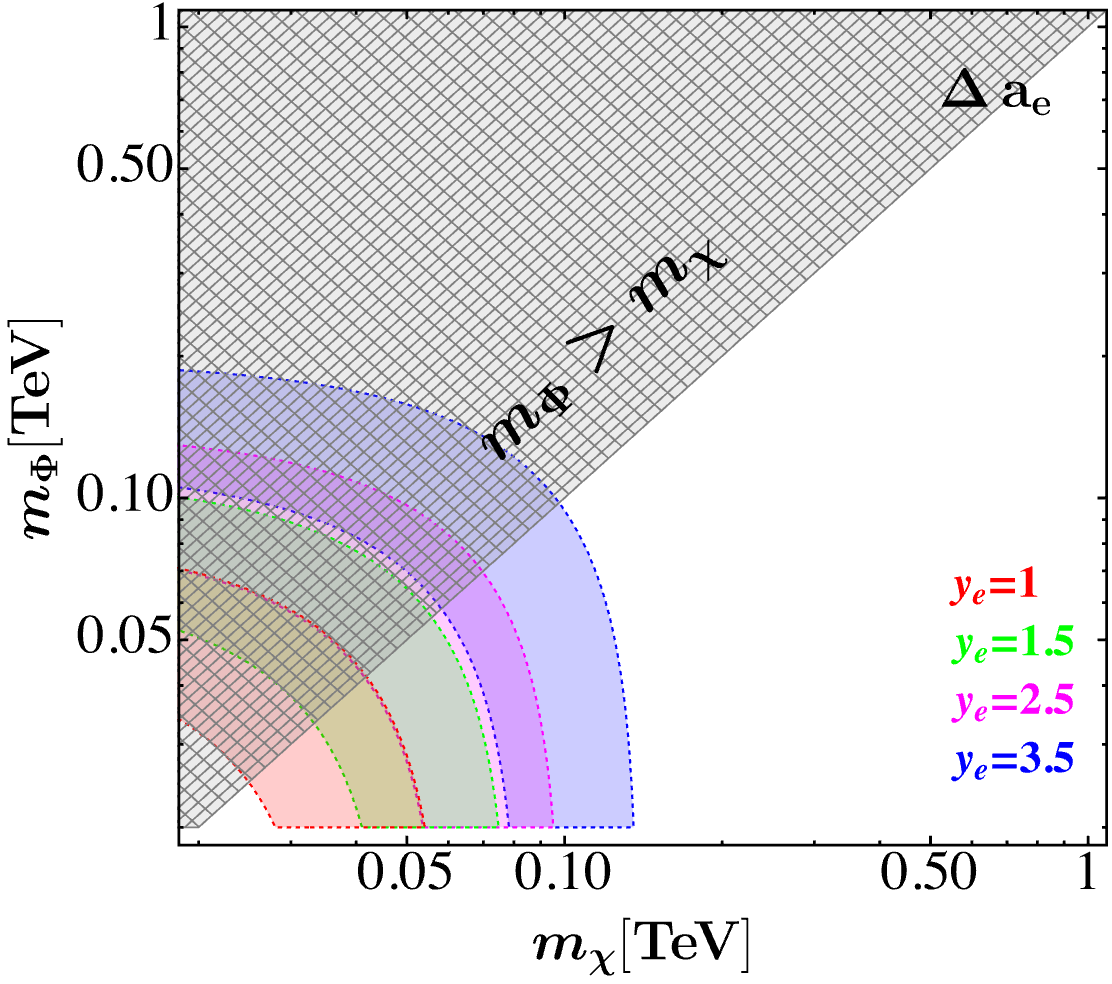}\label{fig:elec_dipole}}\quad
\subfloat[]{\includegraphics[width=0.475\linewidth]{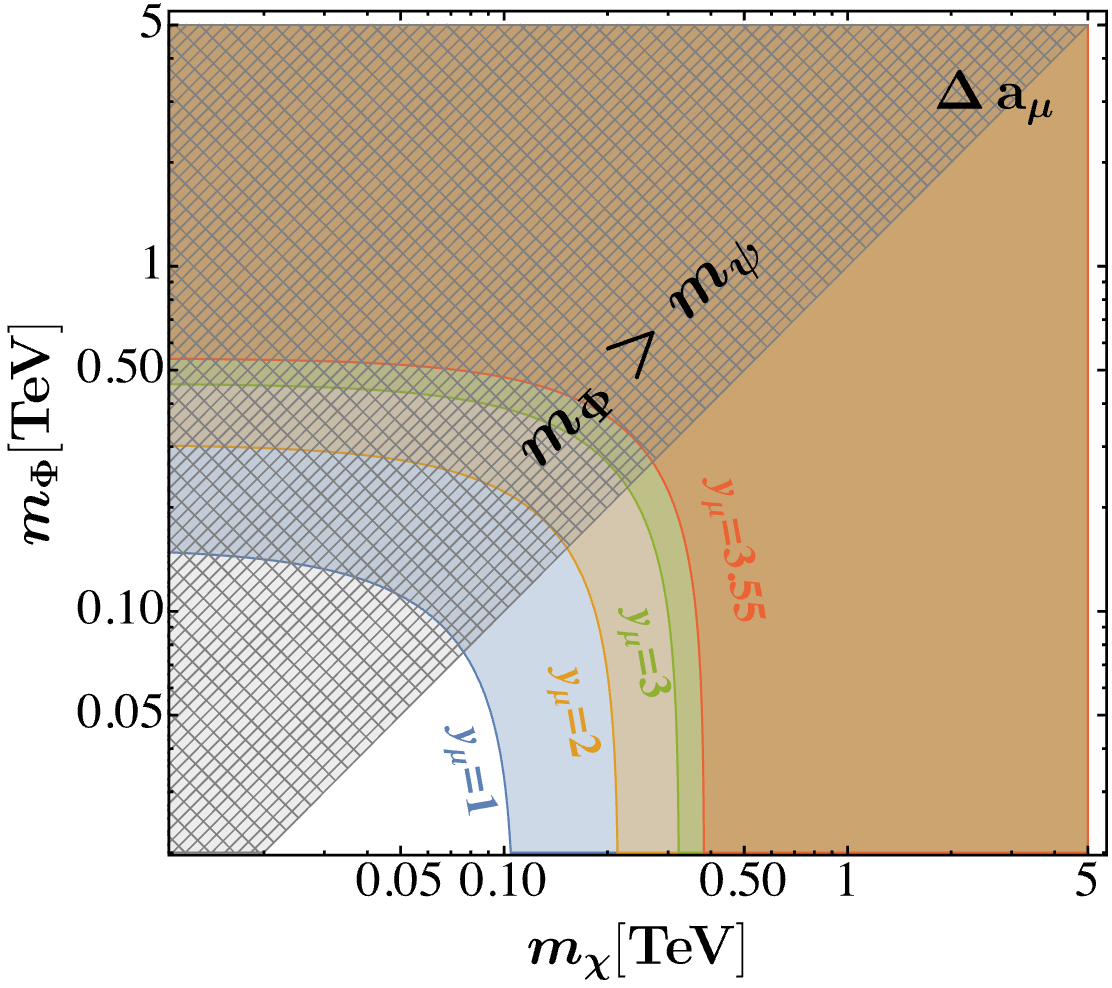}\label{fig:muon_dipole}}
\caption{Allowed regions (in colored bands) of the $(\mchi,\,\mphi)$ parameter space at the $1\sigma$ confidence level from the magnetic dipole moments of charged leptons ($e$ and $\mu$) for the benchmark points specified in the corresponding legends. For details, please see the text.}
\label{fig:lepton_dipole}
\end{figure}

Figure\,.~\ref{fig:lepton_dipole} shows the allowed parameter space in the DM-VLL mass plane for a few benchmark values of the lepton Yukawa coupling $\yl$, up to the perturbative unitarity bound, $\yl \sim \sqrt{4\pi}$. The allowed parameter space from the electron AMM is shown in \Fig\ref{fig:elec_dipole} as bands, obtained by considering the $1\sigma$ uncertainty of the observable. Similarly, \Fig\ref{fig:muon_dipole} shows the constraints from the muon AMM for the $1\sigma$ uncertainty of $\Delta a_\mu$. For the $1\sigma$ uncertainty, the electron AMM is inconsistent with zero, whereas the muon AMM is consistent with zero. Consequently, the former leads to a band in the parameter space, while the latter provides only a lower bound. As discussed in the text, the allowed band from the electron AMM corresponds to the region that can explain the observed discrepancy. However, the higher-mass region remains allowed even if the model does not fully account for the experimental result. We find that, even for Yukawa couplings as large as the perturbative unitarity bound, regions with $\mphi\gtrsim 0.5$~TeV and $\mchi\gtrsim 1$~TeV remain allowed by $\Delta a_e$ and are also consistent with the constraint from $\Delta a_\mu$. We do not show the constraints from $\Delta a_\tau$, since only an upper limit is currently available for this observable. Except for very low masses, $(\mchi\,,~\mphi)<\mathcal{O}(1\,\GeV)$, the entire parameter space remains allowed by the $\tau$ AMM.
\subsubsection{Charged Lepton Flavor Violating Decays}
In the present NP framework, charged LFV decays are induced at the one-loop level through the Yukawa interactions involving the VLL and the dark matter particle, as shown in the Feynman diagrams of the last three panels of \Fig\ref{feyn:LFV}. These processes can broadly be classified into two categories: radiative LFV decays, $\ell_{\alpha}\to\ell_{\beta}\gamma$, and three-body charged LFV decays, $\ell_{\alpha}\to\ell_{\beta}\ell_{\gamma}\ell_{\delta}$, where the final-state leptons may belong to the same or different generations. In the SM, charged LFV decays are highly suppressed, making them excellent probes of NP. To date, no experimental evidence for such processes has been observed, and the absence of any signal places stringent 90\% C.L. upper limits on their branching ratios, as summarized in \autoref{tab:LFV_upper_limits}. The most stringent upper bound is obtained for the $\mathcal{B}(\mu \to e \gamma)$. The MEG-II experiment is expected to achieve an improved upper limit $ \mathcal{B}( \mu \to e \gamma) < 6 \times 10^{-14}$. Also, for the three-body LFV decay $\mu \to 3e$, the projected sensitivity of the Mu3e experiment is several orders of magnitude more stringent, with expected upper limits of $\mathcal{B}(\mu \to 3e) < 2.0\,(1.0)\times10^{-15}$ from the Mu3e Phase-I (Phase-II) projections \cite{Mu3e:2020gyw, COMET:2025sdw}. In our analysis, we have mainly used the current experimental upper bounds, and in the final summary plots (Figs.~\ref{fig:relic-summary-1sigma} and \ref{fig:relic-summary-2sigma}), which combine results from all three sectors, namely, flavor, dark matter, and direct collider searches, we have also shown the impact of the projected MEG-II sensitivity. These bounds, therefore, provide powerful constraints on the parameter space of the present model. The recent update on the branching ratio of the $\mathcal{B}(\tau \to \mu \gamma) $ from Belle-II \cite{Belle-II:2026usi} provides a more relaxed bound compared to the previous Belle \cite{Belle:2021ysv} result at $90 \%$ C.L, hence we have considered the most stringent bound as the constraint. 
\begin{table}[htb!]
\centering
\begin{tabular}{|c|c|}\hline
\rowcolor{gray!40}Branching Ratios of Processes & Upper Bounds\\\hline\hline
\rowcolor{red!10}$\mu^+ \to e^+ \gamma$& $ 1.5\times 10^{-13}$ \cite{MEGII:2025gzr}\\ 
\rowcolor{red!10}$\tau^\pm \to \mu^\pm \gamma $ & $  4.2\times 10^{-8}$ \cite{Belle:2021ysv} \\
\rowcolor{red!10}$\tau^\pm \to e^\pm \gamma  $ & $ 3.3\times 10^{-8}$ \cite{BaBar:2009hkt} \\
\rowcolor{green!10}$\mu^+ \to e^+\, e^- \, e^+$ & $ 1.0 \times 10^{-12}$ \cite{SINDRUM:1987nra}\\
\rowcolor{green!10}$\tau^+ \to e^+\, e^- \, e^+$ & $ 2.5 \times 10^{-8}$ \cite{Belle-II:2025urb}\\
\rowcolor{green!10}$\tau^- \to \mu^- \, \mu^+ \, \mu^-$ & $1.9 \times 10^{-8}$ \cite{Belle-II:2024sce,LHCb:2026eod,ATLAS:2026kkw,CMS:2026fwo} \\
\rowcolor{cyan!10}$\tau^- \to e^- \, e^+\, \mu^-$ & $ 1.6 \times 10^{-8}$ \cite{Belle-II:2025urb}\\ 
\rowcolor{cyan!10}$\tau^- \to e^-\, \mu^+ \, e^- $ & $1.5 \times 10^{-8}$ \cite{Hayasaka:2010np}\\
\rowcolor{cyan!10}$\tau^- \to \mu^-\, \mu^+ \, e^- $ & $2.4 \times 10^{-8}$ \cite{Belle-II:2025urb}\\
\rowcolor{cyan!10}$\tau^- \to \mu^-\, e^+ \, \mu^- $ & $1.3 \times 10^{-8}$ \cite{Belle-II:2025urb}\\\hline
\end{tabular}
\caption{ Experimental upper bounds at 90\% C.L. on charged LFV decays.}
\label{tab:LFV_upper_limits}
\end{table}

\begin{figure}[htb!]
\centering
\includegraphics[width=0.32\linewidth]{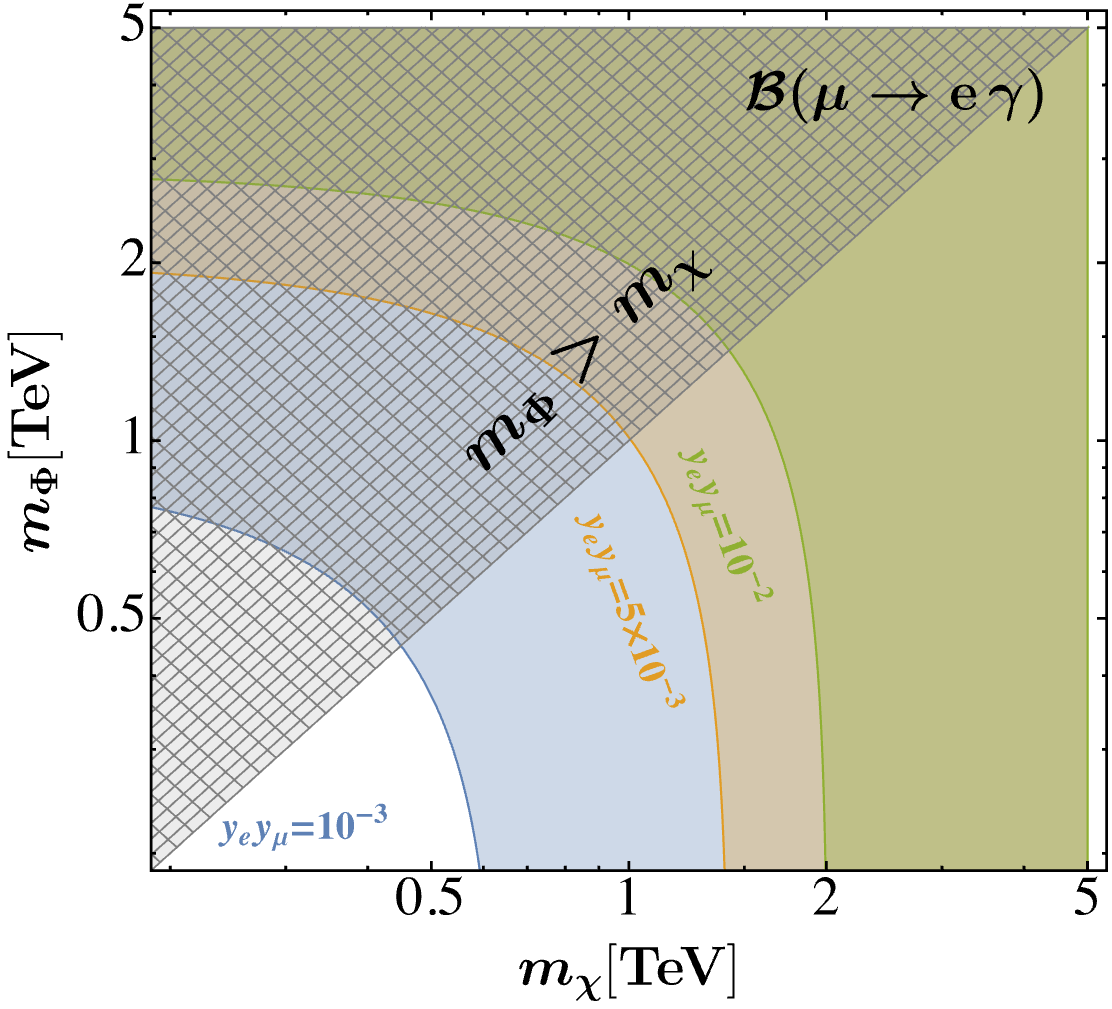}~
\includegraphics[width=0.32\linewidth]{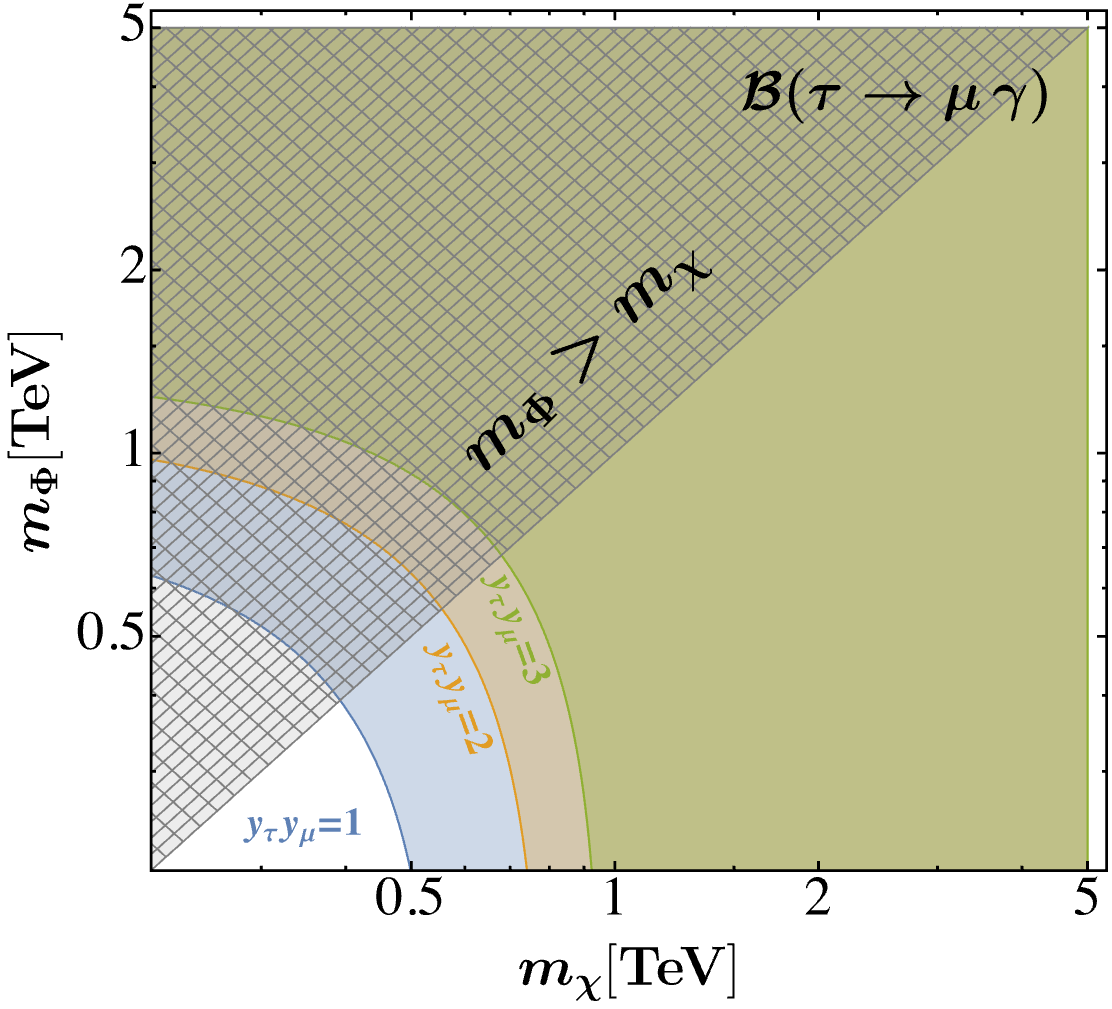}~
\includegraphics[width=0.32\linewidth]{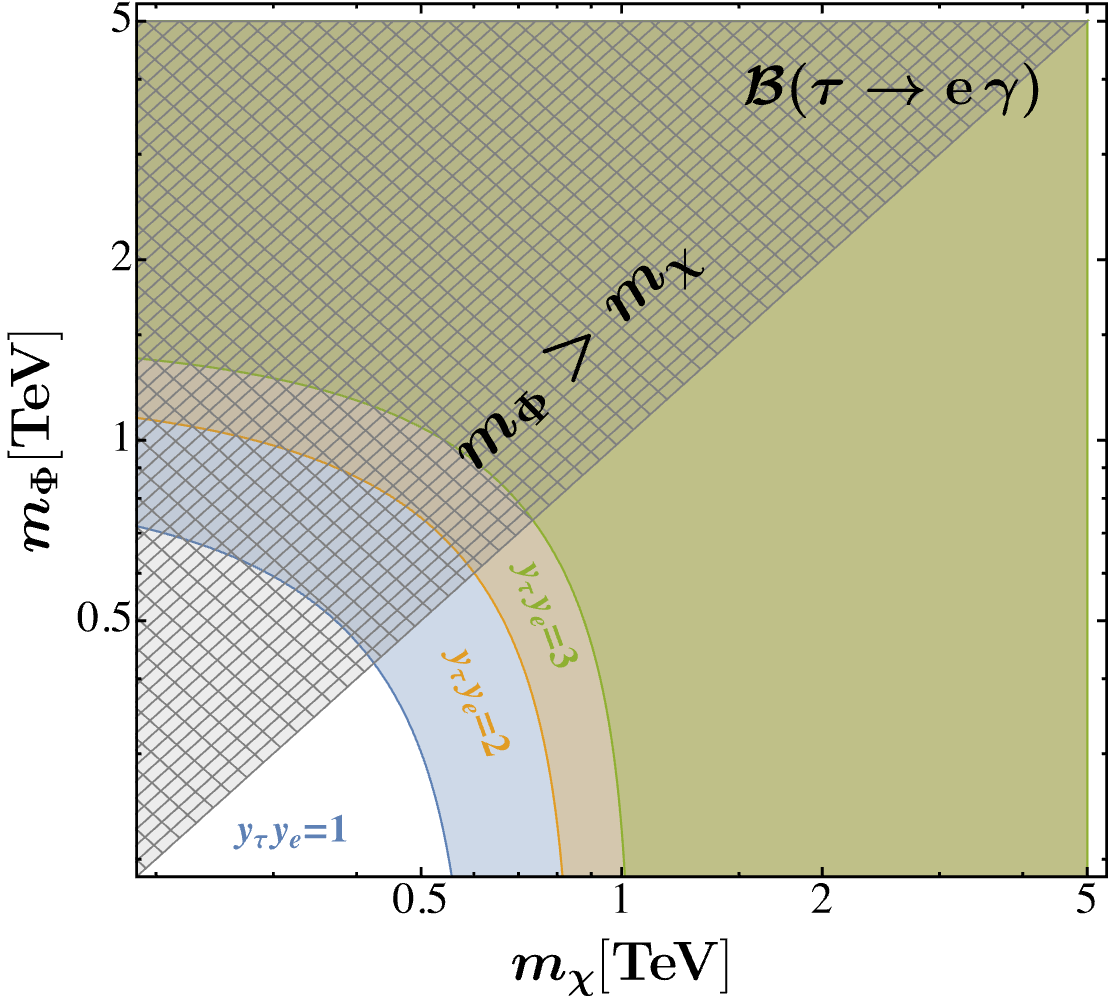}
\caption{Allowed regions (in colored bands) of the parameter space in the $(\mchi,\,\mphi)$ plane, from the radiative LFV decays of charged leptons ($\mu$ and $\tau$) for the benchmark points specified in the corresponding legends.}
\label{fig:cLFV_radiative}
\end{figure}

\paragraph{\underline{Radiative charged LFV Decays :}} The effective Lagrangian in \eq\eqref{eq:gen_lagragian_magnetic_moment} also describes the radiative lepton-flavor-violating decays $\ell_i\to\ell_j\gamma$ ($i\neq j$). Neglecting the mass of the final-state lepton, the corresponding branching ratio is given by
\begin{equation}
\mathcal{B}(\ell_{i} \to \ell_{j}\gamma) = \frac{\alpha_{em} m_{i}^5}{8 \Gamma_{\ell_i}}\left(|A_{ij}^{M}|^2 + |A_{ij}^{E}|^2 \right) = \frac{m_{i}^3}{8 \pi \Gamma_{\ell_i}}\left(|\mu_{ij}^{M}|^2 + |\mu_{ij}^{E}|^2 \right) \,. 
\end{equation}

The allowed parameter space in the $(\mchi\,-\,\mphi)$ plane obtained from the radiative charged-LFV (cLFV) decays is shown in \Fig\ref{fig:cLFV_radiative}. Among these processes, $\mu\to e\gamma$ provides by far the most stringent constraint, with an experimental upper limit of $\mathcal{O}(10^{-13})$, whereas the current bounds on the radiative tau decays are only of $\mathcal{O}(10^{-8})$. Although the dipole amplitude in our model scales with the mass of the decaying lepton, $\mu_{ij}^{M}\propto m_i$, the much shorter lifetime of the tau lepton, $\tau_\tau/\tau_\mu\sim10^{-7}$, partially compensates for the enhancement from the larger tau mass. Assuming all other parameters to be of comparable size, the branching-ratio scaling is approximately
\begin{equation}
\frac{\mathcal{B}(\tau\to\ell_j\gamma)}{\mathcal{B}(\mu\to e\gamma)}
\sim
10^{-7}\frac{m_\tau^5}{m_\mu^5}
\sim
\mathcal{O}(10^{-2})\,.
\end{equation}
However, the experimental upper limits on $\mathcal{B}(\tau\to\ell_j\gamma)$ are roughly five orders of magnitude weaker than that on $\mathcal{B}(\mu\to e\gamma)$. Consequently, for comparable masses of the dark matter particle and the VLL, the decay $\mu\to e\gamma$ imposes much stronger constraints on the Yukawa coupling combination $\ymu \ye$ than the tau decays do on $\yta \yl$. In particular, TeV-scale values of $\mchi$ and $\mphi$ require $\ymu \ye \lesssim0.003$, whereas values as large as $\yta \yl\lesssim2$ remain allowed by the current bounds on the radiative tau decays.

\paragraph{\underline{$\mu^- \to e^-$ conversion in nuclei:}}
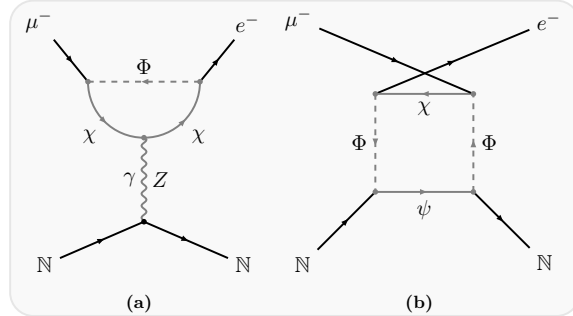
\begin{figure}[htb!]
\centering
\begin{adjustbox}{width=0.5\textwidth}
\begin{tcolorbox}[colback=gray!5, colframe=black!10, boxrule=1pt, arc=4mm, boxsep=0pt, left=0pt, right=0pt, top=0pt, bottom=0pt, width=0.675\textwidth, halign=center]
\subfloat[]{\begin{tikzpicture}
\begin{feynman}
\vertex (a);
\vertex[below left=0.5cm and 1.5cm  of a](a1){\(\mathbb{N}\)};
\vertex[below right=0.5cm and 1.5cm  of a](a2){\(\mathbb{N}\)};
\vertex[above = 1.5cm of a](b);
\vertex[above right=1.0cm and 1.0cm  of b](c1);
\vertex[above left=1.0cm and 1.0cm  of b](b1);
\vertex[above right=1.75cm and 1.5cm of b](c2){\(e^-\)};
\vertex[above left=1.75cm and 1.5cm of b](b2){\(\mu^-\)};
\diagram*{
(b2) --[fermion,line width=0.35mm, arrow size=0.8pt](b1), 
(b1) --[fermion, style=black!50, quarter right,line width=0.35mm,edge label'=\(\color{black}\chi\), arrow size=0.8pt](b),
(b) --[fermion, quarter right,line width=0.35mm,arrow size=0.8pt,style=black!50,edge label'=\(\color{black}\chi\)](c1),
(c1) --[fermion,line width=0.35mm,arrow size=0.8pt,style=black](c2),
(c1) --[charged scalar, line width=0.35mm,arrow size=0.8pt,style=black!50,edge label'=\(\color{black}\Phi\)](b1), 
(a) --[boson,line width=0.35mm,style=black!50,arrow size=0.8pt,edge label={\(\color{black}\gamma\)},edge label'={\(\color{black}Z\)}](b),
(a1) --[fermion,line width=0.35mm, arrow size=0.8pt](a),
(a) --[fermion,line width=0.35mm, arrow size=0.8pt](a2)};
\node at (a)[circle,fill,style=black,inner sep=1pt]{};
\node at (b)[circle,fill,style=gray,inner sep=1pt]{};
\node at (b1)[circle,fill,style=gray,inner sep=1pt]{};
\node at (c1)[circle,fill,style=gray,inner sep=1pt]{};
\end{feynman}
\end{tikzpicture}\label{feyn:muN_eN1}}\,
\subfloat[]{\begin{tikzpicture}
\begin{feynman}
\vertex (a1){\(\mathbb{N}\)};
\vertex[above left = 1cm and 0.25cm  of a1](c);
\vertex[above right = 1.3cm and 1.3cm of a1](a2);
\vertex[above = 1.75cm of a2](a3);
\vertex[above left = 1cm and 1cm of a3](a4){\(\mu^-\)};
\vertex[right = 1.75cm of a2](a5);
\vertex[right = 1.75cm of a3](a6);
\vertex[above right = 1cm and 1cm of a6](a7){\(e^-\)};
\vertex[below right = 1cm and 1cm of a5](a8){\(\mathbb{N}\)};
\vertex[above right = 1cm and 0.25cm of a8](d);
\diagram*{
(a1) --[fermion,line width=0.35mm, arrow size=0.8pt](a2), 
(a3) --[charged scalar, line width=0.35mm,arrow size=0.8pt,style=black!50,edge label'=\(\color{black}\Phi\)](a2), 
(a3) --[fermion, line width=0.35mm,arrow size=0.8pt](a7),
(a2) --[fermion,line width=0.35mm,arrow size=0.8pt,style=black!50,edge label'=\(\color{black}\psi\)](a5),
(a6) --[fermion,line width=0.35mm,style=black!50,arrow size=0.8pt,edge label={\(\color{black}\chi\)}](a3),
(a4) --[fermion,line width=0.35mm, arrow size=0.8pt](a6),
(a5) --[charged scalar,line width=0.35mm, arrow size=0.8pt,style=black!50,edge label'=\(\color{black}\Phi\)](a6),

(a5) --[fermion,line width=0.35mm, arrow size=0.8pt](a8)};
\node at (a2)[circle,fill,style=gray,inner sep=1pt]{};
\node at (a3)[circle,fill,style=gray,inner sep=1pt]{};
\node at (a5)[circle,fill,style=gray,inner sep=1pt]{};
\node at (a6)[circle,fill,style=gray,inner sep=1pt]{};
\end{feynman}
\end{tikzpicture}\label{feyn:muN_eN2}}\,
\end{tcolorbox}
\end{adjustbox}
\caption{Feynman diagrams contributing to the $\mu$ to $e$ conversion process $\mu \, N \to e \, N$ via one-loop penguin (left) and box diagram (right). } 
\label{feyn:muNtoeN}
\end{figure}

The search for neutrinoless $\mu \to e$ conversion in nuclei, see \fig\ref{feyn:muNtoeN} for the respective processes, provides an important probe of LFV processes. The relevant observable is the conversion rate \cite{Kitano:2002mt},
\begin{align}
\mathcal{R}_{\mu^-\to e^-}^{\mathbb{N}} \equiv
\frac{\Gamma(\mu^- + \mathbb{N}(A,Z) \to e^- + \mathbb{N}(A,Z))}
{\Gamma(\mu^- + \mathbb{N}(A,Z) \to \nu_{\mu} + \mathbb{N}(A,Z-1))}\,.
\end{align}
The most stringent current bound on this observable is obtained by the SINDRUM-II experiment using a gold target \cite{SINDRUMII:2006dvw},
\begin{align}
\mathcal{R}_{\mu^-\to e^-}^{\rm Au} < 7\times 10^{-13}\,.
\end{align}
In our analysis, we find that the $\mu\to e\gamma$ process provides the most stringent constraint on the parameter space. The parameter space allowed by the current $\mu\to e\gamma$ bound is already consistent with the existing constraint from $\mu\to e$ conversion. In particular, the contribution from the box diagrams to $\mu\to e$ conversion is subdominant in the parameter region of interest. Therefore, the $\mu\to e$ conversion constraint does not further restrict the parameter space, and we do not show the corresponding parameter-space constraints explicitly. The future experiments Mu2e and COMET, using aluminum targets, are expected to improve the sensitivity to $\mu\to e$ conversion by approximately two and three orders of magnitude, respectively, compared with the current bound. The projected sensitivities for Phase I and Phase II of these experiments are also provided in Refs\,.~\cite{COMET:2018auw, Moritsu:2022lem, Mu2e:2022ggl, Mu2e-II:2022blh}.
\paragraph{\underline{3-body cLFV Decays:}}
In addition to the radiative cLFV decays, the present model also induces the three-body cLFV processes of the form $\ell_{\alpha}\to\ell_{\beta}\ell_{\delta}\bar{\ell}_{\delta}$ at one-loop order. Owing to charge conservation and the existence of only three charged leptons in the SM, the final state necessarily contains a same-flavor lepton pair. These decays are mediated by the dark matter particle and the vector-like lepton through the penguin and box diagrams shown in Figs.~\ref{feyn:muto3ea} and \ref{feyn:muto3eb}. The resulting one-loop amplitudes generate local four-fermion vector operators with both left- and right-handed chiral structures. The corresponding effective Lagrangian can therefore be written as
\begin{equation}
\label{eq:cLFV_3body_gen}
\mathcal{L}_{\ell_{\alpha} \to \ell_{\beta} \ell_{\delta}\bar{\ell}_{\delta}} = \sum_{i,j=\mathtt{L},\mathtt{R}}\mathcal{C}_{ij}(\bar{\ell}_{\beta} \gamma_{\mu} \mathrm{P}_{i} \ell_{\alpha})(\bar{\ell_{\delta}} \gamma^{\mu}\mathrm{P}_{j}\ell_{\delta})\,,
\end{equation}
where $\mathcal{C}_{ij}$ denote the effective Wilson coefficients of the local four-fermion operators, whose explicit expressions are collected in \autoref{app:loops}. Neglecting the masses of the final-state leptons, the corresponding branching ratio is given by
\begin{equation}
\mathcal{B}(\ell_{\alpha} \to \ell_{\beta} \ell_{\delta}\bar{\ell}_{\delta}) = \tau_{\ell_{\alpha}} \frac{m_{\ell_{\alpha}}^5}{96 \pi^3}(|\mathcal{C}_{LL}|^2 + |\mathcal{C}_{\rm LR}|^2+|\mathcal{C}_{\rm RL}|^2+|\mathcal{C}_{\rm RR}|^2).
\end{equation}

The three-body charged lepton flavor-violating decays considered in this work can be classified according to the flavor assignment of the final-state leptons:
(a) $\ell_\alpha^{\pm}\to\ell_\beta^{\pm}\ell_\beta^{\pm}\ell_\beta^{\mp}$,
(b) $\ell_\alpha^{\pm}\to\ell_\beta^{\mp}\ell_\delta^{\pm}\ell_\delta^{\pm}$, and
(c) $\ell_\alpha^{\pm}\to\ell_\beta^{\pm}\ell_\delta^{\pm}\ell_\delta^{\mp}$.
In the first two cases, the final state contains two identical charged leptons. Consequently, the total amplitude must be antisymmetrized under the exchange of these identical fermions, giving
\begin{equation}
\label{eq:cLFV_3Body_amplitude}
\mathcal{M}=\mathcal{M}(p_1\to p_2\,p_3\,p_4)-\mathcal{M}(p_2\leftrightarrow p_3)\,,
\end{equation}
where $p_2$ and $p_3$ denote the momenta of the two identical final-state leptons. In the third case, all final-state leptons are distinguishable, and therefore no additional contribution from particle exchange is required.

\begin{figure}[htb!]
\centering
\includegraphics[width=0.5\linewidth]{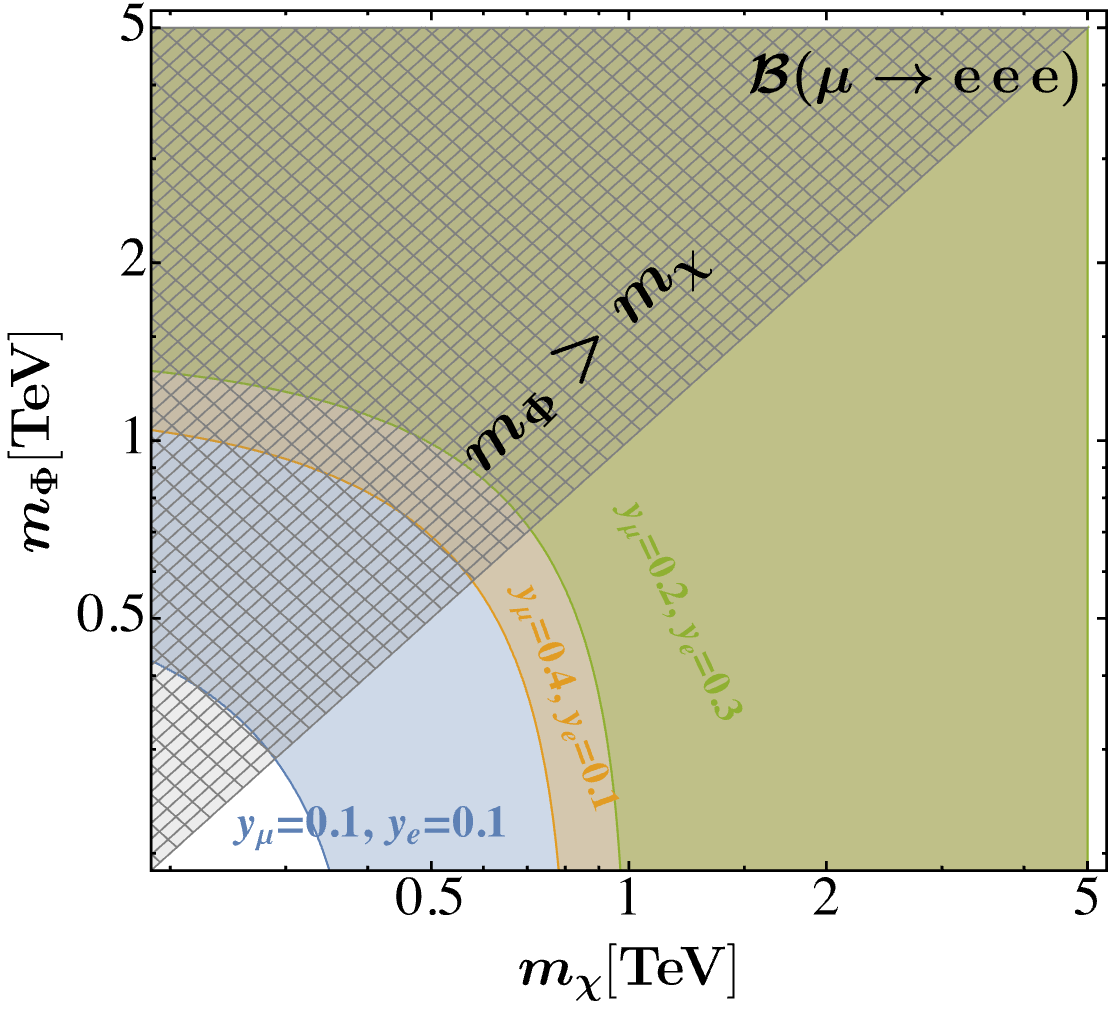}\label{fig:cLFV_3body_1}~
\caption{Allowed regions (in colored bands) in the $(\mchi-\mphi)$ parameter plane from the three-body cLFV decay ($\mu \to e \, e \, e$) at $90 \%$ C.L. for several BPs specified in the plot legends.}
\label{fig:cLFV_3body}
\end{figure}
Figure\,.~\ref{fig:cLFV_3body} presents the allowed regions of the $(\mchi\,-\,\mphi)$ parameter space obtained from the three-body charged lepton flavor-violating decays. The corresponding experimental $90\%$ C.L. upper limits on the branching ratios are summarized in \autoref{tab:LFV_upper_limits}. Figure\,.~\ref{fig:cLFV_3body} displays the allowed parameter space in the $(\mchi\,,~\mphi)$ plane for the decay $\mu^\pm\to e^\pm e^\pm e^\mp$, considering several benchmark values of the relevant Yukawa couplings. Since all couplings in our analysis are taken to be real, the charge-conjugate process yields identical results and is therefore not shown separately. All the one-loop diagrams shown in \Fig\ref{feyn:LFV} contribute to this decay. Numerically, however, the penguin contributions dominate over the box contributions, leading to an approximate scaling of the branching ratio as $(\ymu \ye)^2$. Consequently, accommodating VLL and dark matter masses in the TeV range requires the product of Yukawa couplings to satisfy $\ye\ymu\lesssim \mathcal{O}(10^{-1})$, i.e., the corresponding observable $\mathcal{B}(\mu \to e \gamma)$, will put a stronger bound on the parameter space than this process.

Similarly, for other decays such as $\tau \to 3e$ and $\tau \to 3\mu$, using the coupling orders obtained from the radiative decays, we find that the entire parameter space remains allowed. Hence, we do not show these constraints explicitly.  We also obtain contributions to processes such as $\tau^\pm \to \mu^\pm e^\pm e^\mp$ (and $e \leftrightarrow \mu$) solely through the box diagrams. Since the box contributions are suppressed, these decay modes allow Yukawa couplings of $\mathcal{O}(1)$ and therefore do not impose any significant constraints on the parameter space. Consequently, we do not show them explicitly. Overall, we conclude that the decay $\mu^\pm \to e^\pm e^\pm e^\mp$ provides the most stringent three-body cLFV constraint, requiring the electron and muon Yukawa couplings to be of the order of $\mathcal{O}(10^{-2})$.

\subsubsection{Hadronic $\tau$ LFV Decays}
\begin{figure}[htb!]
\centering
\begin{adjustbox}{width=0.65\linewidth}
\begin{tcolorbox}[colback=gray!5, colframe=black!10, boxrule=0.8pt, arc=4mm, boxsep=0pt, left=0pt, right=0pt, top=0pt, bottom=0pt, width=0.85\linewidth, halign=center]
\subfloat[]{\begin{tikzpicture}
\begin{feynman}
\vertex (a){\(\tau\)};
\vertex[right=1.5cm of a](b);
\vertex[above right=1.25cm and 1.25cm of b](b1);
\vertex[below right=1.25cm and 1.25cm of b](b3);
\vertex[above right=1.5cm and 3.5cm of b](b2){\(\ell\)};
\vertex[below right=1.5cm and 3.5cm of b](b4){\(d_i\)};
\vertex[right=2.5cm of b](b5);
\vertex[right=1.0cm of b5](b6){\(d_j\)};
\diagram* { 
(a) --[fermion,line width=0.35mm, arrow size=0.8pt]
(b) --[fermion, line width=0.35mm,arrow size=0.8pt,style=black!50,edge label=\(\color{black}\chi\)](b1) --[fermion, line width=0.35mm,arrow size=0.8pt](b2),
(b1) --[charged scalar,line width=0.35mm,arrow size=0.8pt,style=black!50,edge label=\(\color{black}\Phi\)](b5),
(b3) --[charged scalar,line width=0.35mm,arrow size=0.8pt,style=black!50,edge label=\(\color{black}\Phi\)](b),
(b6) --[fermion,line width=0.35mm, arrow size=0.8pt]
(b5) --[fermion,line width=0.35mm, arrow size=0.8pt,style=black!50,edge label=\(\color{black}\psi\)](b3) --[fermion,line width=0.35mm, arrow size=0.8pt](b4)};
\node at (b)[circle,fill,style=gray,inner sep=1pt]{};
\node at (b1)[circle,fill,style=gray,inner sep=1pt]{};
\node at (b5)[circle,fill,style=gray,inner sep=1pt]{};
\node at (b3)[circle,fill,style=gray,inner sep=1pt]{};
\end{feynman}
\end{tikzpicture}\label{feyn:tautol1}}\quad
\subfloat[]{\begin{tikzpicture}
\begin{feynman}
\vertex (a){\(\tau\)};
\vertex [right=1.5cm of a](b);
\vertex[above right=0.75cm and 1.5cm of b](b1);
\vertex[below right=0.75cm and 1.5cm of b](c0);
\vertex[right=1.5cm of c0](c1);
\vertex[above right=0.75cm and 2.25cm of b1](b2){\(\ell\)};
\vertex[above right=0.75cm and 0.75cm of c1](c2){\(d_i\)};
\vertex[below right=0.75cm and 0.75cm of c1](c3){\(d_i\)};
\diagram* {
(a) --[fermion,line width=0.35mm,arrow size=1pt,style=black](b),
(b) --[fermion,quarter right,line width=0.35mm,arrow size=1pt,style=black!50,edge label'=\(\color{black}\chi\)](c0),
(c0) --[fermion,line width=0.35mm,arrow size=1pt,style=black!50,edge label'=\(\color{black}\chi\)](b1),
(b1) --[fermion,line width=0.35mm,arrow size=1pt,style=black](b2),
(b1) --[charged scalar,quarter right,line width=0.35mm,arrow size=1pt,style=black!50,edge label'=\(\color{black}\Phi\)](b),
(c0) --[boson,line width=0.35mm,style=black!50,edge label'=\(\color{black}\gamma/Z\)](c1),
(c2) --[fermion,line width=0.35mm,arrow size=1pt,style=black](c1),
(c1) --[fermion,line width=0.35mm,arrow size=1pt,style=black](c3)};
\end{feynman}
\node at (b)[circle,fill,style=gray,inner sep=1pt]{};
\node at (b1)[circle,fill,style=gray,inner sep=1pt]{};
\node at (c0)[circle,fill,style=gray,inner sep=1pt]{};
\node at (c1)[circle,fill,style=black,inner sep=1pt]{};
\end{tikzpicture}\label{feyn:tautol2}}
\end{tcolorbox}
\end{adjustbox}
\caption{Feynman diagram contributing to the LFV hadronic decays of $\tau$ such as: $\tau \to \ell M$, with $M$ being neutral pseudoscalar and vector mesons.}
\label{fig:Feyn_LFV_hadronic_tau}
\end{figure}
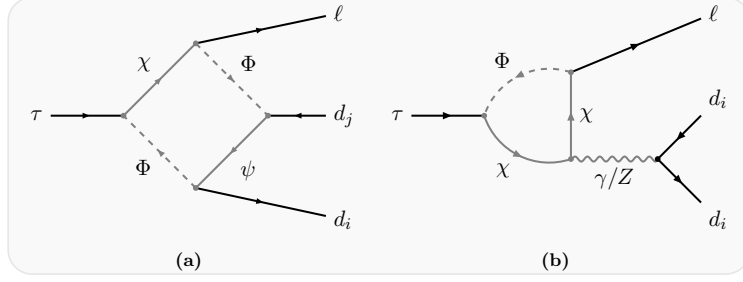

Another important class of LFV processes is the hadronic tau decays, $\tau \to \ell M$, where $M$ denotes either a pseudoscalar or a vector meson. Since the tau is sufficiently heavy, many such decay channels are kinematically allowed. These decays provide a unique connection between quark flavor-changing neutral currents and charged lepton flavor violation. In the SM, both flavor-changing neutral currents and charged lepton flavor-violating processes are highly suppressed, resulting in branching ratios of $\mathcal{B}\sim10^{-50}$ for these decays~\cite{Celis:2013xja, Urquia-Calderon:2025wjx}. Therefore, the observation of any such decay would provide clear evidence for the presence of NP. The current $90\%$ C.L. upper limits on the relevant decay modes are summarized in \autoref{tab:tau_hadron_lepton-limits}. For the processes with final state $\pi^0$ and $\phi$, the HFLAV\cite{HeavyFlavorAveragingGroupHFLAV:2024ctg} average values are considered. The Feynman diagrams contributing to these processes are shown in \Fig\ref{fig:Feyn_LFV_hadronic_tau}. The left panel contributes to all decay modes of the type $\tau \to \ell M$, whereas the right panel contributes only to final-state mesons with quark content $d_i\bar{d}_i$, such as $\pi^0$, $\rho$, and $\omega$.

The general Lagrangian obtained from the Feynman diagrams shown in \fig\ref{fig:Feyn_LFV_hadronic_tau}, can be written as:
\begin{equation}
\mathcal{L} = C_{L(R)L(R)} \, (\bar \ell \gamma_{\mu} \mathrm{P}_{\mathtt{L}(\mathtt{R})} \tau) \, (\bar d_j \gamma^\mu \mathrm{P}_{\mathtt{L}(\mathtt{R})} d_i )\,.
\end{equation}
In the limit of negligible mass of the external particles, we obtain $C_{LL(R)} = 0$. The expressions for the non-zero contributions to the Wilson coefficients are given in \Eq\eqref{eq:LFV_boxLoop} in \autoref{app:loops}. Note that, for $d_i \neq d_j$, only the box-loop contribution is present. Hence, only $C_{\rm RR}$ is non-zero. For the mesonic transition part, the hadronic currents are parametrized in terms of decay constants given by \cite{deMelo:2018hfw}:
\begin{equation}
\label{eq:Pdecay_const}
\langle 0 | \bar d_j \gamma_{\mu} \gamma_5 d_i | P(k)\rangle = i f_{P} k_{\mu}, \, \ \ \ \langle 0 | \bar d_j \gamma_{\mu} d_i | V(k, \epsilon)\rangle = f_{V} m_{V} \epsilon_{\mu} \,. 
\end{equation}
For neutral vector mesons such as $\rho$ and $\omega$, the decay constants are defined as:
\begin{equation}
\langle 0 | \frac{\bar u \gamma_{\mu} u \pm \bar d \gamma_\mu d}{2} | V(k, \epsilon)\rangle = f_{V} m_{V} \epsilon_{\mu}\,. 
\end{equation}
We use the updated values of the decay constants from Refs\,.~\cite{FlavourLatticeAveragingGroupFLAG:2024oxs, Bharucha:2015bzk}. The decay width for the decay of tau to the pseudoscalar and vector mesons can be written as follows:
\begin{subequations}
\begin{align}
\Gamma(\tau \to \ell P) &= \frac{ f_P^2 \left|C_{\rm RL}-C_{\rm RR}\right|^2}{128\pi m_\tau^3} \mathtt{\lambda}^{1/2}(m_\tau^2, m_\ell^2, m_P^2)\left( m_\ell^4 - m_\ell^2\left(m_P^2+2m_\tau^2\right) -  m_P^2m_\tau^2 + m_\tau^4 \right)  \,, \\
\Gamma(\tau \to \ell V) &= \frac{ f_V^2  \left|C_{\rm RL}+C_{\rm RR}\right|^2}{ 128\pi m_\tau^3 } \mathtt{\lambda}^{1/2}(m_\tau^2, m_V^2, m_\ell^2) \left( m_\ell^4 + m_\ell^2\left(m_V^2-2m_\tau^2\right) + m_\tau^4 - 2m_V^4 + m_\tau^2m_V^2 \right) .
\end{align}
\end{subequations}
Here $\mathtt{\lambda}$ is the K\"all\'en function.

\begin{table}[htb!]
\centering
\begin{tabular}{|c|c|}\hline
\rowcolor{gray!40} $\tau$ Hadronic LFVs & Branching Ratio Upper Limits \\\hline\hline
\rowcolor{red!10}$\tau \rightarrow \mu\,K_S^0$  &$ 1.2 \times 10^{-8}$\,\quad \text{\cite{Belle:2025iff}}\\ 
\rowcolor{red!10}$\tau \rightarrow e\,K_S^0$  &$0.8 \times 10^{-8}$\,\quad \text{\cite{Belle:2025iff}}\\
\rowcolor{cyan!10}$\tau \rightarrow \mu\,\pi^0$ &   $8.1 \times 10^{-8}$\,\quad \text{\cite{HeavyFlavorAveragingGroupHFLAV:2024ctg}}\\
\rowcolor{cyan!10}$\tau \rightarrow e\,\pi^0$ & $6.4 \times 10^{-8}$\,\quad \text{\cite{HeavyFlavorAveragingGroupHFLAV:2024ctg}}\\
\rowcolor{blue!10}$\tau\rightarrow \mu\,K^{*0}$  &$ 3.3 \times 10^{-8}$\,\quad \text{\cite{Belle:2023ziz}}\\ 
\rowcolor{blue!10}$\tau \rightarrow e\,K^{*0}$  &$ 1.9 \times 10^{-8}$\,\quad \text{\cite{Belle:2023ziz}}\\ 
\rowcolor{brown!10}$\tau \rightarrow \mu\,\phi$ &   $1.6 \times 10^{-8}$\,\quad \text{\cite{HeavyFlavorAveragingGroupHFLAV:2024ctg}}\\
\rowcolor{brown!10}$\tau \rightarrow e\,\phi$ & $1.3 \times 10^{-8}$\,\quad \text{\cite{HeavyFlavorAveragingGroupHFLAV:2024ctg}}\\
\rowcolor{green!10}$\tau \rightarrow \mu\,\rho^0$ &  $1.7 \times 10^{-8}$\, \quad\text{\cite{Belle:2023ziz}}\\
\rowcolor{green!10}$\tau \rightarrow e\,\rho^0$ & $2.0 \times 10^{-8}$\,\quad\text{\cite{Belle:2023ziz}}\\
\rowcolor{lime!10}$\tau \rightarrow \mu\,\omega$ &  $3.9 \times 10^{-8}$\, \quad\text{\cite{Belle:2023ziz}}\\
\rowcolor{lime!10}$\tau \rightarrow e\,\omega$ & $2.4 \times 10^{-8}$\,\quad\text{\cite{Belle:2023ziz}}\\\hline
\end{tabular}
\caption{Upper limit on the branching ratios of the hadronic LFV decays of $\tau$ at $90\%$ C.L.}
\label{tab:tau_hadron_lepton-limits}
\end{table}

\begin{figure}[htb!]
\centering
\subfloat[]{\includegraphics[width=0.475\linewidth]{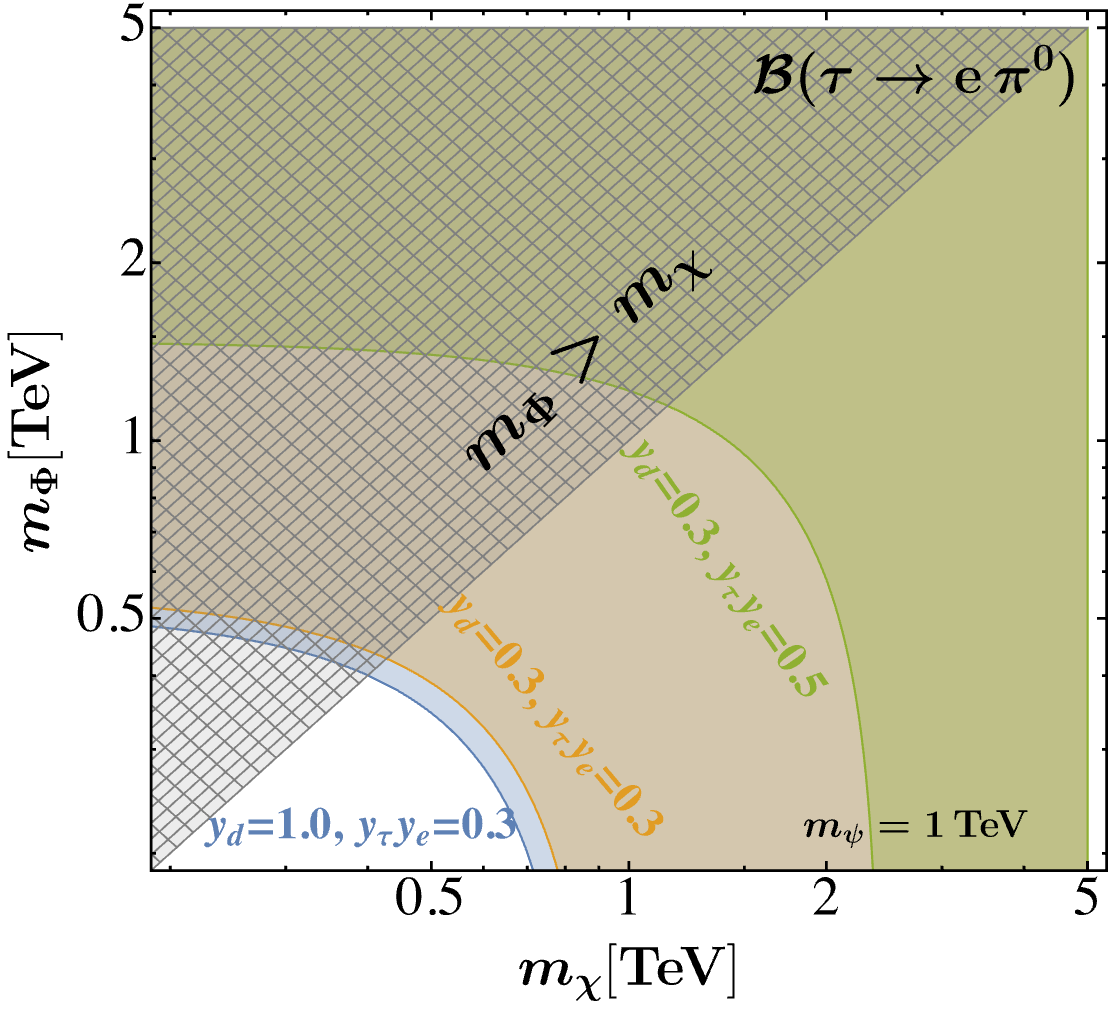}\label{fig:cLFV_tau_1}}\quad
\subfloat[]{\includegraphics[width=0.475\linewidth]{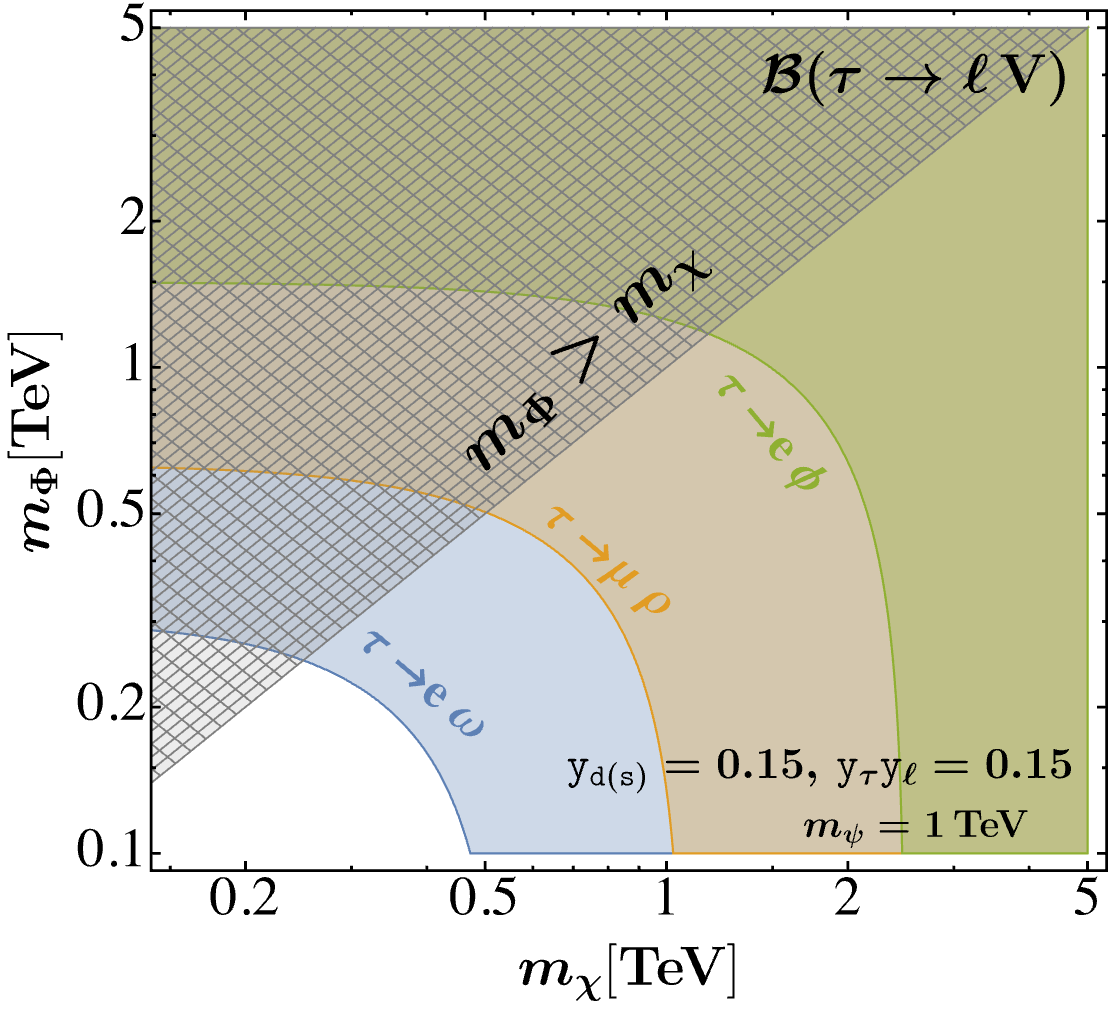}\label{fig:cLFV_tau_2}}
\caption{Allowed regions (in colored bands) in the $(\mchi-\mphi)$ parameter plane from the hadronic LFV decays of the $\tau$ for several benchmark points specified in the plot legends. We fix $\mpsi=1~\TeV$ for all the plots.}
\label{fig:cLFV_tau}
\end{figure}

Figure\,.~\ref{fig:cLFV_tau} shows the allowed parameter space from the most stringent hadronic LFV decays of the $\tau$ lepton. The decays $\tau \to \ell K_{S}$ and $\tau \to \ell K^*$ are generated only through the box diagram, whose contribution is very small. For these decay modes, with $(\mpsi\,,~\mchi\,,~\mphi)\sim 1$ TeV and all the relevant Yukawa couplings fixed to $\mathtt{y}_{f}=1$, we obtain branching ratios of $\mathcal{B}\sim \mathcal{O}(10^{-10})$. Since the current experimental upper bounds on these branching ratios are of the order of $10^{-8}$, the entire parameter space remains allowed. Therefore, we do not show these channels explicitly. In the figure above, we show only the processes that receive contributions from both the box and penguin diagrams.

Figure\,.~\ref{fig:cLFV_tau_1} shows the pseudoscalar meson final state $\tau \to e \pi^0$. Since the dependence on the final-state lepton mass is negligible, similar bounds are obtained for both the electron and muon channels for the same Yukawa couplings. As the experimental upper bound is more stringent for the electron channel, we show only this case. We find that the TeV mass region is allowed for couplings $\yd \sim 0.3$ and $\yta \ye \sim 0.5$. Note that the box contribution scales as $\mathcal{M}_{\rm box} \propto (\mathtt{y}_{d}^2\,\yta\ye)$, whereas the penguin contribution scales as $\mathcal{M}_{\rm peng} \propto (\yta\ye)$.

Figure\,.~\ref{fig:cLFV_tau_2} shows the decays of the $\tau$ lepton into vector mesons. We present the bounds for three representative decay channels corresponding to a BP specified in the legend. The dependence on the couplings is similar to that shown in the previous figure. As before, for each vector meson, we display only the decay channel having the most stringent experimental upper bound for either of the final-state leptons. In this case, the TeV mass region is allowed only for smaller couplings compared to the pseudoscalar case. Here, we fix the couplings to $\yd(\ys)=\yta\ye=0.15$. Among the three decay channels, $\tau \to e \phi$ provides the strongest constraint. Therefore, this process gives the most stringent bound on the Yukawa couplings associated with the tau lepton.
\subsubsection{Meson LFV Decays}
We now consider LFV processes involving mesons, which are induced by flavor-changing quark transitions accompanied by lepton flavor violation, namely $d_i \to d_j \ell_1 \bar{\ell}_2$, where $\ell_1 \neq \ell_2$. These transitions give rise to both purely leptonic decays, $P \to \ell_1 \bar{\ell}_2$, and semileptonic decays, $P \to M \ell_1 \bar{\ell}_2$, where $P$ denotes a pseudoscalar meson and $M$ represents either a pseudoscalar or a vector meson. Several experimental collaborations have reported stringent $90\%$ C.L. upper bounds on these LFV decay modes, which are summarised in \autoref{tab:meson_LFV_exp_bounds}. In the SM, these processes are highly suppressed, leading to branching ratios far below the current experimental sensitivities. Consequently, the observation of any such decay would constitute an unambiguous signal of NP. In what follows, we first examine the constraints arising from two-body leptonic LFV decays of pseudoscalar mesons within the model under our consideration.
\begin{table}[htb!]
\centering
\begin{tabular}{|c|c|}\hline
\rowcolor{gray!40} Meson LFV Decays & Branching Ratio Upper Limits \\\hline\hline
\rowcolor{red!10}$B_s^0 \rightarrow e^\pm \mu^\mp$  &$ 5.4 \times 10^{-9}$\quad \cite{LHCb:2017hag}\\ 
\rowcolor{red!10}$B_s^0 \rightarrow e^\pm \tau^\mp$  &$1.4 \times 10^{-3}$\,\quad \cite{Belle:2023jwr}\\
\rowcolor{red!10}$B_s^0 \rightarrow \mu^\pm \tau^\mp$  &$4.5 \times 10^{-5}$\,\quad \cite{LHCb:2019ujz}\\
\rowcolor{cyan!10}$B^0 \rightarrow e^\pm \mu^\mp$ &   $1.0 \times 10^{-9}$\,\quad \cite{LHCb:2017hag} \\
\rowcolor{cyan!10}$B^0 \rightarrow e^\pm \tau^\mp$ & $1.6 \times 10^{-5}$\,\quad \cite{Belle:2021rod} \\
\rowcolor{cyan!10}$B^0 \rightarrow \mu^\pm \tau^\mp$ & $1.4 \times 10^{-5}$\,\quad \cite{LHCb:2019ujz}\\
\rowcolor{blue!10}$K_L\rightarrow e^\pm \mu^\mp $  &$4.7 \times 10^{-12}$\,\quad \cite{BNL:1998apv}\\ 
\rowcolor{brown!10}$\pi^0 \to e^+ \mu^- $ &   $3.3 \times 10^{-10}$\,\quad \cite{NA62:2021zxl} \\
\rowcolor{green!10}$\phi \rightarrow e^\pm \mu^\mp $ &  $2.0 \times 10^{-6}$\, \quad \cite{Achasov:2009en}\\
\rowcolor{lime!10}$\eta \rightarrow e^+ \mu^-  +  e^- \mu^+ $ &  $6.0 \times 10^{-6}$\, \quad \cite{White:1995jc}\\\hline
\end{tabular}
\caption{Current experimental upper limits at 90\% C.L. on the branching ratios of the meson LFV decays considered in this work.}
\label{tab:meson_LFV_exp_bounds}
\end{table}

For these processes, the relevant Feynman diagrams are shown in \fig\ref{fig:Feyn_didjlmln_box}. In general, the LFV transition is induced solely by the box diagram, in which the VLL, VLQ, and DM particles propagate inside the loop. An exception arises for the decays $\pi^0$, $\phi$, and $\eta$, which also receive contributions from a penguin diagram. In this case, the quark pair annihilates into a $\gamma/Z$, while the LFV vertex is generated through a loop involving the VLL and the DM particle. Evaluating these diagrams yields the following effective Lagrangian:
\begin{equation}\label{eq:loop_P2lilj}
\mathcal{L}_{\rm eff}^{d_{i} \to d_{j} \ell_1 \bar{\ell}_2} = \mathcal{C}_{RL(R)} \, (\bar{d}_{j} \gamma_{\mu} \mathrm{P}_{\mathtt{R}} d_{i})(\bar{\ell}_{2} \gamma^{\mu} \mathrm{P}_{\mathtt{L}(\mathtt{R})} \ell_{1}) \,,
\end{equation}
where the expressions of the Wilson coefficients are collected in \autoref{app:loops}.
The corresponding branching ratio for the leptonic decay $P \to \ell_{1} \bar{\ell}_{2}$ is given by
\begin{equation}
\BR(P\to \ell_{1} \bar{\ell}_{2}) = \tau_{P}\frac{|\mathcal{C}_{\rm RR}-\mathcal{C}_{\rm LR}|^2\, f_P^2}{64\, \pi\, M_P^3} \; \lambda^{1/2}(m_{P}^2, m_{\ell_1}^2, m_{\ell_2}^2) \left[(m_{\ell_1}^2 + m_{\ell_2}^2) M_P^2 - (m_{\ell_1}^2 - m_{\ell_2}^2)^2\right] \,.
\label{eq:br_P2lilj}
\end{equation}
Here, $f_{P}$ denotes the decay constant as previously defined in \eq\eqref{eq:Pdecay_const}.
The numerical values of the decay constants used in our analysis are the same as those adopted in Sec\,.~\ref{subsec:meson_mixing}. As evident from \eq\eqref{eq:br_P2lilj}, the branching ratio is helicity suppressed by the masses of the final-state leptons.
\begin{figure}
\centering
\subfloat[]{\includegraphics[width=0.475\linewidth]{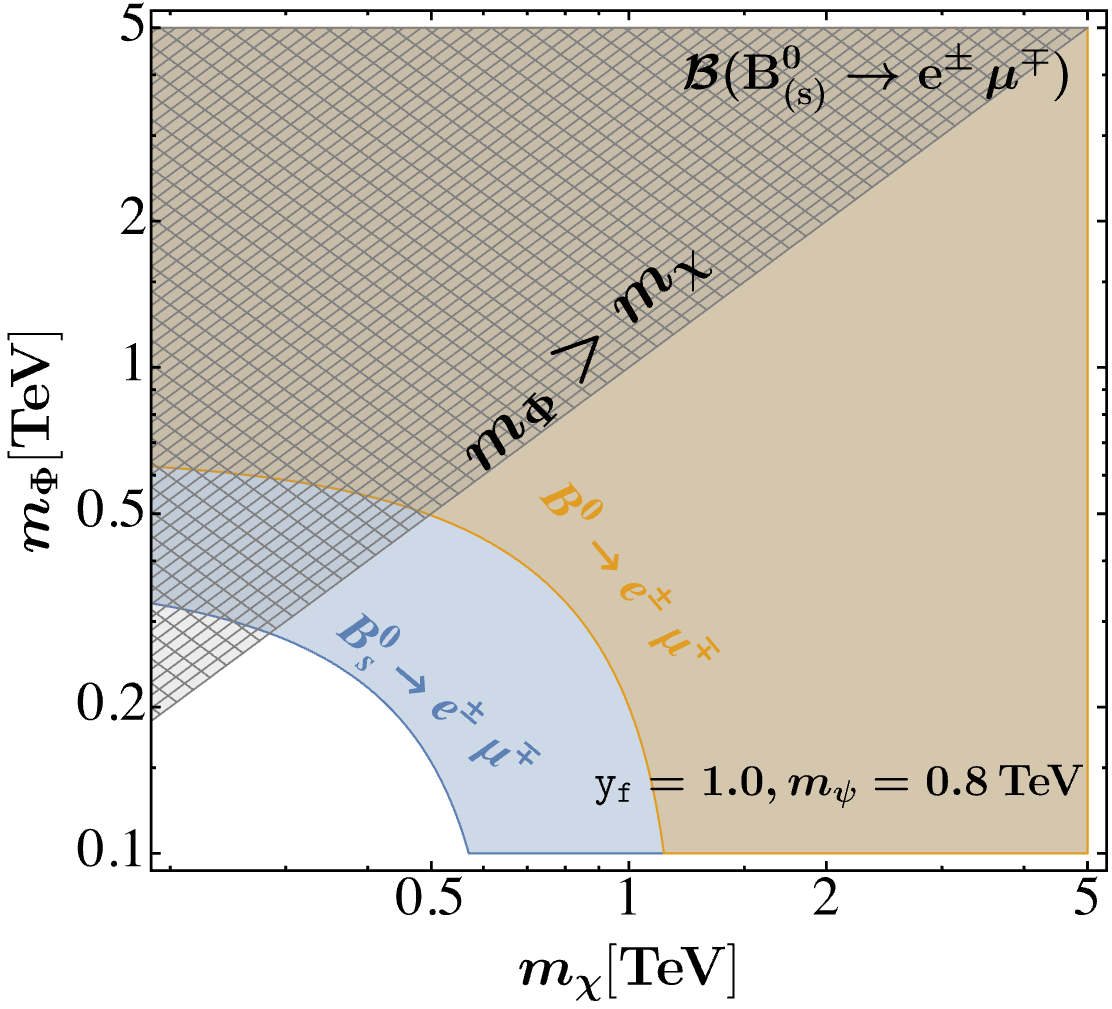}\label{fig:meson_LFV_1}}\quad
\subfloat[]{\includegraphics[width=0.475\linewidth]{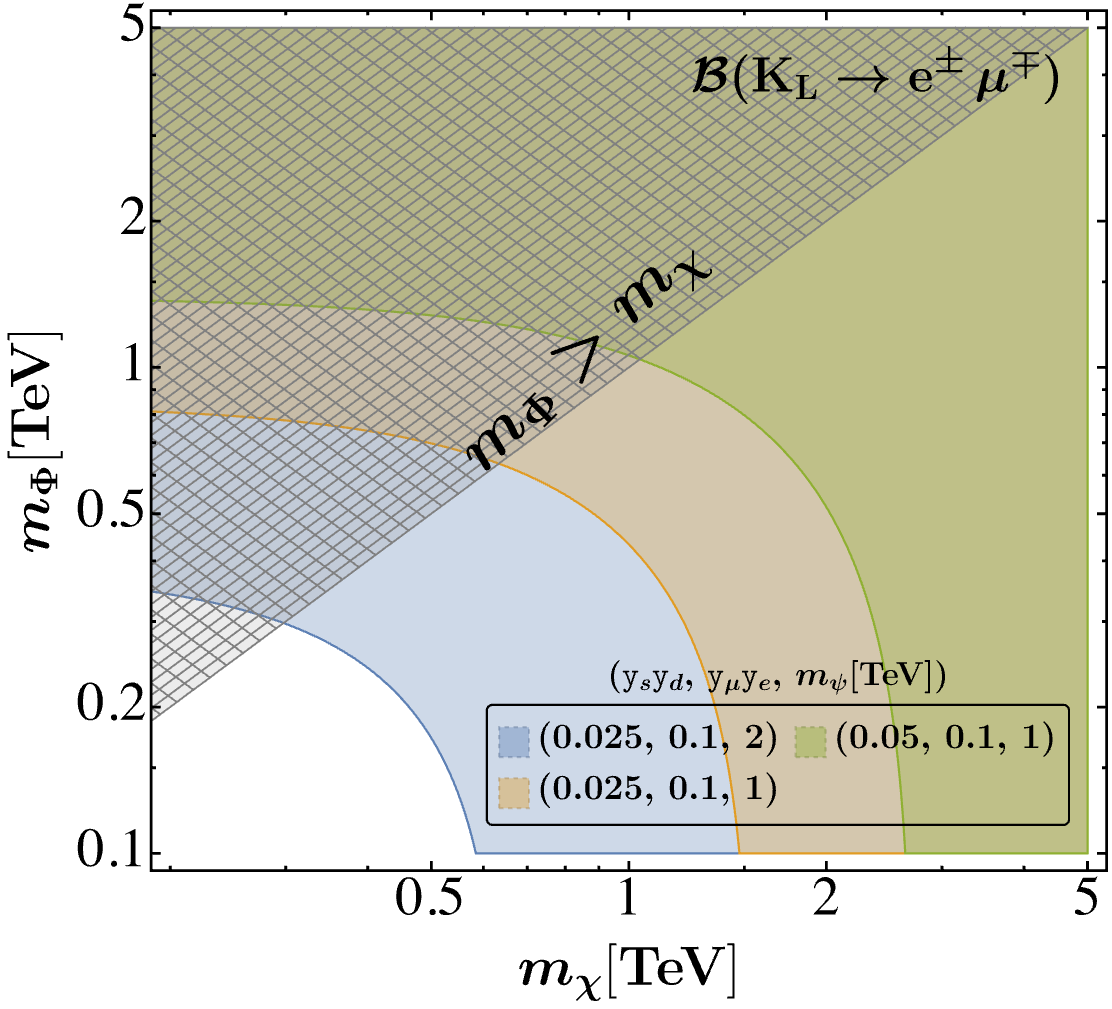}\label{fig:meson_LFV_2}}
\caption{Allowed parameter space (shaded region) in the VLQ-CSDM mass plane ($\mchi-\mphi$) derived from two-body meson LFV decays at $90\%$ C.L. for universal couplings $\mathtt{y}^{}_{\rm f}=\ye=\ymu=\yd=\ys$.}
\label{fig:meson_LFV}
\end{figure}

Figure\,.~\ref{fig:meson_LFV} shows the allowed parameter space in the $(\mchi\,,~\mphi)$ plane for the two-body meson LFV decays for a few benchmark points. 
For the processes, $B_{(s)}^0 \to e^\pm \mu^\mp$, the allowed parameter space (colored regions) is shown in \Fig\ref{fig:meson_LFV_1}. Here, all Yukawa couplings are fixed to unity, while the VLQ mass is set to $\mpsi=0.8~\TeV$. Among these two processes, $B^0 \to e^\pm \mu^\mp$ provides a comparatively stronger constraint due to its stringent experimental upper bound on the branching ratio. The branching ratio scales as $\propto (\mathtt{y}_{d_i}\,\mathtt{y}_{d_j}\,\ye\,\ymu)^2$. As can be seen, even for $\mathcal{O}(1)$ Yukawa couplings, almost the entire parameter space ($\mchi>1$ TeV and $\mphi>0.5 $ TeV) remains allowed. We do not show the corresponding decays involving $\tau$ leptons, as the experimental upper bounds on these channels are several orders of magnitude weaker than those for the light leptons.

Figure\,.~\ref{fig:meson_LFV_2} shows the allowed parameter space for the decay $K_{L}\to e^\pm\mu^\mp$ for three benchmark points. The current experimental upper bound on its branching ratio is of $\mathcal{O}(10^{-12})$, which is approximately three orders of magnitude more stringent than those for the $B$-meson LFV decays. In addition, the comparatively longer lifetime of the $K_L$ meson further enhances the sensitivity of this process. Consequently, $K_L\to e^\pm\mu^\mp$ provides significantly stronger constraints on the parameter space than the corresponding LFV decays of the $B$ mesons. Since this decay is generated solely through the one-loop box diagram, its dependence on the Yukawa couplings is similar to that discussed for the $B$-meson case. We find that masses in the TeV range remain allowed for quark coupling combinations satisfying $\yd\ys\lesssim0.025$ and lepton coupling combinations satisfying $\ye\ymu\lesssim0.1$.

For the decays $(\pi^0,\rho^0,\eta)\to e^\pm\mu^\mp$, an additional penguin diagram contributes, analogous to the hadronic LFV decays of the $\tau$ lepton. Among these channels, the most stringent experimental limit is obtained from $\pi^0\to e^\pm\mu^\mp$, whereas the bounds on the $\eta$ and $\rho^0$ decays are approximately four orders of magnitude weaker. For $\mathcal{O}(1)$ Yukawa couplings and $\mpsi=0.8~\TeV$, the predicted branching ratios are typically of $\mathcal{O}(10^{-14})$. Therefore, these channels impose no additional constraints on the parameter space. Experimental upper bounds are also available for three-body LFV meson decays of the form $P\to M\ell_1^\pm\ell_2^\mp$, where $P$ and $M$ denote pseudoscalar or vector mesons. However, with the exception of the $K_L$ channel, these decays do not provide constraints stronger than those already discussed. Therefore, we do not present their results explicitly.
\subsection{Electroweak Observables}
Unlike sequential chiral fermions, whose masses originate entirely from EWSB through Yukawa interactions with the Higgs field, the VLFs considered in this work possess a gauge-invariant Dirac mass term, $m_{\VLF}$ ($\mpsi\,,~\mchi$), which is independent of the Higgs vacuum expectation value and is invariant under the $\rm SU(2)_\mathtt{L} \times U(1)_\mathtt{Y}$ gauge symmetry. Consequently, the heavy VLFs satisfy the Appelquist-Carazzone decoupling theorem~\cite{Appelquist:1974tg}, implying that their contributions to the renormalizable electroweak sector become progressively suppressed as their mass increases. Since the VLFs do not mix with the SM fermions, their contributions to the electroweak gauge sector arise solely through one-loop vacuum-polarization corrections to the gauge-boson propagators, encoded in the self-energies $\Pi_{VV}(q^2)$ ($V=Z,~\gamma$). Although the unrenormalized self-energies contain ultraviolet-divergent and heavy-mass-dependent terms, these are absorbed into the renormalization of the electroweak parameters through the standard on-shell renormalization procedure, leaving finite physical corrections that are suppressed by inverse powers of the heavy mass. Equivalently, after integrating out the VLFs, the resulting effective theory contains higher-dimensional operators with Wilson coefficients proportional to $1/m_{\VLF}^2$. Consequently, the residual corrections to the gauge-boson self-energies scale as $\mathcal{O}(q^2/m_{\VLF}^2)$ or, at the electroweak scale, as $\mathcal{O}(v^2/m_{\VLF}^2)$. Furthermore, because the VLFs form an isolated $\rm SU(2)_\mathtt{L}$ singlet and do not induce weak-isospin mass splittings, their contribution to the custodial-symmetry-breaking parameter $T$ is negligible, while the $U$ parameter is further suppressed by $m_Z^2/m_{\VLF}^2$. Accordingly, the resulting oblique corrections remain highly suppressed in the heavy-mass limit. Therefore, for $m_{\VLF}\sim\mathcal{O}(\TeV)$, the corrections to electroweak precision observables, including the $Z$-boson mass, remain well below the current experimental sensitivity. We have also implemented the model in the Mathematica package \texttt{SARAH}~\cite{Staub:2008uz} and found that its contributions to the $S$ and $U$ parameters are smaller than $\mathcal{O}(10^{-17})$.
\subsubsection{$Z$-pole Observables}\label{sec:Z_pole}
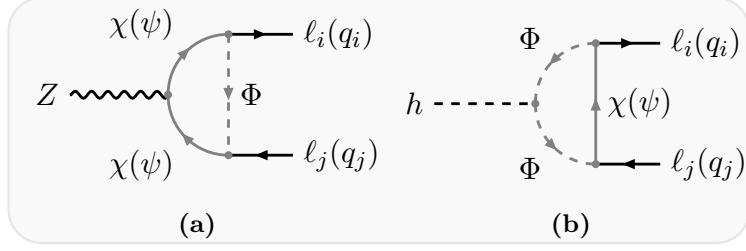
\begin{figure}[htb!]
\centering
\begin{adjustbox}{width=0.65\textwidth}
\begin{tcolorbox}[colback=gray!5, colframe=black!10, boxrule=0.8pt, arc=4mm, boxsep=0pt, left=0pt, right=0pt, top=0pt, bottom=0pt, width=0.61\textwidth, halign=center]
\subfloat[]{
\begin{tikzpicture}
\begin{feynman}
\vertex (a1){\(Z\)};
\vertex[right=1.5cm of a1](a2);
\vertex[above right=0.75cm and 0.75cm of a2](a3);
\vertex[below right=0.75cm and 0.75cm of a2](a4);
\vertex[right= 0.8cm of a3](a5){\(\ell_i(q_i)\)};
\vertex[right= 0.8cm of a4](a6){\(\ell_{j}(q_j)\)};
\diagram*{(a1) --[boson,line width=0.35mm,arrow size=1pt,style=black] (a2) --[fermion,quarter left,line width=0.35mm,arrow size=1pt,style=black!50,edge label=\(\color{black}\chi (\psi)\)] (a3) --[fermion,line width=0.35mm,arrow size=1pt,style=black](a5),
(a6) --[fermion,line width=0.35mm,arrow size=1pt,style=black](a4) --[fermion,quarter left,line width=0.35mm,arrow size=1pt,style=black!50,edge label=\(\color{black}\chi (\psi)\)](a2), 
(a3) --[charged scalar,arrow size=1pt,line width=0.35mm,style=black!50,edge label=\(\color{black}\Phi\)](a4)};
\end{feynman}
\node at (a2)[circle,fill,style=gray,inner sep=1pt]{};
\node at (a3)[circle,fill,style=gray,inner sep=1pt]{};
\node at (a4)[circle,fill,style=gray,inner sep=1pt]{};
\end{tikzpicture}
\label{fig:Feynman_Zfifj}}
\subfloat[]{\begin{tikzpicture}
\begin{feynman}
\vertex (a1){\(h\)};
\vertex[right=1.5cm of a1](a2);
\vertex[above right=0.75cm and 0.75cm of a2](a3);
\vertex[below right=0.75cm and 0.75cm of a2](a4);
\vertex[right= 0.8cm of a3](a5){\(\ell_i(q_i)\)};;
\vertex[right= 0.8cm of a4](a6){\(\ell_{j}(q_j)\)};;
\diagram*{
(a1) --[scalar,line width=0.35mm,arrow size=1pt,style=black](a2),
(a3) --[charged scalar,quarter right,line width=0.35mm,arrow size=1pt,style=black!50,edge label'=\(\color{black}\Phi\)](a2),
(a3) --[fermion,line width=0.35mm,arrow size=1pt,style=black](a5),
(a6) --[fermion,line width=0.35mm,arrow size=1pt,style=black](a4),
(a2) --[charged scalar,quarter right,line width=0.35mm,arrow size=1pt,style=black!50,edge label'=\(\color{black}\Phi\)](a4),
(a4) --[fermion,arrow size=1pt,line width=0.35mm,style=black!50,edge label'=\(\color{black}\chi(\psi)\)](a3)};
\end{feynman}
\node at (a2)[circle,fill,style=gray,inner sep=1pt]{};
\node at (a3)[circle,fill,style=gray,inner sep=1pt]{};
\node at (a4)[circle,fill,style=gray,inner sep=1pt]{};
\end{tikzpicture}\label{fig:Feynman_H2fifj}}
\end{tcolorbox}
\end{adjustbox}
\caption{Feynman diagrams contributing to the processes $Z \to f_{i} \bar{f}_{j}$ (left) and $h \to f_i \bar{f}_j$ (right). Decays to leptons are mediated by the VLL $\chi$, and decays to quarks are mediated by the VLQ $\psi$.}
\label{fig:Feyn_Zpole}
\end{figure}

The electroweak precision observables (EWPOs) measured at the $Z$ pole provide an important probe of the NP model. Although the contributions to the oblique electroweak parameters are negligible, the VLFs induce one-loop corrections to the effective $Zf\bar{f}$ couplings, thereby modifying several $Z$-pole observables. High-precision measurements of these observables were performed at LEP-I and SLAC at the $Z$ pole ($\sqrt{s}=m_Z$). The main observables include the partial and total decay widths of the $Z$ boson into fermions, the ratio observables ($R_f$), and the forward-backward asymmetries ($A_{\mathrm{FB}}^f$). Since these measurements are in excellent agreement with the SM predictions, they provide stringent constraints on new physics through its loop-induced effects. The ratio observables are defined as
\begin{equation}\label{eq:def_Zpole}
R_{q} = \frac{\Gamma_{Z\to q \bar q}}{\Gamma_{\rm had}}, \, \quad\quad R_{\ell} = \frac{\Gamma_{\rm had}}{\Gamma_{Z\to \ell \bar \ell}}\,.   
\end{equation}
Here $Z_{\rm had}$ signifies the total decay width of $Z$ decaying to all the quarks (kinematically available). In our model, the process $Z \to f \bar f$ is modified at one loop via the VLQ (VLL) $\psi \, (\chi)$ and the DM $\Phi$ as shown in \fig\ref{fig:Feynman_Zfifj}. This will modify the vector and axial vector interaction strength of the $Z$ boson to the fermion pairs as:
\begin{equation}
\mathcal{L}_{\rm eff} = - i \frac{g}{4 \cos \theta}\bar{f} \, \gamma^{\mu} \left(v_{f}^{\rm tot} + \gamma_{5} \,  a_{f}^{\rm tot} \right)\, f \, Z_{\mu} \,,  
\end{equation}
with, 
\begin{equation} \label{eq:Z_eff_coup}
v_{f}^{\rm tot} \to v_{f}^{\rm SM} + \delta \, v_{f},  \text{\ \ and \ \ } a_{f}^{\rm tot} \to a_{f}^{\rm SM} +\delta \,  a_{f},
\end{equation}
In SM, we have \cite{ALEPH:2005ab, ParticleDataGroup:2024cfk}: 
\begin{subequations}
\begin{eqnarray}
& v_{f}^{\rm SM} = 2 \sqrt{\rho_{f}} (I_{3} - 2 \, \kappa_{f} Q_{f} s_{W}^2) \,,  \\ 
& a_{f}^{\rm SM} = 2 \sqrt{\rho_{f}} \,  I_{3} \,. 
\end{eqnarray}\end{subequations}
At tree-level, $\rho_{f} = \kappa_{f} = 1$ whereas the higher-order corrections result in $\rho_{\ell} = 0.9977, \, \rho_{b} = 0.9866$ and $\kappa_{\ell} = 1.0014, \, \kappa_{b} = 1.0068$\,. 
The total decay width of $Z$ decaying to the fermion pair is given by \cite {Kolay:2024wns, Kolay:2025jip, Soni:2010xh}:
\begin{eqnarray}
\Gamma_{\rm tot} \left(Z \to f \bar f \right) = &  \frac{N_{c}}{48} \frac{\alpha}{s_{W}^2 c_{W}^2}m_{Z} \sqrt{1-\mu_{f}^2} \left( |a_{f}^{\rm tot}|^2 (1-\mu_f^2) + |v_{f}^{\rm tot}|^2 (1+\displaystyle\frac{\mu_{f}^2}{2}) \right) (1+\delta_{b}^{(0)}) \nonumber \\ & (1+\delta_{\rm QED}) (1+\delta_{\rm QCD})(1+\delta_{f})\,. 
\end{eqnarray}
Here $\mu_{f}$ is the factor that takes care of the non-negligible mass of the fermions and is given by: $\mu_{f} = m_{f}/m_{Z}$, relevant for the heavy fermions like the $b$ quark and $\tau$ lepton. The parameters $\delta_{b}^{(0)}, \, \delta_{\rm QED}, \, \delta_{\rm QCD}, \delta_{f}$ are the corrections to the tree-level SM contribution to the decays. The explicit expression of the corrections is discussed in detail in \cite{Soni:2010xh}. We have collected the expressions of $\delta v_f,\,\delta a_f$ generated in the model of our consideration in \autoref{app:loops}. The SM expectations and the experimental measurements of the $R_{f}$ observables are given in \autoref{tab:EWPO_Zpole}. As shown in the table, all measurements are consistent with the SM within $1\sigma$. 

In our scenario, since the VLQ couples only to the down-type quarks, it gives rise to non-zero contributions to the decays of the $Z$ boson to all down-type quarks. The total hadronic decay width in the presence of NP can be written as: 
\begin{equation}
\Gamma_{\rm had}^{\rm tot} = \Gamma_{\rm had}^{\rm SM} + \delta \Gamma_{\rm had}\,.
\end{equation}
$\delta \Gamma_{\rm had}$ contains the pure NP contribution and the SM-NP interference term to the $Z$ decays to all the down quarks. Similarly, the total decay width to a particular fermion can also be written as:
\begin{equation}
\Gamma_{Z\to q \bar q (\ell \bar \ell)}^{\rm tot} = \Gamma_{Z\to q \bar q (\ell \bar{\ell})}^{\rm SM} + \delta \, \Gamma_{q (\ell)} \,.
\end{equation}
The ratio observable for the quarks, in the presence of the NP, and assuming that the effect is NP is smaller than the SM, will be given by:
\begin{eqnarray}
R_{q}^{\rm tot} = \frac{ \Gamma_{Z \to q \bar{q}}^{\rm SM} + \delta \, \Gamma_{q} }{ \Gamma_{\rm had}^{\rm SM}+ \delta \Gamma_{\rm had} } = R_{q}^{\rm SM} \left(1 + \frac{\delta \Gamma_{q}}{\Gamma^{\rm SM}_{Z \to q \bar q}} - \frac{\delta \Gamma_{\rm had}}{\Gamma_{\rm had}^{\rm SM}} \right) \,.
\end{eqnarray}
If the non-negligible NP contribution only arises for the decay to $b-$quark, the expression can be found from \cite{Kolay:2024wns}. On the other hand, the ratio observables for the leptons can be written in the presence of the NP as:
\begin{equation}\label{eq:Def_Rl}
R_{\ell}^{\rm tot} = R_{\ell}^{\rm SM} \left( 1 - \frac{\delta \, \Gamma_{\ell}}{\Gamma_{\rm Z \to \ell \ell }^{\rm SM} } + \frac{\delta \, \Gamma_{\rm had}}{\Gamma_{\rm had}^{\rm SM}} \right) \,. 
\end{equation}

The $Z$-pole asymmetry observables constitute a crucial class of precision electroweak measurements that probe the chiral structure of the $Zf\bar{f}$ interaction in $e^+e^- \to Z \to f\bar{f}$ processes near the $Z$ resonance, $\sqrt{s}\simeq m_Z$ \cite{ParticleDataGroup:2024cfk}. They were measured with high precision at LEP~\cite{ALEPH:2005ab} and SLC~\cite{SLD:2000leq}, with the latter benefiting from a longitudinally polarised electron beam. These observables arise from the parity-violating nature of the neutral weak current and are sensitive to the unequal couplings of the $Z$ boson to left- and right-handed fermions. They therefore provide stringent tests of the SM and sensitive probes of NP contributions to the effective $Zf\bar{f}$ couplings, as well as precise determinations of the effective weak mixing angle, $\sin^2\theta_{\rm eff}$. The relevant quantities include the asymmetry parameter $A_f$, the forward-backward asymmetry $A_{\mathrm{FB}}^{0,f}$, the left-right asymmetry $A_{\mathrm{LR}}$, and the left-right forward-backward asymmetry $A_{\mathrm{FB}}^{\mathrm{LR}}(f)$.

The basic parity-violating asymmetry parameter $A_f$ for a given fermion flavor $f$ characterizes the relative strength of the left- and right-handed $Zf\bar{f}$ couplings and is defined as
\begin{equation}
A_f = \frac{2g_V^f g_A^f}{(g_V^f)^2+(g_A^f)^2},
\label{eq:Af}
\end{equation}
where $g_{V,A}^f$ denote the vector and axial-vector, $Zf\bar{f}$ couplings, respectively. For charged leptons, the corresponding asymmetry parameter is denoted by $A_\ell$. The asymmetry parameter $A_f$ enters the various $Z$-pole asymmetry observables and provides a direct measure of the chiral structure of the $Zf\bar{f}$ interaction.

The left-right asymmetry $A_{\mathrm{LR}}$ was measured at the SLC by the SLD experiment using a longitudinally polarised electron beam and provides a direct probe of the initial-state electron asymmetry parameter $A_e$. It is defined as
\begin{equation}
A_{\mathrm{LR}} =\frac{\sigma_L-\sigma_R}{\sigma_L+\sigma_R},
\label{eq:Z_pole_ALR_AFB}
\end{equation}
where $\sigma_{L,R}$ denote the total cross sections for left- and right-handed incident electrons. At the $Z$ pole, the asymmetry is related to $A_e$ through
\begin{equation}
A_{\mathrm{LR}}=P_e A_e\,,
\label{eq:ALRPeAe}
\end{equation}
where $P_e$ denotes the longitudinal polarisation of the electron beam. The large beam polarisation achieved at SLD \cite{SLD:2000ujp}, with $P_e \simeq 75\%-80\%$, enabled a precise determination of $A_e$ with minimal dependence on the couplings of the final-state fermions.

At LEP-I, where the electron beam was unpolarised, the forward-backward asymmetry $A_{\mathrm{FB}}^f$ provides a probe of both the initial- and final-state fermion couplings. It is defined as
\begin{equation}
A_{\mathrm{FB}}^f=\frac{\sigma_F^f-\sigma_B^f}{\sigma_F^f+\sigma_B^f},
\label{eq:AFB}
\end{equation}
where $\sigma_F^f$ and $\sigma_B^f$ denote the cross sections for the final-state fermion $f$ produced in the forward and backward hemispheres, respectively, with the forward direction defined relative to the incoming electron. For a longitudinally polarised electron beam, the asymmetry is given by
\begin{equation}
A_{\mathrm{FB}}^f=\frac{3}{4}\frac{A_e+P_e}{1+A_eP_e} A_f,
\label{eq:AFB_polarised}
\end{equation}
where $P_e$ denotes the electron-beam polarisation. In the absence of beam polarisation ($P_e=0$), as at LEP-I, this expression reduces at the $Z$ pole to
\begin{equation}
A_{\mathrm{FB}}^{0,f}=\frac{3}{4}A_eA_f\,.
\label{eq:asymmetry_obs_relation}
\end{equation}
For heavy-quark final states, $A_{\mathrm{FB}}^{0,b}$ and $A_{\mathrm{FB}}^{0,c}$ probe the products $A_eA_b$ and $A_eA_c$, respectively. Thus, unlike $A_{\mathrm{LR}}$, which provides a direct probe of $A_e$, the forward-backward asymmetry is sensitive to both the initial- and final-state asymmetry parameters.

For a longitudinally polarized electron beam, the forward-backward asymmetry can be combined with beam polarization to construct the polarized left-right forward-backward asymmetry, $A_{\rm LR}^{\rm FB}(f)$. It is defined experimentally as
\begin{equation}
A_{\mathrm{FB}}^{\mathrm{LR}}(f)=\frac{\sigma_{\rm LF}^f-\sigma_{\rm LB}^f-\sigma_{\rm RF}^f+\sigma_{\rm RB}^f}{\sigma_{\rm LF}^f+\sigma_{\rm LB}^f+\sigma_{\rm RF}^f+\sigma_{\rm RB}^f}\,,
\label{eq:AFBLR}
\end{equation}
where the first and second subscripts denote the helicity of the incident electron and the forward or backward direction of the final-state fermion, respectively. At the $Z$ pole, assuming full beam polarization ($P_e = 1$) or normalizing by $P_e$, this observable isolates the final-state parameter:
\begin{equation}
A_{\mathrm{FB}}^{\rm LR}(f)=\frac{3}{4}A_f\,.
\label{eq:Z_pole_forbck_lftrt}
\end{equation}
Thus, unlike the unpolarised forward-backward asymmetry $A_{\mathrm{FB}}^{0,f}$, which depends on the product $A_eA_f$, $A_{\mathrm{FB}}^{\rm LR}(f)$ provides a direct probe of the final-state asymmetry parameter $A_f$. The SLD measurements of this observable for heavy-quark final states, in particular $b$ and $c$ quarks, therefore provide direct constraints on the corresponding chiral $Zf\bar{f}$ couplings.

The $\tau$-lepton polarisation measured at LEP provides a complementary probe of the leptonic asymmetry parameters. Reconstructed from the kinematic distributions of the $\tau$ decay products, the polarisation is given by
\begin{equation}
\mathcal{P}_{\tau}(\theta)=-\frac{A_{\tau}(1+\cos^2\theta)+2A_e\cos\theta}{    (1+\cos^2\theta)+2A_{\tau}A_e\cos\theta}\,,
\label{eq:Ptau}
\end{equation}
where $\theta$ denotes the angle between the outgoing $\tau^-$ and the incoming electron. The angular-averaged polarisation directly determines the $\tau$ asymmetry parameter,
\begin{equation}
\langle\mathcal{P}_{\tau}\rangle=-A_{\tau}\,,
\label{eq:Ptau_avg}
\end{equation}
while its angular dependence provides sensitivity to $A_e$. Thus, the $\tau$-polarisation measurements constrain both $A_{\tau}$ and $A_e$ through the polarisation and its angular dependence.

It is important to emphasize that the aforementioned $Z$-pole asymmetry observables are not mutually independent, as they are all functions of the fundamental parity-violating parameters $A_e$ and $A_f$. In our analysis, we therefore adopt $A_f$ as the primary observable rather than treating the individual experimental asymmetries-$A_{\rm LR}$, $A_{\rm FB}^{0,f}$, and $A_{\rm LR}^{\rm FB}(f)$-as separate inputs. This choice is motivated by several key advantages: $A_f$ directly characterizes the intrinsic chiral structure of the $Zf\bar{f}$ vertex, can be computed straightforwardly from modified couplings in NP scenarios, and eliminates explicit dependencies on experimental factors such as beam polarization. Furthermore, framing the constraints in terms of $A_f$ provides a clean, flavor-specific parameterization that avoids double-counting correlated experimental inputs while cleanly mapping NP contributions onto the corresponding $Zf\bar{f}$ coupling modifications.

In the presence of NP modifying the SM $Z f \bar{f}$ couplings by small shifts $\bar{g}_V^f = g_V^{\text{SM}} + \delta v_f$ and $\bar{g}_A^f = g_A^{\text{SM}} + \delta a_f$, the resulting asymmetry parameter $A_f$ is parameterized linearly (neglecting higher-order NP terms) as \cite{Kolay:2024wns, Kolay:2025jip, Kala:2025srq}:
\begin{align}
A_f^{\text{tot}} &= \frac{2(g_A^{\text{SM}} + \delta a_f)(g_V^{\text{SM}} + \delta v_f)}{(g_A^{\text{SM}} + \delta a_f)^2 + (g_V^{\text{SM}} + \delta v_f)^2}\, \nonumber \\&
\approx A_f^{\text{SM}}\left(1 + \frac{\delta v_f}{g_V^{\text{SM}}} + \frac{\delta a_f}{g_A^{\text{SM}}} - \frac{2 g_V^{\text{SM}} \delta v_f}{(g_V^{\text{SM}})^2 + (g_A^{\text{SM}})^2} - \frac{2 g_A^{\text{SM}} \delta a_f}{(g_V^{\text{SM}})^2 + (g_A^{\text{SM}})^2}\right)\,.
\end{align}
We have neglected the higher-order contribution of the NP. The mass effect of the quarks can be embedded in asymmetry observables as shown in \cite{Novikov:1999af}.
\begin{table}[htb!]
\begin{center}
\begin{tabular}{|c|c|c|}\hline
\rowcolor{gray!25}{\bf Observables}& {\bf Experiment} \cite{ParticleDataGroup:2024cfk,ALEPH:2005ab} & {\bf Standard Model}  \cite{ParticleDataGroup:2024cfk, Reina:2025suh} \\\hline\hline
\rowcolor{green!10} $R_{b}$   & $0.21629 \pm 0.00066$ & $0.21588 \pm 0.000010$ \\
\rowcolor{lime!15} $R_{\tau}$ & $20.767 \pm 0.025$   & $20.794 \pm 0.006$  \\ 
\rowcolor{cyan!7} $R_{\mu}$   & $20.767 \pm 0.025$   & $20.749 \pm 0.006$  \\ 
\rowcolor{magenta!10} $R_{e}$ & $20.767 \pm 0.025$   & $20.749 \pm 0.006$  \\\hline \hline

\rowcolor{orange!10}$A_{b}$ & $0.923 \pm  0.020$ & $0.934724 \pm  0.000040$  \\
\rowcolor{red!10} $A_{s}$ & $0.895 \pm  0.091$ & $0.935631 \pm 0.000040$ \\
\rowcolor{purple!12} $A_{\tau}$ & $ 0.1465 \pm 0.0033 $ & $0.14684 \pm 0.00048$  \\
\rowcolor{violet!15} $A_{\mu}$ & $0.1465 \pm 0.0033$& $0.14684 \pm 0.00048$  \\
\rowcolor{blue!10} $A_{e}$ & $0.1465 \pm 0.0033$ & $0.14684 \pm 0.00048$  \\ \hline 
\end{tabular}
\end{center}
\caption{Updated SM values \cite{Dubovyk:2019szj, Freitas:2014hra, Dubovyk:2018rlg, Reina:2025suh} and measurements\cite{ALEPH:2005ab,ParticleDataGroup:2024cfk} with $1\sigma$ uncertainties, of the electroweak $Z$-pole ratio and asymmetry observables.}
\label{tab:EWPO_Zpole}
\end{table}

The SM predictions and experimental measurements of the $Z$-pole observables are given in \autoref{tab:EWPO_Zpole}. The SM predictions are updated to the two-loop level (whenever applicable) \cite{Reina:2025suh}. All measurements are in agreement with the SM predictions within $1\sigma$ uncertainty.

For the asymmetry observables, various quantities of the form $A_{\mathrm{FB}}$ and $A_f$ can be constructed, as shown in \Eqs\eqref{eq:Af}-\eqref{eq:Z_pole_forbck_lftrt}. These observables are not completely independent, and their relations can be derived from the above expressions, as given in \eq\eqref{eq:asymmetry_obs_relation}. In our analysis, we consider only the observables $A_f$, for which precise measurements are available from SLD~\cite{SLD:2000leq}, LEP-I~\cite{ALEPH:2005ab}, and CMS~\cite{CMS:2023mgq}. For $A_\tau$, we use the most recent measurement from CMS~\cite{CMS:2023mgq}, which has a smaller uncertainty. For the electron and muon, the SM predictions are identical~\cite{ParticleDataGroup:2024cfk, Reina:2025suh}, as the charged-lepton masses are negligible compared to the $Z$-boson mass. The experimental values of $A_e$ and $A_\mu$ are available from both LEP-I~\cite{ALEPH:2005ab} and SLD~\cite{SLD:2000ujp}. Although the SLD measurements have smaller uncertainties, we use the LEP-I measurements in our analysis, as they are consistent with the SM predictions within $1\sigma$, whereas the SLD measurements deviate from the SM predictions by approximately $2\sigma$.
\begin{figure}[htb!]
\centering
\subfloat[]{\includegraphics[width=0.32\linewidth]{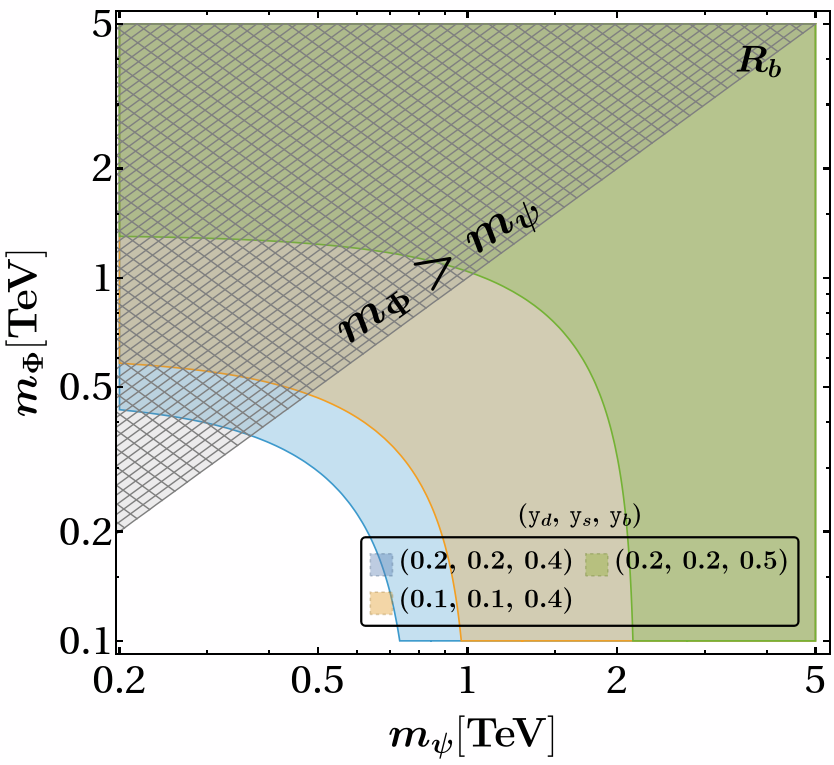}\label{fig:Zpole_ratio1}}~
\subfloat[]{\includegraphics[width=0.32\linewidth]{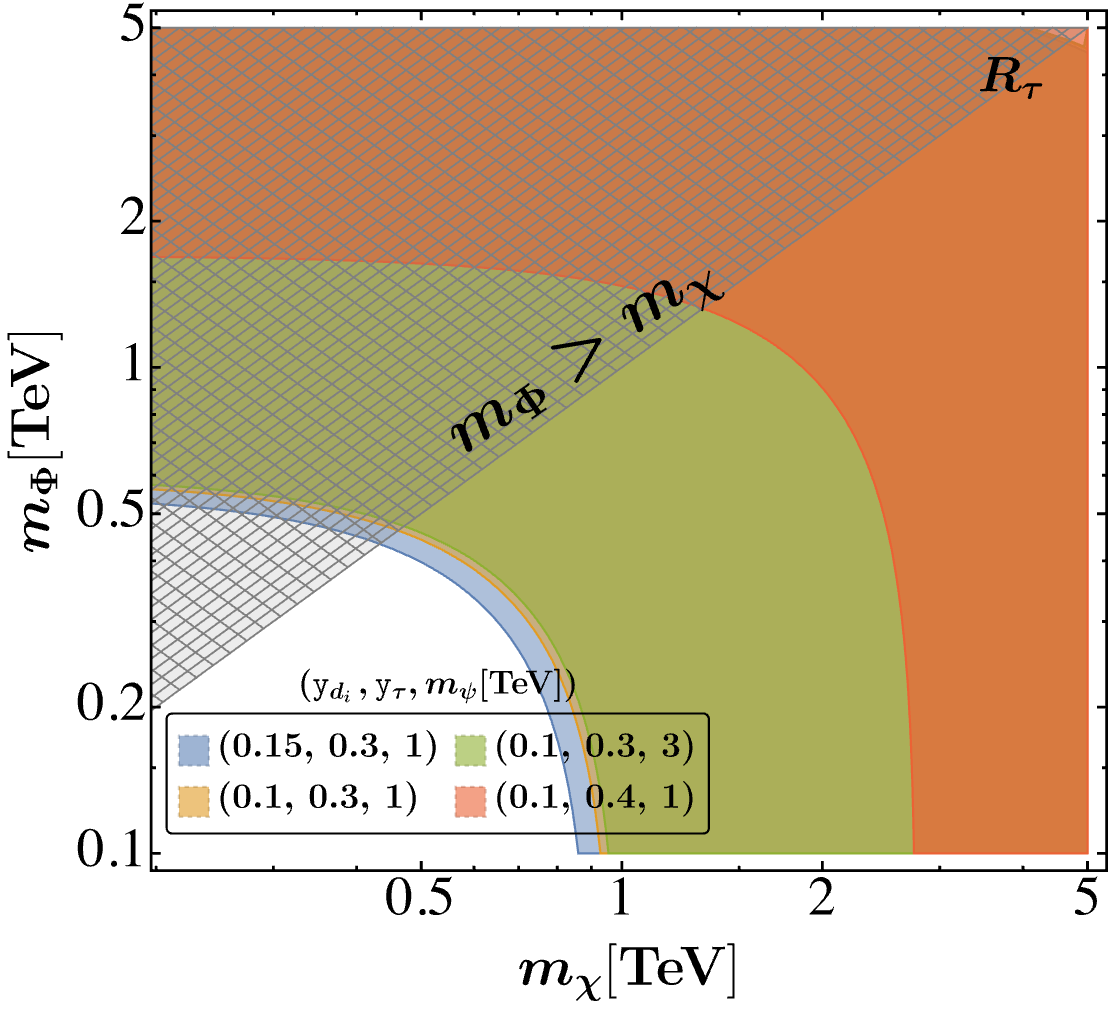}\label{fig:Zpole_ratio2}}~
\subfloat[]{\includegraphics[width=0.32\linewidth]{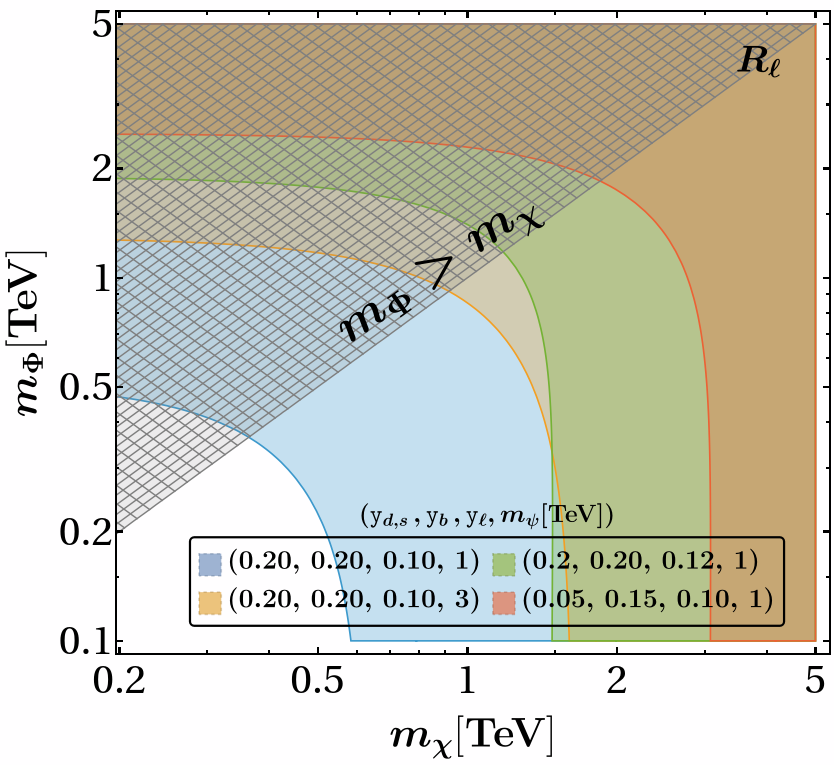}\label{fig:Zpole_ratio3}}
\caption{Allowed parameter space (colored band region) in the (VLF-DM) mass plane, from the $Z$-pole ratio observables $R_{f}$, varied within $1\sigma$ uncertainties of the observables. The observables are $R_b$ (left), $R_\tau$ (middle) and $R_{\ell}$ (right).}
\label{fig:Z-pole_ratio}
\end{figure}

\paragraph{\underline{Bounds on the Parameter Space from Ratio Observables}:}

Figure\,.~\ref{fig:Z-pole_ratio} shows the allowed parameter space from the $Z$-pole ratio observables defined in \eq\eqref{eq:def_Zpole}. Here, we present the constraints from the observables $R_{b}$, $R_{\tau}$, and $R_{\ell}$, with $\ell=(e\,,~\mu)$. The experimental measurements of all these observables are consistent with the SM predictions at the $1\sigma$ level. Figure\,.~\ref{fig:Zpole_ratio1} shows the allowed parameter space from the observable $R_{b}$ for a few benchmark points. Since $R_b$ depends on $\Gamma_{\rm had}$, it is also sensitive to variations in the other quark Yukawa couplings in addition to $\yb$. The higher-mass region remains allowed even for relatively large values of $\yb \sim 0.5$. As the values of the other couplings, $\yd$ and $\ys$, decrease, the allowed parameter space expands towards lower masses. In particular, for $(\yd\,,~\ys)<0.2$ and $\yb<0.4$, the region with $\mphi>0.4~\text{TeV}$ and $\mpsi>0.7~\text{TeV}$ is allowed. 

Figure\,.~\ref{fig:Zpole_ratio2} shows the allowed parameter space from the lepton $Z$-pole observable $R_{\tau}$. This observable depends on both the lepton and quark Yukawa couplings, as seen from \eq\eqref{eq:Def_Rl}, as well as on the VLQ mass. The dependence on the VLQ mass and quark couplings is relatively mild. The plots are shown for a few benchmark points with $\yd=\ys=\yb$, denoted by $\mathtt{y}_{d_i}$ in the plot legend. Although the variation is primarily driven by $\yta$, the quark Yukawa couplings also affect the allowed region. In contrast to the case of $R_b$, increasing $\mathtt{y}_{d_i}$ extends the allowed parameter space towards lower values of both $\mchi$ and $\mphi$, while increasing $\mpsi$ restricts it to larger values of $\mchi$ and $\mphi$. The opposite dependence on the quark Yukawa couplings arises because $R_{\tau}$ is defined as the inverse of $R_b$. On the other hand, larger values of $\yta$ favor larger $\mphi$ and $\mchi$. This behavior can be understood from the fact that the SM prediction for $R_{\tau}$ is larger than the experimental measurement: increasing the quark couplings enhances $R_{\tau}$, whereas increasing $\yta$ suppresses it. Consequently, for smaller quark couplings and comparatively larger $\yta$, lower DM masses lead to smaller values of $R_{\tau}$, favoring lower $\mchi$ and lighter VLLs. Nevertheless, for $\yta \lesssim 0.15$, the TeV region of the $(\mchi\,,~\mphi)$ plane remains allowed.

Figure\,.~\ref{fig:Zpole_ratio3} shows the allowed parameter space from the observable $R_{\ell}$, where $\ell=(e,\mu)$. Since the electron and muon masses are negligible compared to the $Z$-boson mass, both observables exhibit identical dependence on the model parameters. Consequently, the SM predictions and the corresponding experimental measurements are identical in both cases. Therefore, the dependence of $R_{\ell}$ on the relevant parameters is similar to that of $R_{\tau}$. For $\yq=0.2$ and $\yl\leq 0.1$, the TeV region of the $(\mchi\,,~\mphi)$ mass plane is allowed.

\paragraph{\underline{Bounds on the Parameter Space from Asymmetry Observables}:}

\begin{figure}[htbp]
\centering
\subfloat[]{\includegraphics[width=0.32\linewidth]{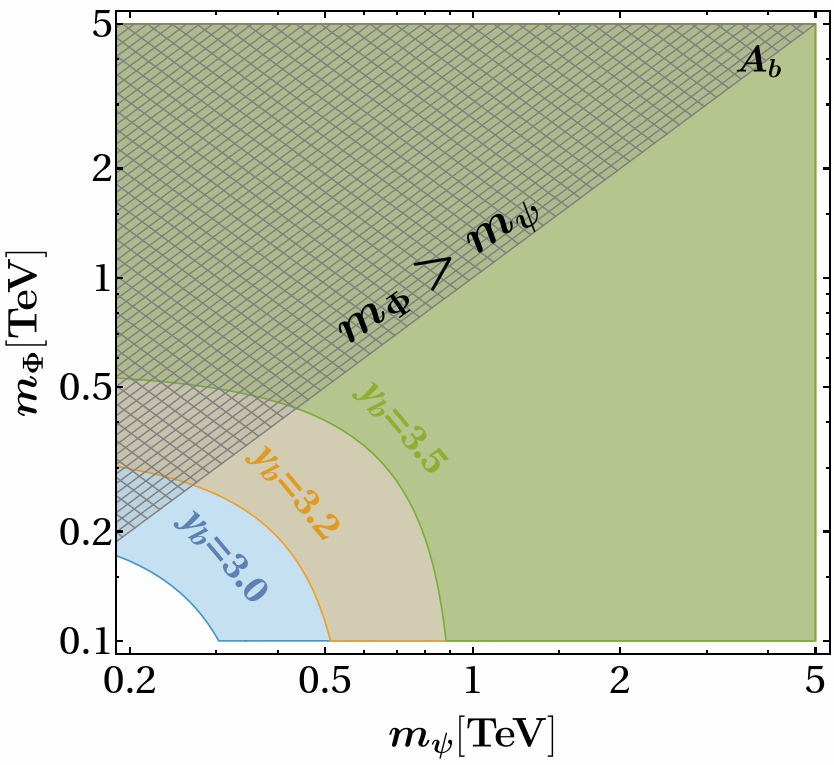}\label{fig:Zpole_asymm1}}~
\subfloat[]{\includegraphics[width=0.32\linewidth]{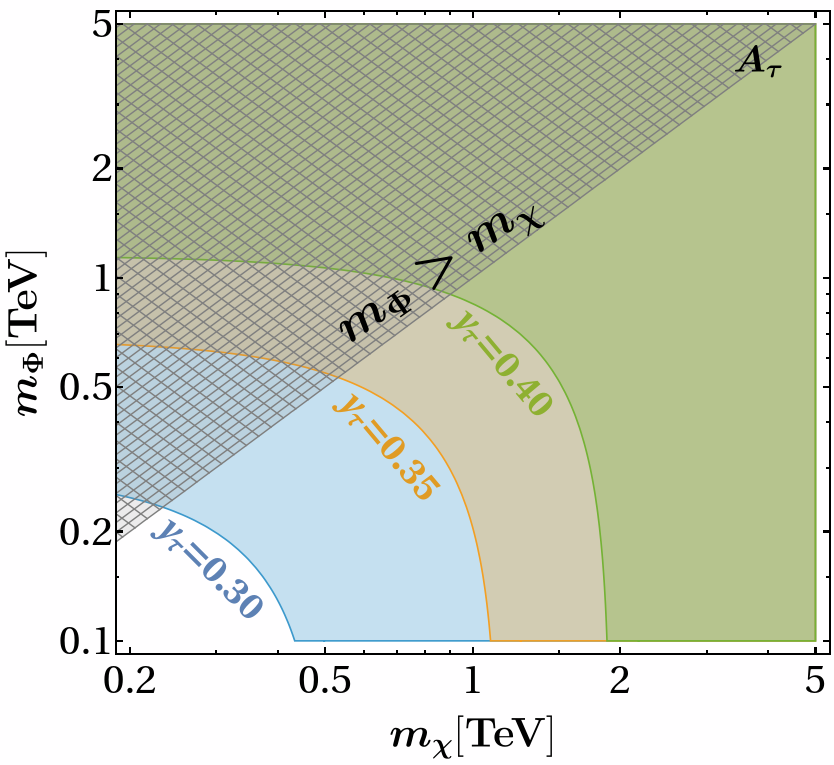}\label{fig:Zpole_asymm2}}~
\subfloat[]{\includegraphics[width=0.32\linewidth]{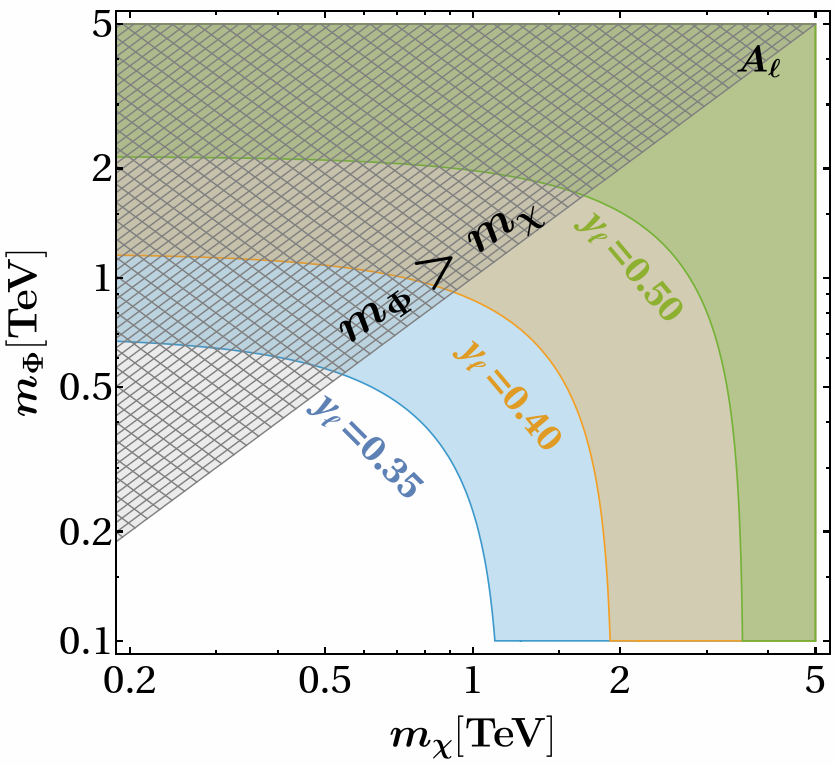}\label{fig:Zpole_asymm3}} ~
\caption{Allowed parameter space (colored band region) in the (VLF-DM) mass plane, from the $Z$-pole asymmetry observables $A_{f}$, varied within $1\sigma$ error of the observables. The observables are $A_b$, $A_\tau$, $A_{\ell}$ (from left to right).}
\label{fig:Zpole_asymm}
\end{figure}

The asymmetry observables, $A_f$, constitute another important set of $Z$-pole electroweak precision observables. The corresponding SM predictions and experimental measurements are listed in \autoref{tab:EWPO_Zpole}, while the allowed parameter space obtained from these observables is shown in \Fig\ref{fig:Zpole_asymm} for a few representative benchmark points. The constraints from $A_b$ are presented in \Fig\ref{fig:Zpole_asymm1} for $\yb=(3.0,\,3.2,\,3.5)$. Since $A_b$ depends only on the modifications to the vector and axial-vector couplings of the $Zb\bar b$ vertex, it is sensitive only to $\yb$, $\mpsi$, and $\mphi$, and, unlike $R_b$, does not provide any meaningful constraint for $\yb\lesssim 3$. The observable $A_s$ does not impose any additional constraints on the parameter space, and the whole region remains allowed even for very large Yukawa coupling; hence, we have not shown it explicitly. The constraints from $A_\tau$ are shown in the $(\mchi\,,~\mphi)$ plane in \Fig\ref{fig:Zpole_asymm2}, while those from $A_\ell$ is presented separately in \Fig\ref{fig:Zpole_asymm3}. Similar to $R_\ell$, the observables $A_\mu$ and $A_e$ are denoted collectively as $A_\ell$, since their SM predictions and experimental values are identical due to the negligible charged-lepton masses compared to the $Z$-boson mass. Here, the experimental value of $A_\ell$ is taken from the LEP-I \cite{ALEPH:2005ab} experiment, as it agrees with the SM prediction within $1\sigma$. Overall, these asymmetry observables yield weaker constraints than the ratio observables, allowing TeV-scale DM and mediator masses for $\yta\lesssim0.18$, $\ymu\lesssim0.35$, and $\ye\lesssim0.30$. Consequently, they do not impose any additional constraints on the parameter space beyond those already obtained from the ratio observables. 

\paragraph{\underline{flavor Violating Decays of $Z$ Boson}:}

Flavor-violating decays of the $Z$ boson, involving both quarks and charged leptons, provide important probes of physics beyond the SM. In the SM, these processes are absent at the tree level and arise only through higher-order loop corrections, resulting in highly suppressed branching ratios. The quark flavor changing neutral current (FCNC) decays, $Z\to d_i\bar{d}_j$, are induced through one-loop diagrams mediated by the $W$ boson and up-type quarks, yielding branching ratios of $\mathcal{O}(10^{-7}$-$10^{-9})$~\cite{Aranda:2020tqw, Eilam:2002as, Bernabeu:1986pk}. In contrast, the charged LFV decays are further suppressed by the tiny neutrino masses, resulting in branching ratios of $\mathcal{O}(10^{-54})$~\cite{Illana:2000ic}. In the present model, the vector-like fermions generate additional loop-level contributions to the flavor-violating $Zf_i\bar{f}_j$ vertices, giving rise to both quark FCNC decays, $Z\to q_i\bar{q}_j$, and charged LFV decays, $Z\to \ell_i\bar{\ell}_j$, as illustrated in \fig\ref{fig:Feynman_Zfifj}. Experimentally, stringent upper limits are available for the LFV decays of the $Z$ boson from several experiments and are summarized in \autoref{tab:Z_flavor_violating}. In contrast, no updated direct searches exist for the quark FCNC modes, and the only direct limits are those from the LEP and SLD experiment~\cite{Atwood:2002ke, Fuster:1999dj}, which reported the relatively weak bound $\mathcal{B}(Z\to b\,\bar{d}_j)<1.8\times10^{-3}$. In the following, we study the cLFV decays $Z\to \ell_i \bar{\ell}_j$ and derive the corresponding constraints on the model parameter space.
\begin{table}[htb!]
\centering
\begin{tabular}{|c|c|}\hline
\rowcolor{gray!40} Rare $Z$ and $H$ Decays & Branching Ratio Upper Limits \\\hline\hline
\rowcolor{lime!10}$Z \rightarrow e^\pm \mu^\mp $ & $1.9 \times 10^{-7}$\, \quad \cite{CMS:2025wqy} \\ 
\rowcolor{brown!10}$Z\to e^{\pm}\tau^{\mp}$ &   $5.0 \times 10^{-6}$\,\quad \cite{ATLAS:2021bdj}  \\ 
\rowcolor{green!10}$Z \rightarrow \mu^\pm \tau^\pm $ &  $6.5 \times 10^{-6}$\, \quad \cite{ATLAS:2021bdj} \\
\rowcolor{blue!10}$h\to e^{\pm}\mu^{\mp} $  &$4.4\times 10^{-5}$ \,\cite{CMS:2023pte}\\ 
\rowcolor{cyan!10}$h\to e^{\pm}\tau^{\mp} $  & $2.0 \times 10^{-3}\,$ \cite{ATLAS:2023mvd} \\ 
\rowcolor{olive!10}$h\to \mu^{\pm}\tau^{\mp} $& $1.5\times 10^{-3} $ \cite{ATLAS:2023mvd} \\ 
\hline
\end{tabular}
\caption{Current experimental upper limits at 95\% C.L. on the branching ratios of the LFV decays of $Z$- and Higgs-boson considered in this work.}
\label{tab:Z_flavor_violating}
\end{table}

\begin{figure}[htbp]
\centering
\subfloat[]{\includegraphics[width=0.33\linewidth]{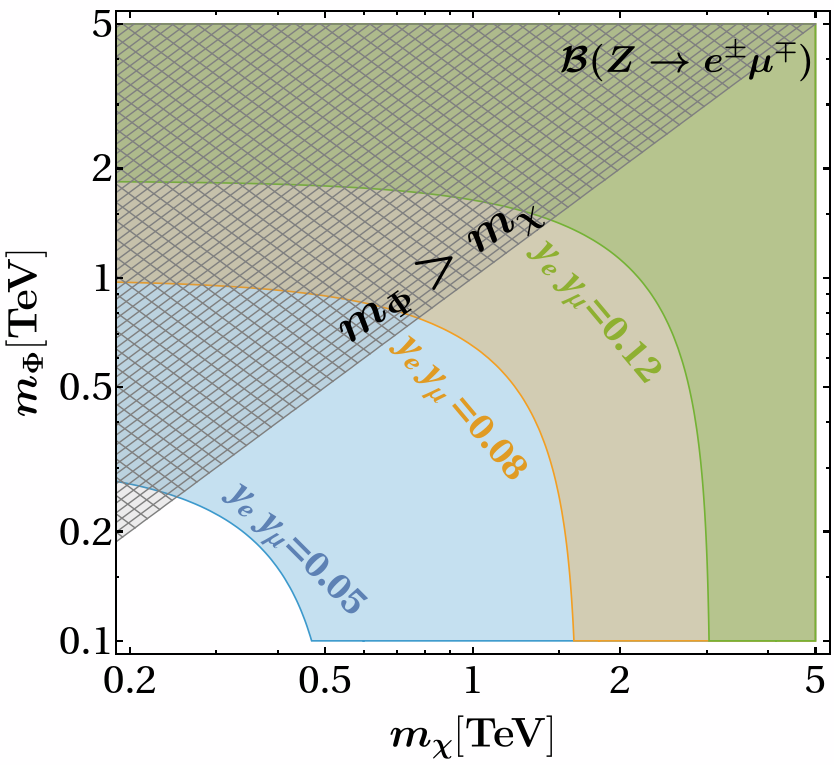}\label{fig:Z2fifj1}}~~
\subfloat[]{\includegraphics[width=0.33\linewidth]{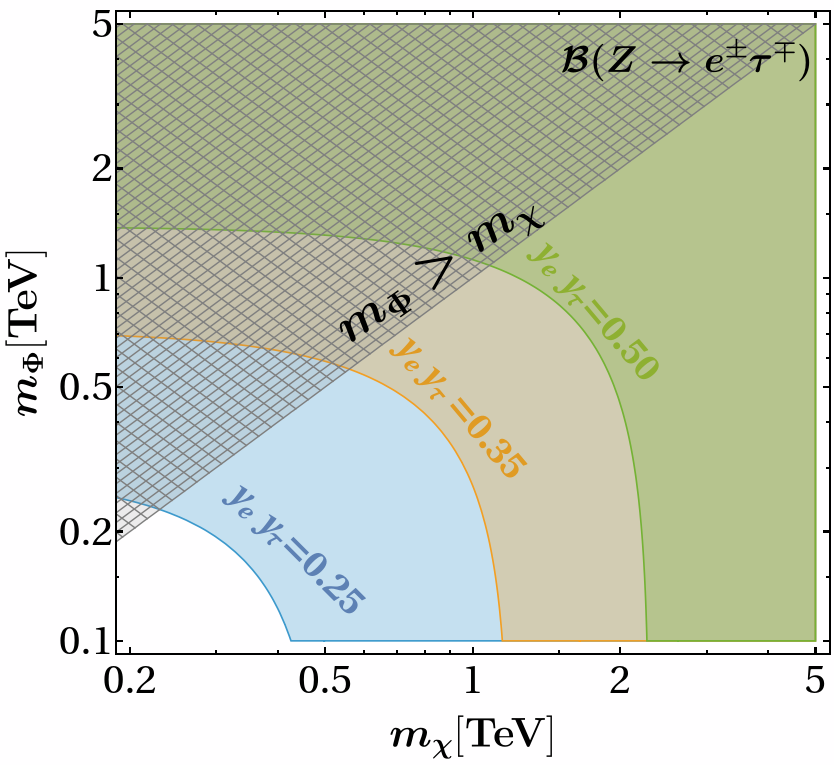}\label{fig:Z2fifj2}}~~
\subfloat[]{\includegraphics[width=0.33\linewidth]{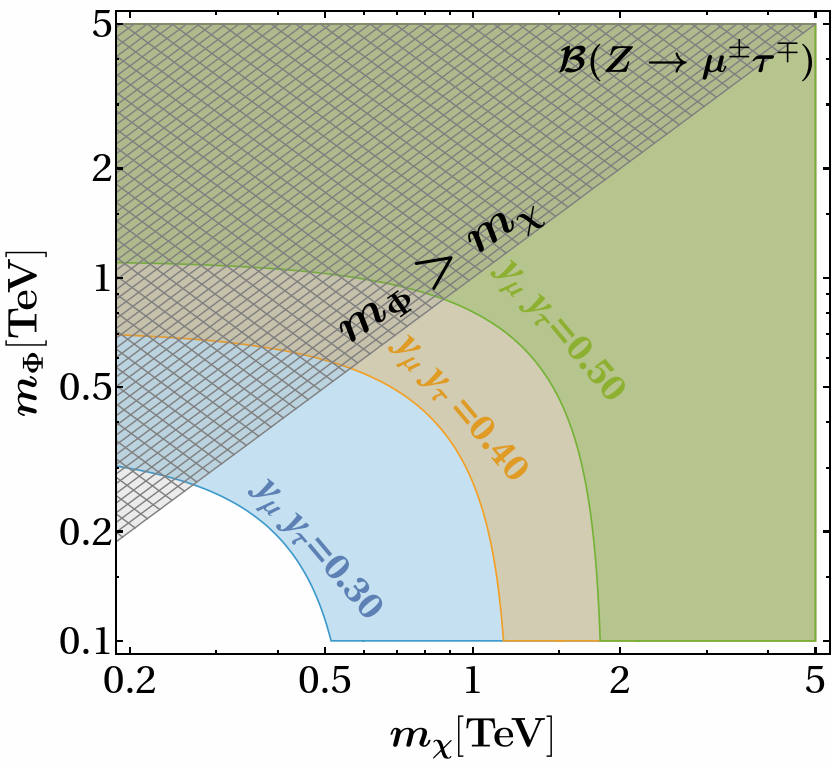}\label{fig:Z2fifj3}}
\caption{Allowed parameter space (colored band region) at $95\%$ C.L. in the (VLF-DM) mass plane, from the flavor-violating decays of the $Z$-boson to the charged leptons. The processes are shown (from left to right) $Z\to (e^\pm\mu^\mp,\tau^\pm e^\mp,\tau^\pm\mu^\mp)$.}
\label{fig:Z2fifj}
\end{figure}
Figure\,.~\ref{fig:Z2fifj} shows the allowed parameter space obtained from the LFV decays $Z\to \ell_i^\pm \ell_j^\mp$. Among these processes, $Z\to e^\pm \mu^\mp$ provides the strongest constraint, as it has the most stringent experimental upper bound on its branching ratio. The experimental upper limits on the other LFV decays are approximately one order of magnitude weaker, leading to comparatively weaker constraints on the parameter space. We find that TeV-scale VLL and DM masses are allowed for $\mathtt{y}_{\ell_i}^{}\mathtt{y}_{\ell_j}^{}\lesssim (0.07,\,0.35,\,0.40)$ for the $(e^\pm\mu^\mp,\tau^\pm e^\mp,\tau^\pm\mu^\mp)$ channels, respectively. We do not show the allowed parameter space from flavor-changing quark decays because the current experimental upper limits on their branching ratios are too weak to provide meaningful constraints. Therefore, even for $\yq\sim 1$, the entire parameter space considered in this analysis remains allowed.
\subsubsection{Higgs decays to Fermions}
The presence of the VLFs also induces loop-level contributions to the Higgs decays $h \to f_i \bar{f}_j$, including both the flavor-conserving ($i=j$) and flavor-violating ($i\neq j$) channels. The corresponding Feynman diagram is shown in \fig\ref{fig:Feynman_H2fifj}. In the SM, the flavor-violating quark decays arise at the one-loop level through flavor-changing neutral current interactions, leading to branching ratios of $\mathcal{O}(10^{-7}-10^{-8})$ for the $h\to d_i\bar{d}_j$ channels \cite{Benitez-Guzman:2015ana}. The corresponding charged lepton flavor-violating Higgs decays are negligible in the SM. Experimentally, the $95\%$ C.L. upper bounds on the flavor-violating Higgs decays into charged leptons are of $\mathcal{O}(10^{-3}-10^{-5})$ \cite{CMS:2023pte, ATLAS:2023mvd, CMS:2021rsq}, as mentioned in \autoref{tab:Z_flavor_violating}.

In our scenario, the loop contribution is suppressed by the external fermion mass. Furthermore, as evident from the Feynman diagram in \Fig\ref{fig:Feynman_H2fifj}, the amplitude is proportional to the Higgs-portal coupling, $\lphiH$, which is taken to be of $\mathcal{O}(10^{-3})$ in our analysis to satisfy the dark-matter direct detection constraints, which will be discussed in the subsequent section, introducing an additional suppression. Consequently, both the flavor-conserving and flavor-violating Higgs decays receive negligible NP contributions. Even by taking all the new couplings to be of $\mathcal{O}(1)$ and assuming $\lambda_{\Phi H}\sim\mathcal{O}(1)$, we obtain $\BR(h\to f_i\bar{f}_j)\sim\mathcal{O}(10^{-13})$ for the flavor-violating channels ($i\neq j$), with both the DM and VLF masses below $1~\TeV$, which remains several orders of magnitude below the current experimental sensitivities. Therefore, neither the flavor-conserving nor the flavor-violating Higgs decays provide any meaningful constraints on the parameter space of the model, and we do not present dedicated numerical analyses or parameter-space plots for these observables.
\subsection{Combined Flavor Constraints}\label{sec:all_flav_ewpos}
\begin{figure}[htb!]
\centering
\includegraphics[width=0.8\linewidth]{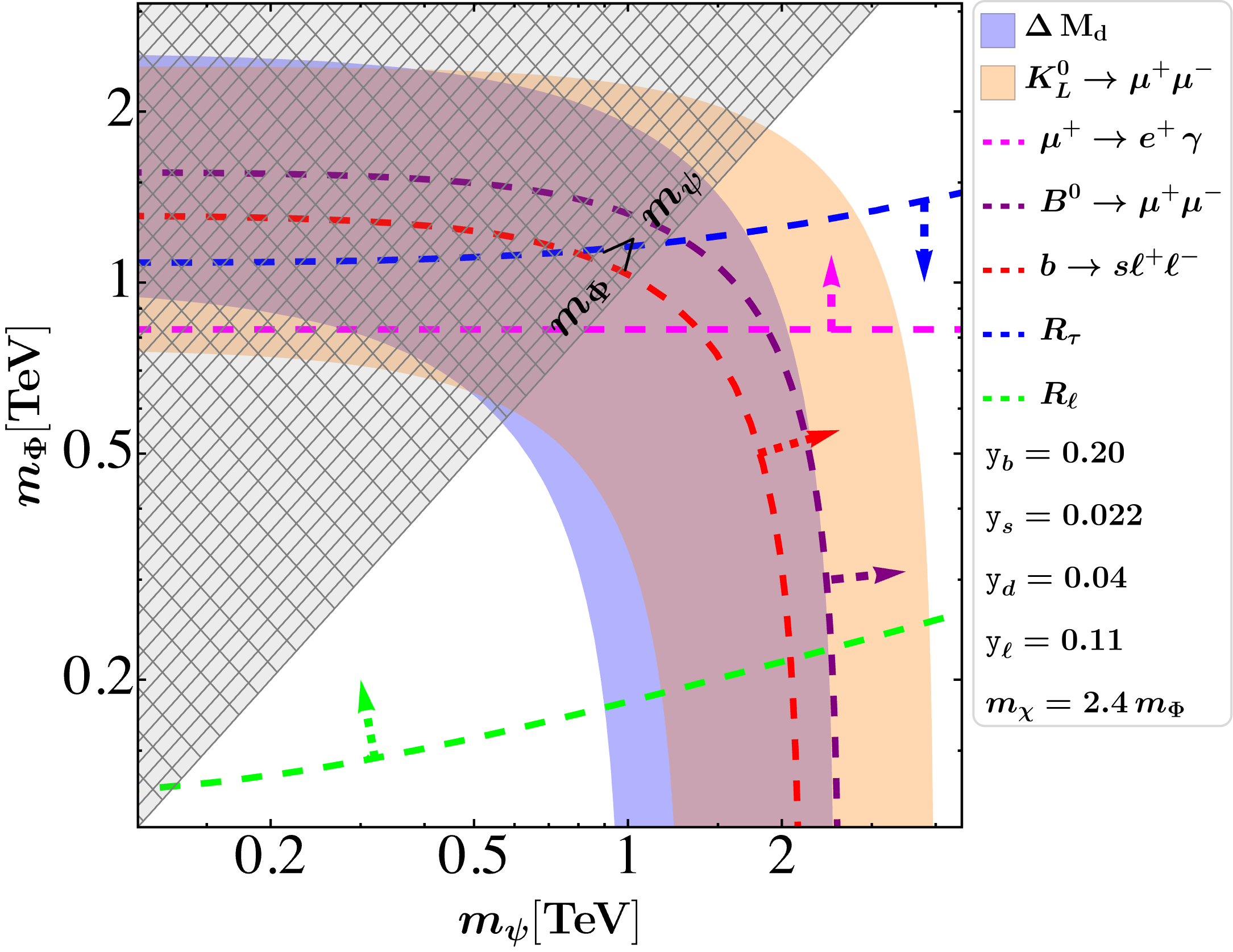}
\caption{
Shaded regions and arrows indicate the parameter space allowed by individual flavor and electroweak precision constraints, which significantly restrict the available space. Results correspond to the benchmark points shown on the right side of the figure with $\ell=(\tau, \mu, e)$. Observables with measured central values include $1\sigma$ uncertainties, while upper bounds are applied at $90\%$ or $95\%$ C.L., as noted. Observables imposing no constraints are omitted.}
\label{fig:all_flav_ewpo_combined}
\end{figure}
In the previous sections, we have discussed the relevant processes in detail and examined the contributions of the present model to these processes and their resulting constraints on the parameter space. For VLF and DM masses in the $\sim$TeV range, the different processes constrain the Yukawa couplings $\mathtt{y}_f^{}$ to be within the range $\sim\mathcal{O}(0.1-1.0)$. To identify a common parameter space that is simultaneously allowed by all the considered processes, it is therefore important to identify a suitable benchmark point for which the corresponding constraints admit an overlapping region of the mass parameters. Such a benchmark point allows us to consistently study the combined constraints from all the processes and identify the surviving parameter space.

Figure\,.~\ref{fig:all_flav_ewpo_combined} shows the allowed parameter space obtained from the relevant flavor and electroweak constraints for the benchmark point $\yb = 0.20$, $\ys=0.022$, $\yd = 0.04$, $\yl=\yta =\ymu =\ye = 0.11$, and $\mchi = 2.4\mphi$. The purple dashed line represents the lower bound on the mass parameter from the dileptonic decay of the $B^0$ meson, with the arrow indicating that the region towards larger masses is allowed. Similarly, the red dashed line and arrow denote the constraint from the combined analysis of $b \to s\ell^+\ell^-$ processes, which is discussed in detail in the corresponding section. The blue dashed line and arrow show the constraint from the $R_{\tau}$ observable. As discussed previously, $R_{\tau}$ depends predominantly on the coupling $\yta$ and the mass of the VLL $\mchi$, with a subleading dependence on the VLQ parameters. Among the considered observables, $\mu \to e\gamma$ provides the strongest constraint on the parameter space. The stringent upper bound from the recent MEG-II result~\cite{MEGII:2025gzr} excludes a significant portion of the parameter space, as indicated by the magenta line, while the region with $\mphi > 0.82$~TeV remains allowed by this constraint. In addition, meson mixing and the decay $K_L \to \mu^+\mu^-$ impose both lower and upper bounds on the parameter space when the corresponding data are considered with their $1\sigma$ uncertainties. 
It is also worth noting that, unlike the other observables, $R_\tau$ places an upper bound on the allowed mass parameter space, as discussed in Sec\,.~\ref{sec:Z_pole}. In our analysis, the quark couplings are chosen to be sufficiently small to satisfy the low-energy flavor constraints. For relatively small $\yta=0.11$, lower DM masses ($\mphi\lesssim 1.0~\TeV$) are therefore preferred to remain within the experimentally allowed $1\sigma$ range of $R_\tau$.
Combining all these constraints, we find that, for this benchmark point, the final allowed region is restricted to $\mpsi \leq 1.5$~TeV and, along the $y$-axis, to the region above the purple and magenta boundary, corresponding to the overlap of the orange and blue shaded regions.
\section{Dark Matter Phenomenology}
\label{sec:darkmatter}
The viability of a BSM scenario is not determined solely by its ability to reproduce the observed dark matter relic abundance. It must also satisfy the constraints from current DM search experiments while predicting regions of parameter space that can be probed by future DM detection experiments. We recall that the complex scalar $\Phi$ is the dark matter candidate in the present model, as it is the lightest $\Zthree$-odd field and is therefore stable.
\begin{itemize}
\item \textbf{Dark Matter relic density \\}
The Planck measurements of the CMB anisotropies, analyzed within the framework of the standard $\Lambda\rm CDM$ cosmological model, yield a present-day (temperature $\sim 2.7255 \rm~ K$) cold dark matter relic density is \cite{Planck:2018vyg}:
\begin{align}
\omgdm h^2=0.1200\pm 0.0012\,,
\end{align}
where $h=0.6736\pm 0.0054$ is a dimensionless Hubble parameter.

In this study, we restrict ourselves to dark matter masses up to $4~\TeV$, so that freeze-out always occurs after EWSB, $\TFO\simeq \dfrac{\mphi}{25}\lesssim\TEW$, after which SM particles acquire mass through the spontaneous symmetry breaking of the SM Higgs field. Furthermore, the DM relic density is evaluated using the \texttt{micrOMEGAs} package \cite{Alguero:2023zol}.
\item \textbf{Invisible decay \\}
Precision measurements of the total decay widths of the $Z$ and Higgs boson provide stringent constraints on possible invisible decay modes, which are particularly important for low-mass dark matter candidates with $m_{\rm DM} \lesssim m_Z/2~(m_h/2)$, respectively. The LEP measurement of the invisible $Z$-boson width, $\Gamma_{\rm inv}^{\rm exp}=499.3\pm1.5~\MeV$~\cite{ATLAS:2023ynf,ALEPH:2005ab,CMS:2022ett}, places stringent constraints on any additional invisible decay channels predicted by new physics. For the Higgs boson, a combined study of Run I+II results provides an upper limit on the invisible Higgs boson branching ratio of ${\BR}(h \to {\rm inv}) < 0.107$ is observed at 95\% CL \cite{ATLAS:2023tkt}. However, in our focused analysis regime, since the direct detection limits are more stringent for scalar dark matter, we do not incorporate these constraints further.
\item \textbf{Dark Matter direct detection constraints \\}
The most stringent constraints on the DM–nucleon scattering cross section are obtained from the LUX-ZEPLIN (LZ) experiment \cite{LZ:2024zvo}, while the proposed XLZD experiment \cite{XLZD:2024nsu}, with higher sensitivity, is expected to improve this limit by up to two orders of magnitude compared to the present observational bounds.

In the present model, the CSDM interacts with down-type quarks inside the nucleon through the Yukawa interaction (see \fig\ref{feyn:wimp-dd1}), in addition to the Higgs portal interaction (see \fig\ref{feyn:wimp-dd2}). Consequently, for sufficiently large Yukawa couplings ($\yd$), CSDM-nucleon scattering dominates the DM-nucleus recoil rate.
Since the total DM relic abundance consists of equal contributions from DM particles and antiparticles, only half of the local DM density is attributed to the CSDM particle. Accordingly, the effective spin-independent direct detection cross section is given by,
\begin{align}
\sphiN = \frac{1}{2}\,\sigma_{\Phi \mathbb{N}}^{\rm SI}=\frac{1}{2}\,\sigma_{\Phi^* \mathbb{N}}^{\rm SI}\,.
\end{align}
One should note that the DM-nucleon scattering cross section, mediated through both the $s$- and $t$-channel processes, is directly proportional to the Yukawa coupling ($\mathtt{y}_d^4$) and the Higgs portal coupling ($\lambda_{\Phi\rm  H}^2$), while being inversely proportional to the mediator mass ($m_\psi^{-4}$).

\begin{figure}[htb!]
\centering
\begin{adjustbox}{width=\textwidth}
\begin{tcolorbox}[colback=gray!5, colframe=black!10, boxrule=0.8pt, arc=4mm, boxsep=0pt, left=0pt, right=0pt, top=0pt, bottom=0pt, width=0.46\textwidth, halign=center]
\subfloat[]{
\begin{tikzpicture}
\begin{feynman}
\vertex (a);
\vertex[above left=0.9cm and 0.9cm  of a] (a1){\(\Phi\)};
\vertex[below left=0.9cm and 0.9cm  of a] (a2){\(d\)}; 
\vertex[right=0.9cm of a] (b); 
\vertex[above right=0.9cm and 0.9cm of b] (c1){\(\Phi\)};
\vertex[below right=0.9cm and 0.9cm of b] (c2){\(d\)};
\diagram*{
(a1) -- [line width=0.35mm, charged scalar, arrow size=0.8pt,style=black] (a),
(a2) -- [line width=0.35mm,fermion, arrow size=0.8pt,style=black] (a),
(a) -- [line width=0.35mm,fermion,arrow size=0.8pt, edge label'={\(\color{black}{\psi}\)}, style=black!50] (b),
(b) -- [line width=0.35mm,charged scalar, arrow size=0.8pt] (c1),
(b) -- [line width=0.35mm,fermion, arrow size=0.8pt] (c2)};
\node at (a)[circle,fill,style=black,inner sep=1pt]{};
\node at (b)[circle,fill,style=black,inner sep=1pt]{};
\end{feynman}
\end{tikzpicture}
\label{feyn:wimp-dd1}}
\subfloat[]{\begin{tikzpicture}
\begin{feynman}
\vertex (a);
\vertex[above left=0.45cm and 0.9cm of a] (a1){\(\Phi\)};
\vertex[above right=0.45cm and 0.9cm  of a] (a2){\(\Phi\)}; 
\vertex[below = 0.9cm of a] (b); 
\vertex[below left=0.45cm and 0.9cm of b] (c1){\(\rm \mathbb{N}\)};
\vertex[below right=0.45cm and 0.9cm of b] (c2){\(\rm \mathbb{N}\)};
\diagram*{
(a1) -- [line width=0.35mm,charged scalar, arrow size=0.8pt,style=black] (a) -- [line width=0.35mm,charged scalar, arrow size=0.8pt,style=black] (a2),
(b) -- [line width=0.35mm, scalar, arrow size=0.8pt, edge label'={\(\color{black}{h}\)}, style=black!50] (a) ,
(c1) -- [line width=0.35mm,fermion, arrow size=0.8pt] (b),
(b) -- [line width=0.35mm,fermion, arrow size=0.8pt] (c2)};
\node at (a)[circle,fill,style=black,inner sep=1pt]{};
\node at (b)[circle,fill,style=black,inner sep=1pt]{};
\end{feynman}
\end{tikzpicture}
\label{feyn:wimp-dd2}}
\end{tcolorbox}
\begin{tcolorbox}[colback=gray!5, colframe=black!10, boxrule=0.8pt, arc=4mm, boxsep=0pt, left=0pt, right=0pt, top=0pt, bottom=0pt, width=0.5\textwidth, halign=center]
\subfloat[]{\begin{tikzpicture}
\begin{feynman}
\vertex (a);
\vertex[above left=0.9cm and 0.9cm  of a] (a1){\(\Phi\)};
\vertex[below left=0.9cm and 0.9cm  of a] (a2){\(\Phi\)}; 
\vertex[right=0.9cm of a] (b); 
\vertex[above right=0.9cm and 0.9cm of b] (c1){\(\tau(b)\)};
\vertex[below right=0.9cm and 0.9cm of b] (c2){\(\tau(b)\)};
\diagram*{
(a1) -- [line width=0.35mm, charged scalar, arrow size=0.8pt,style=black] (a),
(a) -- [line width=0.35mm, charged scalar, arrow size=0.8pt,style=black] (a2),
(a) -- [line width=0.35mm, scalar,arrow size=0.8pt, edge label'={\(\color{black}{\rm h}\)}, style=black!50] (b),
(c1) -- [line width=0.35mm,fermion, arrow size=0.8pt] (b),
(b) -- [line width=0.35mm,fermion, arrow size=0.8pt] (c2)};
\node at (a)[circle,fill,style=black,inner sep=1pt]{};
\node at (b)[circle,fill,style=black,inner sep=1pt]{};
\end{feynman}
\end{tikzpicture}
\label{feyn:wimp-id1}}
\subfloat[]{\begin{tikzpicture}
\begin{feynman}
\vertex (a);
\vertex[above left=0.45cm and 0.9cm of a] (a1){\(\Phi\)};
\vertex[above right=0.45cm and 0.9cm  of a] (a2){\(\tau(b)\)}; 
\vertex[below = 0.9cm of a] (b); 
\vertex[below left=0.45cm and 0.9cm of b] (c1){\(\Phi\)};
\vertex[below right=0.45cm and 0.9cm of b] (c2){\(\tau(b)\)};
\diagram*{
(a1) -- [line width=0.35mm, charged scalar, arrow size=0.8pt,style=black] (a),
(a2) -- [line width=0.35mm, fermion, arrow size=0.8pt,style=black] (a),
(a) -- [line width=0.35mm, fermion, arrow size=0.8pt, edge label={\(\color{black}{\chi(\psi)}\)}, style=black!50] (b) ,
(b) -- [line width=0.35mm, charged scalar, arrow size=0.8pt] (c1),
(b) -- [line width=0.35mm, fermion, arrow size=0.8pt] (c2)};
\node at (a)[circle,fill,style=black,inner sep=1pt]{};
\node at (b)[circle,fill,style=black,inner sep=1pt]{};
\end{feynman}
\end{tikzpicture}
\label{feyn:wimp-id2}}
\end{tcolorbox}
\end{adjustbox}
\caption{The Feynman diagrams are related to the direct (\ref{feyn:wimp-dd1}, \ref{feyn:wimp-dd2}) and indirect (\ref{feyn:wimp-id1}, \ref{feyn:wimp-id2}) detection of WIMPs. Similar Feynman diagrams are also possible for $\Phi^*$ in the context of direct detection prospects.}
\label{fig:feynman-dd-id}
\end{figure}
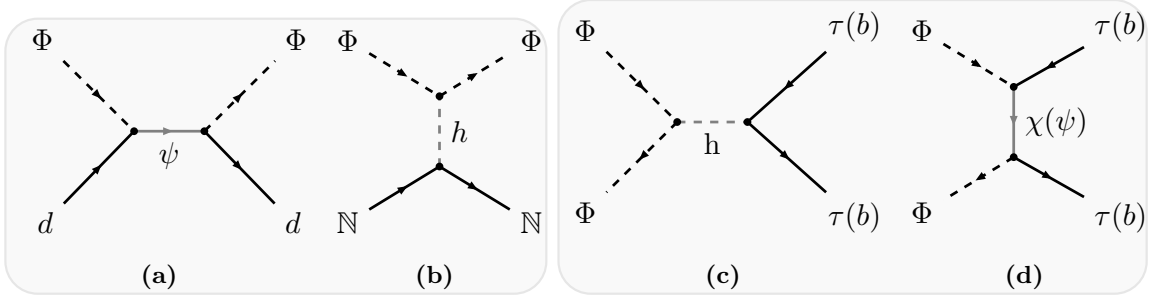
\item \textbf{Dark Matter Indirect detection constraints \\}
The observation of an excess of positrons, gamma rays, antiprotons, or neutrinos over the corresponding SM background predictions may provide evidence for dark matter annihilation or decay into these particles. These signals can originate not only from direct annihilation or decay but also from prompt particle production, cascade decays of intermediate states, and final-state radiation, all of which contribute to the observable cosmic-ray and gamma-ray spectra. In the case of DM annihilation or decay into quarks and gluons, the quarks hadronise into hadrons, which subsequently decay into stable SM particles. In particular, the decay of neutral pions into two photons constitutes the dominant source of continuum gamma rays, while charged pions decay into muons and muon neutrinos, followed by muon decay, producing positrons and neutrinos. Antiprotons are also generated during the hadronisation process. In addition, neutrinos can originate from direct DM self-annihilation, cascade decays of intermediate states, or radiative decay channels. These stable particles serve as the primary messengers in indirect dark matter searches \cite{Conrad:2017pms, Gaskins:2016cha, Strigari:2018utn, Arguelles:2019ouk}.

The CSDM annihilate in the Galactic core and produces an excess of gamma-ray, X-ray, neutrino, and other astrophysical signals. Indirect DM searches using gamma-ray telescopes targeting several dwarf spheroidal galaxies (dSphs) provide excellent probes due to their low astrophysical backgrounds. In this work, we use data \cite{Fermi-LAT:2025gei} from a combined analysis by the most sensitive currently operating gamma-ray instruments: the Fermi-LAT telescope; the ground-based imaging atmospheric Cherenkov telescope arrays H.E.S.S., MAGIC, and VERITAS; and the HAWC water Cherenkov detector. These data provide constraints on the velocity-weighted DM self-annihilation cross section as a function of the DM mass, spanning the range from $5~\GeV$ to $100~\TeV$.

In this model framework, the CSDM can annihilate into bottom-quark or tau-lepton pairs (\figs\ref{feyn:wimp-id1} and \ref{feyn:wimp-id2}) through $s$-channel Higgs mediation and $t$-channel VLQ (VLL) mediation. Consequently, both the Higgs-portal and Yukawa-portal couplings are subject to stringent constraints from the experimental upper limits on the DM self-annihilation cross section. In addition, the CSDM semi-annihilation process into a Higgs boson \cite{Queiroz:2019acr} is also constrained by indirect detection observations, particularly in the resonant region where $\mphi \simeq m_h$. However, our analysis is not limited to the DM self-annihilation channel. Instead, for completeness, we include all relevant DM annihilation, cascade, and radiative processes that contribute to the production of photons and positrons. The corresponding indirect detection constraints are evaluated using the \texttt{micrOMEGAs} package, which incorporates the Fermi-LAT and CMB observational limits.
\end{itemize}
\subsection*{Combined Constraints from Dark Matter Relic Density, Direct Detection, and Indirect Searches:}
\begin{figure}[htb!]
\centering
\includegraphics[width=0.475\linewidth]{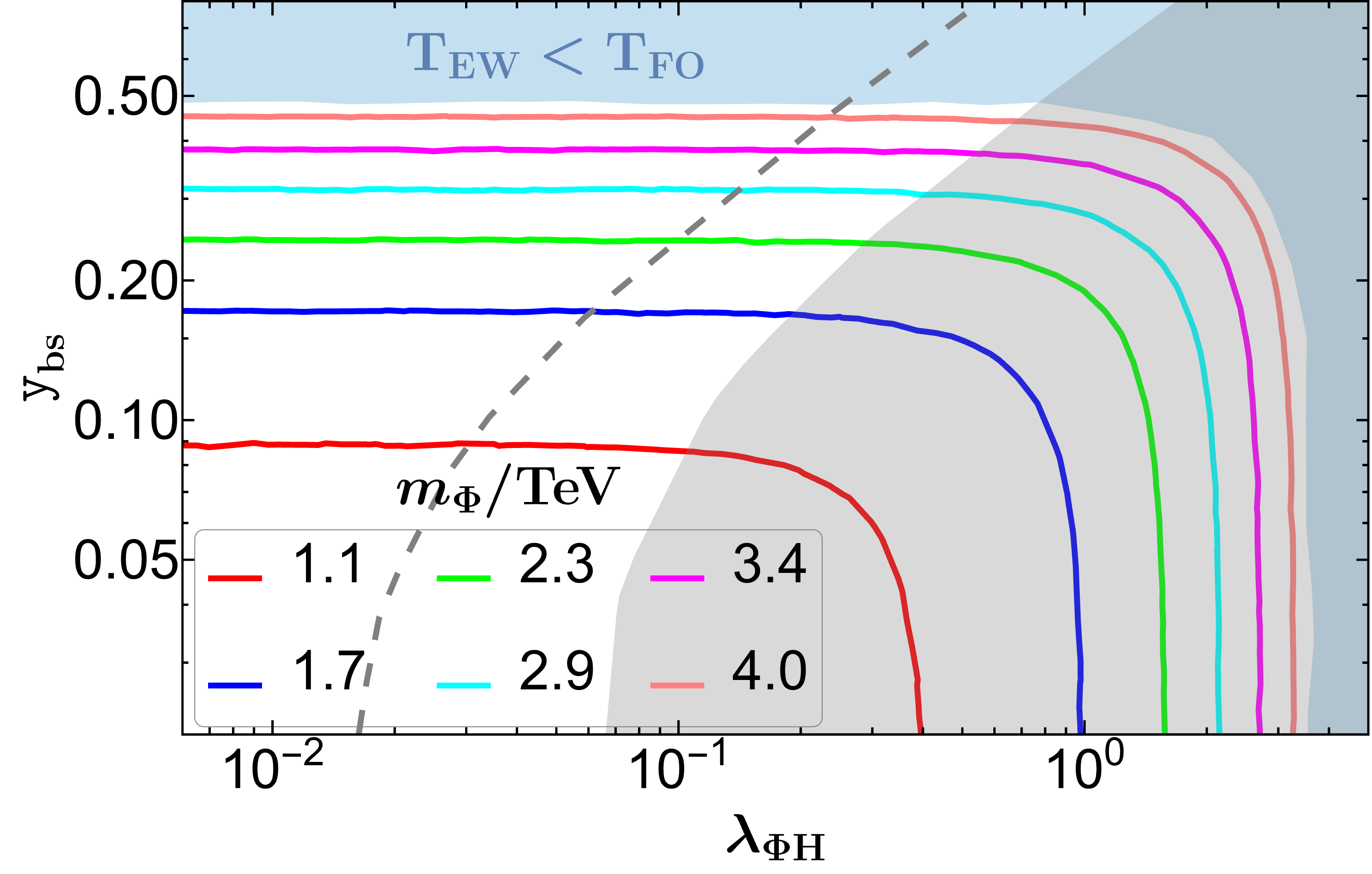}
\includegraphics[width=0.475\linewidth]{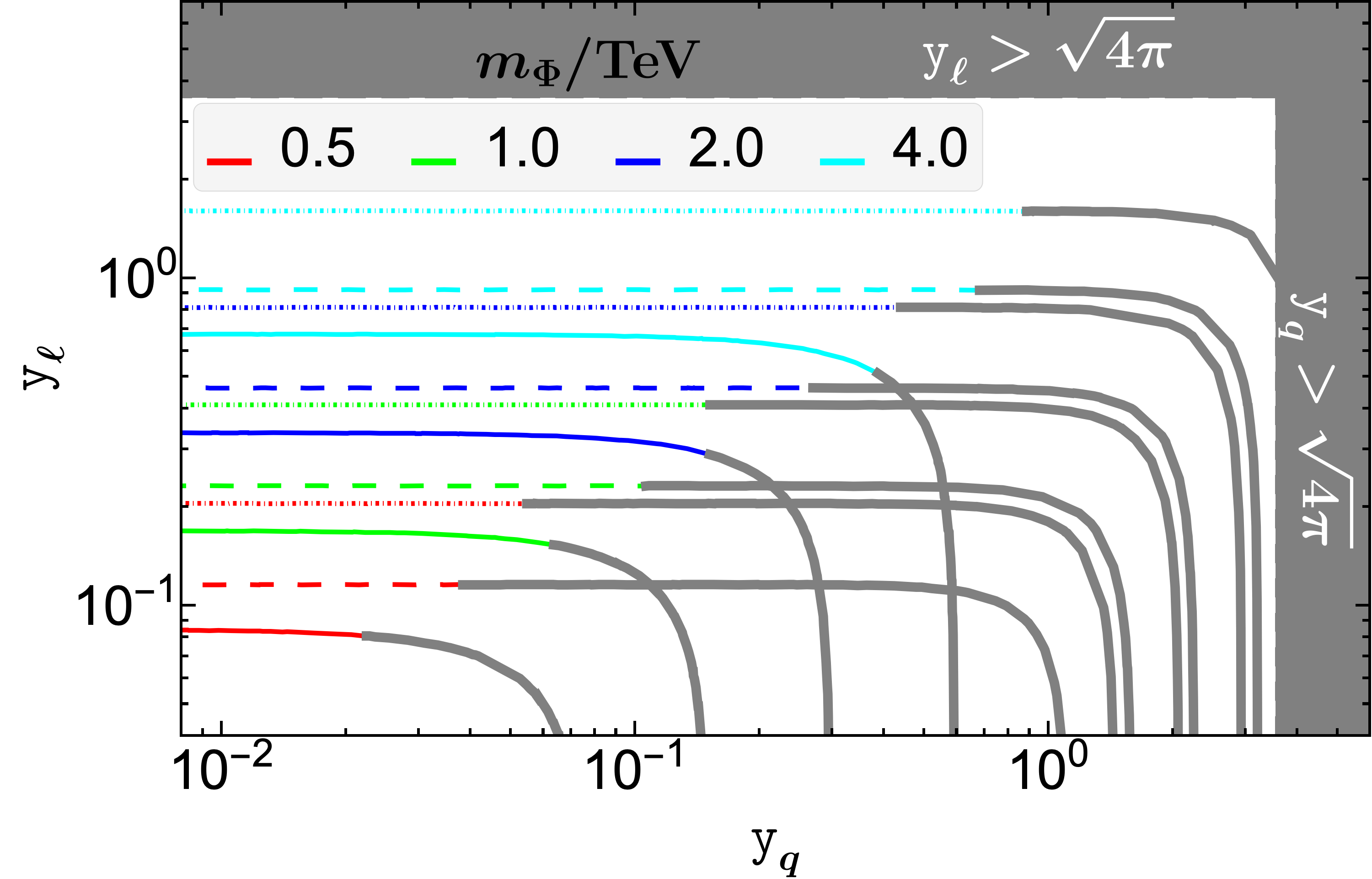}
\caption{
\texttt{Left: }The thick lines represent the regions satisfying the observed DM relic density in the $\lphiH-\ybs$ plane, with $\ybs=\yb=\ys$, for six different DM masses, as indicated in the legend. The remaining parameters are fixed as $\mpsi=1.2\mphi$, $\mchi=1.5\mphi$, $\mu_3=2.5\mphi$, $\mathtt{y}_\ell=0.2$, and $\yd=0.01$. The light-blue shaded region is excluded by the $\TFO<\TEW$ criterion, which requires DM freeze-out to occur after EWSB. The grey shaded region is excluded by the LZ-2025 data, while the grey dashed line corresponds to the future DD projection by the XLZD experiment.
\texttt{Right: }The thick, dashed, and dot-dashed colored lines correspond to $\mpsi = 5\mchi/4 = 25\mphi/16$, $\mpsi = 3\mchi/2 = 9\mphi/4$, and $\mpsi = 7\mchi/4 = 49\mphi/16$, respectively, and represent the relic density-satisfying parameter space in the $\yq-\yl$ plane. The remaining parameters are fixed at $\lphiH = 0.01$ and $\mu_3 = 2\mphi$. Different colored lines correspond to different DM masses, as indicated in the figure legend. The grey thick overlay on the colored lines denotes the regions excluded by the LZ-2025 data.}
\label{fig:relic1}
\end{figure}

\begin{figure}[htb!]
\centering
\includegraphics[width=1\linewidth]{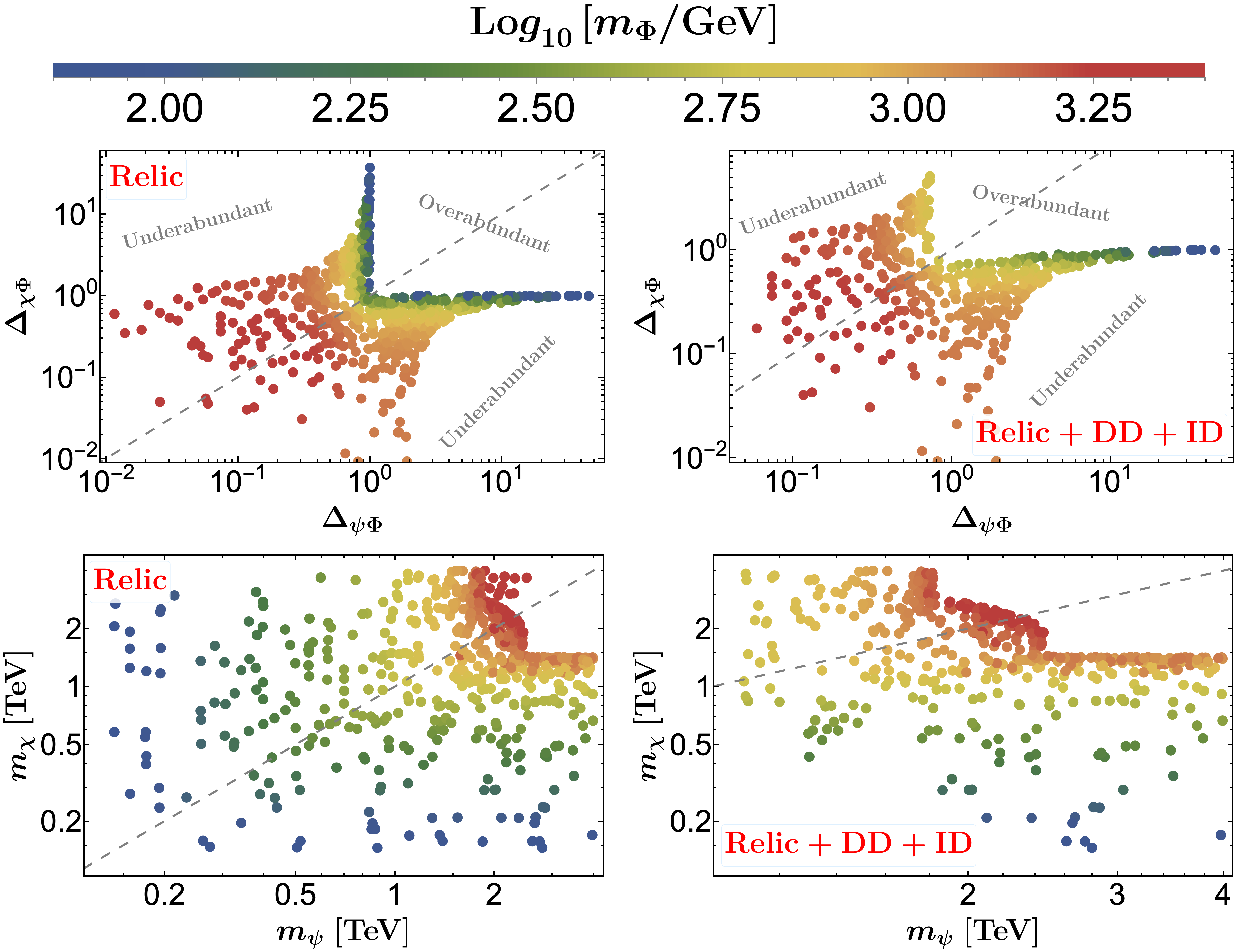}
\caption{The top-left and bottom-left panels show the parameter space allowed by the relic density constraint in the $\Delpsi-\Delchi$ and $\mpsi-\mchi$ planes, respectively. The top-right and bottom-right panels show the corresponding parameter space satisfying the combined relic density, DD, and ID constraints in the same planes. The grey dashed lines in the top and bottom panels correspond to $\Delpsi=\Delchi$ and $\mpsi=\mchi$, respectively. The dark rainbow colour bar indicates the variation of ${\rm Log_{10}}[\mphi/\GeV]$. We always restrict ourselves to the region satisfying $\TFO < \TEW$ to ensure consistency, such that freeze-out occurs after EWSB. For these figures, we fix the other parameters as $\lphiH=0.001$, $\mu_3=3\mphi$, $\yd=0.03$, $\ys=0.02$, $\yb=0.20$, $\ye=0.10$, $\ymu=0.05$, and $\yta=0.25$.}
\label{fig:relic2}
\end{figure}

As already mentioned, the complex scalar $\Phi$ serves as the DM candidate under the mass hierarchy assumption $\mpsi-m_d>\mphi<\mchi-m_e$. Therefore, the relic density depends not only on the interaction strength of $\Phi$ but also on the masses and interactions of $\psi$ and $\chi$, as these particles are eventually converted into the CSDM through the co-annihilation mechanism. The Feynman diagrams relevant to the CSDM self-annihilation and semi-annihilation processes are shown in \figs\ref{fig:feynman-relic1} and \ref{fig:feynman-relic2}, while \fig\ref{fig:feynman-relic3} depicts all relevant co-annihilation processes. The effective cross section for coannihilation in the presence of two VLFs and a complex scalar, all of which have identical charges under the same $\Zthree$ symmetry, relevant for the DM relic density evaluation, is estimated using the following Boltzmann Equation (BEQ) \cite{Griest:1990kh, Gondolo:2004sc, Edsjo:1997bg},
\begin{align}
\dfrac{dn_{\rm dm}}{dt}+3\mathcal{H}n_{\rm dm}=-\sveff\left[n_{\rm dm}^2-(n^{\rm eq}_{\rm dm})^2\right]\,,
\end{align}
where $\mathcal{H}$ is the Hubble expansion rate of the universe, $n_{\rm dm}=2(n_\psi+n_\chi+n_\Phi)$, and
\begin{align}
\sveff=\sum_{ij}\langle\sigma_{ij}v_{ij}\rangle\dfrac{n_i^{\rm eq}}{n_{\rm dm}^{\rm eq}}\dfrac{n_j^{\rm eq}}{n_{\rm dm}^{\rm eq}}\,,
\end{align}
with $v_{ij}=\sqrt{(p_i.p_j)^2-m_i^2m_j^2}/(E_iE_j)$. The equilibrium number density ratio, assuming the non-relativistic limit, can be expressed as,
\begin{align}
\dfrac{n_i^{\rm eq}}{n_{\rm dm}^{\rm eq}}=\dfrac{g_i(1+\Delta_{i\Phi})^{3/2}e^{-x\Delta_{i\Phi}}}{\sum_kg_k(1+\Delta_{k\Phi})^{3/2}e^{-x\Delta_{k\Phi}}}\,,
\label{eq:beq1}
\end{align}
where $x=\mphi/T$, $g_i$ denotes the degrees of freedom of the $i^{\text{th}}$ species and 
\begin{equation}
\label{eq:Delta_def}
\Delta_{i\Phi}=\dfrac{m_i-\mphi}{\mphi},
\end{equation}
denoting the degree of mass splitting between the BSM particles. Finally, by solving this BEQ, we obtain the present-day DM relic density as
\begin{align}
\omgdm h^2=9.49 \times 10^4 (\mphi/\GeV) (n_{\rm dm}[T_0]/{\rm cm^{-3}})\,,
\end{align}
where $T_0=2.348\times 10^{-13}~\GeV$ is the present day bath temperature \cite{ParticleDataGroup:2024cfk}.
However, we have numerically calculated the DM relic density, $\omgdm h^2$, the spin-independent DM-nucleon scattering cross section, $\sphiN$, and the indirect detection constraints using the \texttt{micrOMEGAs} package. The corresponding results are illustrated in \figs\ref{fig:relic1}, and \ref{fig:relic2}.

In \fig\ref{fig:relic1}, we show the relic-density contours, indicated by the colored lines, in the $\lphiH-\ybs$ (left) and $\yq-\yl$ (right) parameter planes. In the left panel, the six relic-density contours (red, blue, green, cyan, magenta and pink) correspond to $\mphi=\{1.1,~1.7,~2.3,~2.9,~3.4$, $\text{and}~4.0\}~\TeV$, while the remaining parameters are fixed to $\yl=0.2$, $\yd=0.01$, $\mpsi=\mchi=1.2\mphi$, and $\mu_3=2\mphi$. The light-blue shaded region is excluded since DM freeze-out occurs before EWSB phase transition, i.e., $\TFO>\TEW$, while still reproducing the observed DM relic density. In this regime, the relic-density calculation requires before-EWSB processes, yielding a parameter space that generally differs from that obtained in the after-EWSB framework; see Ref\,.~\cite{Bhattacharya:2021rwh}. Physically, a larger Yukawa coupling lowers the DM relic density, requiring a heavier CSDM to reproduce the observed abundance. Beyond a certain mass ($\gtrsim 4~\TeV$), however, the condition $\TFO<\TEW$ is violated, i.e., $\TFO>\TEW$, placing the model in the regime before EWSB and thereby forbidding that parameter region. Decreasing the Yukawa coupling, $\yq$, requires a smaller $\mphi$ to reproduce the observed relic density. The same argument also applies to the Higgs-portal coupling $\lphiH$. The grey shaded region is excluded by the latest LZ-2025 upper limit on the spin-independent CSDM-nucleon scattering cross section, $\sphiN$. As $\sphiN$ depends not only on $\lphiH$ and $\mphi$ but also simultaneously on $\yd$ and $\mpsi$. For a fixed value of $\yd$, the $\mpsi$ is varied according to the relation $\mpsi = 1.2\mphi$ as $\mphi$ changes. For smaller values of $\mphi$ (i.e., smaller $\mpsi$), the dominant contribution to $\sphiN$ arises from the Higgs portal processes, resulting in a relatively larger upper limit on $\lphiH$ of approximately $0.07$. However, in the larger $\mphi$ (i.e., larger $\mpsi$) regime, the VLQ-mediated contribution to $\sphiN$ becomes suppressed compared to the Higgs portal contribution. Consequently, as $\mphi$ increases, a larger value of $\lphiH$ is allowed from the direct detection.

In the right panel of \fig\ref{fig:relic1}, we also show the relic density-satisfied regions, identified by the red, green, blue, and cyan colored lines, corresponding to $\mphi=\{0.5,~1.0,~2.0,~4.0\}$ TeV, respectively. The thick, dashed, and dot-dashed lines correspond to $\mpsi = 5\mchi/4 = 25\mphi/16$, $\mpsi = 3\mchi/2 = 9\mphi/4$, and $\mpsi = 7\mchi/4 = 49\mphi/16$, respectively. The grey overlay on the colored lines indicates the regions excluded by the direct detection limits from the LZ-2025 experiment. The grey-shaded regions at the top and on the right correspond to the parameter space excluded by the perturbativity limits on the quark and lepton Yukawa couplings $\yq$ and $\yl$, respectively. However, the explanation of the relic density and direct detection limits are very similar to that of the left figure. The only difference is that, in this case, we vary the lepton Yukawa coupling, $\yl$, which does not contribute to the $\sphiN$ cross section but does contribute to the DM relic density. With increasing $\mphi$, a larger value of $\yl$ is generally required to reproduce the observed relic density through the co-annihilation process $\Phi\Phi \to \chibar \ell$. However, when $\yq$ becomes sufficiently large compared to $\yl$, the process $\Phi\Phi \to \psibar q_i$ (with $q_i=d\,,s\,,b$) starts to dominate, making the relic density essentially independent of $\yl$. Consequently, unlike in the left panel, the grey coating over the thick, dashed, and dot-dashed lines of the same colour does not follow a continuous curve, although it remains continuous for different colours corresponding to the same line style. This is because the thick, dashed, and dot-dashed lines correspond to different values of $\mpsi$, as mentioned earlier, resulting in distinct direct-detection exclusion regions. The most stringent exclusion limit on the relic density-allowed parameter space comes from CSDM–nucleon scattering (with the dominant contribution arising from neutron rather than interactions with the proton), as measured by the LZ-2025 experiment. However, the exclusion limit derived from indirect searches for DM annihilation into SM final states, based on recent Fermi-LAT observations, is negligible compared to that obtained from direct detection experiments.

To gain further insight into these dynamics, we present four plots in \fig\ref{fig:relic2}. The left panels show the relic density in the $\mpsi-\mchi$ and $\Delpsi-\Delchi$ (defined in \eq\eqref{eq:Delta_def}) planes, while the right panels display the parameter space satisfying the relic density together with the direct and indirect detection constraints in the same planes. The top and bottom rows correspond to the $\Delpsi-\Delchi$ and $\mpsi-\mchi$ planes, respectively. The dark rainbow colour bar indicates the variation of the CSDM mass, shown on a log scale. All other parameters are fixed at $\lphiH=0.001$, $\mu_3=3\mphi$, $\yd=0.03$, $\ys=0.02$, $\yb=0.20$, $\ye=0.10$, $\ymu=0.05$, and $\yta=0.25$. The colour gradient exhibits a linear correlation among the parameters $\mpsi$, $\mchi$, and $\mphi$ in the parameter space satisfying the observed DM relic density (left panel) and, altogether i.e., the relic density, the direct and indirect detection constraints (right panel). Importantly, since $\lphiH = 10^{-3}$ is required to evade the DD constraints, the contribution of the Higgs-portal processes to the DM relic density becomes subdominant. However, near the Higgs resonance ($\mphi \simeq m_h/2$), these processes can become dominant, although the corresponding parameter space may be subject to indirect detection constraints. Therefore, the dominant contributions to the DM relic density arise from processes involving the Yukawa couplings $\yl$ and $\yq$, as well as from gauge boson-induced processes (involving gluons, photons, and $Z$ boson). Hence, for fixed values of $\yq$ and $\yl$, the direct dependence on the parameters $\mpsi$, $\mchi$, and $\mphi$ arises primarily from the coannihilation processes (see \figs\ref{fig:feynman-relic2} and \ref{fig:feynman-relic3} for the corresponding Feynman diagrams), as the CSDM self-annihilation processes into SM particles are Higgs-mediated. For the benchmark parameters chosen in \fig\ref{fig:relic2}, one of the coannihilation channels involving either $\psi$ or $\chi$ dominates the DM relic density. Specifically, for $\mpsi > \mchi$, the relic density is dominated by the coannihilation channels of $\chi$, whereas for $\mpsi < \mchi$, it is dominated by those of $\psi$. In the case of $\psi$, the allowed parameter space extends slightly to $\mphi \gtrsim 1~\TeV$, even though $\yq$ is not significantly larger than $\yl$ but is instead of the same order, owing to the gluon-mediated strong annihilation of $\psi$ into quarks and gluons, which contributes to the CSDM relic density through coannihilation and is absent in the annihilation processes of $\chi$.

In the top-left panel of \fig\ref{fig:relic2}, we show the relic density-allowed rainbow-colored points in the $\Delpsi-\Delchi$ plane. The colour gradient, ranging from dark red to deep blue, indicates the dependence of $\mphi$ from high to low mass range. The top-left and bottom-right regions of the plot correspond to the completely underabundant regime, whereas the top-right region is entirely overabundant and can never achieve the correct relic density for this benchmark choice. However, as $\Delpsi$ ($\Delchi$) decreases from $\Delpsi = \Delchi$ dashed line, which corresponds to $\mpsi=2\mphi$ ($\mchi=2\mphi$), the coannihilation channels ($\Phi\Phi^*\to \psi(\chi)\,q\,(\ell)$ with $q=d,s,b$) become increasingly effective. For $\Delpsi$ ($\Delchi$) greater than $1.0$, these coannihilation processes are kinematically inaccessible, making them incapable of reducing the dark matter relic density. As a result, the relic density remains overabundant in this region. With decreasing $\Delpsi$ ($\Delchi$), the enhanced coannihilation cross section reduces the relic density, which can be compensated by increasing $\mphi$. Moreover, if $\Delpsi>1$ while $\Delchi\ll 1$, or vice versa, the significantly enhanced coannihilation cross section drives the dark matter relic density into the underabundant regime. The plot exhibits a nearly symmetric behaviour about the grey dashed line, which corresponds to $\mpsi=\mchi$, as the relevant coupling values are of the same order. The same explanation also applies to the right panels of \fig\ref{fig:relic2}; however, in this case, we have additionally imposed the direct detection and indirect detection constraints along with the relic density constraint. The recent DD exclusion limits from LZ-2025 on $\sphiN$ exclude some points in the low dark matter mass regime, i.e., where lower values of $\mpsi$ or $\mchi$ are required to achieve the correct relic density through coannihilation. Concurrently, such lower values of the VLF masses enhance $\sphiN$, even though $\lphiH$ is very small. We note that the indirect detection constraints are significantly weaker than those from direct detection. This is because the Higgs-portal coupling is required to be very small in order to satisfy the stringent direct detection bounds, rendering the Higgs-mediated annihilation channels negligible. Consequently, the dominant contribution to the dark matter annihilation cross section arises from $t$-channel exchange of the heavy vector-like fermions (VLFs). Owing to the large VLF masses and the moderate values of the Yukawa couplings, the present-day annihilation cross section remains suppressed, resulting in comparatively weak indirect detection constraints. However, for other benchmark points, these constraints may become important, as elaborated in the discussion of the summary plot later in Sec\,.~\ref{sec:combined}.

In the bottom panels of \fig\ref{fig:relic2}, we illustrate the same scenario but in the $\mpsi-\mchi$ plane using the same benchmark point. As discussed in the previous paragraph, the effectiveness of the coannihilation processes in determining the dark matter relic density depends on the mass splitting. Smaller values of $\Delpsi$ ($\Delchi$) correspond to a smaller mass separation between $\mpsi$ ($\mchi$) and $\mphi$. For a fixed value of $\mpsi$ ($\mchi$), an increase in $\mphi$ requires a corresponding increase in $\mchi$ ($\mpsi$) to achieve the correct relic density through coannihilation, as the self-annihilation processes are suppressed. This characteristic is also visible in the bottom-left plot. Similar to the top-left plot, it exhibits a nearly symmetric distribution of the coloured points about the grey dashed line for the same reason. Similarly, we have also imposed the direct detection (mostly effective) and indirect detection (less stringent) constraints on the relic density-allowed points, as shown in the bottom-right panel of \fig\ref{fig:relic2}. Since the benchmark value $\yd =0.03$ is relatively large and the spin-independent DM--nucleon scattering cross section, $\sphiN$, scales inversely with $\mpsi$, the LZ-2025 limits exclude VLF masses below approximately $1.0~\TeV$. In contrast, VLL masses as low as $\sim 0.5~\TeV$ remain allowed, since $\sphiN$ does not depend on $\mchi$.

\section{Collider Robustness of Dark Matter parameter space}
\label{sec:collider}
From a collider perspective, the simultaneous presence of a VLL and VLQ yields a rich phenomenology across strong and electroweak production channels, leading to decays into SM particles and exotic states. These processes produce distinctive signatures—energetic leptons, heavy-flavor jets, boosted electroweak bosons, and missing transverse momentum—offering multiple complementary avenues to probe the underlying model. While existing LHC searches place stringent constraints on vector-like fermions, sizeable regions of parameter space remain viable, especially in scenarios with non-standard decay patterns, compressed spectra, or suppressed branching fractions into conventional channels. Beyond the LHC, next-generation hadron colliders (e.g., FCC-hh and SPPC) and lepton colliders (such as the ILC, CLIC, and muon colliders) will substantially extend the sensitivity to heavier vector-like fermions through higher-energy reaches and cleaner experimental environments.

In this article, however, we focus on the HL-LHC operating at $\sqrt{s}=14~\TeV$ with an integrated luminosity of $3~\mathrm{ab}^{-1}$. As the next stage of the LHC program, the HL-LHC combines unprecedented luminosity with a mature experimental infrastructure, offering the most realistic near-term opportunity to discover or constrain the VLL–VLQ parameter space. It therefore provides an ideal framework for assessing the collider phenomenology and discovery prospects of this model.
\begin{figure}[htb!]
\centering
\begin{adjustbox}{width=1\linewidth}
\begin{tcolorbox}[colback=gray!5, colframe=black!10, boxrule=1pt, arc=4mm, boxsep=0pt, left=0pt, right=0pt, top=0pt, bottom=0pt, width=0.9\linewidth, halign=center]
\begin{tikzpicture}[baseline=(current bounding box.center)]
\begin{feynman}
\vertex (u) {\(d\)};
\vertex[right=1.25cm of u] (v_in);
\vertex[below=2.5cm of u] (ubar) {\(\bar{d}\)};
\vertex[right=1.25cm of ubar] (v_out);
\vertex[right=1.25cm of v_in] (psi) {\(\psi\)};
\vertex[right=1.25cm of v_out] (psibar) {\(\bar\psi\)};
\diagram*{
(u) -- [fermion,line width=0.35mm,arrow size=0.8pt,style=black] (v_in),
(v_in) -- [fermion,line width=0.35mm,arrow size=0.8pt,style=black] (psi),
(v_out) -- [charged scalar, edge label=\({\color{black}\Phi}\),line width=0.35mm,arrow size=0.8pt,style=gray] (v_in),
(v_out) -- [fermion,line width=0.35mm,arrow size=0.8pt,style=black] (ubar),
(v_out) -- [anti fermion,line width=0.35mm,arrow size=0.8pt,style=black] (psibar)};
\end{feynman}
\node at (v_in)[circle,fill,style=black,inner sep=1pt]{};
\node at (v_out)[circle,fill,style=black,inner sep=1pt]{};
\end{tikzpicture}
\begin{tikzpicture}[baseline=(current bounding box.center)]
\begin{feynman}
\vertex (u) {\(u\,,d\)};
\vertex [below=2.5cm of u] (ubar) {\(\bar{u}\,,\bar{d}\)};
\vertex [right=1.25cm of u, yshift=-1.25cm] (v_in);
\vertex [right=2.5cm of v_in, yshift=+1.25cm] (psi) {\(\psi\,,\chi\)};
\vertex [right=2.5cm of v_in, yshift=-1.25cm] (psibar) {\(\bar\psi\,,\bar\chi\)};
\vertex [right=1.25cm of v_in] (v_out);
\diagram*{
(u) -- [fermion,line width=0.35mm,arrow size=0.8pt,style=black] (v_in),
(v_in) -- [fermion,line width=0.35mm,arrow size=0.8pt,style=black] (ubar),
(v_in) -- [photon, edge label=\({\color{black}\gamma}\), edge label'=\({\color{black}Z}\),line width=0.35mm,arrow size=0.8pt,style=gray] (v_out),
(v_out) -- [fermion,line width=0.35mm,arrow size=0.8pt,style=black] (psi),
(v_out) -- [anti fermion,line width=0.35mm,arrow size=0.8pt,style=black] (psibar)};
\end{feynman}
\node at (v_in)[circle,fill,style=black,inner sep=1pt]{};
\node at (v_out)[circle,fill,style=black,inner sep=1pt]{};
\end{tikzpicture}
\begin{tikzpicture}[baseline=(current bounding box.center)]
\begin{feynman}
\vertex (u) {\(u\,,d\)};
\vertex [below=2.5cm of u] (ubar) {\(\bar{u}\,,\bar{d}\)};
\vertex [right=1.25cm of u, yshift=-1.25cm] (v_in);
\vertex [right=2.5cm of v_in, yshift=+1.25cm] (psi) {\(\psi\)};
\vertex [right=2.5cm of v_in, yshift=-1.25cm] (psibar) {\(\bar\psi\)};
\vertex [right=1.25cm of v_in] (v_out);
\diagram*{
(u) -- [fermion,line width=0.35mm,arrow size=0.8pt,style=black] (v_in),
(v_in) -- [fermion,line width=0.35mm,arrow size=0.8pt,style=black] (ubar),
(v_out) -- [gluon, edge label'=\({\color{black}g}\),line width=0.35mm,arrow size=0.8pt,style=gray] (v_in),
(v_out) -- [fermion,line width=0.35mm,arrow size=0.8pt,style=black] (psi),
(v_out) -- [anti fermion,line width=0.35mm,arrow size=0.8pt,style=black] (psibar)};
\end{feynman}
\node at (v_in)[circle,fill,style=black,inner sep=1pt]{};
\node at (v_out)[circle,fill,style=black,inner sep=1pt]{};
\end{tikzpicture}
\end{tcolorbox}
\end{adjustbox}
\caption{Feynman diagrams contributing to $\psi\psibar$ and $\chi\chibar$ pair production.}
\label{fig:s_channel_production}
\end{figure}
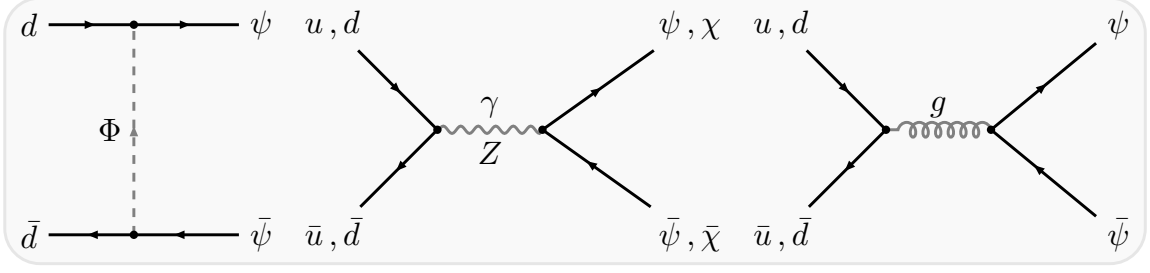


\subsection{LHC constraints on singlet Vector-Like Fermions with Dark Matter}
\label{LHC_constraints}
In this section, we summarize the most relevant collider bounds obtained by recasting existing analyses with \textsc{CheckMATE.v2}~\cite{Dercks:2016npn}.
For the VLQ sector, we consider pair production,
$$
pp \rightarrow \psi\; \psibar\,,
$$
with the corresponding Feynman diagrams shown in \Fig\ref{fig:s_channel_production}. This is followed by the dominant decay mode,
$$
\psi \rightarrow \Phi\; b\,,
$$
resulting in a final state with two $b$-jets and missing transverse energy from the invisible scalar $\Phi$, the dark matter candidate in the present model. The analysis is performed for two benchmark scenarios defined as (\texttt{BP1}) and (\texttt{BP2}) mentioned in \autoref{tab:benchmark}, which satisfy the flavor constraints at the $1\sigma$ and $2\sigma$ levels, respectively.
We have chosen the quark Yukawa couplings ($\yq$) to satisfy low-energy flavor observables, while keeping equal couplings for all charged leptons is sufficient to remain consistent with current experimental bounds. However, there are other potential benchmarks one could explore.
\begin{table}[htb!]
\centering
\begin{adjustbox}{width=1\linewidth}
\begin{tabular}{|c|c|c|c|c|c|c|c|c|}\hline
Benchmark Points & $\mchi/\mphi$ & $\lphiH$ & $\yd$ & $\ys$ & $\yb$ & $\ye$ & $\ymu$ & $\yta$ \\\hline
\texttt{BP1} & $2.4$ & \multirow{2}{*}{$10^{-3}$} & \multirow{2}{*}{$0.04$} & $0.022$ & \multirow{2}{*}{$0.20$} & $0.11$ & $0.11$ & $0.11$ \\
\texttt{BP2} & $2.2$ &  &  & $0.023$ &  & $0.15$ & $0.15$ & $0.15$ \\\hline
\end{tabular}
\end{adjustbox}
\caption{\texttt{BP1} and \texttt{BP2} are considered for a combined interpretation of the model parameters in the $\mpsi–\mphi$ plane in \figs\ref{fig:relic-summary-1sigma} and \ref{fig:relic-summary-2sigma}, which reflect $1\sigma$ and $2\sigma$ flavor constraints, respectively, and include all current exclusion limits as well as projected sensitivities from the relevant experiments.}
\label{tab:benchmark}
\end{table}

Owing to its similarity to the supersymmetric $b$-jets+$E_T^{\rm miss}$ signature, the model can be constrained by existing SUSY searches. The exclusion contours for the benchmark scenarios \texttt{BP1} and \texttt{BP2}, obtained from a scan over the $(\mpsi\,,~\mphi)$ parameter plane, are shown as the red and blue shaded regions, respectively, in the right panel of \Fig\ref{fig:exclusion_discovery_contours}. The strongest constraints originate from the ATLAS top squark (stop) search in the all-hadronic $t\bar{t}+E_{\text{T}}^{\text{miss}}$ final state~\cite{ATLAS:2020dsf}, the CMS inclusive jets+$E_{\text{T}}^{\text{miss}}$ search~\cite{CMS:2019zmd}, and the ATLAS squark/gluino search~\cite{ATLAS:2020syg}, with each providing the leading sensitivity in different regions of the parameter space.

The resonant production of VLLs, illustrated in \Fig\ref{fig:s_channel_production}, takes place through the pair-production channel:
$$pp\rightarrow \chi\,\chibar\,,$$
followed by the decay,
$$\chi\rightarrow\Phi\,\ell\,,$$
which gives rise to a final state containing two leptons and missing transverse energy. The same benchmark scenarios are adopted, and the exclusion limits are obtained by scanning the $(\mchi\,,~\mphi)$ plane. As shown by the red shaded region in the left panel of \Fig\ref{fig:exclusion_discovery_contours}, the most stringent constraint originates from the ATLAS search for direct chargino and
slepton pair production in final states with two leptons and missing transverse momentum~\cite{ATLAS:2019lff}.

For the VLQ sector, the origin of the exclusion contour is straightforward to understand. As the VLQ mass increases, the pair-production cross section decreases rapidly, eventually falling below the experimental upper limits implemented in \textsc{CheckMATE.v2}. Consequently, the exclusion sensitivity deteriorates at large $\mpsi$. On the other hand, in the compressed region, $\mphi\simeq\mpsi$, the small mass splitting suppresses the transverse momentum of the $b$-jets, causing a significant fraction of the signal events to fail the analysis selection requirements. The interplay of the falling production cross-section at high masses and the reduced acceptance in the compressed regime gives rise to the characteristic shape of the exclusion contour.

A similar behaviour is observed in the VLL sector. The exclusion reach weakens for increasing $\mchi$ due to the rapid decrease of the VLL pair-production cross section. In addition, for low values of $\mchi$, the visible decay products become relatively soft, resulting in leptons and missing transverse energy that frequently fail the kinematic requirements of the ATLAS analysis. This reduction in signal acceptance limits the experimental sensitivity in the low-mass region, while the suppression of the production cross section dominates at high masses. Together, these effects determine the exclusion contour in the $(\mchi\,,~\mphi)$ plane.The black dotted and dashed lines correspond to the benchmark relations $m_{\chi}=2.4,m_{\Phi}$ and $m_{\chi}=2.2,m_{\Phi}$, respectively, adopted for \texttt{BP1} and \texttt{BP2}. Both benchmark scenarios lie outside the current LHC exclusion region over a substantial portion of the parameter space, indicating that the chosen benchmark points are broadly consistent with the existing collider constraints. At the same time, they pass through the projected HL-LHC sensitivity region, demonstrating that a significant portion of the benchmark parameter space can be probed at the HL-LHC.

A comparison of the exclusion reaches in the two sectors reveals a significant difference. While the VLL masses are excluded up to approximately $\mchi\lesssim 380~\GeV$, the VLQ exclusion extends to $\mpsi\lesssim 1.6~\TeV$. This behaviour is primarily driven by the underlying production mechanisms. At hadron colliders, VLL pair production proceeds exclusively through electroweak Drell-Yan processes mediated by an off-shell $\gamma/Z$, resulting in comparatively small production cross sections. In contrast, VLQ pair production receives the dominant QCD contribution through gluon-initiated processes, as illustrated in \fig\ref{fig:s_channel_production}. Depending on the model parameters, an additional $t$-channel contribution mediated by the scalar $\Phi$ can further enhance the production rate. Consequently, the VLQ production cross section is substantially larger than that of the VLL, leading to a considerably stronger exclusion reach at the LHC.
\subsection{Dark Matter Search Prospects at the HL-LHC}
\label{sec:HLLHC}
In this subsection, we assess the discovery reach for the dark matter candidate by analyzing the characteristic missing-energy signatures associated with VLL and VLQ production.
\subsubsection{Prospects for Vector-Like Quark Searches in the $b\bar{b}+\slashed{E}_T$ Channel}
\label{sec:HLLHC_VLQ}
The potential signal topology for probing the VLQ \(\psi\) at the HL-LHC is given by \[pp \rightarrow \psi\,\psibar\,,\qquad \psi \rightarrow \Phi\, b\,;\] which leads to a final state characterised by two \(b\)-jets accompanied by missing transverse energy (\(\slashed{E}_T\)), where the missing energy originates from the invisible scalar \(\Phi\).
\subsection*{Potential Backgrounds}
The dominant SM backgrounds for the \(2b + \slashed{E}_T\) final state arise from several processes. The most significant contribution comes from \(t\bar{t}\) production, which naturally yields two \(b\)-jets along with sizable missing transverse energy originating from the leptonic decays of the \(W\)-bosons. Another important irreducible background is \(Z(\nu\bar{\nu}) + b\bar{b}\), where the invisible decay of the \(Z\)-boson generates genuine missing transverse energy. In addition, processes such as \(b\bar{b}W\) can contribute when the charged leptons are not reconstructed.

Subleading backgrounds also arise from \(t\bar{t}W\) and \(t\bar{t}Z\) production, particularly in leptonic decay channels involving neutrinos. Contributions from \(jjZ\) production, with the \(Z\)-boson decaying invisibly and light jets being mistagged as \(b\)-jets, can also become relevant. Altogether, these backgrounds constitute the primary SM contributions to the \(2b + \slashed{E}_T\) search channel considered in this analysis.

After analysing important kinematic variables for both signal and backgrounds, we have defined the following selection cuts to achieve good discovery significance across the low-to-high $\mpsi-\mphi$ mass plane.

\begin{itemize}
\item \textbf{Cut 1: Jet cleaning (\(p_T^{j} > 50\) GeV):} 
Events are required to contain reconstructed jets with transverse momentum greater than \(50\) GeV to suppress soft QCD activity and detector noise.

\item \textbf{Cut 2: At least two \(b\)-tagged jets (\(N_b \geq 2\)):} 
Events must contain at least two identified \(b\)-jets, consistent with the expected signal topology arising from VLQ pair production.

\item \textbf{Cut 3: Lepton veto:} 
Events containing isolated charged leptons (\(e,\mu\)) are rejected to suppress backgrounds having leptons.

\item \textbf{Cut 4: Leading \(b\)-jet transverse momentum (\(p_T(b_1) > 170\) GeV):} 
The leading \(b\)-jet is required to satisfy a hard transverse momentum cut in order to reduce soft QCD and electroweak backgrounds.

\item \textbf{Cut 5: Subleading \(b\)-jet transverse momentum (\(p_T(b_2) > 80\) GeV):} 
The second-leading \(b\)-jet is required to have transverse momentum above \(80\) GeV to further enhance the signal significance.

\item \textbf{Cut 11: Missing transverse energy (\(\slashed{E}_T > 250\) GeV):} 
A large missing transverse energy requirement is imposed to efficiently suppress SM backgrounds while retaining events containing invisible particles in the final state.

\item \textbf{Cut 12: Effective mass (\(M_{\mathrm{eff}} > 450\) GeV):} 
The effective mass variable, defined as the scalar sum of the transverse momenta of visible objects and missing transverse energy, is required to be larger than \(450\) GeV to isolate high-energy signal events.

\item \textbf{Cut 13: Transverse mass of the leading \(b\)-jet (\(M_T(b_1) > 300\) GeV):} 
A lower bound on the transverse mass constructed using the leading \(b\)-jet and missing transverse momentum is imposed to further suppress residual SM backgrounds.
\end{itemize}
After imposing the aforementioned cuts, we have used the following formula to obtain the significance of the discovery~\cite{Cowan:2010js}:
\begin{align}
\mathcal{Z} = \sqrt{2\left(N_S+N_B\right)\ln\left(\frac{N_S+N_B}{N_B}\right)-2N_S} \,, 
\end{align}
where $N_S$ and $N_B$ represent the number of signal and background events, respectively. The number of background events is determined using the relation:
\begin{align}
N_B = \left(\sum_{i} \sigma_{B}^{i} \times \epsilon_{B}^{i}\right) \times \mathcal{L}_\text{int.}\,,
\end{align}
where $\sigma_{B}^{i}$ is the cross section of the $i^{\text{th}}$ background process, and $\epsilon_{B}^{i}$ denotes its corresponding cut efficiency. The parameter $\mathcal{L}_\text{int.} = 3~\text{ab}^{-1}$ is the integrated luminosity of the HL-LHC. The total background yield, $N_B$, is obtained by summing the contributions from all background processes considered.

The mass points of significance for the discovery $\mathcal{Z} > 3 \sigma$ are shown in solid red and blue lines in the right panel of the \fig\ref{fig:exclusion_discovery_contours} for \texttt{BP1} and \texttt{BP2}, respectively. Clearly, sensitivities up to $\mpsi < 1.5~\text{TeV}$ are achievable at the HL-LHC, beyond which the reach is constrained by the limited centre-of-mass energy required to produce heavier resonant VLQs. Furthermore, the upper bound on $\mphi$ for a given $\mpsi$ is driven by a drop in selection efficiency for cuts 4 and 5, which impose high-$p_T$ requirements on the $b$-jets. As $\mphi$ approaches $\mpsi$, the kinetic energy available to the produced $b$-jets decreases, significantly reducing the survival probability under these kinematics-based cuts.
\subsubsection{Prospects for Vector-Like Lepton Searches in the $e^{+} e^{-}+E_T^{\rm miss}$ Channel}
\label{sec:HLLHC_VLL}
The discovery prospects for the VLL are investigated through the pair-production process
\begin{align}
pp \rightarrow \chi\,\chibar, \qquad \chi \rightarrow \Phi\,e\,;
\end{align}
which gives rise to a final state containing two electrons and missing transverse energy, where the missing momentum is carried away by the invisible scalar $\Phi$.
\subsection*{Potential Backgrounds}
The dominant SM backgrounds to the $2e+E_T^{\rm miss}$ final state arise from $tW$, $t\bar{t}$ and $WW$ production. Additional contributions from $ZZ$, $WZ$, $WZjj$, $t\bar{t}W$, $t\bar{t}Z$, triboson ($VVV$, with $V=W^\pm,~Z$) and four-top production are also included in the analysis. All relevant background processes are generated and passed through the same detector simulation and event selection as the signal.

To maximize the signal significance over the $(\mchi\,,~\mphi)$ parameter space, the following event selection criteria are imposed:
\begin{itemize}
\item \textbf{Muon veto:} Events containing isolated muons are rejected,

$$
N_{\mu}=0.
$$

\item \textbf{Opposite-sign electron pair:} Events are required to contain at least two electrons with opposite electric charges,

$$
N_{e^+e^-}\geq 2.
$$

\item \textbf{$b$-jet veto:} Events containing one or more tagged $b$ jets are rejected,

$$
N_b=0,
$$

thereby suppressing backgrounds from top-quark production.

\item \textbf{Jet veto:} Events containing one or more reconstructed jets are rejected,

$$
N_j=0.
$$

\item \textbf{Azimuthal separation between the leading electron and missing transverse momentum:} The azimuthal angle between the leading electron and the missing transverse momentum is required to satisfy

$$
\Delta\phi(e_1,E_T^{\rm miss})>1.5.
$$

\item \textbf{$Z$-boson veto:} Events consistent with the decay of a $Z$ boson are rejected using a $Z$-mass veto.

\item \textbf{Leading electron transverse momentum:} The leading electron is required to satisfy

$$
p_T(e_1)>20~\GeV.
$$

\item \textbf{Subleading electron transverse momentum:} The subleading electron is required to satisfy

$$
p_T(e_2)>20~\GeV.
$$

\item \textbf{Missing transverse momentum:} Events are required to have sufficiently large missing transverse momentum,

$$
E_T^{\rm miss}>100~\GeV.
$$

\item \textbf{Transverse mass:} The transverse mass constructed from the leading electron and missing transverse momentum is required to satisfy

$$
M_T(e_1,E_T^{\rm miss})>100~\GeV.
$$

\end{itemize}

\begin{figure}[htb!]
\centering
\includegraphics[width=0.5\linewidth]{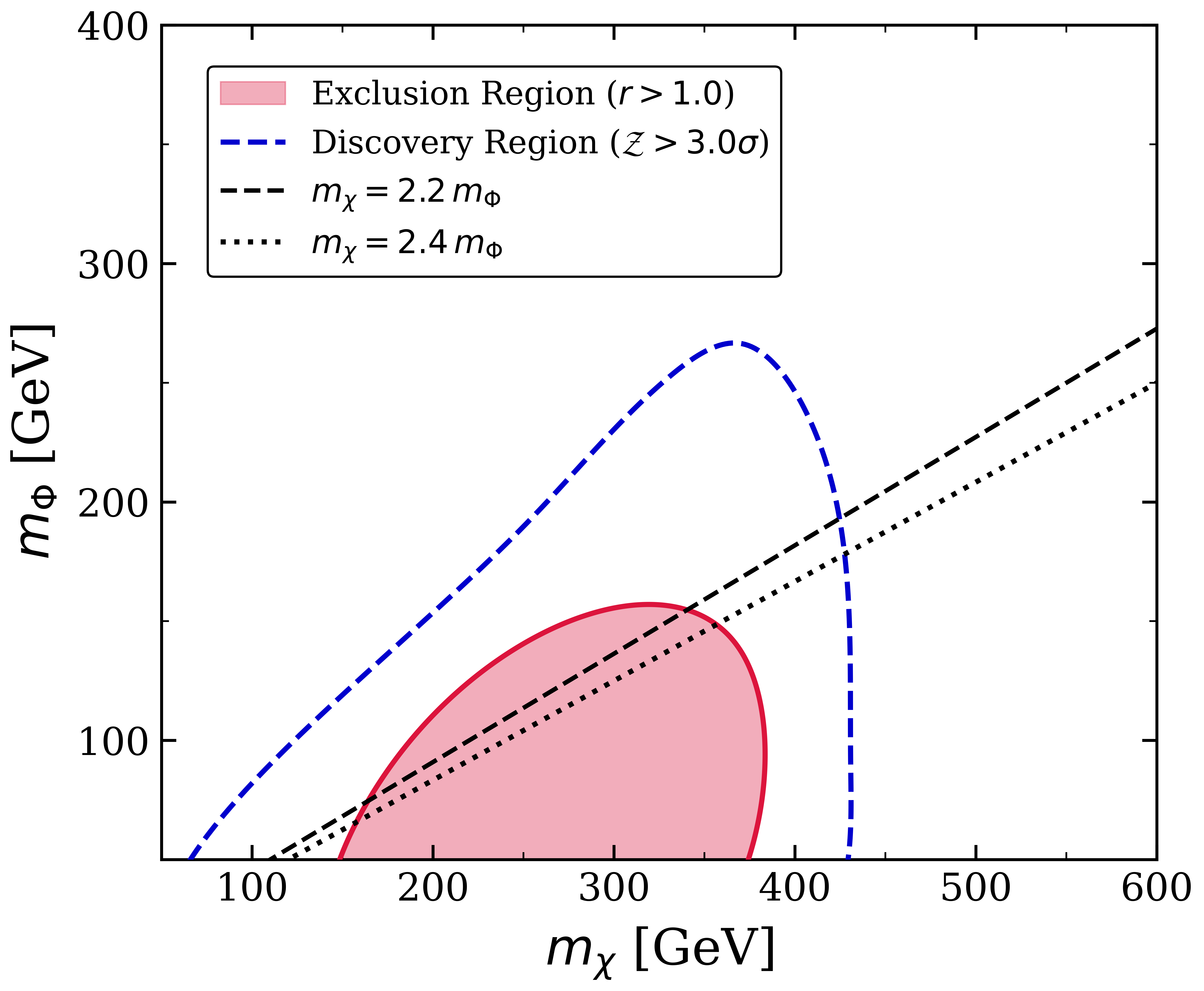}
\includegraphics[width=0.45\linewidth]{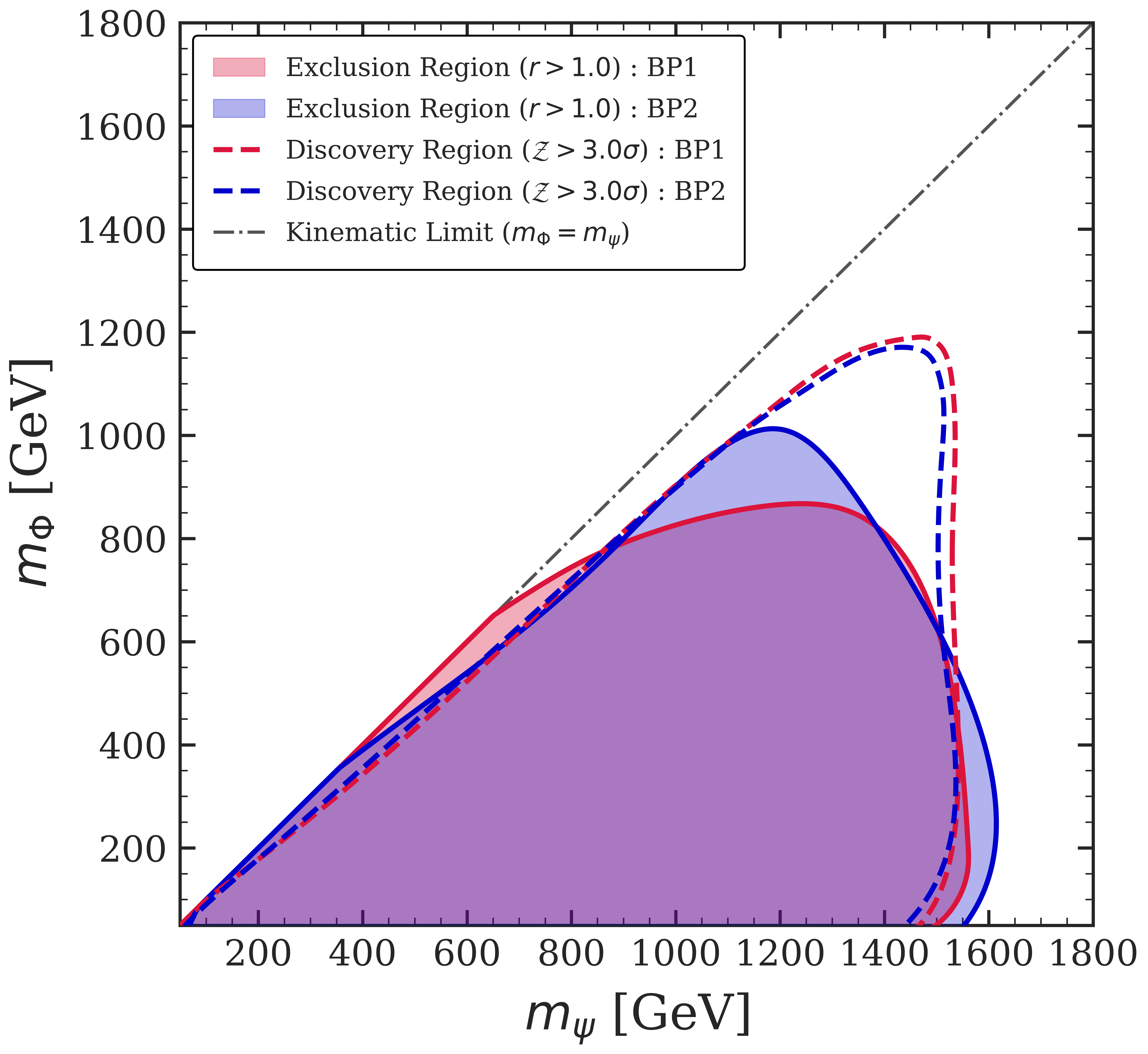}
\caption{Left: Exclusion ($2\sigma$) and projected ($3\sigma$) sensitivity reach for the VLL in the $(\mchi\,,~\mphi)$ plane obtained from the two tau-lepton and missing transverse energy analysis at the HL-LHC with an integrated luminosity of $3~\mathrm{ab}^{-1}$. Right: Exclusion ($2\sigma$) and projected ($3\sigma$) sensitivity reach for the VLQ in the $(\mpsi\,,~\mphi)$ plane obtained from the two b-jets and missing energy analysis for two specific benchmark points discussed in the main text.}
\label{fig:exclusion_discovery_contours}
\end{figure}

These selection criteria significantly suppress the SM backgrounds while retaining good signal efficiency across the $(\mchi\,,~\mphi)$ parameter space. The statistical significance is evaluated using the same procedure and significance estimator described for the VLQ analysis. The projected HL-LHC discovery reach, corresponding to mass points with $\mathcal{Z}\geq3$, is shown by the blue shaded region in the left panel of \fig\ref{fig:exclusion_discovery_contours} for the benchmark scenarios considered. The discovery reach extends up to $\mchi \leq 500\text{ GeV}$ because the process is purely electroweak (QED) mediated, resulting in a suppressed production cross section. 

\section{Combined interpretation}
\label{sec:combined}
In the previous sections, we investigated the individual constraints to identify the allowed parameter space when only exclusion bounds are available, as well as the parameter space consistent with experimental constraints when both upper and lower bounds are imposed. For these two summary plots, we considered two benchmark points (\texttt{BP1} and \texttt{BP2}) listed in \autoref{tab:benchmark}, which differ only in the value of $\ys$, while all other parameters remain identical. However, we leave $\mpsi$ and $\mphi$ free, defining the parameter plane for the summary plots, while $\mu_3$, being relevant only for the DM relic density, is varied up to $2\sqrt{\pi}\mphi$ (corresponding to the maximum value allowed by vacuum stability for $\lambda_\Phi=\pi$) in order to reproduce the observed DM relic density.

In this table, the leptonic Yukawa couplings ($\ymu$ and $\ye$) are chosen to satisfy the stringent constraints from the radiative decay $\mu\to e\gamma$, while $\lphiH$ and $\yd$, being highly sensitive to $\sphiN$, are taken to be sufficiently small to evade all current exclusion limits. All other parameters are chosen such that a non-zero common allowed parameter space is obtained.
\begin{figure}[htb!]
\centering
\includegraphics[width=0.9\linewidth]{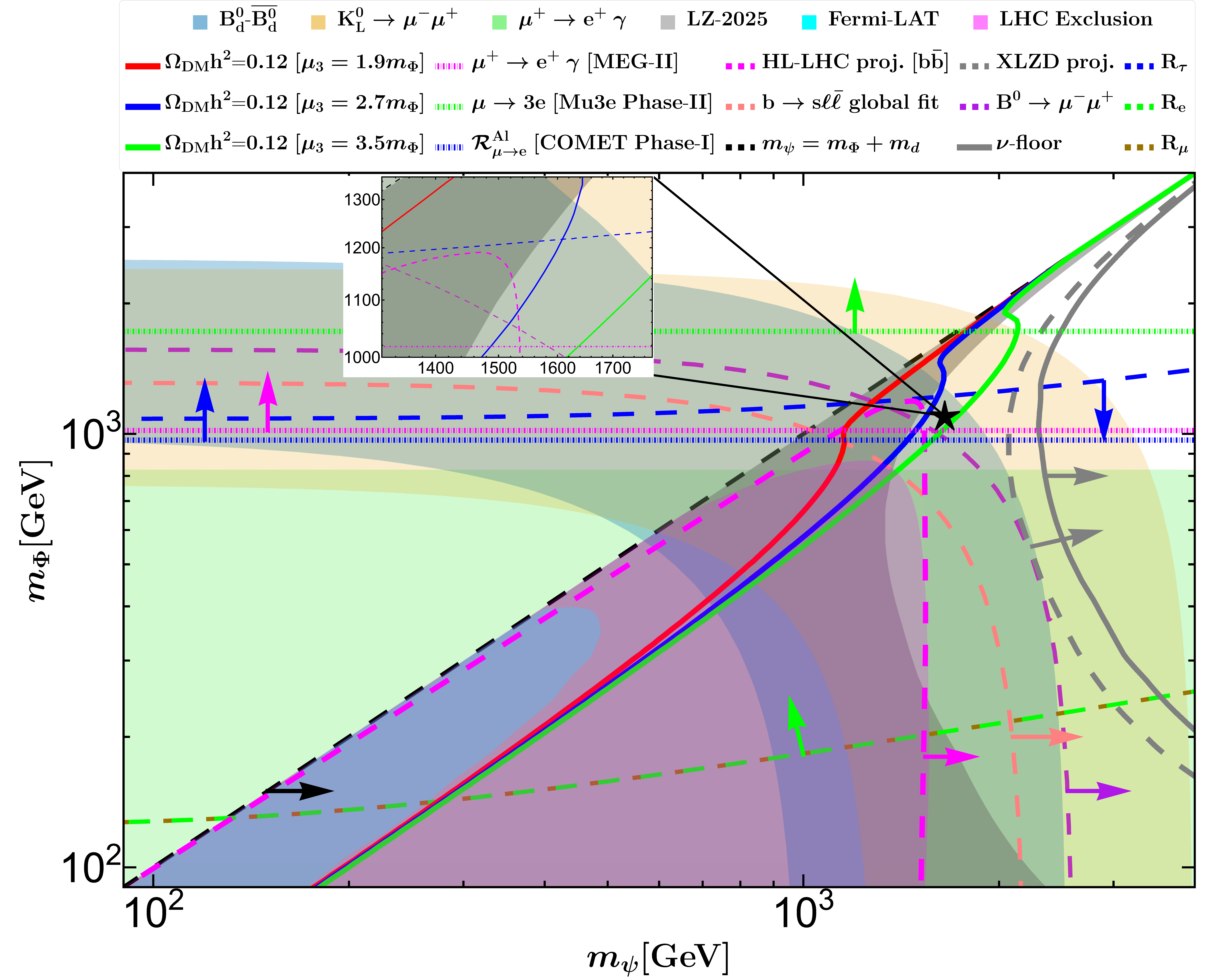}
\caption{Combined phenomenological constraints for benchmark point \texttt{BP1} in the $\mpsi-\mphi$ plane. The flavor constraints are derived from $1\sigma$ experimental bounds. The blue, green, and magenta dot-dashed lines correspond to the projected limits for the respective LFV observables. Other observables, including $A_e$, $A_\mu$, $A_\tau$, $A_b$, $A_s$, $R_b$, $B^0_s-\bar{B}^0_s$, $K^0-\bar{K}^0$, and $B^0_s \to \mu^- \mu^+$, impose no additional constraints in the allowed parameter space. The black dashed line denotes the threshold $\mpsi = \mphi + m_d$, which ensures the stability of the CSDM. The arrows indicate the allowed region for each respective observable.}
\label{fig:relic-summary-1sigma}
\end{figure}

\Fig\ref{fig:relic-summary-1sigma} illustrates that all the relevant observable constraints have been discussed in the previous sections. The remaining observables (which are not included in the figure legend) do not impose any additional constraints on the parameter space in the $\mpsi-\mchi$ plane. Importantly, this plot is obtained using the $1\sigma$ flavor bounds corresponding to the respective observables. For clarity, we summarise the impact of all relevant constraints on the parameter space in the $\mpsi-\mchi$ plane in the following step-by-step discussion.
\begin{itemize}
\item The regions on, above, and below the thick red ($\mu_3 = 1.9~\mphi$), blue ($\mu_3 = 2.7~\mphi$), and green ($\mu_3 = 3.5~\mphi$) lines correspond to the correct, overabundant, and underabundant relic density regimes, respectively. In these relic density contours, the bends appear as a consequence of the dominance of coannihilation processes, $\Phi\Phi \to b \psibar$ and $\Phi\Phi \to \tau\chibar$, owing to the comparatively large corresponding Yukawa couplings, while the processes $\psi\bar\psi \to\rm SM~SM$ are dominant near to the first bend around $\mpsi\sim\mphi+m_Z$. The explanation of the dependence of $\Delpsi$ and $\Delchi$ on the relic density remains the same: a smaller value results in a larger annihilation cross section and, consequently, requires a larger $\mphi$ to reproduce the correct relic density.
\item The grey shaded region is excluded by the current direct detection constraint from the LZ-2025 experiment. For our choice of $\lphiH = 10^{-3}$ and $\yd = 0.04$, VLQ masses up to approximately $1.3~\TeV$ are excluded. Since $\mpsi$ is strongly correlated with $\mphi$, smaller mass splittings are subject to tighter constraints. For $\mpsi \gtrsim 1.3~\TeV$, the contribution to $\sphiN$ from the VLQ-mediated interaction is always subdominant. However, the Higgs portal interaction can become dominant if $\lphiH$ is increased beyond $10^{-3}$. We also present the projected sensitivity in this plane, using the projected direct-detection limit from the XLZD collaboration, shown as the grey dashed line. Future experiments are expected to probe the parameter space down to approximately $\mpsi \approx 2.0~\TeV$ for $\mphi \approx 1.0~\TeV$. In addition, we also include the neutrino floor on the parameter-space plane, as below this limit the neutrino background becomes irreducible compared to the dark matter signal \cite{Billard:2013qya, Monroe:2007xp, OHare:2020lva}. Although several studies have proposed methods to distinguish dark matter signals from the neutrino background, the neutrino floor remains an important benchmark for sensitivity.
\item Further, we added indirect detection constraints using Fermi-LAT data only. As the CMB limits are less stringent, we do not show them here. In this setup, CSDM self-annihilation, cascade decays of intermediate states, and final-state radiation contribute to the total gamma-ray flux. We calculated the indirect detection exclusion limits from gamma-ray observations of the Galactic Center using the \texttt{micrOMEGAs} package, accounting for all aforementioned contributions; the excluded region is indicated by cyan shading. As seen in the cyan region, although the Yukawa couplings are small, $\mathcal{O}(0.1)$, the indirect detection bounds remain strong for $\mphi\lesssim~450\,\GeV$. Consequently, the Fermi-LAT constraints exclude at $95\% \rm C.L.$ the region with $\mpsi \leq 2\mphi$ for $\mphi \leq 450\,\GeV$. Notably, increasing $\mphi$ enhances the dominance of the leptonic-mode coannihilation, $\Phi\Phi\to\chibar\ell$, whereas increasing $\mpsi$ suppresses the contribution from the bottom-mode coannihilation, $\Phi\Phi\to\psibar b$.
\item The current LHC constraints on the VLQ sector are obtained by recasting searches for pair-produced vector-like quarks, $pp\rightarrow\psi\psibar\,,$ followed by the dominant decay mode $\psi\rightarrow\Phi\,b$, as described in Sec\,.~\ref{LHC_constraints}. The resulting excluded regions in the $(\mpsi\,,~\mphi)$ plane are shown by the pink shaded area. The corresponding LHC constraint on the VLL sector, obtained from the pair production of $\chi\bar{\chi}$ followed by $\chi\rightarrow\Phi\,\ell$, is presented in the $(\mchi,~\mphi)$ plane in the left panel of \fig\ref{fig:exclusion_discovery_contours}. These constraints, for benchmark \texttt{BP1}, also exclude regions of the model parameter space in \fig\ref{fig:relic-summary-1sigma} that are already disallowed by other constraints considered in this analysis and thus have not been shown explicitly.

The projected HL-LHC sensitivity is obtained using the cut-based analyses described in Secs\,.~\ref{sec:HLLHC_VLQ} and \ref{sec:HLLHC_VLL}. The regions enclosed by the pink dashed contours in \figs\ref{fig:relic-summary-1sigma} correspond to the parameter space that can be probed through VLQ pair production with a discovery significance of $\mathcal{Z}\geq3$. Similarly, the projected HL-LHC discovery reach for the VLL sector in the $(\mchi,~\mphi)$ plane is shown in the left panel of \fig\ref{fig:exclusion_discovery_contours}. These results demonstrate that the HL-LHC can probe a substantial region of the parameter space that remains unconstrained by current LHC searches.

\item The dashed lines in \Fig\ref{fig:relic-summary-1sigma} show the bounds from different low-energy flavor observables and electroweak observables. The arrows indicate the direction of the allowed parameter space. The bounds from the various observables are obtained by considering the corresponding $1\sigma$ uncertainties of the measurements and upper bounds at $90\%$ ($95\%$) C.L., wherever applicable. The constraints for this benchmark point are shown in detail in \Fig\ref{fig:all_flav_ewpo_combined}, including only those observables that have a significant impact on the allowed parameter space. The remaining observables, which are allowed over the entire parameter space, are not shown in the plot. The upper boundary of the allowed parameter space is determined by the observables $\Delta M_d$ and $K_{L} \to \mu^+ \mu^-$. On the other hand, the most stringent lower boundary is imposed by the observable $\mathcal{B}(\mu \to e \gamma)$. Details of the constraints from the individual observables are discussed in Sec\,.~\ref{sec:all_flav_ewpos}. 
\end{itemize}
Finally, after imposing all relevant constraints applicable to this mass range of $\mpsi$ and $\mphi$, we obtain a very small region of the parameter space around $(\mpsi,~\mphi) \sim (1.65,~1.1)\text{ TeV}$ in the $(\mpsi,~\mphi)$ plane of \fig\ref{fig:relic-summary-1sigma} that simultaneously satisfies all the aforementioned constraints and remains to be probed by future direct, indirect, collider (HL-LHC), and flavor searches. Therefore, any positive signal in one of these searches may hint at the existence of such a model. However, the viability of this model can only be established through multiple complementary experimental observations.
\begin{figure}[htb!]
\centering
\includegraphics[width=1\linewidth]{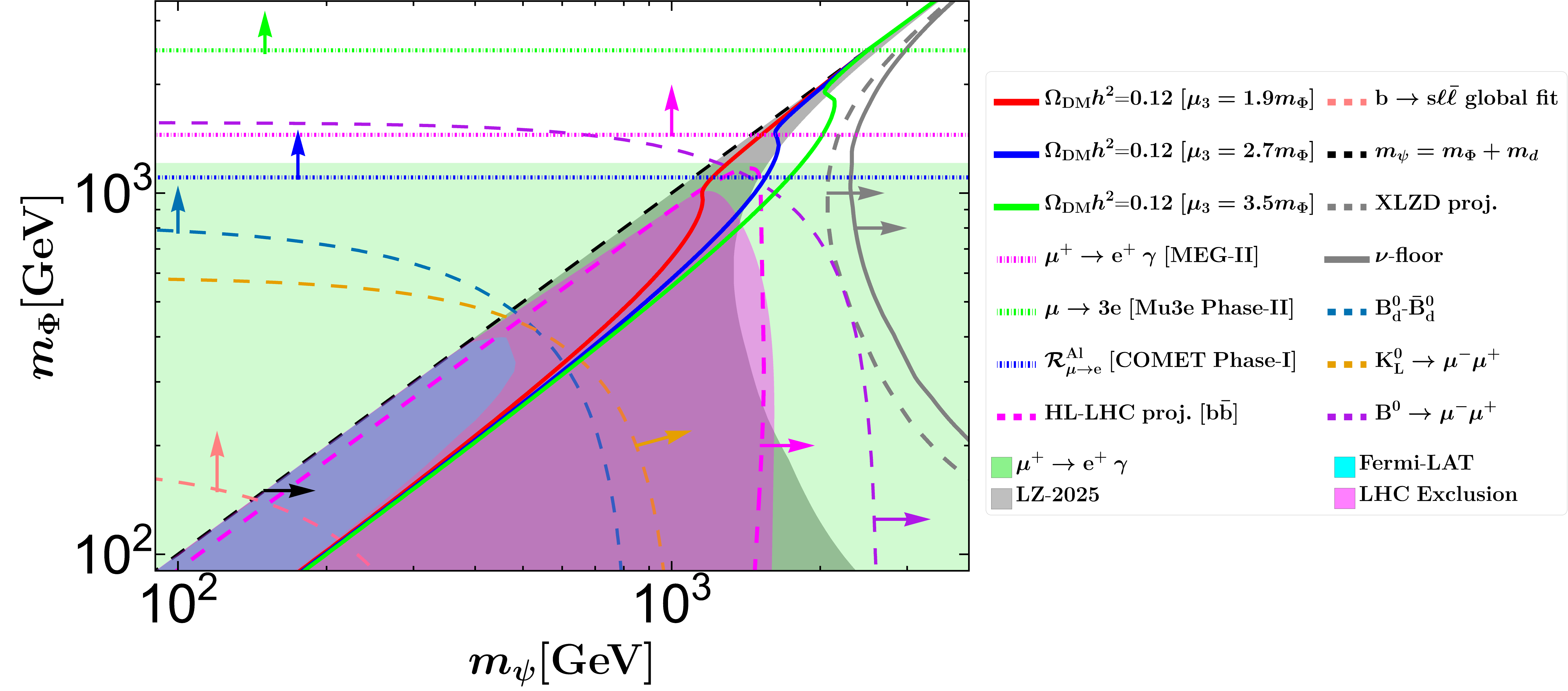}
\caption{Same as \fig\ref{fig:relic-summary-1sigma}, but for \texttt{BP2}, with flavor constraints derived using the $2\sigma$ experimental bounds. Observables not shown in the plot are allowed across the entire parameter space.}
\label{fig:relic-summary-2sigma}
\end{figure}

Figure\,.~\ref{fig:relic-summary-2sigma} presents the corresponding combined analysis for \texttt{BP2}, i.e., the benchmark value $\ys=0.03$, while keeping all other parameters fixed. As in the previous analysis, bounds providing only upper limits are taken at the $90\%$ or $95\%$ confidence level, while the experimental measurements of flavor and electroweak observables are considered with their $2\sigma$ uncertainties. Compared to the previous figure, the excluded regions in the $(\mpsi\,,~\mphi)$ plane are modified, whereas all other features remain essentially unchanged. In particular, the $2\sigma$ treatment weakens the constraints from $\Delta M_{B_d}$ and $K_L^0\to\mu^+\mu^-$, which no longer produce the allowed bands observed in the $1\sigma$ analysis but instead impose only lower bounds on the masses. Consequently, the upper bound on the parameter space is removed, and the entire high-mass region remains allowed by the current flavor and electroweak constraints considered in our analysis.
A substantial region of the parameter space, therefore, remains viable and accessible to future dark matter searches. The upper boundary of the mass plane in this scenario is set by the requirement that freeze-out occurs before EWSB. In particular, the region $800~\GeV \lesssim \mphi \lesssim 4~\TeV$ and $1.5~\TeV \lesssim \mpsi \lesssim 4.0~\TeV$ satisfies all current constraints. This viable region can be further enlarged for other benchmark choices by appropriately adjusting the model parameters to simultaneously satisfy the flavor, collider, and dark matter constraints.
\section{Summary and Conclusion}
\label{sec:summary}
We have investigated a minimal extension of the SM consisting of a $\Zthree$-stabilized complex scalar dark matter candidate, a VLQ, and a VLL, where the new fermions are singlets under $\rm SU(2)_\mathtt{L}$ but carry non-zero $\rm U(1)_\mathtt{Y}$ hypercharge. Besides providing a viable dark matter candidate, the model predicts rich phenomenology at low energies through loop-induced flavor observables, making it testable at Intensity Frontier experiments. At the same time, the requirement of reproducing the observed dark matter relic abundance, together with the stringent constraints from direct and indirect detection experiments, establishes a strong interplay between the Cosmic and Intensity Frontiers, significantly reducing the viable parameter space. The surviving regions are further probed by searches for the new vector-like states at the LHC and the High-Luminosity LHC, thereby linking the Cosmic, Intensity, and Energy Frontiers within a single phenomenological framework.

In addition to the masses of the three new particles, the phenomenology of the model is governed by the Yukawa couplings $\yq$ and $\yl$, which control the interactions of the dark matter particle with the SM quarks and charged leptons through the VLQ and VLL, respectively. These couplings play a central role in determining the flavor, dark matter, and collider phenomenology, leading to strong correlations among the different experimental probes. We have performed a comprehensive phenomenological analysis of this framework, including flavor, electroweak precision, dark matter, and collider observables.

Neutral meson mixing ($B_s^0$, $B^0$, and $K^0$) receives new-physics contributions exclusively from box diagrams involving the complex scalar dark matter particle and the VLQ, with amplitudes determined by the Yukawa couplings $\yq$ and the masses $\mpsi$ and $\mphi$. In contrast, rare leptonic and semileptonic meson decays receive contributions from both box and electroweak penguin diagrams. While the penguin amplitudes depend only on the VLQ sector, the box contributions also involve the VLL through the Yukawa couplings $\yl$ and the mass $\mchi$. Depending on the parameter region, the box diagrams can dominate the new-physics contributions, thereby highlighting the important role of the VLL in flavor observables. On the other hand, rare meson decays with neutrinos in the final state depend exclusively on the VLQ sector and are therefore insensitive to the VLL parameters.

Leptonic observables, including the charged-lepton anomalous magnetic moments and charged lepton flavor-violating processes ($\ell_\alpha \to \ell_\beta \gamma$, $\ell_\alpha \to \ell_\beta \ell_\rho \bar{\ell}_\delta$, and $\mu$--$e$ conversion, arise at one-loop level and depend exclusively on the VLL sector through the parameters $\mchi$, $\mphi$, and $\yl$. The only exception is provided by the semileptonic tau decays ($\tau \to \ell + \text{meson}$), which receive additional box-diagram contributions involving the VLQ and are therefore sensitive to the Yukawa couplings $\yq$ and the mass $\mpsi$. Similar to the charged leptonic meson decays, these processes exhibit a non-trivial interplay between the VLQ and VLL sectors.

Electroweak precision observables, particularly the high-precision $Z$-pole measurements, impose stringent constraints on the model through one-loop corrections to the $Zf\bar{f}$ couplings. Rather than considering the individual partial decay widths, we analyze the experimentally measured observables $R_\ell$, $R_q$, and the forward--backward asymmetries $A_{\mathrm{FB}}^f$, which provide robust and complementary probes of the parameter space. Our analysis shows that the electroweak precision observables place significant constraints on the masses and couplings of the new particles, complementing the constraints obtained from flavor observables.

Beyond the flavor sector, reproducing the observed DM relic abundance constitutes a central motivation for the present framework. Owing to the $\Zthree$ charge assignments, the complex scalar $\Phi$ is stable and serves as the dark matter candidate. Its thermal freeze-out is governed by Higgs-, Yukawa-, and gauge-portal interactions, with self-annihilation, semi-annihilation, and co-annihilation processes involving the vector-like fermions $\psi$ and $\chi$ playing important roles. While direct detection experiments place stringent constraints on the Higgs-portal coupling $\lphiH$ and the Yukawa coupling $\yd$, and indirect detection probes the annihilation channels predominantly controlled by $\yb$, the co-annihilation channels provide sufficient flexibility to reproduce the observed relic abundance without violating the flavor or dark matter constraints. In particular, the mass splittings between the complex scalar and the vector-like fermions, together with the Yukawa couplings and the scalar trilinear coupling $\mu_3$—which affects only the relic abundance—are found to be the key parameters governing the dark matter phenomenology. Our analysis shows that the direct detection bounds require $\lphiH\lesssim5\times10^{-3}$ and $\yd\lesssim 0.05$ for $\mpsi\sim\mathcal{O}(\TeV)$, whereas the current indirect detection limits do not impose any additional constraints on the viable parameter space.

Combining the flavor, electroweak precision, and dark matter constraints, we identify the regions of the parameter space that simultaneously reproduce the observed dark matter relic abundance while satisfying all current experimental measurements at the $1\sigma$ ($2\sigma$) confidence level together with the relevant experimental bounds at $90\%$ C.L., as shown in \fig\ref{fig:relic-summary-1sigma} (\ref{fig:relic-summary-2sigma}). These combined constraints significantly restrict the allowed masses $(\mpsi \,,~\mchi \,,\text{ and }~\mphi)$ and Yukawa couplings. For our benchmark scenario, the $1\sigma$ analysis yields a very narrow region of the parameter space near the $(1.65-1.1)$ TeV point in the $(\mpsi-\mphi)$ plane, whereas the $2\sigma$ analysis leads to a substantially less constrained parameter space, allowing $0.8 \leq \mphi/\TeV \leq 4$ and $1.5 \leq \mpsi/\TeV \leq 4$. These results motivate direct searches for the new particles at high-energy colliders.
At hadron colliders, the vector-like fermions are pair-produced through $s$-channel gauge interactions, with additional $t$-channel contributions arising from the Yukawa interactions. Motivated by the current null results from the LHC, we performed a dedicated projection of the discovery prospects at the High-Luminosity LHC (HL-LHC). Our analysis shows that the VLQ is subject to considerably stronger exclusion limits and correspondingly larger discovery reaches than the VLL in the $b\text{-jets} + E_T^{\text{miss}}$ search channel. This difference originates from the QCD production of the VLQ, supplemented by an additional $t$-channel contribution, whereas VLL pair production proceeds purely through electroweak interactions. These results demonstrate the complementarity of flavor, dark matter, and collider searches, and establish this framework as a well-motivated target for future precision and high-energy experiments.

\subsubsection*{Acknowledgment}
L.K. and R.M. acknowledge support from the DAE-BRNS YSRP grant No.~57/20/02/2024. R.M.\ further acknowledges support from the SERB grant SPG/2022/001238. M.M. further
acknowledges support from the IPPP DIVA Programme. M.M., S.S. and D.P. also acknowledge the use of the SAMKHYA High-Performance Computing Facility at the Institute
of Physics (IOP), Bhubaneswar, as well as the two workstations provided by the Institute
of Physics, Bhubaneswar through the DAE APEX project, which were extensively used
for the numerical computations presented in this work. 

\newpage
\appendix
\section{Relevant Higher-Dimensional Operators in CSDM-EFT}
\label{app:EFT}
The effective field theory (EFT) approach provides a useful framework for studying low-energy BSM physics, particularly in DM searches via direct, indirect, and collider experiments \cite{Alanne:2017oqj, Arcadi:2019lka, Arcadi:2024mli, Criado:2021trs}. Moreover, it is the most effective tool for investigating low-energy flavor phenomenology within the SM framework. The underlying motivation is that a given low-energy experimental observable may receive contributions from various high-energy processes. Since the characteristic energy scale of the experiment is much lower than the masses of the heavy degrees of freedom, these particles cannot be produced on-shell, and their effects are encoded in local, point-like interactions. Consequently, the underlying high-energy dynamics can be systematically captured by effective operators with corresponding Wilson coefficients suppressed by a cut-off scale, which collectively encode the effects of the ultraviolet theory. In the following, we discuss a few of the higher-dimensional operators that can generate the interactions considered in our analysis. We focus only on the possible operator structures and do not discuss the corresponding bounds or phenomenology.

Let's construct the CSDM-EFT operators using a top-down approach from our UV model. For this, we assume that the characteristic energy scale is $\mathcal{E} \ll \{\mpsi\,,~\mchi\}$. In this model framework, the UV Lagrangian corresponding to $\psi$ and $\chi$ is:
\begin{align}
\mathcal{L}_{\psi\chi}\,\equiv\,
\overline{\psi}\left(i\slashed{\mathcal{D}}^{(\psi)}-\mpsi\right)\psi+
\overline{\chi}\left(i\slashed{\mathcal{D}}^{(\chi)}-\mchi\right)\chi-
\left(\mathtt{y}_q^j\overline{\psi}q^j_\mathtt{R}\Phi+
\mathtt{y}_\ell^r\overline{\chi}\ell^{r}_\mathtt{R}\Phi+hc.\right)\,.
\label{eq:psichi}
\end{align}
where $\mathtt{y}_{\ell}=\{\ye,\ymu,\yta\}$ and $\mathtt{y}_{q}=\{\yd,\ys,\yb\}$.

From the equation of motion (EOM) for $\overline\psi$ (or equivalently, for $\psi$), we get,
\begin{subequations}
\begin{gather}
\left(i\slashed{\D}^{(\psi)}-\mpsi\right)\psi=\mathtt{y}_q^j q^j_\mathtt{R}\Phi\\
\implies\psi=\dfrac{1}{i\slashed{\D}^{(\psi)}-\mpsi}\mathtt{y}_q^j q^j_\mathtt{R}\Phi\,.
\end{gather}
\end{subequations}
A series expansion of the inverse operator gives us,
\begin{gather}
\psi=-\dfrac{1}{\mpsi}\left(1+\dfrac{i\slashed{\D}^{(\psi)}}{\mpsi}+\left(\dfrac{i\slashed{\D}^{(\psi)}}{\mpsi}\right)^2+...\right)\mathtt{y}_q^j q^j_\mathtt{R}\Phi\,,
\end{gather}
and
\begin{gather}
\overline\psi=-(\mathtt{y}_q^{k^*} \Phi^*\bar{q}_\mathtt{R}^k)\dfrac{1}{\mpsi}\left(1-\dfrac{i\slashed{\D}^{(\psi)}}{\mpsi}+\left(\dfrac{i\slashed{\D}^{(\psi)}}{\mpsi}\right)^2+...\right)\,.
\end{gather}
In a similar way, we also get for $\chi$,
\begin{gather}
\chi=-\dfrac{1}{\mchi}\left(1+\dfrac{i\slashed{\D}^{(\chi)}}{\mchi}+\left(\dfrac{i\slashed{\D}^{(\chi)}}{\mchi}\right)^2+...\right)\mathtt{y}_\ell^r\ell^r_{\mathtt{R}}\Phi\,,
\end{gather}
and
\begin{gather}
\overline\chi=-(\mathtt{y}_\ell^{p^*} \Phi^*\bar{\ell}^p_\mathtt{R})\dfrac{1}{\mpsi}\left(1-\dfrac{i\slashed{\D}^{(\chi)}}{\mchi}+\left(\dfrac{i\slashed{\D}^{(\chi)}}{\mchi}\right)^2+...\right)\,.
\end{gather}
Substituting the above values of $\{\psi\,,~\bar\psi\}$ and $\{\chi\,,~\bar\chi\}$ in the UV-lagrangian \eq\eqref{eq:psichi}, we get the following tree level effective lagrangian \cite{Fuentes-Martin:2020udw, Alte:2019iug, Crivellin:2022fdf}:
\begin{gather}
\mathcal{L}_{\rm eff}^{\rm dim=5}\equiv 
\dfrac{\mathtt{y}_q^j(\mathtt{y}_q^k)^*}{\mpsi}\boldsymbol{(\Phi^*\Phi)(\bar{d}^j_\mathtt{R}d^k_\mathtt{R})}
\,+\,\dfrac{\mathtt{y}^r_\ell(\mathtt{y}^p_\ell)^*}{\mchi}\boldsymbol{(\Phi^*\Phi)(\bar{\ell}^r_\mathtt{R}\ell^p_\mathtt{R})}\,,
\label{eq:dim5tree}
\end{gather}
and
\begin{eqnarray}
\begin{split}
\mathcal{L}_{\rm eff}^{\rm dim=6}\bigg|_{\rm tree}\equiv
& \dfrac{\mathtt{y}_q^j(\mathtt{y}_q^k)^*}{2m_\psi^2}\boldsymbol{(\Phi^*i\overleftrightarrow{\partial_\mu} \Phi)(\bar{d}^j_{\mathtt{R}}\gamma^\mu d^k_{\mathtt{R}})}
\,+\,\dfrac{\mathtt{y}_q^p(\mathtt{y}_q^s)^*Y_d^{rs}}{2m_\psi^2}\boldsymbol{\left[(\Phi^*\Phi)(\bar{q}^r_{\mathtt{L}}{\rm H}d^p_{\mathtt{R}})\,+\,\text{h.c.}\right]}\\
\,+\, & \dfrac{\mathtt{y}^j_\ell(\mathtt{y}^k_\ell)^*}{2m_\chi^2}\boldsymbol{(\Phi^*i\overleftrightarrow{\partial_\mu}\Phi)(\bar{\ell}_{\mathtt{R}}^{j}\gamma^\mu \ell_{\mathtt{R}}^{k})}
\,+\,\dfrac{\mathtt{y}_\ell^p(\mathtt{y}_\ell^s)^*Y_e^{rs}}{2m_\chi^2}\boldsymbol{\left[(\Phi^*\Phi)(\bar{\ell}^r_{\mathtt{L}}{\rm H} e^p_{\mathtt{R}})\,+\,\text{h.c.}\right]}\,,
\end{split}
\label{eq:dim6tree}
\end{eqnarray}

In the following, we have listed some of the 1-loop level operators using the \textsc{Matchete} package \cite{Fuentes-Martin:2022jrf}.
\begin{eqnarray}
\begin{split}
\mathcal{L}_{\rm eff}^{\rm dim=5}\bigg|^{(0)}_{\rm 1-loop}\equiv & \dfrac{\mu_3}{64\pi^2}\left[\dfrac{|\mathtt{y}^p_\ell|^2}{m_\chi^2}(2\lambda_\Phi-|\mathtt{y}^r_\ell|^2)+3\dfrac{|\mathtt{y}^p_q|^2}{m_\psi^2}(2\lambda_\Phi-|\mathtt{y}^r_q|^2)\right]\boldsymbol{|\Phi|^2[\Phi^3+(\Phi^*)^3]}\\
\,+\,& \dfrac{\mu_3}{64\pi^2}\biggl[
\dfrac{1}{2m_\chi^2}\left(|\mathtt{y}_\ell^p|^2(2\lphiH+3g_Y^2)-15\mathtt{y}_\ell^r(\mathtt{y}_\ell^s)^*(Y_e^{pr})^*Y_e^{ps}\right)\\
\,+\,&\dfrac{1}{m_\chi^2}\left(g_Y^2|\mathtt{y}_\ell^p|^2-3\mathtt{y}_\ell^r(\mathtt{y}_\ell^s)^*(Y_e^{pr})^*Y_e^{ps}\right)\\
\,+\,&\dfrac{3}{2m_\psi^2}\left(|\mathtt{y}_q^p|^2(2\lphiH+g_Y^2)-15\mathtt{y}_q^r(\mathtt{y}_q^s)^*(Y_e^{pr})^*Y_e^{ps}\right)\\
\,+\,&\dfrac{1}{m_\psi^2}\left(g_Y^2|\mathtt{y}_q^p|^2-9\mathtt{y}_q^r(\mathtt{y}_q^s)^*(Y_e^{pr})^*Y_e^{ps}\right)
\biggr]\boldsymbol{(H^\dagger H)[\Phi^3+(\Phi^*)^3]}\,, 
\end{split}
\label{eq:dim6loop0}
\end{eqnarray}

\begin{eqnarray}
\begin{split}
\mathcal{L}_{\rm eff}^{\rm dim=6}\bigg|^{(1)}_{\rm 1-loop}\equiv
& \dfrac{|\mathtt{y}^i_q|^2}{96\pi^2m_\psi^2}\boldsymbol{(\Phi^*\Phi)\textbf{G}_{\mu\nu}\textbf{G}^{\mu\nu}}
\,+\, \dfrac{1}{144\pi^2}\left(\dfrac{|\mathtt{y}^i_q|^2}{m_\psi^2}\,+\,3\dfrac{|\mathtt{y}^i_\ell|^2}{m_\chi^2}\right)\boldsymbol{(\Phi^*\Phi)\textbf{B}_{\mu\nu}\textbf{B}^{\mu\nu}}\,,
\end{split}
\label{eq:dim6loop1}
\end{eqnarray}

\begin{eqnarray}
\begin{split}
\mathcal{L}_{\rm eff}^{\rm dim=6}\bigg|^{(2)}_{\rm 1-loop}\equiv &\dfrac{1}{1152\pi^2}\biggl[\dfrac{9}{m_\psi^2}(\mathtt{y}_q^j)^* (Y_d^{pk})^*\mathtt{y}_q^kY_d^{rj}\left(1-2\ln\dfrac{\mu^2}{m_\psi^2}\right)-2g_Y^2\biggl(\dfrac{(\mathtt{y}_\ell^j)^*\mathtt{y}_\ell^j}{m_\chi^2}\left(3+2\ln\dfrac{\mu^2}{m^2_\chi}\right)\\
\,+\, &\dfrac{(\mathtt{y}_q^j)^*\mathtt{y}_q^j}{m_\psi^2}\left(3+2\ln\dfrac{\mu^2}{m^2_\psi}\right)\biggr)\delta_{pr}\biggr]\boldsymbol{(\Phi^*i\overleftrightarrow{\partial_\mu}\Phi)(\bar{q}^r_{\mathtt{L}}\gamma^\mu q^p_{\mathtt{L}})}\\
\,+\, &\dfrac{g_Y^2}{144\pi^2}\left[\dfrac{\mathtt{y}^i_\chi(\mathtt{y}^i_\ell)^*}{m_\chi^2}\left(3\,+\,2\ln\dfrac{\mu^2}{m^2_\chi}\right)
\,+\,\dfrac{\mathtt{y}^i_\psi(\mathtt{y}^i_\psi)^*}{m_\psi^2}\left(3\,+\,2\ln\dfrac{\mu^2}{m^2_\psi}\right)\right]
\boldsymbol{(\Phi^*i\overleftrightarrow{\partial_\mu}\Phi)(\bar{u}^j_{\mathtt{R}}\gamma^\mu u^j_{\mathtt{R}})}\,\\
\,+\, & \biggl[\dfrac{(\mathtt{y}_q^r)^*\mathtt{y}_q^p}{2m_\psi^2}
\,+\, \dfrac{1}{576\pi^2}\dfrac{(\mathtt{y}_\ell^s)^*\mathtt{y}_\ell^s}{m_\chi^2m_\psi^2}\biggl(-9m_\chi^2(\mathtt{y}_q^r)^*\mathtt{y}_q^p\left(1+2\ln\dfrac{\mu^2}{m_\chi^2}\right)\\
\,+\, &2g_Y^2m_\psi^2\left(3+2\ln\dfrac{\mu^2}{m_\chi^2}\right)\delta_{pr}\biggr)
\,+\, \dfrac{g_Y^2(\mathtt{y}_q^s)^*\mathtt{y}_q^s}{268\pi^2m_\psi^2}\left(3+2\ln\dfrac{\mu^2}{m_\psi^2}\right)\delta_{pr}\\
\,-\, & \dfrac{(\mathtt{y}_q^r)^*\mathtt{y}_q^p}{128\pi^2m_\psi^4}\left(9\mu_3^2\left(3+2\ln\dfrac{\mu^2}{m_\psi^2}\right)+m_\psi^2(\mathtt{y}_q^s)^*\mathtt{y}_q^s\left(13+14\ln\dfrac{\mu^2}{m_\psi^2}\right)\right)\biggr]
\boldsymbol{(\Phi^*i\overleftrightarrow{\partial_\mu}\Phi)(\bar{d}^r_{\mathtt{R}}\gamma^\mu d^p_{\mathtt{R}})}\,,
\end{split}
\label{eq:dim6loop2}
\end{eqnarray}

\begin{eqnarray}
\begin{split}
\mathcal{L}_{\rm eff}^{\rm dim=6}\bigg|^{(3)}_{\rm 1-loop}\equiv &
\dfrac{1}{384\pi^2}\biggl[\dfrac{3}{m_\chi^2}(\mathtt{y}_\ell^j)^* \mathtt{y}_\ell^k(Y_e^{pk})^*Y_e^{rj}\left(1-2\ln\dfrac{\mu^2}{m_\chi^2}\right)+2g_Y^2\biggl(\dfrac{(\mathtt{y}_\ell^j)^*\mathtt{y}_\ell^j}{m_\chi^2}\left(3+2\ln\dfrac{\mu^2}{m^2_\chi}\right)\\
\,+\, &\dfrac{(\mathtt{y}_q^j)^*\mathtt{y}_q^j}{m_\psi^2}\left(3+2\ln\dfrac{\mu^2}{m^2_\psi}\right)\biggr)\delta_{pr}\biggr]\boldsymbol{(\Phi^*i\overleftrightarrow{\partial_\mu}\Phi)(\bar{\ell}^r_{\mathtt{L}}\gamma^\mu \ell^p_{\mathtt{L}})}\\
\,+\, & \biggl[\dfrac{(\mathtt{y}_\ell^r)^*\mathtt{y}_\ell^p}{2m_\chi^2}
\,-\, \dfrac{3}{128\pi^2}\dfrac{(\mathtt{y}_\ell^r)^*\mathtt{y}_\ell^p}{m_\chi^4}\biggl((m_\chi^2(\mathtt{y}_\ell^s)^*\mathtt{y}_\ell^s+3\mu_3^2)\left(3+2\ln\dfrac{\mu^2}{m_\chi^2}\right)\\
\,+\, & 2m_\chi^2(\mathtt{y}_q^s)^*\mathtt{y}_q^s\left(3+2\ln\dfrac{\mu^2}{m_\psi^2}\right)\biggr)
\,+\,\dfrac{1}{96\pi^2}g_Y^2\biggl(\dfrac{(\mathtt{y}_\ell^s)^*\mathtt{y}_\ell^s}{m_\chi^2}\left(3+2\ln\dfrac{\mu^2}{m_\chi^2}\right)\\
\,+\, &\dfrac{(\mathtt{y}_q^s)^*\mathtt{y}_q^s}{m_\psi^2}\left(3+2\ln\dfrac{\mu^2}{m_\psi^2}\right)\biggr)\delta_{pr}\biggr]\boldsymbol{(\Phi^*i\overleftrightarrow{\partial_\mu}\Phi)(\bar{\ell}^r_{\mathtt{R}}\gamma^\mu \ell^p_{\mathtt{R}})}\,,
\end{split}
\label{eq:dim6loop3}
\end{eqnarray}

\begin{eqnarray}
\begin{split}
\mathcal{L}_{\rm eff}^{\rm dim=6}\bigg|^{(4)}_{\rm 1-loop}\equiv &\dfrac{1}{96\pi^2}\biggl[-\dfrac{g_Y^2}{m_\chi^2}(\mathtt{y}_\ell^j)^*\mathtt{y}_\ell^j\left(3+2\ln\dfrac{\mu^2}{m_\chi^2}\right)\,+\,\dfrac{3}{m_\chi^2}(\mathtt{y}_\ell^s)^*\mathtt{y}_\ell^r(Y_e^{pr})^\dagger Y_e^{ps}\left(3+2\ln\dfrac{\mu^2}{m_\chi^2}\right)\\
\,-\, &\dfrac{1}{m_\psi^2}\left(g_Y^2(\mathtt{y}_q^p)^*\mathtt{y}_q^p-9(\mathtt{y}_q^s)^*\mathtt{y}_q^r (Y^{pr}_d)^*Y_d^{ps}\right)\left(3+2\ln\dfrac{\mu^2}{m_\psi^2}\right)\biggr]
\boldsymbol{(\Phi^*\overleftrightarrow{\partial^\mu}\Phi)(H\overleftrightarrow{\mathcal{D}_\mu}H^\dagger)}\,\\&
\,+\,\#_1~(\bar{\ell}^r_\mathtt{L}He_R^p)(\Phi^*\Phi)
\,+\,\#_2~(\bar{q}^r_\mathtt{L}Hd_R^p)(\Phi^*\Phi)
\,\,.\,.\,.
\end{split}
\label{eq:dim6loop4}
\end{eqnarray}
where $\qR = \{\dR, \sR, \bR\}$, $\lR = \{\eR, \muR, \tauR\}$, and $\#_{1,2}$ are loop-generated coefficients similar to the ones above. If we relax the assumption of a common matching scale $\mu\equiv \sqrt{\mpsi\mchi}$ for $\mpsi\sim \mchi$ and instead introduce two distinct scales, $\mu_\psi$ and $\mu_\chi$ for the regimes where $\mpsi\gg\mchi$ or $\mpsi\ll\mchi$, associated with integrating out the heavy VLFs $\psi$ and $\chi$ respectively, we must account for the renormalization group evolution between them. It is worth noting that while the dimension-6 DM-EFT operators remain invariant under the full SM gauge symmetry, the dimension-5 LEFT operators are valid only below the EWSB scale. 

\section{Loop contributions for the flavor and electroweak observables}
\label{app:loops}
\paragraph{\underline{Neutral Meson Mixing:}}
The loop contribution for the neutral meson mixing process of $P^0 - \bar P^0$, with $P^0 = (d_i \bar d_j)$, as introduced in \eq\eqref{eq:mixing_loop}, given by : 
\begin{equation}\label{eq:mixing_loop_eff}
\mathcal{C}_{\rm RR} = \frac{|\mathtt{y}_{d_i}\,\mathtt{y}_{d_j}|^2}{32\pi^2}\frac{m_{\Phi}^4-m_{\psi}^4-2m_{\Phi}^2m_{\psi}^2\log\!\left(\frac{m_{\Phi}^2}{m_{\psi}^2}\right)}{\left(m_{\Phi}^2-m_{\psi}^2\right)^3}\,.
\end{equation}
\paragraph{\underline{Semileptonic transition: $d_i \to d_j \, \ell_\alpha \, \bar{\ell}_\beta $}}
The contribution to the semileptonic transition arises from three-loop diagrams: box and penguins, with $\gamma$ and $Z$ mediation. This will contribute to various processes like semileptonic decays $P \to M \, \ell \,\bar\ell$, where $\alpha = \beta$, and LFV decays such as $P^0 \to \ell_\alpha \, \bar{\ell}_\beta$. Subsequently, we will provide the loop contributions in the massless external-particle limit.
\begin{subequations}
\begin{equation}\begin{split}
\mathcal{C}_{\rm peng}^\gamma & = \frac{Q_{f} e^2 \, \mathtt{y}_{d_i} \mathtt{y}_{d_j}}{1728 \pi^2 \, (m_{\Phi}^2 - m_{\psi}^2)^4}\Big[ 2m_{\Phi}^{6}+25m_{\Phi}^{4}m_{\psi}^{2}+27m_{\Phi}^{2}m_{\psi}^{2}\left(m_{\Phi}^{2}-2m_{\psi}^{2}\right)  \\ &  +6m_{\Phi}^{2}\left(2m_{\Phi}^{4}-9m_{\Phi}^{2}m_{\psi}^{2}+6m_{\psi}^{4}\right)\log\!\left(\frac{m_{\Phi}^{2}}{m_{\psi}^{2}}\right) \Big]\,,
\end{split}
\end{equation}
\begin{equation}
\begin{split}
\mathcal{C}_{\rm peng}^{Z} & = \frac{ Q_{f} e\,g\,\sin\theta_W\, \mathtt{y}_{d_i}\, \mathtt{y}_{d_j}  }{ 384\,\cos^{2}\theta_W\, m_Z^{2}\, \left(m_{\Phi}^{2}-m_{\psi}^{2}\right)^{2} \pi^{2}}\Bigg[ 2m_{\Phi}^{2}\left(m_{\Phi}^{2}-2m_{\psi}^{2}\right)\log\!\left(\frac{m_{\Phi}^{2}}{m_{\psi}^{2}}\right)  \\ & 
 - \left(m_{\Phi}^{2}-m_{\psi}^{2}\right) \left( m_{\Phi}^{2}-3m_{\psi}^{2} 
\right) \Bigg]\,,
\end{split}
\end{equation}
\begin{equation}\begin{split}
\mathcal{C_{\rm box}} & = \frac{
\mathtt{y}_{d_i}\,\mathtt{y}_{d_j}\,\mathtt{y}_{\ell_\alpha}\, \mathtt{y}_{\ell_\beta}}{32\pi^{2}
\left(m_{\Phi}^{2}-m_{\chi}^{2}\right)^{2}
\left(m_{\Phi}^{2}-m_{\psi}^{2}\right)^{2}
\left(m_{\chi}^{2}-m_{\psi}^{2}\right)}
\left[m_{\chi}^{4}\left(m_{\Phi}^{2}-m_{\psi}^{2}\right)^{2}
\log\!\left(\frac{m_{\Phi}^{2}}{m_{\psi}^{2}}\right) \right.\\&\left.  
 +\left(m_{\Phi}^{2}-m_{\chi}^{2}\right)\left(m_{\Phi}^{2}\left(m_{\Phi}^{2}
 -m_{\psi}^{2}\right)\left(m_{\psi}^{2}-m_{\chi}^{2}\right)
 -\left(m_{\chi}^{2}-m_{\Phi}^{2}\right)m_{\psi}^{4}
\log\!\left(\frac{m_{\Phi}^{2}}{m_{\psi}^{2}}\right)
\right)\right]\,.
\end{split}\end{equation}
\end{subequations}

The above expressions arise from the different loop calculations. For the transition $d_i \to d_j \ell\bar\ell$, the charge of the VLF will be $Q_f = -1/3$. The corresponding Wilson coefficients (as used in eq.~\eqref{eq:rare_BR_formula}) can be written in terms of the above functions: 
\begin{equation}
\begin{split}
\mathcal{C}_{9}^\prime &=\frac{1}{K_{c}}\left(\mathcal{C_{\rm box}}  + \mathcal{C}_{\rm peng}^\gamma + c_{V_{\ell}} \mathcal{C}_{\rm peng}^{Z} \right)\,, \\
\mathcal{C}_{10}&=\frac{c_{V_{\ell}}^{} }{K_{c}}\mathcal{C}_{\rm peng}^{Z}\,, \\
\mathcal{C}_{10}^\prime&=\frac{1}{K_{c}}\left(\mathcal{C_{\rm box}}-c_{A_{\ell}} \mathcal{C}_{\rm peng}^{Z}\right)\,.
\end{split}
\label{eq:wilson_expression}
\end{equation}
where $c_{V_{f}}\,,~c_{A_{f}}$ being the vector and axial vector coupling of $Z$ boson with the SM fermion pairs. The common constant to all the Wilson coefficients can be given by:
\begin{equation}
K_{c}=\frac{\sqrt{2\,\pi}}{\alpha\, G_F\, |V_{ti}\,V_{tj}^*|}\,.
\end{equation}
In our case, the Wilson coefficient $C_9$ is suppressed by the external particle mass, hence we have not shown the expression here. 

For the process, with quark transition $d_i \to d_j \, \ell_\alpha \, \ell_\beta$ (with $i \neq j$ and $\alpha \neq \beta$), only the box diagram will contribute. The expressions can be obtained by replacing the VLQ mass and coupling with VLL and the appropriate replacement of the couplings, along with $Q_f = -1$. In the text, we have written this expression in terms of the chiral basis, which will be given by (as given by Eq.~\eqref{eq:loop_P2lilj}):
\begin{align}\label{eq:LFV_boxLoop}
C_{\rm RR} = 2\,\mathcal{C_{\rm box}}\,. 
\end{align}

For the transition $\tau \to \ell \, d_{i} \, \bar d_i $, an additional contribution will arise to both $C_{\rm RR}$ and $C_{\rm RL}$, which are given in the following with the suitable replacement of $\psi$ by $\chi$:
\begin{subequations}\begin{align}
C_{\rm RR} &= 2 \mathcal{C}_{\rm box} + \mathcal{C}_{\rm peng}^{\gamma} + (c_{V_{d_i}} - c_{A_{d_i}}) \mathcal{C}_{\rm penguin}^Z \,,  \\
C_{\rm RL} &= \mathcal{C}_{\rm peng}^{\gamma} + (c_{V_{d_i}} + c_{A_{d_i}}) \mathcal{C}_{\rm penguin}^Z \,,
\end{align}\end{subequations}
\paragraph{\underline{ Radiative lepton decays and anomalous dipole Moments:}}

The coefficients for $\ell_i \to \ell_j \gamma$ transition introduced in Eq.~\eqref{eq:gen_lagragian_magnetic_moment} are obtained as
\begin{subequations}
\begin{eqnarray}\label{eq:magnetic_moment_loop}
& \mu_{ij}^M = \frac{e \,( m_{i} + m_{j}) \mathtt{y}_{\ell_i} \mathtt{y}^*_{\ell_j} }{192 \pi^2 (m_{\Phi}^2 - m_{\psi}^2)^4} \left[ m_{\psi}^6 + 2 m_{\Phi}^6 + 3 m_{\Phi}^2 m_{\psi}^2 (m_{\Phi}^2 - 2 m_{\psi}^2) - 6 m_{\Phi}^4 m_{\psi}^2 \, \log\, \frac{m_{\Phi^2}}{m_\psi^2} \right] \,, \\ 
&\mu_{ij}^E = \frac{e \,( m_{i} - m_{j}) \mathtt{y}_{\ell_i} \mathtt{y}^*_{\ell_j}}{192 \pi^2 (m_{\Phi}^2 - m_{\psi}^2)^4} \left[ m_{\psi}^6 + 2 m_{\Phi}^6 + 3 m_{\Phi}^2 m_{\psi}^2 (m_{\Phi}^2 - 2 m_{\psi}^2) - 6 m_{\Phi}^4 m_{\psi}^2 \, \log\, \frac{m_{\Phi^2}}{m_\psi^2} \right] \,. 
\end{eqnarray}
\end{subequations}

\paragraph{\underline{Electroweak observables:}} 
Although the penguin diagram of the $Z \to f_i \bar{f}_j$ decay has the same structure, in this case, we have non-negligible transfer momentum, and hence, the loop function in eq.~\eqref{eq:Z_eff_coup} will be given as :
\begin{equation}
\delta v_f = \delta a_f   = - \frac{Q_{\psi/\chi} \, s_W^2 \, \mathtt{y}_i \mathtt{y}_j}{32\,\pi^2\,m_{Z}^2} \, L_{Z}  \,,
\end{equation}
with the loop function $L_{Z}$ given by:
\begin{align}
L_{Z}\,=\, & m_Z^2 + 2m_{\Phi}^2 - 2m_\psi^2 + 2m_{\Phi}^2 \log\left(\frac{m_{\Phi}^2}{m_\psi^2}\right) \nonumber \\
+& 2\left[ m_{\Phi}^2\left(m_{\Phi}^2-2m_\psi^2\right) +m_\psi^2\left(m_\psi^2-m_Z^2\right) \right] C_{0}( 0,0,-m_Z^2,\mpsi,\mphi,\mpsi)  \nonumber  \\
-& 2\left(m_Z^2+2m_{\Phi}^2-2m_\psi^2\right) \sqrt{1+4\frac{m_\psi^2}{m_Z^2}} \log\left[ \frac{m_Z^2+2m_\psi^2+ \sqrt{m_Z^4+4m_Z^2m_\psi^2}}{ 2m_\psi^2 } \right]\,,
\end{align}
with $C_0$ is the scalar three-point Passarino–Veltman loop function.
\section{Feynman diagram relevant for Dark Matter phenomenology}
\label{app:feynman}

The Feynman diagrams for DM annihilation, semi-annihilation and co-annihilation channels are shown in Figs.~\ref{fig:feynman-relic1}, \ref{fig:feynman-relic2} and \ref{fig:feynman-relic3}.

\begin{figure}[htb!]
\centering
\begin{adjustbox}{width=1.0\textwidth, keepaspectratio}
\subfloat[]{
\begin{tikzpicture}
\begin{feynman}
\vertex (a);
\vertex[above left=0.45cm and 0.9cm of a] (a1){\(\Phi\)};
\vertex[above right=0.45cm and 0.9cm  of a] (a2){\(h\)}; 
\vertex[below = 0.9cm of a] (b); 
\vertex[below left=0.45cm and 0.9cm of b] (c1){\(\Phi\)};
\vertex[below right=0.45cm and 0.9cm of b] (c2){\(h\)};
\diagram*{
(a1) -- [line width=0.35mm,charged scalar, arrow size=0.8pt,style=black] (a),
(a2) -- [line width=0.35mm, scalar, arrow size=0.8pt,style=black] (a),
(a) -- [line width=0.35mm, charged scalar, arrow size=0.8pt, edge label={\(\color{black}{\Phi}\)}, style=black!50] (b),
(b) -- [line width=0.35mm,charged scalar, arrow size=0.8pt] (c1),
(b) -- [line width=0.35mm,scalar, arrow size=0.8pt] (c2)};
\node at (a)[circle,fill,style=black,inner sep=1pt]{};
\node at (b)[circle,fill,style=black,inner sep=1pt]{};
\end{feynman}
\end{tikzpicture}\label{fig:feyn01}}
\subfloat[]{
\begin{tikzpicture}
\begin{feynman}
\vertex (a);
\vertex[above left=0.9cm and 0.9cm of a] (a1){\(\Phi\)};
\vertex[above right=0.9cm and 0.9cm  of a] (a2){\(h\)}; 
\vertex[below left=0.9cm and 0.9cm of a] (c1){\(\Phi\)};
\vertex[below right=0.9cm and 0.9cm of a] (c2){\(h\)};
\diagram*{
(a1) -- [line width=0.35mm,charged scalar, arrow size=0.8pt,style=black] (a),
(a2) -- [line width=0.35mm, scalar, arrow size=0.8pt,style=black] (a),
(a) -- [line width=0.35mm,charged scalar, arrow size=0.8pt] (c1),
(a) -- [line width=0.35mm,scalar, arrow size=0.8pt] (c2)};
\node at (a)[circle,fill,style=black,inner sep=1pt]{};
\end{feynman}
\end{tikzpicture}\label{fig:feyn02}}
\subfloat[]{\begin{tikzpicture}
\begin{feynman}
\vertex (a);
\vertex[above left=0.9cm and 0.9cm  of a] (a1){\(\Phi\)};
\vertex[below left=0.9cm and 0.9cm  of a] (a2){\(\Phi\)}; 
\vertex[right=0.9cm of a] (b); 
\vertex[above right=0.9cm and 0.9cm of b] (c1){\(\rm SM\)};
\vertex[below right=0.9cm and 0.9cm of b] (c2){\(\rm SM\)};
\diagram*{
(a1) -- [line width=0.35mm, charged scalar, arrow size=0.8pt,style=black] (a),
(a) -- [line width=0.35mm,charged scalar, arrow size=0.8pt,style=black] (a2),
(a) -- [line width=0.35mm,scalar,arrow size=0.8pt, edge label={\(\color{black}{h}\)}, style=black!50] (b),
(b) -- [line width=0.35mm,plain, arrow size=0.8pt] (c1),
(b) -- [line width=0.35mm,plain, arrow size=0.8pt] (c2)};
\node at (a)[circle,fill,style=black,inner sep=1pt]{};
\node at (b)[circle,fill,style=black,inner sep=1pt]{};
\end{feynman}
\end{tikzpicture}\label{fig:feyn03}}
\subfloat[]{
\begin{tikzpicture}
\begin{feynman}
\vertex (a);
\vertex[above left=0.45cm and 0.9cm of a] (a1){\(\Phi\)};
\vertex[above right=0.45cm and 0.9cm  of a] (a2){\(\ell_i(q_i)\)}; 
\vertex[below = 0.9cm of a] (b); 
\vertex[below left=0.45cm and 0.9cm of b] (c1){\(\Phi\)};
\vertex[below right=0.45cm and 0.9cm of b] (c2){\(\ell_j(q_j)\)};
\diagram*{
(a1) -- [line width=0.35mm, charged scalar, arrow size=0.8pt,style=black] (a),
(a) -- [line width=0.35mm, fermion, arrow size=0.8pt,style=black] (a2),
(b) -- [line width=0.35mm, fermion, arrow size=0.8pt, edge label'={\(\color{black}{\chi(\psi)}\)}, style=black!50] (a) ,
(c2) -- [line width=0.35mm, fermion, arrow size=0.8pt] (b),
(b) -- [line width=0.35mm, charged scalar, arrow size=0.8pt] (c1)};
\node at (a)[circle,fill,style=black,inner sep=1pt]{};
\node at (b)[circle,fill,style=black,inner sep=1pt]{};
\end{feynman}
\end{tikzpicture}\label{fig:feyn04}}
\end{adjustbox}

\begin{adjustbox}{width=0.9\textwidth, keepaspectratio}
\subfloat[]{
\begin{tikzpicture}
\begin{feynman}
\vertex (a);
\vertex[above left=0.45cm and 0.9cm of a] (a1){\(\Phi\)};
\vertex[above right=0.45cm and 0.9cm  of a] (a2){\(\Phi\)}; 
\vertex[below = 0.9cm of a] (b); 
\vertex[below left=0.45cm and 0.9cm of b] (c1){\(\Phi\)};
\vertex[below right=0.45cm and 0.9cm of b] (c2){\(h\)};
\diagram*{
(a1) -- [line width=0.35mm,charged scalar, arrow size=0.8pt,style=black] (a),
(a2) -- [line width=0.35mm,charged scalar, arrow size=0.8pt,style=black] (a),
(b) -- [line width=0.35mm, charged scalar, arrow size=0.8pt, edge label={\(\color{black}{\Phi}\)}, style=black!50] (a) ,
(c1) -- [line width=0.35mm,charged scalar, arrow size=0.8pt] (b),
(b) -- [line width=0.35mm,scalar, arrow size=0.8pt] (c2)};
\node at (a)[circle,fill,style=black,inner sep=1pt]{};
\node at (b)[circle,fill,style=black,inner sep=1pt]{};
\end{feynman}
\end{tikzpicture}\label{fig:feyn05}}
\subfloat[]{
\begin{tikzpicture}
\begin{feynman}
\vertex (a);
\vertex[above left=0.9cm and 0.9cm  of a] (a1){\(\Phi\)};
\vertex[below left=0.9cm and 0.9cm  of a] (a2){\(\Phi\)}; 
\vertex[right=0.9cm of a] (b); 
\vertex[above right=0.9cm and 0.9cm of b] (c1){\(\Phi\)};
\vertex[below right=0.9cm and 0.9cm of b] (c2){\(h\)};
\diagram*{
(a1) -- [line width=0.35mm, charged scalar, arrow size=0.8pt,style=black] (a),
(a2) -- [line width=0.35mm,charged scalar, arrow size=0.8pt,style=black] (a),
(b) -- [line width=0.35mm,charged scalar,arrow size=0.8pt, edge label={\(\color{black}{\Phi}\)}, style=black!50] (a),
(c1) -- [line width=0.35mm,charged scalar, arrow size=0.8pt] (b),
(b) -- [line width=0.35mm,scalar, arrow size=0.8pt] (c2)};
\node at (a)[circle,fill,style=black,inner sep=1pt]{};
\node at (b)[circle,fill,style=black,inner sep=1pt]{};
\end{feynman}
\end{tikzpicture}\label{fig:feyn06}}
\subfloat[]{\begin{tikzpicture}
\begin{feynman}
\vertex (a);
\vertex[above left=0.9cm and 0.9cm  of a] (a1){\(\Phi\)};
\vertex[below left=0.9cm and 0.9cm  of a] (a2){\(\Phi\)}; 
\vertex[right=0.9cm of a] (b); 
\vertex[above right=0.9cm and 0.9cm of b] (c1){\(\chi(\psi)\)};
\vertex[below right=0.9cm and 0.9cm of b] (c2){\(\ell_i(q_i)\)};
\diagram*{
(a1) -- [line width=0.35mm, charged scalar, arrow size=0.8pt,style=black] (a),
(a2) -- [line width=0.35mm,charged scalar, arrow size=0.8pt,style=black] (a),
(b) -- [line width=0.35mm,charged scalar,arrow size=0.8pt, edge label={\(\color{black}{\Phi}\)}, style=black!50] (a),
(c1) -- [line width=0.35mm,fermion, arrow size=0.8pt] (b),
(b) -- [line width=0.35mm,fermion, arrow size=0.8pt] (c2)};
\node at (a)[circle,fill,style=black,inner sep=1pt]{};
\node at (b)[circle,fill,style=black,inner sep=1pt]{};
\end{feynman}
\end{tikzpicture}\label{fig:feyn07}}
\end{adjustbox}
\caption{Dark Matter self ($\Phi\Phi^*\to\rm SM ~SM$ [\ref{fig:feyn01},~\ref{fig:feyn02},~\ref{fig:feyn03},~\ref{fig:feyn04}]) and semi-annihilation ($\Phi\Phi\to\Phi^* h$ [\ref{fig:feyn05},~\ref{fig:feyn06}] and $\Phi\Phi\to\chi(\psi)\ell_i(q_i)$ [\ref{fig:feyn07}]) where SM $=\{h,~W,~Z,~\text{quarks and leptons}\}$, $\ell_{i,j}=\{e,~\mu,~\tau\}$ and $q_{i,j}=\{d,~s,~b\}$.}
\label{fig:feynman-relic1}
\end{figure}
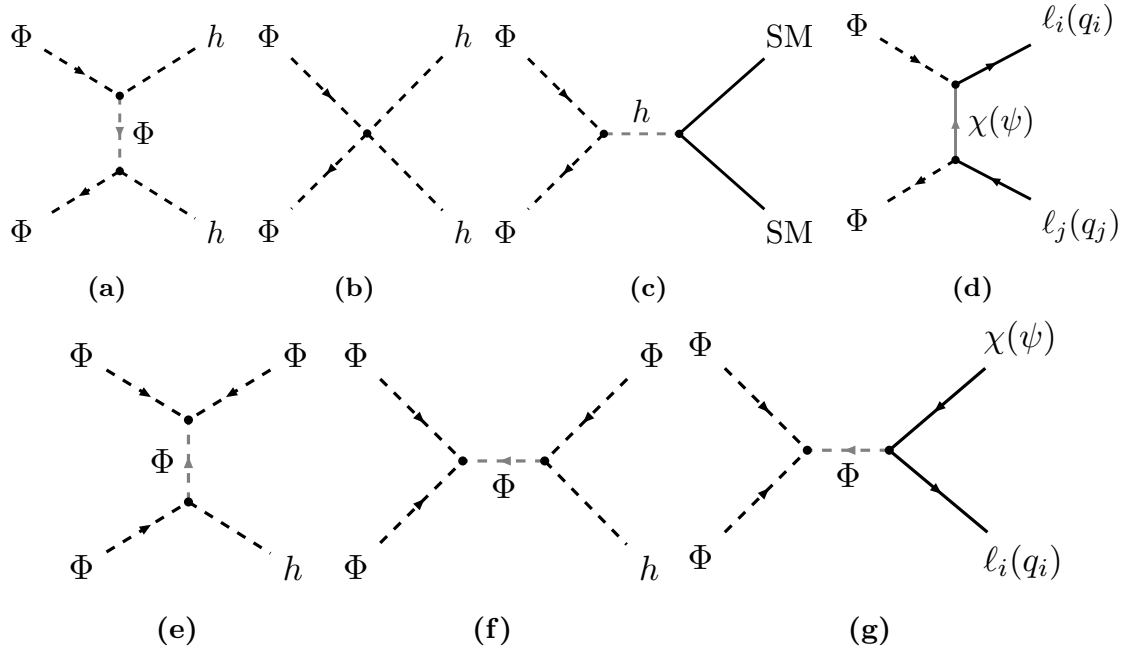

\begin{figure}[htb!]
\centering
\begin{adjustbox}{width=1.0\linewidth, keepaspectratio}
\subfloat[]{
\begin{tikzpicture}
\begin{feynman}
\vertex (a);
\vertex[above left=0.45cm and 0.9cm of a] (a1){\(\psi\)};
\vertex[above right=0.45cm and 0.9cm  of a] (a2){\(q_i\)}; 
\vertex[below = 0.9cm of a] (b); 
\vertex[below left=0.45cm and 0.9cm of b] (c1){\(\chi\)};
\vertex[below right=0.45cm and 0.9cm of b] (c2){\(\ell_j\)};
\diagram*{
(a1) -- [line width=0.35mm, fermion, arrow size=0.8pt,style=black] (a),
(a) -- [line width=0.35mm, fermion, arrow size=0.8pt,style=black] (a2),
(a) -- [line width=0.35mm, charged scalar, arrow size=0.8pt, edge label'={\(\color{black}{\Phi}\)}, style=black!50] (b) ,
(c2) -- [line width=0.35mm, fermion, arrow size=0.8pt] (b),
(b) -- [line width=0.35mm, fermion, arrow size=0.8pt] (c1)};
\node at (a)[circle,fill,style=black,inner sep=1pt]{};
\node at (b)[circle,fill,style=black,inner sep=1pt]{};
\end{feynman}
\end{tikzpicture}\label{fig:feyn08}}
\subfloat[]{\begin{tikzpicture}
\begin{feynman}
\vertex (a);
\vertex[above left=0.45cm and 0.9cm of a] (a1){\(\chi(\psi)\)};
\vertex[above right=0.45cm and 0.9cm  of a] (a2){\(\ell_i(q_j)\)}; 
\vertex[below = 0.9cm of a] (b); 
\vertex[below left=0.45cm and 0.9cm of b] (c1){\(\Phi\)};
\vertex[below right=0.45cm and 0.9cm of b] (c2){\(\Phi\)};
\diagram*{
(a1) -- [line width=0.35mm, fermion, arrow size=0.8pt,style=black] (a),
(a) -- [line width=0.35mm, fermion, arrow size=0.8pt,style=black] (a2),
(a) -- [line width=0.35mm, charged scalar, arrow size=0.8pt, edge label'={\(\color{black}{\Phi}\)}, style=black!50] (b) ,
(c2) -- [line width=0.35mm, charged scalar, arrow size=0.8pt] (b),
(c1) -- [line width=0.35mm, charged scalar, arrow size=0.8pt] (b)};
\node at (a)[circle,fill,style=black,inner sep=1pt]{};
\node at (b)[circle,fill,style=black,inner sep=1pt]{};
\end{feynman}
\end{tikzpicture}\label{fig:feyn09}}
\subfloat[]{\begin{tikzpicture}
\begin{feynman}
\vertex (a);
\vertex[above left=0.9cm and 0.9cm  of a] (a1){\(\chi(\psi)\)};
\vertex[below left=0.9cm and 0.9cm  of a] (a2){\(\Phi\)}; 
\vertex[right=0.9cm of a] (b); 
\vertex[above right=0.9cm and 0.9cm of b] (c1){\(\ell_i(q_j)\)};
\vertex[below right=0.9cm and 0.9cm of b] (c2){\(h\)};
\diagram*{
(a1) -- [line width=0.35mm, fermion, arrow size=0.8pt,style=black] (a),
(a) -- [line width=0.35mm,charged scalar, arrow size=0.8pt,style=black] (a2),
(a) -- [line width=0.35mm,fermion,arrow size=0.8pt, edge label={\(\color{black}{\ell_i(q_j)}\)}, style=black!50] (b),
(b) -- [line width=0.35mm,fermion, arrow size=0.8pt] (c1),
(b) -- [line width=0.35mm,scalar, arrow size=0.8pt] (c2)};
\node at (a)[circle,fill,style=black,inner sep=1pt]{};
\node at (b)[circle,fill,style=black,inner sep=1pt]{};
\end{feynman}
\end{tikzpicture}\label{fig:feyn10}}
\end{adjustbox}

\begin{adjustbox}{width=1.0\linewidth, keepaspectratio}
\subfloat[]{
\begin{tikzpicture}
\begin{feynman}
\vertex (a);
\vertex[above left=0.45cm and 0.9cm of a] (a1){\(\chi(\psi)\)};
\vertex[above right=0.45cm and 0.9cm  of a] (a2){\(\ell_i(q_j)\)}; 
\vertex[below = 0.9cm of a] (b); 
\vertex[below left=0.45cm and 0.9cm of b] (c1){\(\Phi\)};
\vertex[below right=0.45cm and 0.9cm of b] (c2){\(h\)};
\diagram*{
(a1) -- [line width=0.35mm, fermion, arrow size=0.8pt,style=black] (a),
(a) -- [line width=0.35mm, fermion, arrow size=0.8pt,style=black] (a2),
(a) -- [line width=0.35mm, charged scalar, arrow size=0.8pt, edge label'={\(\color{black}{\Phi}\)}, style=black!50] (b) ,
(c2) -- [line width=0.35mm, scalar, arrow size=0.8pt] (b),
(b) -- [line width=0.35mm, charged scalar, arrow size=0.8pt] (c1)};
\node at (a)[circle,fill,style=black,inner sep=1pt]{};
\node at (b)[circle,fill,style=black,inner sep=1pt]{};
\end{feynman}
\end{tikzpicture}\label{fig:feyn11}}
\subfloat[]{\begin{tikzpicture}
\begin{feynman}
\vertex (a);
\vertex[above left=0.9cm and 0.9cm  of a] (a1){\(\chi(\psi)\)};
\vertex[below left=0.9cm and 0.9cm  of a] (a2){\(\Phi\)}; 
\vertex[right=0.9cm of a] (b); 
\vertex[above right=0.9cm and 0.9cm of b] (c1){\(\ell_i(q_j)\)};
\vertex[below right=0.9cm and 0.9cm of b] (c2){\(B\)};
\diagram*{
(a1) -- [line width=0.35mm, fermion, arrow size=0.8pt,style=black] (a),
(a) -- [line width=0.35mm,charged scalar, arrow size=0.8pt,style=black] (a2),
(a) -- [line width=0.35mm,fermion,arrow size=0.8pt, edge label={\(\color{black}{\ell_i(q_j)}\)}, style=black!50] (b),
(b) -- [line width=0.35mm,fermion, arrow size=0.8pt] (c1),
(b) -- [line width=0.35mm,boson, arrow size=0.8pt] (c2)};
\node at (a)[circle,fill,style=black,inner sep=1pt]{};
\node at (b)[circle,fill,style=black,inner sep=1pt]{};
\end{feynman}
\end{tikzpicture}\label{fig:feyn12}}
\subfloat[]{
\begin{tikzpicture}
\begin{feynman}
\vertex (a);
\vertex[above left=0.45cm and 0.9cm of a] (a1){\(\chi(\psi)\)};
\vertex[above right=0.45cm and 0.9cm  of a] (a2){\(B\)}; 
\vertex[below = 0.9cm of a] (b); 
\vertex[below left=0.45cm and 0.9cm of b] (c1){\(\Phi\)};
\vertex[below right=0.45cm and 0.9cm of b] (c2){\(\ell_i(q_j)\)};
\diagram*{
(a1) -- [line width=0.35mm, fermion, arrow size=0.8pt,style=black] (a),
(a) -- [line width=0.35mm, boson, arrow size=0.8pt,style=black] (a2),
(a) -- [line width=0.35mm, fermion, arrow size=0.8pt, edge label'={\(\color{black}{\chi(\psi)}\)}, style=black!50] (b) ,
(b) -- [line width=0.35mm, fermion, arrow size=0.8pt] (c2),
(b) -- [line width=0.35mm, charged scalar, arrow size=0.8pt] (c1)};
\node at (a)[circle,fill,style=black,inner sep=1pt]{};
\node at (b)[circle,fill,style=black,inner sep=1pt]{};
\end{feynman}
\end{tikzpicture}\label{fig:feyn15}}
\end{adjustbox}

\begin{adjustbox}{width=1.0\linewidth, keepaspectratio}
\subfloat[]{\begin{tikzpicture}
\begin{feynman}
\vertex (a);
\vertex[above left=0.9cm and 0.9cm  of a] (a1){\(\chi(\psi)\)};
\vertex[below left=0.9cm and 0.9cm  of a] (a2){\(\Phi\)}; 
\vertex[right=0.9cm of a] (b); 
\vertex[above right=0.9cm and 0.9cm of b] (c1){\(u_i\)};
\vertex[below right=0.9cm and 0.9cm of b] (c2){\(W\)};
\diagram*{
(a1) -- [line width=0.35mm, fermion, arrow size=0.8pt,style=black] (a),
(a) -- [line width=0.35mm, charged scalar, arrow size=0.8pt,style=black] (a2),
(a) -- [line width=0.35mm, fermion,arrow size=0.8pt, edge label={\(\color{black}{q_i}\)}, style=black!50] (b),
(b) -- [line width=0.35mm, fermion, arrow size=0.8pt] (c1),
(b) -- [line width=0.35mm, charged boson, arrow size=0.8pt] (c2)};
\node at (a)[circle,fill,style=black,inner sep=1pt]{};
\node at (b)[circle,fill,style=black,inner sep=1pt]{};
\end{feynman}
\end{tikzpicture}\label{fig:feyn14}}
\subfloat[]{
\begin{tikzpicture}
\begin{feynman}
\vertex (a);
\vertex[above left=0.45cm and 0.9cm of a] (a1){\(\psi\)};
\vertex[above right=0.45cm and 0.9cm  of a] (a2){\(g\)}; 
\vertex[below = 0.9cm of a] (b); 
\vertex[below left=0.45cm and 0.9cm of b] (c1){\(\Phi\)};
\vertex[below right=0.45cm and 0.9cm of b] (c2){\(q_i\)};
\diagram*{
(a1) -- [line width=0.35mm, fermion, arrow size=0.8pt,style=black] (a),
(a) -- [line width=0.35mm, gluon, arrow size=0.8pt,style=black] (a2),
(a) -- [line width=0.35mm, fermion, arrow size=0.8pt, edge label'={\(\color{black}{\psi}\)}, style=black!50] (b) ,
(b) -- [line width=0.35mm, fermion, arrow size=0.8pt] (c2),
(b) -- [line width=0.35mm, charged scalar, arrow size=0.8pt] (c1)};
\node at (a)[circle,fill,style=black,inner sep=1pt]{};
\node at (b)[circle,fill,style=black,inner sep=1pt]{};
\end{feynman}
\end{tikzpicture}\label{fig:feyn155}}
\subfloat[]{\begin{tikzpicture}
\begin{feynman}
\vertex (a);
\vertex[above left=0.9cm and 0.9cm  of a] (a1){\(\chi(\psi)\)};
\vertex[below left=0.9cm and 0.9cm  of a] (a2){\(\Phi\)}; 
\vertex[right=0.9cm of a] (b); 
\vertex[above right=0.9cm and 0.9cm of b] (c1){\(\nu_{\ell_i}\)};
\vertex[below right=0.9cm and 0.9cm of b] (c2){\(W\)};
\diagram*{
(a1) -- [line width=0.35mm, fermion, arrow size=0.8pt,style=black] (a),
(a) -- [line width=0.35mm, charged scalar, arrow size=0.8pt,style=black] (a2),
(a) -- [line width=0.35mm, fermion,arrow size=0.8pt, edge label={\(\color{black}{\ell_i}\)}, style=black!50] (b),
(b) -- [line width=0.35mm, fermion, arrow size=0.8pt] (c1),
(b) -- [line width=0.35mm, charged boson, arrow size=0.8pt] (c2)};
\node at (a)[circle,fill,style=black,inner sep=1pt]{};
\node at (b)[circle,fill,style=black,inner sep=1pt]{};
\end{feynman}
\end{tikzpicture}\label{fig:feyn13}}
\end{adjustbox}
\caption{Dark Matter coannihilation channels, $\psi\chibar\to q_i\bar{\ell}_j$ [\ref{fig:feyn08}], $\chi(\psi)\Phi\to \ell_j(q_j)\Phi^*$ [\ref{fig:feyn09}], $\chi(\psi)\Phi^*\to \ell_i(q_j)h$ [\ref{fig:feyn10},~\ref{fig:feyn11}], $\chi(\psi)\Phi^*\to \ell_i(q_j)B$ [\ref{fig:feyn12},~\ref{fig:feyn15}],
$\chi(\psi)\Phi^*\to u_iW$ [\ref{fig:feyn14}],
$\chi(\psi)\Phi^*\to q_i g$ [\ref{fig:feyn155}],
$\chi(\psi)\Phi^*\to \nu_\ell W$ [\ref{fig:feyn13}]
where $B=\{\gamma,~Z\}$, $u_i=\{u,~c,~t\}$, $\ell_{i,j}=\{e,~\mu,~\tau\}$ and $q_{i,j}=\{d,~s,~b\}$.}
\label{fig:feynman-relic2}
\end{figure}
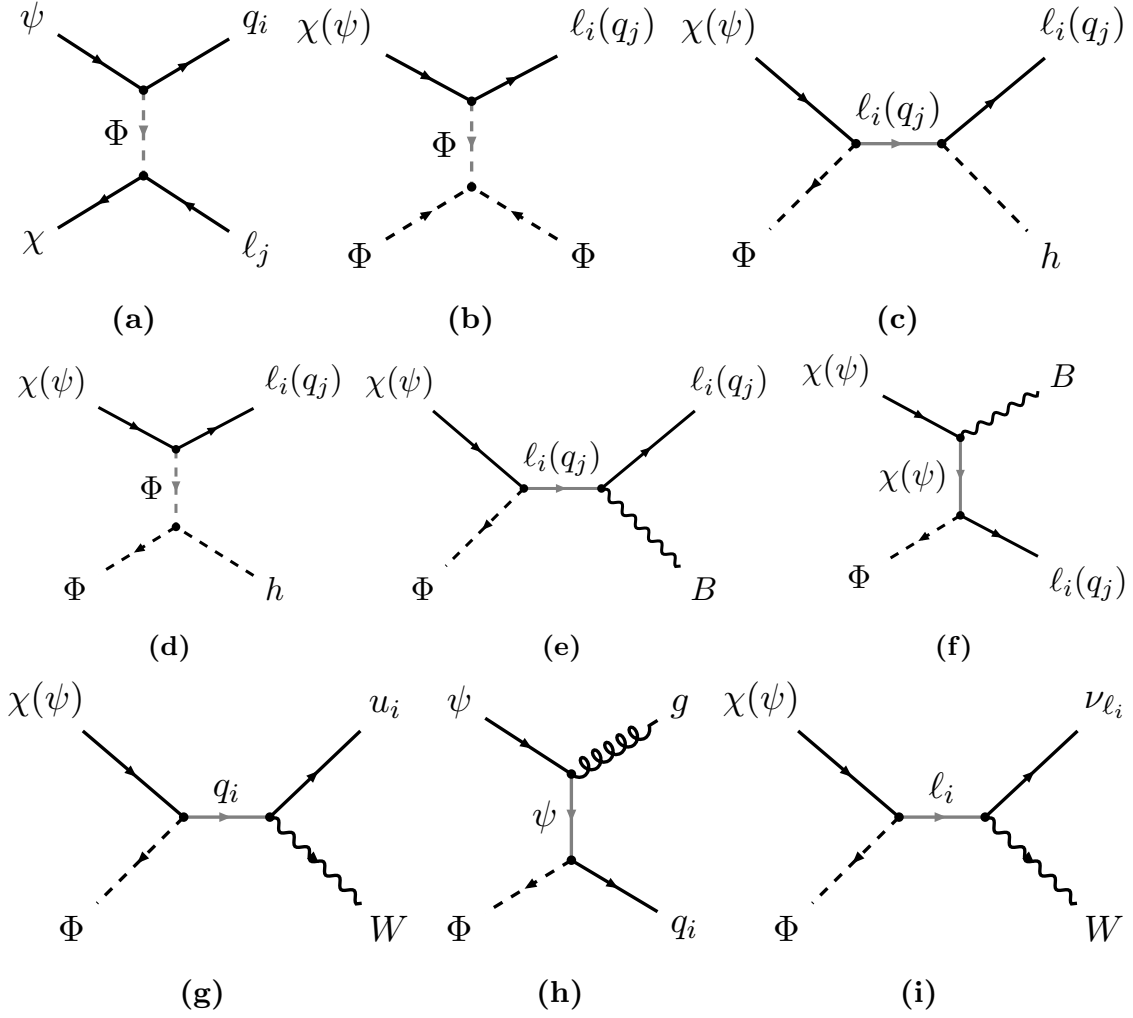

\begin{figure}[htb!]
\centering
\begin{adjustbox}{width=1.0\linewidth, keepaspectratio}
\subfloat[]{\begin{tikzpicture}
\begin{feynman}
\vertex (a);
\vertex[above left=0.9cm and 0.9cm  of a] (a1){\(\mathcal{F}\)};
\vertex[below left=0.9cm and 0.9cm  of a] (a2){\(\mathcal{F}\)}; 
\vertex[right=0.9cm of a] (b); 
\vertex[above right=0.9cm and 0.9cm of b] (c1){\(h\)};
\vertex[below right=0.9cm and 0.9cm of b] (c2){\(Z\)};
\diagram*{
(a1) -- [line width=0.35mm, fermion, arrow size=0.8pt,style=black] (a),
(a) -- [line width=0.35mm, fermion, arrow size=0.8pt,style=black] (a2),
(a) -- [line width=0.35mm, boson,arrow size=0.8pt, edge label={\(\color{black}{Z}\)}, style=black!50] (b),
(b) -- [line width=0.35mm,scalar, arrow size=0.8pt] (c1),
(b) -- [line width=0.35mm,boson, arrow size=0.8pt] (c2)};
\node at (a)[circle,fill,style=black,inner sep=1pt]{};
\node at (b)[circle,fill,style=black,inner sep=1pt]{};
\end{feynman}
\end{tikzpicture}\label{fig:feyn16}}
\subfloat[]{
\begin{tikzpicture}
\begin{feynman}
\vertex (a);
\vertex[above left=0.45cm and 0.9cm of a] (a1){\(\psi(\chi)\)};
\vertex[above right=0.45cm and 0.9cm  of a] (a2){\(q_i(\ell_i)\)}; 
\vertex[below = 0.9cm of a] (b); 
\vertex[below left=0.45cm and 0.9cm of b] (c1){\(\psi(\chi)\)};
\vertex[below right=0.45cm and 0.9cm of b] (c2){\(q_j(\ell_j)\)};
\diagram*{
(a1) -- [line width=0.35mm, fermion, arrow size=0.8pt,style=black] (a),
(a) -- [line width=0.35mm, fermion, arrow size=0.8pt,style=black] (a2),
(a) -- [line width=0.35mm, charged scalar, arrow size=0.8pt, edge label'={\(\color{black}{\Phi}\)}, style=black!50] (b) ,
(c2) -- [line width=0.35mm, fermion, arrow size=0.8pt] (b),
(b) -- [line width=0.35mm, fermion, arrow size=0.8pt] (c1)};
\node at (a)[circle,fill,style=black,inner sep=1pt]{};
\node at (b)[circle,fill,style=black,inner sep=1pt]{};
\end{feynman}
\end{tikzpicture}\label{fig:feyn17}}
\subfloat[]{\begin{tikzpicture}
\begin{feynman}
\vertex (a);
\vertex[above left=0.9cm and 0.9cm  of a] (a1){\(\mathcal{F}\)};
\vertex[below left=0.9cm and 0.9cm  of a] (a2){\(\mathcal{F}\)}; 
\vertex[right=0.9cm of a] (b); 
\vertex[above right=0.9cm and 0.9cm of b] (c1){\(q_i(\ell_j)\)};
\vertex[below right=0.9cm and 0.9cm of b] (c2){\(q_i(\ell_j)\)};
\diagram*{
(a1) -- [line width=0.35mm, fermion, arrow size=0.8pt,style=black] (a),
(a) -- [line width=0.35mm, fermion, arrow size=0.8pt,style=black] (a2),
(a) -- [line width=0.35mm, boson,arrow size=0.8pt, edge label={\(\color{black}{\gamma/Z}\)}, style=black!50] (b),
(c1) -- [line width=0.35mm,fermion, arrow size=0.8pt] (b),
(b) -- [line width=0.35mm,fermion, arrow size=0.8pt] (c2)};
\node at (a)[circle,fill,style=black,inner sep=1pt]{};
\node at (b)[circle,fill,style=black,inner sep=1pt]{};
\end{feynman}
\end{tikzpicture}\label{fig:feyn18}}
\end{adjustbox}

\begin{adjustbox}{width=1.0\linewidth, keepaspectratio}
\subfloat[]{\begin{tikzpicture}
\begin{feynman}
\vertex (a);
\vertex[above left=0.9cm and 0.9cm  of a] (a1){\(\psi\)};
\vertex[below left=0.9cm and 0.9cm  of a] (a2){\(\psi\)}; 
\vertex[right=0.9cm of a] (b); 
\vertex[above right=0.9cm and 0.9cm of b] (c1){\(q_i\)};
\vertex[below right=0.9cm and 0.9cm of b] (c2){\(q_i\)};
\diagram*{
(a1) -- [line width=0.35mm, fermion, arrow size=0.8pt,style=black] (a),
(a) -- [line width=0.35mm, fermion, arrow size=0.8pt,style=black] (a2),
(a) -- [line width=0.35mm, gluon,arrow size=0.8pt, edge label={\(\color{black}{g}\)}, style=black!50] (b),
(c1) -- [line width=0.35mm, fermion, arrow size=0.8pt] (b),
(b) -- [line width=0.35mm, fermion, arrow size=0.8pt] (c2)};
\node at (a)[circle,fill,style=black,inner sep=1pt]{};
\node at (b)[circle,fill,style=black,inner sep=1pt]{};
\end{feynman}
\end{tikzpicture}\label{fig:feyn19}}
\subfloat[]{\begin{tikzpicture}
\begin{feynman}
\vertex (a);
\vertex[above left=0.9cm and 0.9cm  of a] (a1){\(\mathcal{F}\)};
\vertex[below left=0.9cm and 0.9cm  of a] (a2){\(\mathcal{F}\)}; 
\vertex[right=0.9cm of a] (b); 
\vertex[above right=0.9cm and 0.9cm of b] (c1){\(W^+\)};
\vertex[below right=0.9cm and 0.9cm of b] (c2){\(W^-\)};
\diagram*{
(a1) -- [line width=0.35mm, fermion, arrow size=0.8pt,style=black] (a),
(a) -- [line width=0.35mm, fermion, arrow size=0.8pt,style=black] (a2),
(a) -- [line width=0.35mm, boson,arrow size=0.8pt, edge label={\(\color{black}{\gamma/Z}\)}, style=black!50] (b),
(c1) -- [line width=0.35mm, charged boson, arrow size=0.8pt] (b),
(b) -- [line width=0.35mm, charged boson, arrow size=0.8pt] (c2)};
\node at (a)[circle,fill,style=black,inner sep=1pt]{};
\node at (b)[circle,fill,style=black,inner sep=1pt]{};
\end{feynman}
\end{tikzpicture}\label{fig:feyn20}}
\subfloat[]{\begin{tikzpicture}
\begin{feynman}
\vertex (a);
\vertex[above left=0.9cm and 0.9cm  of a] (a1){\(\psi\)};
\vertex[below left=0.9cm and 0.9cm  of a] (a2){\(\psi\)}; 
\vertex[right=0.9cm of a] (b); 
\vertex[above right=0.9cm and 0.9cm of b] (c1){\(g\)};
\vertex[below right=0.9cm and 0.9cm of b] (c2){\(g\)};
\diagram*{
(a1) -- [line width=0.35mm, fermion, arrow size=0.8pt,style=black] (a),
(a) -- [line width=0.35mm, fermion, arrow size=0.8pt,style=black] (a2),
(a) -- [line width=0.35mm, gluon,arrow size=0.8pt, edge label={\(\color{black}{g}\)}, style=black!50] (b),
(c1) -- [line width=0.35mm, gluon, arrow size=0.8pt] (b),
(b) -- [line width=0.35mm, gluon, arrow size=0.8pt] (c2)};
\node at (a)[circle,fill,style=black,inner sep=1pt]{};
\node at (b)[circle,fill,style=black,inner sep=1pt]{};
\end{feynman}
\end{tikzpicture}\label{fig:feyn200}}
\end{adjustbox}

\begin{adjustbox}{width=0.8\linewidth, keepaspectratio}
\subfloat[]{
\begin{tikzpicture}
\begin{feynman}
\vertex (a);
\vertex[above left=0.45cm and 0.9cm of a] (a1){\(\psi\)};
\vertex[above right=0.45cm and 0.9cm  of a] (a2){\(g\)}; 
\vertex[below = 0.9cm of a] (b); 
\vertex[below left=0.45cm and 0.9cm of b] (c1){\(\psi\)};
\vertex[below right=0.45cm and 0.9cm of b] (c2){\(g\)};
\diagram*{
(a1) -- [line width=0.35mm, fermion, arrow size=0.8pt,style=black] (a),
(a) -- [line width=0.35mm, gluon, arrow size=0.8pt,style=black] (a2),
(a) -- [line width=0.35mm, fermion, arrow size=0.8pt, edge label'={\(\color{black}{\psi}\)}, style=black!50] (b) ,
(c2) -- [line width=0.35mm, gluon, arrow size=0.8pt] (b),
(b) -- [line width=0.35mm, fermion, arrow size=0.8pt] (c1)};
\node at (a)[circle,fill,style=black,inner sep=1pt]{};
\node at (b)[circle,fill,style=black,inner sep=1pt]{};
\end{feynman}
\end{tikzpicture}\label{fig:feyn21}}
\subfloat[]{
\begin{tikzpicture}
\begin{feynman}
\vertex (a);
\vertex[above left=0.45cm and 0.9cm of a] (a1){\(\mathcal{F}\)};
\vertex[above right=0.45cm and 0.9cm  of a] (a2){\(B\)}; 
\vertex[below = 0.9cm of a] (b); 
\vertex[below left=0.45cm and 0.9cm of b] (c1){\(\mathcal{F}\)};
\vertex[below right=0.45cm and 0.9cm of b] (c2){\(B\)};
\diagram*{
(a1) -- [line width=0.35mm, fermion, arrow size=0.8pt,style=black] (a),
(a) -- [line width=0.35mm, boson, arrow size=0.8pt,style=black] (a2),
(a) -- [line width=0.35mm, fermion, arrow size=0.8pt, edge label'={\(\color{black}{\mathcal{F}}\)}, style=black!50] (b) ,
(c2) -- [line width=0.35mm, boson, arrow size=0.8pt] (b),
(b) -- [line width=0.35mm, fermion, arrow size=0.8pt] (c1)};
\node at (a)[circle,fill,style=black,inner sep=1pt]{};
\node at (b)[circle,fill,style=black,inner sep=1pt]{};
\end{feynman}
\end{tikzpicture}\label{fig:feyn22}}
\subfloat[]{\begin{tikzpicture}
\begin{feynman}
\vertex (a);
\vertex[above left=0.9cm and 0.9cm  of a] (a1){\(\mathcal{F}\)};
\vertex[below left=0.9cm and 0.9cm  of a] (a2){\(\mathcal{F}\)}; 
\vertex[right=0.9cm of a] (b); 
\vertex[above right=0.9cm and 0.9cm of b] (c1){\(\nu_{\ell_i}^{}\)};
\vertex[below right=0.9cm and 0.9cm of b] (c2){\(\nu_{\ell_i}^{}\)};
\diagram*{
(a1) -- [line width=0.35mm, fermion, arrow size=0.8pt,style=black] (a),
(a) -- [line width=0.35mm, fermion, arrow size=0.8pt,style=black] (a2),
(a) -- [line width=0.35mm, boson, arrow size=0.8pt, edge label={\(\color{black}{Z}\)}, style=black!50] (b),
(c1) -- [line width=0.35mm, fermion, arrow size=0.8pt] (b),
(b) -- [line width=0.35mm, fermion, arrow size=0.8pt] (c2)};
\node at (a)[circle,fill,style=black,inner sep=1pt]{};
\node at (b)[circle,fill,style=black,inner sep=1pt]{};
\end{feynman}
\end{tikzpicture}\label{fig:feyn23}}
\end{adjustbox}
\caption{Dark Matter coannihilation channels, 
$\mathcal{F}\bar{\mathcal{F}}\to \rm h~Z$ [\ref{fig:feyn16}],
$\psi\psibar(\chi\bar\chi)\to q_i\bar{q}_j~(\ell_i\bar{\ell}_j)$ [\ref{fig:feyn17}],
$\psi\psibar(\chi\bar\chi)\to q_i\bar{q}_i~(\ell_i\bar{\ell}_i)$ [\ref{fig:feyn17}, \ref{fig:feyn18}, \ref{fig:feyn19} (\ref{fig:feyn17}, \ref{fig:feyn18})],
$\mathcal{F}\bar{\mathcal{F}}\to \rm W^+~W^-$ [\ref{fig:feyn20}],
$\psi\psibar\to g~g$ [\ref{fig:feyn200}, \ref{fig:feyn21}],
$\mathcal{F}\bar{\mathcal{F}}\to B~B$ [\ref{fig:feyn16}],
$\mathcal{F}\bar{\mathcal{F}}\to \nu_{\ell_i}\bar{\nu}_{\ell_i}$ [\ref{fig:feyn23}],
where $\mathcal{F}=\{\psi,~\chi\}$,~$B=\{\gamma,~Z\}$, $\ell_{i,j}=\{e,~\mu,~\tau\}$ and $q_{i,j}=\{d,~s,~b\}$.}
\label{fig:feynman-relic3}
\end{figure}

\bibliographystyle{JHEP} 
\bibliography{biblio.bib}

\providecommand{\href}[2]{#2}\begingroup\raggedright\begin{thebibliography}{100}

\bibitem{Barger:2008jx}
V.~Barger, P.~Langacker, M.~McCaskey, M.~Ramsey-Musolf and G.~Shaughnessy,
  \emph{{Complex Singlet Extension of the Standard Model}},
  \href{https://doi.org/10.1103/PhysRevD.79.015018}{\emph{Phys. Rev. D}
  {\bfseries 79} (2009) 015018}
  [\href{https://arxiv.org/abs/0811.0393}{{\ttfamily 0811.0393}}].

\bibitem{Rindler-Daller:2013zxa}
T.~Rindler-Daller and P.R.~Shapiro, \emph{{Complex scalar field dark matter on
  galactic scales}},
  \href{https://doi.org/10.1142/S021773231430002X}{\emph{Mod. Phys. Lett. A}
  {\bfseries 29} (2014) 1430002}
  [\href{https://arxiv.org/abs/1312.1734}{{\ttfamily 1312.1734}}].

\bibitem{McDonald:1993ex}
J.~McDonald, \emph{{Gauge singlet scalars as cold dark matter}},
  \href{https://doi.org/10.1103/PhysRevD.50.3637}{\emph{Phys. Rev. D}
  {\bfseries 50} (1994) 3637}
  [\href{https://arxiv.org/abs/hep-ph/0702143}{{\ttfamily hep-ph/0702143}}].

\bibitem{Belanger:2012zr}
G.~Belanger, K.~Kannike, A.~Pukhov and M.~Raidal, \emph{{$Z_3$ Scalar Singlet
  Dark Matter}},
  \href{https://doi.org/10.1088/1475-7516/2013/01/022}{\emph{JCAP} {\bfseries
  01} (2013) 022} [\href{https://arxiv.org/abs/1211.1014}{{\ttfamily
  1211.1014}}].

\bibitem{Hektor:2019ote}
A.~Hektor, A.~Hryczuk and K.~Kannike, \emph{{Improved bounds on
  $\mathbb{Z}_{3}$ singlet dark matter}},
  \href{https://doi.org/10.1007/JHEP03(2019)204}{\emph{JHEP} {\bfseries 03}
  (2019) 204} [\href{https://arxiv.org/abs/1901.08074}{{\ttfamily
  1901.08074}}].

\bibitem{Giacchino:2015hvk}
F.~Giacchino, A.~Ibarra, L.~Lopez~Honorez, M.H.G.~Tytgat and S.~Wild,
  \emph{{Signatures from Scalar Dark Matter with a Vector-like Quark
  Mediator}}, \href{https://doi.org/10.1088/1475-7516/2016/02/002}{\emph{JCAP}
  {\bfseries 02} (2016) 002}
  [\href{https://arxiv.org/abs/1511.04452}{{\ttfamily 1511.04452}}].

\bibitem{Acaroglu:2022hrm}
H.~Acaro{\u{g}}lu, P.~Agrawal and M.~Blanke, \emph{{Lepton-flavoured scalar
  dark matter in Dark Minimal Flavour Violation}},
  \href{https://doi.org/10.1007/JHEP05(2023)106}{\emph{JHEP} {\bfseries 05}
  (2023) 106} [\href{https://arxiv.org/abs/2211.03809}{{\ttfamily
  2211.03809}}].

\bibitem{Bai:2014osa}
Y.~Bai and J.~Berger, \emph{{Lepton Portal Dark Matter}},
  \href{https://doi.org/10.1007/JHEP08(2014)153}{\emph{JHEP} {\bfseries 08}
  (2014) 153} [\href{https://arxiv.org/abs/1402.6696}{{\ttfamily 1402.6696}}].

\bibitem{Biondini:2026ryb}
S.~Biondini, A.~Greljo, X.~Ponce~D{\'\i}az and A.~Valenti, \emph{{Charting the
  flavor structure of dark matter}},
  \href{https://doi.org/10.1103/p45x-l7qb}{\emph{Phys. Rev. D} {\bfseries 113}
  (2026) 095021} [\href{https://arxiv.org/abs/2601.07921}{{\ttfamily
  2601.07921}}].

\bibitem{Ishiwata:2013gma}
K.~Ishiwata and M.B.~Wise, \emph{{Phenomenology of heavy vectorlike leptons}},
  \href{https://doi.org/10.1103/PhysRevD.88.055009}{\emph{Phys. Rev. D}
  {\bfseries 88} (2013) 055009}
  [\href{https://arxiv.org/abs/1307.1112}{{\ttfamily 1307.1112}}].

\bibitem{Falkowski:2013jya}
A.~Falkowski, D.M.~Straub and A.~Vicente, \emph{{Vector-like leptons: Higgs
  decays and collider phenomenology}},
  \href{https://doi.org/10.1007/JHEP05(2014)092}{\emph{JHEP} {\bfseries 05}
  (2014) 092} [\href{https://arxiv.org/abs/1312.5329}{{\ttfamily 1312.5329}}].

\bibitem{Barducci:2018esg}
D.~Barducci, A.~Deandrea, S.~Moretti, L.~Panizzi and H.~Prager,
  \emph{{Characterizing dark matter interacting with extra charged leptons}},
  \href{https://doi.org/10.1103/PhysRevD.97.075006}{\emph{Phys. Rev. D}
  {\bfseries 97} (2018) 075006}
  [\href{https://arxiv.org/abs/1801.02707}{{\ttfamily 1801.02707}}].

\bibitem{CMS:2024bni}
{\scshape CMS} collaboration, \emph{{Review of searches for vector-like quarks,
  vector-like leptons, and heavy neutral leptons in proton{\textendash}proton
  collisions at {\ensuremath{\sqrt{}}}s=13 TeV at the CMS experiment}},
  \href{https://doi.org/10.1016/j.physrep.2024.09.012}{\emph{Phys. Rept.}
  {\bfseries 1115} (2025) 570}
  [\href{https://arxiv.org/abs/2405.17605}{{\ttfamily 2405.17605}}].

\bibitem{Mitra:2025cmo}
M.~Mitra, D.~Pradhan and S.~Saha, \emph{{Complex scalar dark matter with
  effective Higgs portals beyond radiation domination}},
  \href{https://doi.org/10.1103/245r-nqyk}{\emph{Phys. Rev. D} {\bfseries 113}
  (2026) 115047} [\href{https://arxiv.org/abs/2512.23228}{{\ttfamily
  2512.23228}}].

\bibitem{Bhattacharya:2024nla}
S.~Bhattacharya, L.~Kolay and D.~Pradhan, \emph{{Multiparticle scalar dark
  matter with $\mathbb{Z}_N$ symmetry}},
  \href{https://arxiv.org/abs/2410.16275}{{\ttfamily 2410.16275}}.

\bibitem{Aguilar-Saavedra:2013qpa}
J.A.~Aguilar-Saavedra, R.~Benbrik, S.~Heinemeyer and M.~P{\'e}rez-Victoria,
  \emph{{Handbook of vectorlike quarks: Mixing and single production}},
  \href{https://doi.org/10.1103/PhysRevD.88.094010}{\emph{Phys. Rev. D}
  {\bfseries 88} (2013) 094010}
  [\href{https://arxiv.org/abs/1306.0572}{{\ttfamily 1306.0572}}].

\bibitem{Acaroglu:2021qae}
H.~Acaro{\u{g}}lu and M.~Blanke, \emph{{Tasting flavoured Majorana dark
  matter}}, \href{https://doi.org/10.1007/JHEP05(2022)086}{\emph{JHEP}
  {\bfseries 05} (2022) 086}
  [\href{https://arxiv.org/abs/2109.10357}{{\ttfamily 2109.10357}}].

\bibitem{Acaroglu:2023phy}
H.~Acaro{\u{g}}lu, M.~Blanke, J.~Heisig, M.~Kr{\"a}mer and L.~Rathmann,
  \emph{{Flavoured Majorana Dark Matter then and now: from freeze-out scenarios
  to LHC signatures}},
  \href{https://doi.org/10.1007/JHEP06(2024)179}{\emph{JHEP} {\bfseries 06}
  (2024) 179} [\href{https://arxiv.org/abs/2312.09274}{{\ttfamily
  2312.09274}}].

\bibitem{Benincasa:2023vyp}
N.~Benincasa, A.~Hryczuk, K.~Kannike and M.~Laletin, \emph{{Phase transitions
  and gravitational waves in a model of~$\mathbb{Z}_{3}$ scalar dark matter}},
  \href{https://doi.org/10.1007/JHEP02(2024)207}{\emph{JHEP} {\bfseries 02}
  (2024) 207} [\href{https://arxiv.org/abs/2312.04627}{{\ttfamily
  2312.04627}}].

\bibitem{Aranda:2019vda}
A.~Aranda, D.~Hern{\'a}ndez-Otero, J.~Hern{\'a}ndez-Sanchez, V.~Keus,
  S.~Moretti, D.~Rojas-Ciofalo et~al., \emph{{Z$_3$ symmetric inert ( 2+1
  )-Higgs-doublet model}},
  \href{https://doi.org/10.1103/PhysRevD.103.015023}{\emph{Phys. Rev. D}
  {\bfseries 103} (2021) 015023}
  [\href{https://arxiv.org/abs/1907.12470}{{\ttfamily 1907.12470}}].

\bibitem{Bhattacharya:2025mlg}
S.~Bhattacharya, L.~Kolay, D.~Pradhan and A.~Sarkar, \emph{{Up-type FCNC in
  presence of Dark Matter}},
  \href{https://doi.org/10.1007/JHEP02(2026)239}{\emph{JHEP} {\bfseries 02}
  (2026) 239} [\href{https://arxiv.org/abs/2504.20045}{{\ttfamily
  2504.20045}}].

\bibitem{DiazSaez:2022nhp}
B.~D{\'\i}az~S{\'a}ez and K.~Ghorbani, \emph{{Z $_{3}$ scalar dark matter with
  strong positron fluxes}},
  \href{https://doi.org/10.1088/1475-7516/2023/02/002}{\emph{JCAP} {\bfseries
  02} (2023) 002} [\href{https://arxiv.org/abs/2203.09282}{{\ttfamily
  2203.09282}}].

\bibitem{Lahiri:2024rxc}
J.~Lahiri, D.~Pradhan and A.~Sarkar, \emph{{The Influence of Lepton Portal on
  the WIMP-pFIMP framework}},
  \href{https://arxiv.org/abs/2410.19734}{{\ttfamily 2410.19734}}.

\bibitem{Burgess:2000yq}
C.P.~Burgess, M.~Pospelov and T.~ter Veldhuis, \emph{{The Minimal model of
  nonbaryonic dark matter: A Singlet scalar}},
  \href{https://doi.org/10.1016/S0550-3213(01)00513-2}{\emph{Nucl. Phys. B}
  {\bfseries 619} (2001) 709}
  [\href{https://arxiv.org/abs/hep-ph/0011335}{{\ttfamily hep-ph/0011335}}].

\bibitem{Lerner:2009xg}
R.N.~Lerner and J.~McDonald, \emph{{Gauge singlet scalar as inflaton and
  thermal relic dark matter}},
  \href{https://doi.org/10.1103/PhysRevD.80.123507}{\emph{Phys. Rev. D}
  {\bfseries 80} (2009) 123507}
  [\href{https://arxiv.org/abs/0909.0520}{{\ttfamily 0909.0520}}].

\bibitem{Athron:2018ipf}
P.~Athron, J.M.~Cornell, F.~Kahlhoefer, J.~Mckay, P.~Scott and S.~Wild,
  \emph{{Impact of vacuum stability, perturbativity and XENON1T on global fits
  of $\mathbb {Z}_2$ and $\mathbb {Z}_3$ scalar singlet dark matter}},
  \href{https://doi.org/10.1140/epjc/s10052-018-6314-y}{\emph{Eur. Phys. J. C}
  {\bfseries 78} (2018) 830}
  [\href{https://arxiv.org/abs/1806.11281}{{\ttfamily 1806.11281}}].

\bibitem{Kannike:2012pe}
K.~Kannike, \emph{{Vacuum Stability Conditions From Copositivity Criteria}},
  \href{https://doi.org/10.1140/epjc/s10052-012-2093-z}{\emph{Eur. Phys. J. C}
  {\bfseries 72} (2012) 2093}
  [\href{https://arxiv.org/abs/1205.3781}{{\ttfamily 1205.3781}}].

\bibitem{Heisig:2024mwr}
J.~Heisig, \emph{{Conversion-Driven Leptogenesis: A Testable Theory of Dark
  Matter and Baryogenesis at the Electroweak Scale}},
  \href{https://doi.org/10.1103/PhysRevLett.133.191803}{\emph{Phys. Rev. Lett.}
  {\bfseries 133} (2024) 191803}
  [\href{https://arxiv.org/abs/2404.12428}{{\ttfamily 2404.12428}}].

\bibitem{Belfatto:2026uze}
B.~Belfatto, M.~Blanke, J.~Heisig, L.~Rathmann and F.~Wilsch,
  \emph{{Conversion-Driven Baryogenesis in Flavored Dark Matter Models}},
  \href{https://arxiv.org/abs/2607.11147}{{\ttfamily 2607.11147}}.

\bibitem{Kolay:2024wns}
L.~Kolay and S.~Nandi, \emph{{Exploring constraints on Simplified Dark Matter
  model through flavour and electroweak observables}},
  \href{https://doi.org/10.1007/JHEP10(2024)008}{\emph{JHEP} {\bfseries 10}
  (2024) 008} [\href{https://arxiv.org/abs/2403.20303}{{\ttfamily
  2403.20303}}].

\bibitem{Kolay:2025jip}
L.~Kolay and S.~Nandi, \emph{{Flavour and Electroweak Precision Constraints on
  a Simplified Dark Matter Model with a Light Spin-0 Mediator}},
  \href{https://arxiv.org/abs/2503.15609}{{\ttfamily 2503.15609}}.

\bibitem{Kala:2025srq}
S.~Kala, L.~Kolay, L.~Mukherjee and S.~Nandi, \emph{{Constraining anomalous Wtb
  and related SMEFT couplings using low-energy and electroweak precision
  observables}}, \href{https://doi.org/10.1007/JHEP11(2025)071}{\emph{JHEP}
  {\bfseries 11} (2025) 071}
  [\href{https://arxiv.org/abs/2505.07926}{{\ttfamily 2505.07926}}].

\bibitem{Albrecht:2024oyn}
J.~Albrecht, F.~Bernlochner, A.~Lenz and A.~Rusov, \emph{{Lifetimes of
  b-hadrons and mixing of neutral B-mesons: theoretical and experimental
  status}}, \href{https://doi.org/10.1140/epjs/s11734-024-01124-3}{\emph{Eur.
  Phys. J. ST} {\bfseries 233} (2024) 359}
  [\href{https://arxiv.org/abs/2402.04224}{{\ttfamily 2402.04224}}].

\bibitem{Wang:2022lfq}
B.~Wang, \emph{{Calculating $\Delta m_K$ with lattice QCD}},
  \href{https://doi.org/10.22323/1.396.0141}{\emph{PoS} {\bfseries LATTICE2021}
  (2022) 141} [\href{https://arxiv.org/abs/2301.01387}{{\ttfamily
  2301.01387}}].

\bibitem{LHCb:2023sim}
{\scshape LHCb} collaboration, \emph{{Improved Measurement of CP Violation
  Parameters in Bs0{\textrightarrow}J/{\ensuremath{\psi}}K+K- Decays in the
  Vicinity of the {\ensuremath{\phi}}(1020) Resonance}},
  \href{https://doi.org/10.1103/PhysRevLett.132.051802}{\emph{Phys. Rev. Lett.}
  {\bfseries 132} (2024) 051802}
  [\href{https://arxiv.org/abs/2308.01468}{{\ttfamily 2308.01468}}].

\bibitem{Belle-II:2023bps}
{\scshape Belle-II} collaboration, \emph{{Measurement of the B0 lifetime and
  flavor-oscillation frequency using hadronic decays reconstructed in
  2019{\textendash}2021 Belle II data}},
  \href{https://doi.org/10.1103/PhysRevD.107.L091102}{\emph{Phys. Rev. D}
  {\bfseries 107} (2023) L091102}
  [\href{https://arxiv.org/abs/2302.12791}{{\ttfamily 2302.12791}}].

\bibitem{ParticleDataGroup:2024cfk}
{\scshape Particle Data Group} collaboration, \emph{{Review of particle
  physics}}, \href{https://doi.org/10.1103/PhysRevD.110.030001}{\emph{Phys.
  Rev. D} {\bfseries 110} (2024) 030001}.

\bibitem{KTeV:2010sng}
{\scshape KTeV} collaboration, \emph{{Precise Measurements of Direct CP
  Violation, CPT Symmetry, and Other Parameters in the Neutral Kaon System}},
  \href{https://doi.org/10.1103/PhysRevD.83.092001}{\emph{Phys. Rev. D}
  {\bfseries 83} (2011) 092001}
  [\href{https://arxiv.org/abs/1011.0127}{{\ttfamily 1011.0127}}].

\bibitem{FlavourLatticeAveragingGroupFLAG:2024oxs}
{\scshape Flavour Lattice Averaging Group (FLAG)} collaboration, \emph{{FLAG
  Review 2024}},  \href{https://arxiv.org/abs/2411.04268}{{\ttfamily
  2411.04268}}.

\bibitem{Bazavov:2017lyh}
A.~Bazavov et~al., \emph{{$B$- and $D$-meson leptonic decay constants from
  four-flavor lattice QCD}},
  \href{https://doi.org/10.1103/PhysRevD.98.074512}{\emph{Phys. Rev. D}
  {\bfseries 98} (2018) 074512}
  [\href{https://arxiv.org/abs/1712.09262}{{\ttfamily 1712.09262}}].

\bibitem{Bobeth:1999mk}
C.~Bobeth, M.~Misiak and J.~Urban, \emph{{Photonic penguins at two loops and
  $m_t$ dependence of $BR[B \to X_s l^+ l^-]$}},
  \href{https://doi.org/10.1016/S0550-3213(00)00007-9}{\emph{Nucl. Phys. B}
  {\bfseries 574} (2000) 291}
  [\href{https://arxiv.org/abs/hep-ph/9910220}{{\ttfamily hep-ph/9910220}}].

\bibitem{Becirevic:2012fy}
D.~Becirevic, N.~Kosnik, F.~Mescia and E.~Schneider, \emph{{Complementarity of
  the constraints on New Physics from $B_s \to \mu^+ \mu^-$ and from $B \to K
  l^+l^-$ decays}},
  \href{https://doi.org/10.1103/PhysRevD.86.034034}{\emph{Phys. Rev. D}
  {\bfseries 86} (2012) 034034}
  [\href{https://arxiv.org/abs/1205.5811}{{\ttfamily 1205.5811}}].

\bibitem{Altmannshofer:2008dz}
W.~Altmannshofer, P.~Ball, A.~Bharucha, A.J.~Buras, D.M.~Straub and M.~Wick,
  \emph{{Symmetries and Asymmetries of $B \to K^{*} \mu^{+} \mu^{-}$ Decays in
  the Standard Model and Beyond}},
  \href{https://doi.org/10.1088/1126-6708/2009/01/019}{\emph{JHEP} {\bfseries
  01} (2009) 019} [\href{https://arxiv.org/abs/0811.1214}{{\ttfamily
  0811.1214}}].

\bibitem{CMS:2022mgd}
{\scshape CMS} collaboration, \emph{{Measurement of the
  B$^0_\mathrm{S}$$\to$$\mu^+\mu^-$ decay properties and search for the
  B$^0$$\to$$\mu^+\mu^-$ decay in proton-proton collisions at $\sqrt{s}$ = 13
  TeV}}, \href{https://doi.org/10.1016/j.physletb.2023.137955}{\emph{Phys.
  Lett. B} {\bfseries 842} (2023) 137955}
  [\href{https://arxiv.org/abs/2212.10311}{{\ttfamily 2212.10311}}].

\bibitem{Beneke:2019slt}
M.~Beneke, C.~Bobeth and R.~Szafron, \emph{{Power-enhanced leading-logarithmic
  QED corrections to $B_q \to \mu^+\mu^-$}},
  \href{https://doi.org/10.1007/JHEP10(2019)232}{\emph{JHEP} {\bfseries 10}
  (2019) 232} [\href{https://arxiv.org/abs/1908.07011}{{\ttfamily
  1908.07011}}].

\bibitem{LHCb:2020pcv}
{\scshape LHCb} collaboration, \emph{{Search for the Rare Decays $B^0_s\to
  e^+e^-$ and $B^0\to e^+e^-$}},
  \href{https://doi.org/10.1103/PhysRevLett.124.211802}{\emph{Phys. Rev. Lett.}
  {\bfseries 124} (2020) 211802}
  [\href{https://arxiv.org/abs/2003.03999}{{\ttfamily 2003.03999}}].

\bibitem{LHCb:2017myy}
{\scshape LHCb} collaboration, \emph{{Search for the decays
  $B_s^0\to\tau^+\tau^-$ and $B^0\to\tau^+\tau^-$}},
  \href{https://doi.org/10.1103/PhysRevLett.118.251802}{\emph{Phys. Rev. Lett.}
  {\bfseries 118} (2017) 251802}
  [\href{https://arxiv.org/abs/1703.02508}{{\ttfamily 1703.02508}}].

\bibitem{Bobeth:2013uxa}
C.~Bobeth, M.~Gorbahn, T.~Hermann, M.~Misiak, E.~Stamou and M.~Steinhauser,
  \emph{{$B_{s,d} \to l^+ l^-$ in the Standard Model with Reduced Theoretical
  Uncertainty}},
  \href{https://doi.org/10.1103/PhysRevLett.112.101801}{\emph{Phys. Rev. Lett.}
  {\bfseries 112} (2014) 101801}
  [\href{https://arxiv.org/abs/1311.0903}{{\ttfamily 1311.0903}}].

\bibitem{LHCb:2021vsc}
{\scshape LHCb} collaboration, \emph{{Analysis of Neutral B-Meson Decays into
  Two Muons}},
  \href{https://doi.org/10.1103/PhysRevLett.128.041801}{\emph{Phys. Rev. Lett.}
  {\bfseries 128} (2022) 041801}
  [\href{https://arxiv.org/abs/2108.09284}{{\ttfamily 2108.09284}}].

\bibitem{DAmbrosio:1994fgc}
G.~D'Ambrosio, G.~Ecker, G.~Isidori and H.~Neufeld, \emph{{Radiative
  nonleptonic kaon decays}},  11, 1994
  [\href{https://arxiv.org/abs/hep-ph/9411439}{{\ttfamily hep-ph/9411439}}].

\bibitem{BNLE871:1998bii}
{\scshape BNL E871} collaboration, \emph{{First observation of the rare decay
  mode K0(L) ---{\ensuremath{>}} e+ e-}},
  \href{https://doi.org/10.1103/PhysRevLett.81.4309}{\emph{Phys. Rev. Lett.}
  {\bfseries 81} (1998) 4309}
  [\href{https://arxiv.org/abs/hep-ex/9810007}{{\ttfamily hep-ex/9810007}}].

\bibitem{Valencia:1997xe}
G.~Valencia, \emph{{Long distance contribution to K(L) ---{\ensuremath{>}}
  lepton+ lepton-}},
  \href{https://doi.org/10.1016/S0550-3213(98)00116-3}{\emph{Nucl. Phys. B}
  {\bfseries 517} (1998) 339}
  [\href{https://arxiv.org/abs/hep-ph/9711377}{{\ttfamily hep-ph/9711377}}].

\bibitem{LHCb:2020ycd}
{\scshape LHCb} collaboration, \emph{{Constraints on the $K^0_S \rightarrow
  \mu^+ \mu^-$ Branching Fraction}},
  \href{https://doi.org/10.1103/PhysRevLett.125.231801}{\emph{Phys. Rev. Lett.}
  {\bfseries 125} (2020) 231801}
  [\href{https://arxiv.org/abs/2001.10354}{{\ttfamily 2001.10354}}].

\bibitem{DAmbrosio:2017klp}
G.~D'Ambrosio and T.~Kitahara, \emph{{Direct $CP$ Violation in $K \to \mu^+
  \mu^-$}}, \href{https://doi.org/10.1103/PhysRevLett.119.201802}{\emph{Phys.
  Rev. Lett.} {\bfseries 119} (2017) 201802}
  [\href{https://arxiv.org/abs/1707.06999}{{\ttfamily 1707.06999}}].

\bibitem{Buras:2022qip}
A.J.~Buras, \emph{{Standard Model predictions for rare K and B decays without
  new physics infection}},
  \href{https://doi.org/10.1140/epjc/s10052-023-11222-6}{\emph{Eur. Phys. J. C}
  {\bfseries 83} (2023) 66} [\href{https://arxiv.org/abs/2209.03968}{{\ttfamily
  2209.03968}}].

\bibitem{KLOE:2008acb}
{\scshape KLOE} collaboration, \emph{{Search for the $K(S) \to e^+ e^-$ decay
  with the KLOE detector}},
  \href{https://doi.org/10.1016/j.physletb.2009.01.037}{\emph{Phys. Lett. B}
  {\bfseries 672} (2009) 203}
  [\href{https://arxiv.org/abs/0811.1007}{{\ttfamily 0811.1007}}].

\bibitem{Sehgal:1969zok}
L.M.~Sehgal, \emph{{Electromagnetic contribution to the decays K(S)
  ---{\ensuremath{>}} lepton anti-lepton and K(L) ---{\ensuremath{>}} lepton
  anti-lepton}}, \href{https://doi.org/10.1103/PhysRev.183.1511}{\emph{Phys.
  Rev.} {\bfseries 183} (1969) 1511}.

\bibitem{Mandal:2019gff}
R.~Mandal and A.~Pich, \emph{{Constraints on scalar leptoquarks from lepton and
  kaon physics}}, \href{https://doi.org/10.1007/JHEP12(2019)089}{\emph{JHEP}
  {\bfseries 12} (2019) 089}
  [\href{https://arxiv.org/abs/1908.11155}{{\ttfamily 1908.11155}}].

\bibitem{LHCb:2020lmf}
{\scshape LHCb} collaboration, \emph{{Measurement of $CP$-Averaged Observables
  in the $B^{0}\rightarrow K^{*0}\mu^{+}\mu^{-}$ Decay}},
  \href{https://doi.org/10.1103/PhysRevLett.125.011802}{\emph{Phys. Rev. Lett.}
  {\bfseries 125} (2020) 011802}
  [\href{https://arxiv.org/abs/2003.04831}{{\ttfamily 2003.04831}}].

\bibitem{Alguero:2023jeh}
M.~Alguer{\'o}, A.~Biswas, B.~Capdevila, S.~Descotes-Genon, J.~Matias and
  M.~Novoa-Brunet, \emph{{To (b)e or not to (b)e: no electrons at LHCb}},
  \href{https://doi.org/10.1140/epjc/s10052-023-11824-0}{\emph{Eur. Phys. J. C}
  {\bfseries 83} (2023) 648}
  [\href{https://arxiv.org/abs/2304.07330}{{\ttfamily 2304.07330}}].

\bibitem{Alguero:2021anc}
M.~Alguer{\'o}, B.~Capdevila, S.~Descotes-Genon, J.~Matias and M.~Novoa-Brunet,
  \emph{{$b\rightarrow s\ell ^+\ell ^-$ global fits after $R_{K_S}$ and
  $R_{K^{*+}}$}},
  \href{https://doi.org/10.1140/epjc/s10052-022-10231-1}{\emph{Eur. Phys. J. C}
  {\bfseries 82} (2022) 326}
  [\href{https://arxiv.org/abs/2104.08921}{{\ttfamily 2104.08921}}].

\bibitem{Wen:2023pfq}
Q.~Wen and F.~Xu, \emph{{Global fits of new physics in b{\textrightarrow}s
  after the RK(*) 2022 release}},
  \href{https://doi.org/10.1103/PhysRevD.108.095038}{\emph{Phys. Rev. D}
  {\bfseries 108} (2023) 095038}
  [\href{https://arxiv.org/abs/2305.19038}{{\ttfamily 2305.19038}}].

\bibitem{Biswas:2020uaq}
A.~Biswas, S.~Nandi, S.K.~Patra and I.~Ray, \emph{{New physics in
  b{\textrightarrow}s{\ensuremath{\ell}}{\ensuremath{\ell}} decays with complex
  Wilson coefficients}},
  \href{https://doi.org/10.1016/j.nuclphysb.2021.115479}{\emph{Nucl. Phys. B}
  {\bfseries 969} (2021) 115479}
  [\href{https://arxiv.org/abs/2004.14687}{{\ttfamily 2004.14687}}].

\bibitem{Ciuchini:2022wbq}
M.~Ciuchini, M.~Fedele, E.~Franco, A.~Paul, L.~Silvestrini and M.~Valli,
  \emph{{Constraints on lepton universality violation from rare B decays}},
  \href{https://doi.org/10.1103/PhysRevD.107.055036}{\emph{Phys. Rev. D}
  {\bfseries 107} (2023) 055036}
  [\href{https://arxiv.org/abs/2212.10516}{{\ttfamily 2212.10516}}].

\bibitem{Hurth:2023jwr}
T.~Hurth, F.~Mahmoudi and S.~Neshatpour, \emph{{$B$ anomalies in the post
  $R_{K^{(*)}}$ era}},
  \href{https://doi.org/10.1103/PhysRevD.108.115037}{\emph{Phys. Rev. D}
  {\bfseries 108} (2023) 115037}
  [\href{https://arxiv.org/abs/2310.05585}{{\ttfamily 2310.05585}}].

\bibitem{London:2021lfn}
D.~London and J.~Matias, \emph{{$B$ Flavour Anomalies: 2021 Theoretical Status
  Report}},
  \href{https://doi.org/10.1146/annurev-nucl-102020-090209}{\emph{Ann. Rev.
  Nucl. Part. Sci.} {\bfseries 72} (2022) 37}
  [\href{https://arxiv.org/abs/2110.13270}{{\ttfamily 2110.13270}}].

\bibitem{Gubernari:2022hxn}
N.~Gubernari, M.~Reboud, D.~van Dyk and J.~Virto, \emph{{Improved theory
  predictions and global analysis of exclusive $b \to s\mu^+\mu^-$ processes}},
  \href{https://doi.org/10.1007/JHEP09(2022)133}{\emph{JHEP} {\bfseries 09}
  (2022) 133} [\href{https://arxiv.org/abs/2206.03797}{{\ttfamily
  2206.03797}}].

\bibitem{Alguero:2022wkd}
M.~Alguer{\'o}, J.~Matias, B.~Capdevila and A.~Crivellin, \emph{{Disentangling
  lepton flavor universal and lepton flavor universality violating effects in
  b{\textrightarrow}s{\ensuremath{\ell}}+{\ensuremath{\ell}}- transitions}},
  \href{https://doi.org/10.1103/PhysRevD.105.113007}{\emph{Phys. Rev. D}
  {\bfseries 105} (2022) 113007}
  [\href{https://arxiv.org/abs/2205.15212}{{\ttfamily 2205.15212}}].

\bibitem{Biswas:2022lhu}
A.~Biswas, S.~Nandi, S.K.~Patra and I.~Ray, \emph{{Study of the b
  {\textrightarrow} d{\ensuremath{\ell}}{\ensuremath{\ell}} transitions in the
  Standard Model and test of New Physics sensitivities}},
  \href{https://doi.org/10.1007/JHEP03(2023)247}{\emph{JHEP} {\bfseries 03}
  (2023) 247} [\href{https://arxiv.org/abs/2208.14463}{{\ttfamily
  2208.14463}}].

\bibitem{Bause:2022rrs}
R.~Bause, H.~Gisbert, M.~Golz and G.~Hiller, \emph{{Model-independent analysis
  of $b \rightarrow d$ processes}},
  \href{https://doi.org/10.1140/epjc/s10052-023-11586-9}{\emph{Eur. Phys. J. C}
  {\bfseries 83} (2023) 419}
  [\href{https://arxiv.org/abs/2209.04457}{{\ttfamily 2209.04457}}].

\bibitem{Mescia:2006jd}
F.~Mescia, C.~Smith and S.~Trine, \emph{{K(L) ---{\ensuremath{>}} pi0 e+ e- and
  K(L) ---{\ensuremath{>}} pi0 mu+ mu-: A Binary star on the stage of flavor
  physics}}, \href{https://doi.org/10.1088/1126-6708/2006/08/088}{\emph{JHEP}
  {\bfseries 08} (2006) 088}
  [\href{https://arxiv.org/abs/hep-ph/0606081}{{\ttfamily hep-ph/0606081}}].

\bibitem{DAmbrosio:2022kvb}
G.~D'Ambrosio, A.M.~Iyer, F.~Mahmoudi and S.~Neshatpour, \emph{{Anatomy of kaon
  decays and prospects for lepton flavour universality violation}},
  \href{https://doi.org/10.1007/JHEP09(2022)148}{\emph{JHEP} {\bfseries 09}
  (2022) 148} [\href{https://arxiv.org/abs/2206.14748}{{\ttfamily
  2206.14748}}].

\bibitem{DAmbrosio:2024ncc}
G.~D'Ambrosio and M.~Knecht, \emph{{Predictions for the Rare Kaon Decays KS,L
  $\to \pi^{0} \ell^{+} \ell^{-} $ from QCD in the Limit of a Large Number of
  Colours}}, \href{https://doi.org/10.3390/universe10120457}{\emph{Universe}
  {\bfseries 10} (2024) 457}
  [\href{https://arxiv.org/abs/2409.08568}{{\ttfamily 2409.08568}}].

\bibitem{Chen:2003nz}
C.-H.~Chen, C.Q.~Geng and I.-L.~Ho, \emph{{Forward backward asymmetry in $K^{+}
  \to \pi^{+} \ell \ell $}},
  \href{https://doi.org/10.1103/PhysRevD.67.074029}{\emph{Phys. Rev. D}
  {\bfseries 67} (2003) 074029}
  [\href{https://arxiv.org/abs/hep-ph/0302207}{{\ttfamily hep-ph/0302207}}].

\bibitem{Gao:2003wy}
D.-N.~Gao, \emph{{Long distance contribution to the forward backward asymmetry
  in decays K+ ---{\ensuremath{>}} pi+ l+ l-}},
  \href{https://doi.org/10.1103/PhysRevD.69.094030}{\emph{Phys. Rev. D}
  {\bfseries 69} (2004) 094030}
  [\href{https://arxiv.org/abs/hep-ph/0311253}{{\ttfamily hep-ph/0311253}}].

\bibitem{Belle-II:2023esi}
{\scshape Belle-II} collaboration, \emph{{Evidence for
  B+{\textrightarrow}K+{\ensuremath{\nu}}{\ensuremath{\nu}}{\textasciimacron}
  decays}}, \href{https://doi.org/10.1103/PhysRevD.109.112006}{\emph{Phys. Rev.
  D} {\bfseries 109} (2024) 112006}
  [\href{https://arxiv.org/abs/2311.14647}{{\ttfamily 2311.14647}}].

\bibitem{Buras:2024ewl}
A.J.~Buras, J.~Harz and M.A.~Mojahed, \emph{{Disentangling new physics in $
  K\to \pi \nu \overline{\nu} $ and $ B\to K\left({K}^{\ast}\right)\nu
  \overline{\nu} $ observables}},
  \href{https://doi.org/10.1007/JHEP10(2024)087}{\emph{JHEP} {\bfseries 10}
  (2024) 087} [\href{https://arxiv.org/abs/2405.06742}{{\ttfamily
  2405.06742}}].

\bibitem{Chen:2024jlj}
F.-Z.~Chen, Q.~Wen and F.~Xu, \emph{{Correlating $B\rightarrow K^{(*)} \nu
  \bar{\nu }$ and flavor anomalies in SMEFT}},
  \href{https://doi.org/10.1140/epjc/s10052-024-13425-x}{\emph{Eur. Phys. J. C}
  {\bfseries 84} (2024) 1012}
  [\href{https://arxiv.org/abs/2401.11552}{{\ttfamily 2401.11552}}].

\bibitem{Buras:2014fpa}
A.J.~Buras, J.~Girrbach-Noe, C.~Niehoff and D.M.~Straub, \emph{{$ B\to
  {K}^{\left(\ast \right)}\nu \overline{\nu} $ decays in the Standard Model and
  beyond}}, \href{https://doi.org/10.1007/JHEP02(2015)184}{\emph{JHEP}
  {\bfseries 02} (2015) 184} [\href{https://arxiv.org/abs/1409.4557}{{\ttfamily
  1409.4557}}].

\bibitem{Becirevic:2023aov}
D.~Be{\v{c}}irevi{\'c}, G.~Piazza and O.~Sumensari, \emph{{Revisiting
  $B\rightarrow K^{(*)} \nu {\bar{\nu }}$ decays in the Standard Model and
  beyond}}, \href{https://doi.org/10.1140/epjc/s10052-023-11388-z}{\emph{Eur.
  Phys. J. C} {\bfseries 83} (2023) 252}
  [\href{https://arxiv.org/abs/2301.06990}{{\ttfamily 2301.06990}}].

\bibitem{Belle:2017oht}
{\scshape Belle} collaboration, \emph{{Search for $\boldsymbol{B\to
  h\nu\bar{\nu}}$ decays with semileptonic tagging at Belle}},
  \href{https://doi.org/10.1103/PhysRevD.96.091101}{\emph{Phys. Rev. D}
  {\bfseries 96} (2017) 091101}
  [\href{https://arxiv.org/abs/1702.03224}{{\ttfamily 1702.03224}}].

\bibitem{BaBar:2013npw}
{\scshape BaBar} collaboration, \emph{{Search for $B \to K^{(*)} \nu \overline
  \nu$ and invisible quarkonium decays}},
  \href{https://doi.org/10.1103/PhysRevD.87.112005}{\emph{Phys. Rev. D}
  {\bfseries 87} (2013) 112005}
  [\href{https://arxiv.org/abs/1303.7465}{{\ttfamily 1303.7465}}].

\bibitem{Straub:2018kue}
D.M.~Straub, \emph{{flavio: a Python package for flavour and precision
  phenomenology in the Standard Model and beyond}},
  \href{https://arxiv.org/abs/1810.08132}{{\ttfamily 1810.08132}}.

\bibitem{NA62:2021zjw}
{\scshape NA62} collaboration, \emph{{Measurement of the very rare
  K$^{+}${\textrightarrow}$ {\pi}^{+}\nu \overline{\nu} $ decay}},
  \href{https://doi.org/10.1007/JHEP06(2021)093}{\emph{JHEP} {\bfseries 06}
  (2021) 093} [\href{https://arxiv.org/abs/2103.15389}{{\ttfamily
  2103.15389}}].

\bibitem{NA62:2024pjp}
{\scshape NA62} collaboration, \emph{{Observation of the $ {K}^{+}\to
  {\pi}^{+}\nu \overline{\nu} $ decay and measurement of its branching ratio}},
  \href{https://doi.org/10.1007/JHEP02(2025)191}{\emph{JHEP} {\bfseries 02}
  (2025) 191} [\href{https://arxiv.org/abs/2412.12015}{{\ttfamily
  2412.12015}}].

\bibitem{KOTO:2024zbl}
{\scshape KOTO} collaboration, \emph{{Search for the
  KL{\textrightarrow}{\ensuremath{\pi}}0{\ensuremath{\nu}}{\ensuremath{\nu}}{\textasciimacron}
  Decay at the J-PARC KOTO Experiment}},
  \href{https://doi.org/10.1103/PhysRevLett.134.081802}{\emph{Phys. Rev. Lett.}
  {\bfseries 134} (2025) 081802}
  [\href{https://arxiv.org/abs/2411.11237}{{\ttfamily 2411.11237}}].

\bibitem{Volkov:2019phy}
S.~Volkov, \emph{{Calculating the five-loop QED contribution to the electron
  anomalous magnetic moment: Graphs without lepton loops}},
  \href{https://doi.org/10.1103/PhysRevD.100.096004}{\emph{Phys. Rev. D}
  {\bfseries 100} (2019) 096004}
  [\href{https://arxiv.org/abs/1909.08015}{{\ttfamily 1909.08015}}].

\bibitem{Aoyama:2019ryr}
T.~Aoyama, T.~Kinoshita and M.~Nio, \emph{{Theory of the Anomalous Magnetic
  Moment of the Electron}},
  \href{https://doi.org/10.3390/atoms7010028}{\emph{Atoms} {\bfseries 7} (2019)
  28}.

\bibitem{Fan:2022eto}
X.~Fan, T.G.~Myers, B.A.D.~Sukra and G.~Gabrielse, \emph{{Measurement of the
  Electron Magnetic Moment}},
  \href{https://doi.org/10.1103/PhysRevLett.130.071801}{\emph{Phys. Rev. Lett.}
  {\bfseries 130} (2023) 071801}
  [\href{https://arxiv.org/abs/2209.13084}{{\ttfamily 2209.13084}}].

\bibitem{Parker:2018vye}
R.H.~Parker, C.~Yu, W.~Zhong, B.~Estey and H.~M{\"u}ller, \emph{{Measurement of
  the fine-structure constant as a test of the Standard Model}},
  \href{https://doi.org/10.1126/science.aap7706}{\emph{Science} {\bfseries 360}
  (2018) 191} [\href{https://arxiv.org/abs/1812.04130}{{\ttfamily
  1812.04130}}].

\bibitem{Morel:2020dww}
L.~Morel, Z.~Yao, P.~Clad{\'e} and S.~Guellati-Kh{\'e}lifa,
  \emph{{Determination of the fine-structure constant with an accuracy of 81
  parts per trillion}},
  \href{https://doi.org/10.1038/s41586-020-2964-7}{\emph{Nature} {\bfseries
  588} (2020) 61}.

\bibitem{Hanneke:2008tm}
D.~Hanneke, S.~Fogwell and G.~Gabrielse, \emph{{New Measurement of the Electron
  Magnetic Moment and the Fine Structure Constant}},
  \href{https://doi.org/10.1103/PhysRevLett.100.120801}{\emph{Phys. Rev. Lett.}
  {\bfseries 100} (2008) 120801}
  [\href{https://arxiv.org/abs/0801.1134}{{\ttfamily 0801.1134}}].

\bibitem{Aliberti:2025beg}
R.~Aliberti et~al., \emph{{The anomalous magnetic moment of the muon in the
  Standard Model: an update}},
  \href{https://doi.org/10.1016/j.physrep.2025.08.002}{\emph{Phys. Rept.}
  {\bfseries 1143} (2025) 1}
  [\href{https://arxiv.org/abs/2505.21476}{{\ttfamily 2505.21476}}].

\bibitem{Muong-2:2025xyk}
{\scshape Muon g-2} collaboration, \emph{{Measurement of the Positive Muon
  Anomalous Magnetic Moment to 127~ppb}},
  \href{https://doi.org/10.1103/7clf-sm2v}{\emph{Phys. Rev. Lett.} {\bfseries
  135} (2025) 101802} [\href{https://arxiv.org/abs/2506.03069}{{\ttfamily
  2506.03069}}].

\bibitem{Gonzalez-Sprinberg:2000lzf}
G.A.~Gonzalez-Sprinberg, A.~Santamaria and J.~Vidal, \emph{{Model independent
  bounds on the tau lepton electromagnetic and weak magnetic moments}},
  \href{https://doi.org/10.1016/S0550-3213(00)00275-3}{\emph{Nucl. Phys. B}
  {\bfseries 582} (2000) 3}
  [\href{https://arxiv.org/abs/hep-ph/0002203}{{\ttfamily hep-ph/0002203}}].

\bibitem{Eidelman:2007sb}
S.~Eidelman and M.~Passera, \emph{{Theory of the tau lepton anomalous magnetic
  moment}}, \href{https://doi.org/10.1142/S0217732307022694}{\emph{Mod. Phys.
  Lett. A} {\bfseries 22} (2007) 159}
  [\href{https://arxiv.org/abs/hep-ph/0701260}{{\ttfamily hep-ph/0701260}}].

\bibitem{ATLAS:2022ryk}
{\scshape ATLAS} collaboration, \emph{{Observation of the
  {\ensuremath{\gamma}}{\ensuremath{\gamma}}{\textrightarrow}{\ensuremath{\tau}}{\ensuremath{\tau}}
  Process in Pb+Pb Collisions and Constraints on the {\ensuremath{\tau}}-Lepton
  Anomalous Magnetic Moment with the ATLAS Detector}},
  \href{https://doi.org/10.1103/PhysRevLett.131.151802}{\emph{Phys. Rev. Lett.}
  {\bfseries 131} (2023) 151802}
  [\href{https://arxiv.org/abs/2204.13478}{{\ttfamily 2204.13478}}].

\bibitem{Lindner:2016bgg}
M.~Lindner, M.~Platscher and F.S.~Queiroz, \emph{{A Call for New Physics : The
  Muon Anomalous Magnetic Moment and Lepton Flavor Violation}},
  \href{https://doi.org/10.1016/j.physrep.2017.12.001}{\emph{Phys. Rept.}
  {\bfseries 731} (2018) 1} [\href{https://arxiv.org/abs/1610.06587}{{\ttfamily
  1610.06587}}].

\bibitem{Mu3e:2020gyw}
{\scshape Mu3e} collaboration, \emph{{Technical design of the phase I Mu3e
  experiment}}, \href{https://doi.org/10.1016/j.nima.2021.165679}{\emph{Nucl.
  Instrum. Meth. A} {\bfseries 1014} (2021) 165679}
  [\href{https://arxiv.org/abs/2009.11690}{{\ttfamily 2009.11690}}].

\bibitem{COMET:2025sdw}
{\scshape COMET, MEG, Mu2e, Mu3e} collaboration, \emph{{Charged Lepton Flavour
  Violations searches with muons: present and future}},
  \href{https://arxiv.org/abs/2503.22461}{{\ttfamily 2503.22461}}.

\bibitem{Belle-II:2026usi}
{\scshape Belle-II} collaboration, \emph{{Search for the
  lepton-flavor-violating decay $ \tau^{\pm} \to \mu^{\pm} \gamma$ at Belle
  II}},  \href{https://arxiv.org/abs/2608.23009}{{\ttfamily 2608.23009}}.

\bibitem{Belle:2021ysv}
{\scshape Belle} collaboration, \emph{{Search for lepton-flavor-violating
  tau-lepton decays to $\ell\gamma$ at Belle}},
  \href{https://doi.org/10.1007/JHEP10(2021)019}{\emph{JHEP} {\bfseries 10}
  (2021) 19} [\href{https://arxiv.org/abs/2103.12994}{{\ttfamily 2103.12994}}].

\bibitem{MEGII:2025gzr}
{\scshape MEG II} collaboration, \emph{{New limit on the ${\upmu ^+ \rightarrow
  e^+\upgamma }$ decay with the MEG II experiment}},
  \href{https://doi.org/10.1140/epjc/s10052-025-14906-3}{\emph{Eur. Phys. J. C}
  {\bfseries 85} (2025) 1177}
  [\href{https://arxiv.org/abs/2504.15711}{{\ttfamily 2504.15711}}].

\bibitem{BaBar:2009hkt}
{\scshape BaBar} collaboration, \emph{{Searches for Lepton Flavor Violation in
  the Decays $\tau^\pm \to e^\pm \gamma$ and $\tau^\pm \to \mu^\pm \gamma$}},
  \href{https://doi.org/10.1103/PhysRevLett.104.021802}{\emph{Phys. Rev. Lett.}
  {\bfseries 104} (2010) 021802}
  [\href{https://arxiv.org/abs/0908.2381}{{\ttfamily 0908.2381}}].

\bibitem{SINDRUM:1987nra}
{\scshape SINDRUM} collaboration, \emph{{Search for the Decay $\mu^+ \to e^+
  e^+ e^-$}}, \href{https://doi.org/10.1016/0550-3213(88)90462-2}{\emph{Nucl.
  Phys. B} {\bfseries 299} (1988) 1}.

\bibitem{Belle-II:2025urb}
{\scshape Belle-II} collaboration, \emph{{Search for the
  lepton-flavor-violating $\tau^{-} \rightarrow e^{\mp} \ell^{\pm} \ell^{\mp}$
  decays at Belle II}},
  \href{https://doi.org/10.1007/JHEP12(2025)169}{\emph{JHEP} {\bfseries 12}
  (2025) 169} [\href{https://arxiv.org/abs/2507.18236}{{\ttfamily
  2507.18236}}].

\bibitem{Belle-II:2024sce}
{\scshape Belle-II} collaboration, \emph{{Search for lepton-flavor-violating
  $\tau^- \to \mu^- \mu^+ \mu^-$ decays at Belle~II}},
  \href{https://doi.org/10.1007/JHEP09(2024)062}{\emph{JHEP} {\bfseries 09}
  (2024) 062} [\href{https://arxiv.org/abs/2405.07386}{{\ttfamily
  2405.07386}}].

\bibitem{LHCb:2026eod}
{\scshape LHCb} collaboration, \emph{{Search for $\tau^-\to\mu^-\mu^+\mu^-$
  decays at the LHCb experiment with Run 2 data}},
  \href{https://arxiv.org/abs/2601.20785}{{\ttfamily 2601.20785}}.

\bibitem{ATLAS:2026kkw}
{\scshape ATLAS} collaboration, \emph{{A search for lepton-flavour violating
  $\tau\to 3\mu$ decays with the ATLAS detector}},
  \href{https://arxiv.org/abs/2603.18099}{{\ttfamily 2603.18099}}.

\bibitem{CMS:2026fwo}
{\scshape CMS} collaboration, \emph{{Search for the lepton flavor violating
  $\tau^{-}\to\mu^{-}\mu^{+}\mu^{-}$ decay in proton-proton collisions at
  $\sqrt{s}=13.6~\mathrm{TeV}$}}, .

\bibitem{Hayasaka:2010np}
K.~Hayasaka et~al., \emph{{Search for Lepton Flavor Violating Tau Decays into
  Three Leptons with 719 Million Produced Tau+Tau- Pairs}},
  \href{https://doi.org/10.1016/j.physletb.2010.03.037}{\emph{Phys. Lett. B}
  {\bfseries 687} (2010) 139}
  [\href{https://arxiv.org/abs/1001.3221}{{\ttfamily 1001.3221}}].

\bibitem{Kitano:2002mt}
R.~Kitano, M.~Koike and Y.~Okada, \emph{{Detailed calculation of lepton flavor
  violating muon electron conversion rate for various nuclei}},
  \href{https://doi.org/10.1103/PhysRevD.76.059902}{\emph{Phys. Rev. D}
  {\bfseries 66} (2002) 096002}
  [\href{https://arxiv.org/abs/hep-ph/0203110}{{\ttfamily hep-ph/0203110}}].

\bibitem{SINDRUMII:2006dvw}
{\scshape SINDRUM II} collaboration, \emph{{A Search for muon to electron
  conversion in muonic gold}},
  \href{https://doi.org/10.1140/epjc/s2006-02582-x}{\emph{Eur. Phys. J. C}
  {\bfseries 47} (2006) 337}.

\bibitem{COMET:2018auw}
{\scshape COMET} collaboration, \emph{{COMET Phase-I Technical Design Report}},
  \href{https://doi.org/10.1093/ptep/ptz125}{\emph{PTEP} {\bfseries 2020}
  (2020) 033C01} [\href{https://arxiv.org/abs/1812.09018}{{\ttfamily
  1812.09018}}].

\bibitem{Moritsu:2022lem}
{\scshape COMET} collaboration, \emph{{Search for Muon-to-Electron Conversion
  with the COMET Experiment {\textdagger}}},
  \href{https://doi.org/10.3390/universe8040196}{\emph{Universe} {\bfseries 8}
  (2022) 196} [\href{https://arxiv.org/abs/2203.06365}{{\ttfamily
  2203.06365}}].

\bibitem{Mu2e:2022ggl}
{\scshape Mu2e} collaboration, \emph{{Mu2e Run I Sensitivity Projections for
  the Neutrinoless Conversion Search in Aluminum}},
  \href{https://doi.org/10.3390/universe9010054}{\emph{Universe} {\bfseries 9}
  (2023) 54} [\href{https://arxiv.org/abs/2210.11380}{{\ttfamily 2210.11380}}].

\bibitem{Mu2e-II:2022blh}
{\scshape Mu2e-II} collaboration, \emph{{Mu2e-II: Muon to electron conversion
  with PIP-II}},  in \emph{{Snowmass 2021}}, 3, 2022
  [\href{https://arxiv.org/abs/2203.07569}{{\ttfamily 2203.07569}}].

\bibitem{Celis:2013xja}
A.~Celis, V.~Cirigliano and E.~Passemar, \emph{{Lepton flavor violation in the
  Higgs sector and the role of hadronic $\tau$-lepton decays}},
  \href{https://doi.org/10.1103/PhysRevD.89.013008}{\emph{Phys. Rev. D}
  {\bfseries 89} (2014) 013008}
  [\href{https://arxiv.org/abs/1309.3564}{{\ttfamily 1309.3564}}].

\bibitem{Urquia-Calderon:2025wjx}
K.A.~Urqu{\'\i}a-Calder{\'o}n and O.~Ruchayskiy, \emph{{Revisiting lepton
  flavor violation: {\ensuremath{\tau}} and meson decays}},
  \href{https://doi.org/10.1007/JHEP04(2026)067}{\emph{JHEP} {\bfseries 04}
  (2026) 067} [\href{https://arxiv.org/abs/2511.20320}{{\ttfamily
  2511.20320}}].

\bibitem{HeavyFlavorAveragingGroupHFLAV:2024ctg}
{\scshape Heavy Flavor Averaging Group (HFLAV)} collaboration, \emph{{Averages
  of $b$-hadron, $c$-hadron, and $\tau$-lepton properties as of 2023}},
  \href{https://arxiv.org/abs/2411.18639}{{\ttfamily 2411.18639}}.

\bibitem{deMelo:2018hfw}
J.P.B.C.~de~Melo and K.~Tsushima, \emph{{$\rho$-meson properties in medium}},
  \href{https://doi.org/10.1016/j.physletb.2018.10.059}{\emph{Phys. Lett. B}
  {\bfseries 788} (2019) 137}
  [\href{https://arxiv.org/abs/1802.06096}{{\ttfamily 1802.06096}}].

\bibitem{Bharucha:2015bzk}
A.~Bharucha, D.M.~Straub and R.~Zwicky, \emph{{$B\to V\ell^+\ell^-$ in the
  Standard Model from light-cone sum rules}},
  \href{https://doi.org/10.1007/JHEP08(2016)098}{\emph{JHEP} {\bfseries 08}
  (2016) 098} [\href{https://arxiv.org/abs/1503.05534}{{\ttfamily
  1503.05534}}].

\bibitem{Belle:2025iff}
{\scshape Belle, Belle-II} collaboration, \emph{{Search for
  lepton-flavor-violating ${\tau }^{-}\to {{\ell}}^{-}{K}_{s}^{0}$ decays at
  Belle and Belle II}},
  \href{https://doi.org/10.1007/JHEP08(2025)092}{\emph{JHEP} {\bfseries 08}
  (2025) 092} [\href{https://arxiv.org/abs/2504.15745}{{\ttfamily
  2504.15745}}].

\bibitem{Belle:2023ziz}
{\scshape Belle} collaboration, \emph{{Search for lepton-flavor-violating
  {\ensuremath{\tau}} decays into a lepton and a vector meson using the full
  Belle data sample}},
  \href{https://doi.org/10.1007/JHEP06(2023)118}{\emph{JHEP} {\bfseries 06}
  (2023) 118} [\href{https://arxiv.org/abs/2301.03768}{{\ttfamily
  2301.03768}}].

\bibitem{LHCb:2017hag}
{\scshape LHCb} collaboration, \emph{{Search for the lepton-flavour violating
  decays B$_{(s)}^{0} \to e^{\pm}\mu^{\mp}$}},
  \href{https://doi.org/10.1007/JHEP03(2018)078}{\emph{JHEP} {\bfseries 03}
  (2018) 078} [\href{https://arxiv.org/abs/1710.04111}{{\ttfamily
  1710.04111}}].

\bibitem{Belle:2023jwr}
{\scshape Belle} collaboration, \emph{{Search for $B{}^0_s \rightarrow
  \ell^{\mp} \tau^{\pm}$ with the Semi-leptonic Tagging Method at Belle}},
  \href{https://doi.org/10.1007/JHEP08(2023)178}{\emph{JHEP} {\bfseries 08}
  (2023) 178} [\href{https://arxiv.org/abs/2301.10989}{{\ttfamily
  2301.10989}}].

\bibitem{LHCb:2019ujz}
{\scshape LHCb} collaboration, \emph{{Search for the lepton-flavour-violating
  decays $B^{0}_{s}\to\tau^{\pm}\mu^{\mp}$ and $B^{0}\to\tau^{\pm}\mu^{\mp}$}},
  \href{https://doi.org/10.1103/PhysRevLett.123.211801}{\emph{Phys. Rev. Lett.}
  {\bfseries 123} (2019) 211801}
  [\href{https://arxiv.org/abs/1905.06614}{{\ttfamily 1905.06614}}].

\bibitem{Belle:2021rod}
{\scshape Belle} collaboration, \emph{{Search for $B^{0} \to \tau^\pm \ell^\mp$
  ($\ell=e,\mu$) with a hadronic tagging method at Belle}},
  \href{https://doi.org/10.1103/PhysRevD.104.L091105}{\emph{Phys. Rev. D}
  {\bfseries 104} (2021) L091105}
  [\href{https://arxiv.org/abs/2108.11649}{{\ttfamily 2108.11649}}].

\bibitem{BNL:1998apv}
{\scshape BNL} collaboration, \emph{{New limit on muon and electron lepton
  number violation from K0(L) ---{\ensuremath{>}} mu+- e-+ decay}},
  \href{https://doi.org/10.1103/PhysRevLett.81.5734}{\emph{Phys. Rev. Lett.}
  {\bfseries 81} (1998) 5734}
  [\href{https://arxiv.org/abs/hep-ex/9811038}{{\ttfamily hep-ex/9811038}}].

\bibitem{NA62:2021zxl}
{\scshape NA62} collaboration, \emph{{Search for Lepton Number and Flavor
  Violation in $K^+$ and $\pi^0$ Decays}},
  \href{https://doi.org/10.1103/PhysRevLett.127.131802}{\emph{Phys. Rev. Lett.}
  {\bfseries 127} (2021) 131802}
  [\href{https://arxiv.org/abs/2105.06759}{{\ttfamily 2105.06759}}].

\bibitem{Achasov:2009en}
M.N.~Achasov et~al., \emph{{Search for Lepton Flavor Violation Process
  $e^{+}e^{-} \to e\mu$ in the Energy Region $\sqrt{s}=984 - 1060 MeV$ and$
  \phi\to e\mu$ Decay}},
  \href{https://doi.org/10.1103/PhysRevD.81.057102}{\emph{Phys. Rev. D}
  {\bfseries 81} (2010) 057102}
  [\href{https://arxiv.org/abs/0911.1232}{{\ttfamily 0911.1232}}].

\bibitem{White:1995jc}
D.B.~White et~al., \emph{{Search for the decays eta ---{\ensuremath{>}} mu e
  and eta ---{\ensuremath{>}} e+ e-}},
  \href{https://doi.org/10.1103/PhysRevD.53.6658}{\emph{Phys. Rev. D}
  {\bfseries 53} (1996) 6658}.

\bibitem{Appelquist:1974tg}
T.~Appelquist and J.~Carazzone, \emph{{Infrared Singularities and Massive
  Fields}}, \href{https://doi.org/10.1103/PhysRevD.11.2856}{\emph{Phys. Rev. D}
  {\bfseries 11} (1975) 2856}.

\bibitem{Staub:2008uz}
F.~Staub, \emph{{SARAH}},  \href{https://arxiv.org/abs/0806.0538}{{\ttfamily
  0806.0538}}.

\bibitem{ALEPH:2005ab}
{\scshape ALEPH, DELPHI, L3, OPAL, SLD, LEP Electroweak Working Group, SLD
  Electroweak Group, SLD Heavy Flavour Group} collaboration, \emph{{Precision
  electroweak measurements on the $Z$ resonance}},
  \href{https://doi.org/10.1016/j.physrep.2005.12.006}{\emph{Phys. Rept.}
  {\bfseries 427} (2006) 257}
  [\href{https://arxiv.org/abs/hep-ex/0509008}{{\ttfamily hep-ex/0509008}}].

\bibitem{Soni:2010xh}
A.~Soni, A.K.~Alok, A.~Giri, R.~Mohanta and S.~Nandi, \emph{{SM with four
  generations: Selected implications for rare B and K decays}},
  \href{https://doi.org/10.1103/PhysRevD.82.033009}{\emph{Phys. Rev. D}
  {\bfseries 82} (2010) 033009}
  [\href{https://arxiv.org/abs/1002.0595}{{\ttfamily 1002.0595}}].

\bibitem{SLD:2000leq}
{\scshape SLD} collaboration, \emph{{A High precision measurement of the
  left-right Z boson cross-section asymmetry}},
  \href{https://doi.org/10.1103/PhysRevLett.84.5945}{\emph{Phys. Rev. Lett.}
  {\bfseries 84} (2000) 5945}
  [\href{https://arxiv.org/abs/hep-ex/0004026}{{\ttfamily hep-ex/0004026}}].

\bibitem{SLD:2000ujp}
{\scshape SLD} collaboration, \emph{{An Improved direct measurement of leptonic
  coupling asymmetries with polarized Z bosons}},
  \href{https://doi.org/10.1103/PhysRevLett.86.1162}{\emph{Phys. Rev. Lett.}
  {\bfseries 86} (2001) 1162}
  [\href{https://arxiv.org/abs/hep-ex/0010015}{{\ttfamily hep-ex/0010015}}].

\bibitem{Novikov:1999af}
V.A.~Novikov, L.B.~Okun, A.N.~Rozanov and M.I.~Vysotsky, \emph{{Theory of $Z$
  boson decays}}, \href{https://doi.org/10.1088/0034-4885/62/9/201}{\emph{Rept.
  Prog. Phys.} {\bfseries 62} (1999) 1275}
  [\href{https://arxiv.org/abs/hep-ph/9906465}{{\ttfamily hep-ph/9906465}}].

\bibitem{Reina:2025suh}
L.~Reina and L.~Silvestrini, \emph{{Electroweak precision tests}},
  \href{https://arxiv.org/abs/2511.16534}{{\ttfamily 2511.16534}}.

\bibitem{Dubovyk:2019szj}
I.~Dubovyk, A.~Freitas, J.~Gluza, T.~Riemann and J.~Usovitsch,
  \emph{{Electroweak pseudo-observables and Z-boson form factors at two-loop
  accuracy}}, \href{https://doi.org/10.1007/JHEP08(2019)113}{\emph{JHEP}
  {\bfseries 08} (2019) 113}
  [\href{https://arxiv.org/abs/1906.08815}{{\ttfamily 1906.08815}}].

\bibitem{Freitas:2014hra}
A.~Freitas, \emph{{Higher-order electroweak corrections to the partial widths
  and branching ratios of the Z boson}},
  \href{https://doi.org/10.1007/JHEP04(2014)070}{\emph{JHEP} {\bfseries 04}
  (2014) 070} [\href{https://arxiv.org/abs/1401.2447}{{\ttfamily 1401.2447}}].

\bibitem{Dubovyk:2018rlg}
I.~Dubovyk, A.~Freitas, J.~Gluza, T.~Riemann and J.~Usovitsch, \emph{{Complete
  electroweak two-loop corrections to Z boson production and decay}},
  \href{https://doi.org/10.1016/j.physletb.2018.06.037}{\emph{Phys. Lett. B}
  {\bfseries 783} (2018) 86}
  [\href{https://arxiv.org/abs/1804.10236}{{\ttfamily 1804.10236}}].

\bibitem{CMS:2023mgq}
{\scshape CMS} collaboration, \emph{{Measurement of the {\ensuremath{\tau}}
  lepton polarization in Z boson decays in proton-proton collisions at $
  \sqrt{s} $ = 13 TeV}},
  \href{https://doi.org/10.1007/JHEP01(2024)101}{\emph{JHEP} {\bfseries 01}
  (2024) 101} [\href{https://arxiv.org/abs/2309.12408}{{\ttfamily
  2309.12408}}].

\bibitem{Aranda:2020tqw}
J.I.~Aranda, G.~Gonz{\'a}lez-Estrada, J.~Monta{\~n}o, F.~Ram{\'\i}rez-Zavaleta
  and E.S.~Tututi, \emph{{Revisiting the rare $H\to q_iq_j$ decays in the
  standard model}}, \href{https://doi.org/10.1088/1361-6471/abb44d}{\emph{J.
  Phys. G} {\bfseries 47} (2020) 125001}
  [\href{https://arxiv.org/abs/2009.07166}{{\ttfamily 2009.07166}}].

\bibitem{Eilam:2002as}
G.~Eilam, \emph{{Rare Z decays}},
  \href{https://doi.org/10.1016/S0920-5632(03)80189-5}{\emph{Nucl. Phys. B
  Proc. Suppl.} {\bfseries 116} (2003) 306}
  [\href{https://arxiv.org/abs/hep-ph/0211458}{{\ttfamily hep-ph/0211458}}].

\bibitem{Bernabeu:1986pk}
J.~Bernabeu, M.B.~Gavela and A.~Santamaria, \emph{{CP Violation at the $Z^0$
  Peak}}, \href{https://doi.org/10.1103/PhysRevLett.57.1514}{\emph{Phys. Rev.
  Lett.} {\bfseries 57} (1986) 1514}.

\bibitem{Illana:2000ic}
J.I.~Illana and T.~Riemann, \emph{{Charged lepton flavor violation from massive
  neutrinos in Z decays}},
  \href{https://doi.org/10.1103/PhysRevD.63.053004}{\emph{Phys. Rev. D}
  {\bfseries 63} (2001) 053004}
  [\href{https://arxiv.org/abs/hep-ph/0010193}{{\ttfamily hep-ph/0010193}}].

\bibitem{Atwood:2002ke}
D.~Atwood, S.~Bar-Shalom, G.~Eilam and A.~Soni, \emph{{Flavor changing Z decays
  from scalar interactions at a giga Z linear collider}},
  \href{https://doi.org/10.1103/PhysRevD.66.093005}{\emph{Phys. Rev. D}
  {\bfseries 66} (2002) 093005}
  [\href{https://arxiv.org/abs/hep-ph/0203200}{{\ttfamily hep-ph/0203200}}].

\bibitem{Fuster:1999dj}
{\scshape DELPHI} collaboration, \emph{{A Search for FCNC in Z
  --{\ensuremath{>}} b anti-d, b anti-s decays}},  in \emph{{1999 International
  Europhysics Conference on High-Energy Physics}}, 6, 1999.

\bibitem{CMS:2025wqy}
{\scshape CMS} collaboration, \emph{{Search for charged lepton flavor violating
  Z and Z' boson decays in proton-proton collisions at s=13{\,}{\,}TeV}},
  \href{https://doi.org/10.1103/8lw4-6lyk}{\emph{Phys. Rev. D} {\bfseries 112}
  (2025) 112011} [\href{https://arxiv.org/abs/2508.07512}{{\ttfamily
  2508.07512}}].

\bibitem{ATLAS:2021bdj}
{\scshape ATLAS} collaboration, \emph{{Search for lepton-flavor-violation in
  $Z$-boson decays with $\tau$-leptons with the ATLAS detector}},
  \href{https://doi.org/10.1103/PhysRevLett.127.271801}{\emph{Phys. Rev. Lett.}
  {\bfseries 127} (2022) 271801}
  [\href{https://arxiv.org/abs/2105.12491}{{\ttfamily 2105.12491}}].

\bibitem{CMS:2023pte}
{\scshape CMS} collaboration, \emph{{Search for the lepton-flavor violating
  decay of the Higgs boson and additional Higgs bosons in the e$\mu$ final
  state in proton-proton collisions at $\sqrt{s}$ = 13 TeV}},
  \href{https://doi.org/10.1103/PhysRevD.108.072004}{\emph{Phys. Rev. D}
  {\bfseries 108} (2023) 072004}
  [\href{https://arxiv.org/abs/2305.18106}{{\ttfamily 2305.18106}}].

\bibitem{ATLAS:2023mvd}
{\scshape ATLAS} collaboration, \emph{{Searches for lepton-flavour-violating
  decays of the Higgs boson into $e\tau$ and $\mu\tau$ in $\sqrt{s}=13$ TeV
  $pp$ collisions with the ATLAS detector}},
  \href{https://doi.org/10.1007/JHEP07(2023)166}{\emph{JHEP} {\bfseries 07}
  (2023) 166} [\href{https://arxiv.org/abs/2302.05225}{{\ttfamily
  2302.05225}}].

\bibitem{Benitez-Guzman:2015ana}
L.G.~Benitez-Guzm{\'a}n, I.~Garc{\'\i}a-Jim{\'e}nez, M.A.~L{\'o}pez-Osorio,
  E.~Mart{\'\i}nez-Pascual and J.J.~Toscano, \emph{{Revisiting the flavor
  changing neutral current Higgs decays $H\;\to \;{q}_{i}{q}_{j}$ in the
  Standard Model}},
  \href{https://doi.org/10.1088/0954-3899/42/8/085002}{\emph{J. Phys. G}
  {\bfseries 42} (2015) 085002}
  [\href{https://arxiv.org/abs/1506.02718}{{\ttfamily 1506.02718}}].

\bibitem{CMS:2021rsq}
{\scshape CMS} collaboration, \emph{{Search for lepton-flavor violating decays
  of the Higgs boson in the $\mu\tau$ and e$\tau$ final states in proton-proton
  collisions at $\sqrt{s}$ = 13 TeV}},
  \href{https://doi.org/10.1103/PhysRevD.104.032013}{\emph{Phys. Rev. D}
  {\bfseries 104} (2021) 032013}
  [\href{https://arxiv.org/abs/2105.03007}{{\ttfamily 2105.03007}}].

\bibitem{Planck:2018vyg}
{\scshape Planck} collaboration, \emph{{Planck 2018 results. VI. Cosmological
  parameters}},
  \href{https://doi.org/10.1051/0004-6361/201833910}{\emph{Astron. Astrophys.}
  {\bfseries 641} (2020) A6}
  [\href{https://arxiv.org/abs/1807.06209}{{\ttfamily 1807.06209}}].

\bibitem{Alguero:2023zol}
G.~Alguero, G.~Belanger, F.~Boudjema, S.~Chakraborti, A.~Goudelis, S.~Kraml
  et~al., \emph{{micrOMEGAs 6.0: N-component dark matter}},
  \href{https://doi.org/10.1016/j.cpc.2024.109133}{\emph{Comput. Phys. Commun.}
  {\bfseries 299} (2024) 109133}
  [\href{https://arxiv.org/abs/2312.14894}{{\ttfamily 2312.14894}}].

\bibitem{ATLAS:2023ynf}
{\scshape ATLAS} collaboration, \emph{{Measurement of the Z boson invisible
  width at s=13 TeV with the ATLAS detector}},
  \href{https://doi.org/10.1016/j.physletb.2024.138705}{\emph{Phys. Lett. B}
  {\bfseries 854} (2024) 138705}
  [\href{https://arxiv.org/abs/2312.02789}{{\ttfamily 2312.02789}}].

\bibitem{CMS:2022ett}
{\scshape CMS} collaboration, \emph{{Precision measurement of the Z boson
  invisible width in pp collisions at s=13 TeV}},
  \href{https://doi.org/10.1016/j.physletb.2022.137563}{\emph{Phys. Lett. B}
  {\bfseries 842} (2023) 137563}
  [\href{https://arxiv.org/abs/2206.07110}{{\ttfamily 2206.07110}}].

\bibitem{ATLAS:2023tkt}
{\scshape ATLAS} collaboration, \emph{{Combination of searches for invisible
  decays of the Higgs boson using 139 fb{\ensuremath{-}}1 of proton-proton
  collision data at s=13 TeV collected with the ATLAS experiment}},
  \href{https://doi.org/10.1016/j.physletb.2023.137963}{\emph{Phys. Lett. B}
  {\bfseries 842} (2023) 137963}
  [\href{https://arxiv.org/abs/2301.10731}{{\ttfamily 2301.10731}}].

\bibitem{LZ:2024zvo}
{\scshape LZ} collaboration, \emph{{Dark Matter Search Results from
  4.2{\,}{\,}Tonne-Years of Exposure of the LUX-ZEPLIN (LZ) Experiment}},
  \href{https://doi.org/10.1103/4dyc-z8zf}{\emph{Phys. Rev. Lett.} {\bfseries
  135} (2025) 011802} [\href{https://arxiv.org/abs/2410.17036}{{\ttfamily
  2410.17036}}].

\bibitem{XLZD:2024nsu}
{\scshape XLZD} collaboration, \emph{{The XLZD Design Book: towards the
  next-generation liquid xenon observatory for dark matter and neutrino
  physics}}, \href{https://doi.org/10.1140/epjc/s10052-025-14810-w}{\emph{Eur.
  Phys. J. C} {\bfseries 85} (2025) 1192}
  [\href{https://arxiv.org/abs/2410.17137}{{\ttfamily 2410.17137}}].

\bibitem{Conrad:2017pms}
J.~Conrad and O.~Reimer, \emph{{Indirect dark matter searches in gamma and
  cosmic rays}}, \href{https://doi.org/10.1038/nphys4049}{\emph{Nature Phys.}
  {\bfseries 13} (2017) 224}
  [\href{https://arxiv.org/abs/1705.11165}{{\ttfamily 1705.11165}}].

\bibitem{Gaskins:2016cha}
J.M.~Gaskins, \emph{{A review of indirect searches for particle dark matter}},
  \href{https://doi.org/10.1080/00107514.2016.1175160}{\emph{Contemp. Phys.}
  {\bfseries 57} (2016) 496}
  [\href{https://arxiv.org/abs/1604.00014}{{\ttfamily 1604.00014}}].

\bibitem{Strigari:2018utn}
L.E.~Strigari, \emph{{Dark matter in dwarf spheroidal galaxies and indirect
  detection: a review}},
  \href{https://doi.org/10.1088/1361-6633/aaae16}{\emph{Rept. Prog. Phys.}
  {\bfseries 81} (2018) 056901}
  [\href{https://arxiv.org/abs/1805.05883}{{\ttfamily 1805.05883}}].

\bibitem{Arguelles:2019ouk}
C.A.~Arg{\"u}elles, A.~Diaz, A.~Kheirandish, A.~Olivares-Del-Campo, I.~Safa and
  A.C.~Vincent, \emph{{Dark matter annihilation to neutrinos}},
  \href{https://doi.org/10.1103/RevModPhys.93.035007}{\emph{Rev. Mod. Phys.}
  {\bfseries 93} (2021) 035007}
  [\href{https://arxiv.org/abs/1912.09486}{{\ttfamily 1912.09486}}].

\bibitem{Fermi-LAT:2025gei}
{\scshape Fermi-LAT, HAWC, H.E.S.S., MAGIC, VERITAS} collaboration,
  \emph{{Combined dark matter search towards dwarf spheroidal galaxies with
  Fermi-LAT, HAWC, H.E.S.S., MAGIC, and VERITAS}},
  \href{https://arxiv.org/abs/2508.20229}{{\ttfamily 2508.20229}}.

\bibitem{Queiroz:2019acr}
F.S.~Queiroz and C.~Siqueira, \emph{{Search for Semi-Annihilating Dark Matter
  with Fermi-LAT, H.E.S.S., Planck, and the Cherenkov Telescope Array}},
  \href{https://doi.org/10.1088/1475-7516/2019/04/048}{\emph{JCAP} {\bfseries
  04} (2019) 048} [\href{https://arxiv.org/abs/1901.10494}{{\ttfamily
  1901.10494}}].

\bibitem{Griest:1990kh}
K.~Griest and D.~Seckel, \emph{{Three exceptions in the calculation of relic
  abundances}}, \href{https://doi.org/10.1103/PhysRevD.43.3191}{\emph{Phys.
  Rev. D} {\bfseries 43} (1991) 3191}.

\bibitem{Gondolo:2004sc}
P.~Gondolo, J.~Edsjo, P.~Ullio, L.~Bergstrom, M.~Schelke and E.A.~Baltz,
  \emph{{DarkSUSY: Computing supersymmetric dark matter properties
  numerically}},
  \href{https://doi.org/10.1088/1475-7516/2004/07/008}{\emph{JCAP} {\bfseries
  07} (2004) 008} [\href{https://arxiv.org/abs/astro-ph/0406204}{{\ttfamily
  astro-ph/0406204}}].

\bibitem{Edsjo:1997bg}
J.~Edsjo and P.~Gondolo, \emph{{Neutralino relic density including
  coannihilations}},
  \href{https://doi.org/10.1103/PhysRevD.56.1879}{\emph{Phys. Rev. D}
  {\bfseries 56} (1997) 1879}
  [\href{https://arxiv.org/abs/hep-ph/9704361}{{\ttfamily hep-ph/9704361}}].

\bibitem{Bhattacharya:2021rwh}
S.~Bhattacharya, S.~Chakraborti and D.~Pradhan, \emph{{Electroweak symmetry
  breaking and WIMP-FIMP dark matter}},
  \href{https://doi.org/10.1007/JHEP07(2022)091}{\emph{JHEP} {\bfseries 07}
  (2022) 091} [\href{https://arxiv.org/abs/2110.06985}{{\ttfamily
  2110.06985}}].

\bibitem{Dercks:2016npn}
D.~Dercks, N.~Desai, J.S.~Kim, K.~Rolbiecki, J.~Tattersall and T.~Weber,
  \emph{{CheckMATE 2: From the model to the limit}},
  \href{https://doi.org/10.1016/j.cpc.2017.08.021}{\emph{Comput. Phys. Commun.}
  {\bfseries 221} (2017) 383}
  [\href{https://arxiv.org/abs/1611.09856}{{\ttfamily 1611.09856}}].

\bibitem{ATLAS:2020dsf}
{\scshape ATLAS} collaboration, \emph{{Search for a scalar partner of the top
  quark in the all-hadronic $t{\bar{t}}$ plus missing transverse momentum final
  state at $\sqrt{s}=13$ TeV with the ATLAS detector}},
  \href{https://doi.org/10.1140/epjc/s10052-020-8102-8}{\emph{Eur. Phys. J. C}
  {\bfseries 80} (2020) 737}
  [\href{https://arxiv.org/abs/2004.14060}{{\ttfamily 2004.14060}}].

\bibitem{CMS:2019zmd}
{\scshape CMS} collaboration, \emph{{Search for supersymmetry in proton-proton
  collisions at 13 TeV in final states with jets and missing transverse
  momentum}}, \href{https://doi.org/10.1007/JHEP10(2019)244}{\emph{JHEP}
  {\bfseries 10} (2019) 244}
  [\href{https://arxiv.org/abs/1908.04722}{{\ttfamily 1908.04722}}].

\bibitem{ATLAS:2020syg}
{\scshape ATLAS} collaboration, \emph{{Search for squarks and gluinos in final
  states with jets and missing transverse momentum using 139 fb$^{-1}$ of
  $\sqrt{s}$ =13 TeV $pp$ collision data with the ATLAS detector}},
  \href{https://doi.org/10.1007/JHEP02(2021)143}{\emph{JHEP} {\bfseries 02}
  (2021) 143} [\href{https://arxiv.org/abs/2010.14293}{{\ttfamily
  2010.14293}}].

\bibitem{ATLAS:2019lff}
{\scshape ATLAS} collaboration, \emph{{Search for electroweak production of
  charginos and sleptons decaying into final states with two leptons and
  missing transverse momentum in $\sqrt{s}=13$ TeV $pp$ collisions using the
  ATLAS detector}},
  \href{https://doi.org/10.1140/epjc/s10052-019-7594-6}{\emph{Eur. Phys. J. C}
  {\bfseries 80} (2020) 123}
  [\href{https://arxiv.org/abs/1908.08215}{{\ttfamily 1908.08215}}].

\bibitem{Cowan:2010js}
G.~Cowan, K.~Cranmer, E.~Gross and O.~Vitells, \emph{{Asymptotic formulae for
  likelihood-based tests of new physics}},
  \href{https://doi.org/10.1140/epjc/s10052-011-1554-0}{\emph{Eur. Phys. J. C}
  {\bfseries 71} (2011) 1554}
  [\href{https://arxiv.org/abs/1007.1727}{{\ttfamily 1007.1727}}].

\bibitem{Billard:2013qya}
J.~Billard, L.~Strigari and E.~Figueroa-Feliciano, \emph{{Implication of
  neutrino backgrounds on the reach of next generation dark matter direct
  detection experiments}},
  \href{https://doi.org/10.1103/PhysRevD.89.023524}{\emph{Phys. Rev. D}
  {\bfseries 89} (2014) 023524}
  [\href{https://arxiv.org/abs/1307.5458}{{\ttfamily 1307.5458}}].

\bibitem{Monroe:2007xp}
J.~Monroe and P.~Fisher, \emph{{Neutrino Backgrounds to Dark Matter Searches}},
  \href{https://doi.org/10.1103/PhysRevD.76.033007}{\emph{Phys. Rev. D}
  {\bfseries 76} (2007) 033007}
  [\href{https://arxiv.org/abs/0706.3019}{{\ttfamily 0706.3019}}].

\bibitem{OHare:2020lva}
C.A.J.~O'Hare, \emph{{Can we overcome the neutrino floor at high masses?}},
  \href{https://doi.org/10.1103/PhysRevD.102.063024}{\emph{Phys. Rev. D}
  {\bfseries 102} (2020) 063024}
  [\href{https://arxiv.org/abs/2002.07499}{{\ttfamily 2002.07499}}].

\bibitem{Alanne:2017oqj}
T.~Alanne and F.~Goertz, \emph{{Extended Dark Matter EFT}},
  \href{https://doi.org/10.1140/epjc/s10052-020-7999-2}{\emph{Eur. Phys. J. C}
  {\bfseries 80} (2020) 446}
  [\href{https://arxiv.org/abs/1712.07626}{{\ttfamily 1712.07626}}].

\bibitem{Arcadi:2019lka}
G.~Arcadi, A.~Djouadi and M.~Raidal, \emph{{Dark Matter through the Higgs
  portal}}, \href{https://doi.org/10.1016/j.physrep.2019.11.003}{\emph{Phys.
  Rept.} {\bfseries 842} (2020) 1}
  [\href{https://arxiv.org/abs/1903.03616}{{\ttfamily 1903.03616}}].

\bibitem{Arcadi:2024mli}
G.~Arcadi, D.~Cabo-Almeida, S.~Fabian and F.~Goertz, \emph{{Dark particles at
  the LHC: LHC-friendly dark matter characterization via non-linear EFT}},
  \href{https://doi.org/10.1007/JHEP06(2025)126}{\emph{JHEP} {\bfseries 06}
  (2025) 126} [\href{https://arxiv.org/abs/2411.05914}{{\ttfamily
  2411.05914}}].

\bibitem{Criado:2021trs}
J.C.~Criado, A.~Djouadi, M.~Perez-Victoria and J.~Santiago, \emph{{A complete
  effective field theory for dark matter}},
  \href{https://doi.org/10.1007/JHEP07(2021)081}{\emph{JHEP} {\bfseries 07}
  (2021) 081} [\href{https://arxiv.org/abs/2104.14443}{{\ttfamily
  2104.14443}}].

\bibitem{Fuentes-Martin:2020udw}
J.~Fuentes-Martin, M.~K{\"o}nig, J.~Pag{\`e}s, A.E.~Thomsen and F.~Wilsch,
  \emph{{SuperTracer: A Calculator of Functional Supertraces for One-Loop EFT
  Matching}}, \href{https://doi.org/10.1007/JHEP04(2021)281}{\emph{JHEP}
  {\bfseries 04} (2021) 281}
  [\href{https://arxiv.org/abs/2012.08506}{{\ttfamily 2012.08506}}].

\bibitem{Alte:2019iug}
S.~Alte, M.~K{\"o}nig and M.~Neubert, \emph{{Effective Theory for a Heavy
  Scalar Coupled to the SM via Vector-Like Quarks}},
  \href{https://doi.org/10.1140/epjc/s10052-019-6867-4}{\emph{Eur. Phys. J. C}
  {\bfseries 79} (2019) 352}
  [\href{https://arxiv.org/abs/1902.04593}{{\ttfamily 1902.04593}}].

\bibitem{Crivellin:2022fdf}
A.~Crivellin, M.~Kirk, T.~Kitahara and F.~Mescia, \emph{{Large
  t{\textrightarrow}cZ as a sign of vectorlike quarks in light of the W mass}},
  \href{https://doi.org/10.1103/PhysRevD.106.L031704}{\emph{Phys. Rev. D}
  {\bfseries 106} (2022) L031704}
  [\href{https://arxiv.org/abs/2204.05962}{{\ttfamily 2204.05962}}].

\bibitem{Fuentes-Martin:2022jrf}
J.~Fuentes-Mart{\'\i}n, M.~K{\"o}nig, J.~Pag{\`e}s, A.E.~Thomsen and F.~Wilsch,
  \emph{{A proof of concept for matchete: an automated tool for matching
  effective theories}},
  \href{https://doi.org/10.1140/epjc/s10052-023-11726-1}{\emph{Eur. Phys. J. C}
  {\bfseries 83} (2023) 662}
  [\href{https://arxiv.org/abs/2212.04510}{{\ttfamily 2212.04510}}].

\end{thebibliography}\endgroup

\end{document}